\documentclass[a4paper,11pt]{article}
\usepackage[utf8x]{inputenc}

\usepackage[paperwidth=210mm, paperheight=297mm,  text={15cm,22.25cm}]{geometry}

\usepackage{amsmath, amssymb, relsize, xspace}
\usepackage{authblk}
\usepackage{graphicx, color, placeins}
\usepackage[FIGTOPCAP,large,bf,nooneline]{subfigure}
\usepackage[normal,small,bf]{caption}
\usepackage{algpseudocode, algorithm, algorithmicx}

\newcommand{\TODO}[1]{\textcolor{blue}{\textbf{($\blacktriangleright$ #1)}}}

\newcommand{\CH}[1]{\textcolor{red}{#1}}
\newcommand{\SI}{Supplementary Information\xspace}

\newcommand{\PlotsDir}{Plots}
\newcommand{\PlotsSIDir}{Plots/SI}
\newcommand{\SketchesDir}{Sketches}

\renewcommand{\TODO}[1]{}
\renewcommand{\CH}[1]{#1}

\renewcommand{\PlotsDir}{.}
\renewcommand{\PlotsSIDir}{.}
\renewcommand{\SketchesDir}{.}

\newcommand{\unit}[1]{{\rm #1}}
\newcommand{\umsps}{\unit{\mu m^2/s}}
\newcommand{\umps}{\unit{\mu m/s}}

\newcommand{\um}{\unit{\mu m}}

\newcommand{\myvec}[1]{\mathbf{#1}}
\newcommand{\uvec}[1]{\hat{\mathbf{#1}}}
\newcommand{\mat}[1]{\mathcal{#1}}
\newcommand{\Res}{\text{Res}}

\newcommand{\nablaX}{\myvec{\nabla}_X}
\newcommand{\nablaY}{\myvec{\nabla}_Y}
\newcommand{\bigzero}{\mathbb{O}}

\newcommand{\erfc}{\text{erfc}}
\newcommand{\takenat}{\big\rvert}
\newcommand{\pd}{\partial}
\newcommand{\la}{\langle}
\newcommand{\ra}{\rangle}
\newcommand{\nicesum}[2]{\sum\limits_{#1}^{#2}}
\newcommand{\nn}{\nonumber}
\newcommand{\triquad}{\quad\quad\quad}
\newcommand{\quadquad}{\quad\quad\quad\quad}
\newcommand{\undernull}[1]{\underset{=0}{\underbrace{#1}}}
\newcommand{\braceunderset}[2]{\underset{#1}{\underbrace{#2}}}
\newcommand{\person}[1]{\textsc{#1}}
\newcommand{\domaintype}[1]{{\it #1}\xspace}
\newcommand{\AND}{\textbf{and} }
\newcommand{\NOT}{\textbf{not} }
\DeclareFontFamily{U}{rcjhbltx}{}
\DeclareFontShape{U}{rcjhbltx}{m}{n}{<->rcjhbltx}{}
\DeclareSymbolFont{hebrewletters}{U}{rcjhbltx}{m}{n}
\let\aleph\relax\let\beth\relax
\let\gimel\relax\let\daleth\relax
\DeclareMathSymbol{\aleph}{\mathord}{hebrewletters}{39}
\DeclareMathSymbol{\beth}{\mathord}{hebrewletters}{98}
\DeclareMathSymbol{\gimel}{\mathord}{hebrewletters}{103}
\DeclareMathSymbol{\daleth}{\mathord}{hebrewletters}{100}
\DeclareMathSymbol{\lamed}{\mathord}{hebrewletters}{108}
\DeclareMathSymbol{\mem}{\mathord}{hebrewletters}{109}
\DeclareMathSymbol{\ayin}{\mathord}{hebrewletters}{96}
\DeclareMathSymbol{\tsadi}{\mathord}{hebrewletters}{118}
\DeclareMathSymbol{\qof}{\mathord}{hebrewletters}{114}
\DeclareMathSymbol{\shin}{\mathord}{hebrewletters}{152}

\newcommand{\sff}{pdf}

\newcommand{\DomainTypeSketchWidth}{\textwidth}
\newcommand{\GradientPlotWidth}{0.475\textwidth}

\newcommand{\subfigref}[1]{(#1)}
\newcommand{\Eqref}[1]{Eq.~(\ref{#1})}
\newcommand{\GFRD}{GFRD\xspace}
\newcommand{\eGFRD}{eGFRD\xspace}
\newcommand{\eGFRDii}{eGFRD2\xspace}

\newcommand{\Smoldyn}{\textit{Smoldyn}\xspace}
\newcommand{\ChemCell}{\textit{ChemCell}\xspace}
\newcommand{\MCell}{\textit{MCell}\xspace}
\newcommand{\GridCell}{\textit{GridCell}\xspace}
\newcommand{\ReaDDy}{\textit{ReaDDy}\xspace}
\newcommand{\Spatiocyte}{\textit{Spatiocyte}\xspace}
\title{eGFRD in all dimensions}
\author[1]{Thomas R. Sokolowski\footnote{Corresponding author; e-mail: {\tt tsokolowski@ist.ac.at}}\footnote{Present address: Institute of Science and Technology (IST) Austria, Am Campus 1, 3400 Klosterneuburg, Austria}}
\author[1]{Joris Paijmans}
\author[1]{Laurens Bossen}
\author[1]{Martijn Wehrens}
\author[1]{Thomas Miedema}
\author[1]{Nils B. Becker\footnote{Present address: Bioquant Center, University of Heidelberg, 69120 Heidelberg, Germany}}
\author[2]{Kazunari Kaizu}
\author[2]{Koichi Takahashi}
\author[1]{Marileen Dogterom\footnote{Present address: Department of Bionanoscience, Kavli Institute of Nanoscience, Faculty of Applied Sciences, Delft University of Technology, 2628 CJ Delft, The Netherlands}}
\author[1]{Pieter Rein ten Wolde}

\affil[1]{FOM Institute AMOLF, Science Park 104, 1098 XG Amsterdam, The Netherlands}
\affil[2]{RIKEN Quantitative Biology Center (QBIC), RIKEN, 6-2-3 Furuedai, Suita, Osaka 565-0874, Japan}

\date{\today}

\begin{document}

\maketitle

\abstract{\CH{
Biochemical reactions typically occur at low copy numbers, but at once in crowded and diverse environments. Space and stochasticity therefore play an essential role in biochemical networks. Spatial-stochastic simulations have become a prominent tool for understanding how stochasticity at the microscopic level influences the macroscopic behavior of such systems. However, while particle-based models guarantee the level of detail necessary to accurately describe the microscopic dynamics at very low copy numbers, the algorithms used to simulate them oftentimes imply trade-offs between computational efficiency and biochemical accuracy. eGFRD (enhanced Green's Function Reaction Dynamics) is an exact algorithm that evades such trade-offs by partitioning the $N$-particle system into $M\leq N$ analytically tractable one- and two-particle systems; the analytical solutions (Green's functions) then are used to implement an event-driven particle-based scheme that allows particles to make large jumps in time and space while retaining access to their state variables at any moment. Here we present "eGFRD2", a new eGFRD version that implements the principle of eGFRD in all dimensions, thus enabling efficient simulation of biochemical reaction-diffusion processes in the 3D cytoplasm, on 2D planes representing membranes, and on 1D elongated cylinders representative of, e.g., cytoskeletal tracks or DNA; in 1D, it also incorporates convective motion used to model active transport. We find that, for low particle densities, eGFRD2 is up to 3 orders of magnitude faster than optimized Brownian Dynamics. We exemplify the capabilities of eGFRD2 by simulating an idealized model of Pom1 gradient formation, which involves 3D diffusion, active transport on microtubules, and autophosphorylation on the membrane, confirming recent results on this system and demonstrating that it can efficiently operate under genuinely stochastic conditions.}
}

\newpage
\section{Introduction}
Biochemical reactions constitute the basis of all vital functions in biological cells,
ranging from metabolism and gene regulation to environment sensing and intra- and intercellular transport.
\CH{While even the simplest biological cells contain a myriad of different biochemical species,
their individual copy numbers oftentimes only reach numbers as low as thousands, or even dozens \cite{Yu2006,Elf2007,Simicevic2013,Li2014,GarzaDeLeon2017};
this means that specific biochemical reaction pathways usually operate in the extreme low-concentration regime,
while at the same time the cytoplasm is a highly crowded and inhomogenous environment \cite{Zimmerman1993,Ellis2001,Ellis2001-2,Li2009,Zhou2013,Dlugosz2011}.}
These circumstances strongly augment the importance of spatial effects and the inherent stochasticity of biochemical reactions
and at once hinder their direct experimental observation \cite{Mahmutovic2012,Tsimring2014}.
\CH{For example, spatial inhomogeneities can have a strong influence
on the behavior of spatially distributed enzymes \cite{Elf2004,Lawson2015},
even provoking the emergence or destruction of ultrasensitivity \cite{vanAlbada2007,Morelli2008,Takahashi2010,Dushek2011},
and on density-dependent clustering \cite{Jilkine2011,Mugler2013,Wehrens2014};
macromolecular crowding can shift chemical equilibria \cite{Morelli2011,Klann2009,Klann2011} (see \cite{TenWoldeMugler2014} for a review), 
and fast reactant rebindings can significantly enhance the noise in transcription factor and ligand binding \cite{vanZon2006,Mugler2012,Kaizu2014}.
Facilitated diffusion on one-dimensional submanifolds, such as the DNA or cytoskeletal macropolymers, is capable of enhancing the search for target sites \cite{Wang2006,Bonnet2008,Tafvizi2011,Hammar2012,Nguyen2014,Loverdo2008,Loverdo2009-PRL,Loverdo2009-PRE,Benichou2010,Benichou2011,Paijmans2014,Schwarz2016}.
Perhaps most strikingly, spatio-temporal fluctuations at the molecular scale can drastically change the macroscopic behavior on the cellular scale \cite{Takahashi2010,Fange2006,MuglerTenWolde2013}.}

Spatial-stochastic simulations therefore have become an important tool for understanding biochemical mechanisms.
They can roughly be seperated into two classes: lattice- or mesh-based simulation schemes and particle-based schemes \cite{Dobrzynski2007,Sokolowski2017}.
Mesh-based schemes, 
such as {\it MesoRD} \cite{Hattne2005,Fange2006,Wang2013}, {\it URDME} \cite{Drawert2012} and associated techniques \cite{Drawert2010,Lampoudi2009,Isaacson2006}
\CH{(which recently lead to the development of {\it StochSS} \cite{Drawert2016})}, {\it VCell} \cite{Moraru2008} and {\it GMP} \cite{Rodriguez2006,Vigelius2010},
elaborate on the idea of the event-driven (and thus highly efficient) Stochastic Simulation Algorithm by Gillespie \cite{Gillespie1976,Gillespie1977},
by essentially implementing it on a spatial mesh; 
therefore, as a caveat, they have to assume well-mixedness at least locally, 
which---in general---is an inaccurate representation of the real conditions in biological cells.
Particle-based schemes such as \Smoldyn \cite{Andrews2004,Andrews2010}, \MCell \cite{Stiles1996,Stiles2000,Franks2002,Kerr2008,Stefan2014},
\ChemCell \cite{Plimpton2005}, \GridCell \cite{Boulianne2008}, \Spatiocyte \cite{Arjunan2009,Miyauchi2016}, and \ReaDDy \cite{Schoneberg2013}
are traditionally based on the principle of Brownian Dynamics (BD);
here particle diffusion is approximated by a random walk in continuous space with very small propagation steps ($\Delta t\lesssim 10^{-6}s$),
required to render them sufficiently accurate.
At low concentrations these schemes become inefficient, \CH{because most CPU time is spent}
on generating (uninteresting) random movements; 
moreover, since their capability to sample chemical equilibria faithfully depends on how well particle overlaps are resolved, 
their computational efficiency can be only improved at the cost of sacrificing accuracy.

The desire to overcome this antagonism between efficiency and accuracy lead to the development of \eGFRD 
(``enhanced Green's Function Reaction Dynamics'') \cite{vanZon2005-PRL,vanZon2005-JChemPhys,vanZon2006,Takahashi2010}, which is both particle-based and event-driven,
and does not rely on arbitrary definitions of particle contact to sample reactions.
The key idea of \eGFRD is to partition the space filled by the particle cloud into geometrically simple subvolumes (``domains'')
which contain at most two particles; after breaking down the multi-particle problem into a series of
one- and two-particle problems, the full time-dependent analytical solution of the reaction-diffusion problem can be calculated
for each of the domains, and used to sample exact event times and updated particle positions.
This way, large jumps in time and space can be made by each individual particle, rendering \eGFRD 
orders of magnitude more efficient than conventional BD schemes up to $\unit{\mu M}$ concentrations \cite{Takahashi2010}.

However, until now \eGFRD had been limited to simulations of diffusion and particle interactions \CH{in three-dimensional space.}
Yet it is well known that many biochemical reactions occur on finite 1D and 2D submanifolds of the cell, such as the cell membrane,
membranes of intracellular vesicles, and long macropolymers like the DNA or microtubules \CH{\cite{Janmey1998,Kholodenko2006,Moseley2006,Loverdo2008,Li2009,Schwarz2016,Kuchler2016}}.
\CH{In this work we present ``\eGFRDii'', an extended version of \eGFRD that allows for simulations in all dimensions,}
implementing diffusion and particle reactions in 1D and 2D, binding of bulk particles to lower dimensional structures,
and transitions of particles between different structures; 
in 1D, it also features combined diffusive-convective motion with reactions, allowing for simulation of active transport on cytoskeletal tracks.
To accomplish this, we derived and numerically implemented the Green's functions by solving the one- and two-particle reaction-diffusion problems in 1D and 2D,
and integrated them together with the known 3D functions and a BD fallback system into a new user-friendly simulation environment.
In order to exemplify the possiblities of the new \eGFRD we carried out simulations of Pom1 gradient formation, 
which is driven by autophosphorylation on the membrane and active transport, and were able to confirm recent results on this system,
while demonstrating that it can operate efficiently under fundamentally stochastic conditions, owing to low copy numbers.

This paper is organized as follows:
In the first part (``Methods'') we first recapitulate the working principle of \eGFRD,
followed by a description of the new extensions to lower dimensions, 
and a brief presentation of the performance of the new scheme.
In the second part (``Results'') we introduce the studied example system and present our simulation results.
We end by discussing the results and an outlook on further development.

 \newpage
\section{Methods} \label{sec:Methods}
\subsection{The \eGFRD working principle}
\label{sec:eGFRD-Principle}
\begin{figure}[t]
  \centering
  \includegraphics[width=\textwidth]{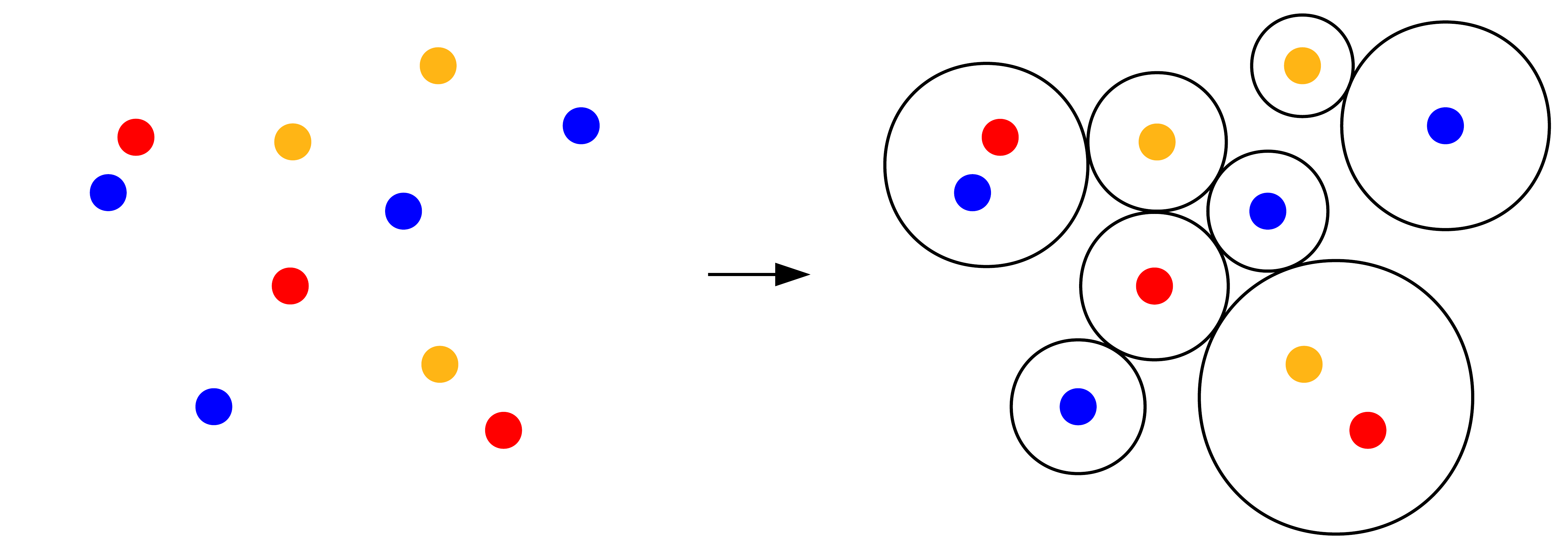}  
\caption{ \label{fig:GFRD-Principle}
  \textbf{Protective domains separate the $N$-particle problem into one- and two-particle problems.}
	  The drawing illustrates how \eGFRD constructs protective domains that contain at most two particles
	  in order to isolate these from the influence of other particles,
	  starting from a random particle configuration.
	  Subsequently, analytical solutions are calculated for each domain individually and used to
	  propagate the domains in an event-driven, asynchronous fashion.
	  We show here a 2D projection for the standard scenario in which particles diffuse and react in unbounded 3D space.
	  In this case protective domains are spherical.
	  Different colors mark different chemical species.
}
\end{figure}

\CH{\eGFRD is an exact algorithm designed to simulate the idealized reaction-diffusion model shown in Figure~\ref{fig:GFRD-Principle},
which is widespread in the field of particle-based stochastic simulation.
In this ``particle-based model'', 
the particles have an idealized, spherical shape with a species-specific radius $R$, 
move by normal free diffusion characterized by a (species-specific) diffusion constant $D$, 
and can interact upon contact with a predefined rate constant $k$;
\eGFRD thus assumes that beyond the contact distance the interaction potential is zero.
In addition, the particles can undergo dissociation, species change or annihilation reactions with predefined rates.
Stochastic simulations of the particle-based model in Fig.~\ref{fig:GFRD-Principle} }
can be straightforwardly carried out using Brownian Dynamics, 
but at low particle density---commonly encountered in biochemical systems---this becomes very inefficient, 
because the vast majority of computation steps is spent on sampling the diffusive random walks of the particles;
it is therefore desirable to skip the particle hops and jump directly between the truly interesting events, i.e., particle encounters and reactions,
employing the known statistics of diffusion \cite{Oppelstrup2006,Oppelstrup2009,Donev2010,Hellander2011,Schwarz2013}.
However---even with the simplifications introduced above---it is generally hard to find an analytical prediction 
for future particle species and positions in an $N$-particle reaction-diffusion system;
nonetheless, as described further below in more detail, exact analytical solutions (Green's functions) can be obtained for the case $N\leq2$.
\eGFRD capitalizes on this fact by dividing the simulation volume into subvolumes, called protective domains,
that contain either one (\domaintype{``Single''} domains) or two (\domaintype{``Pair''} domains) particles, 
in order to isolate the content of each domain from the influence of surrounding particles (and vice versa).
This way the $N$-particle problem is reduced to $M\leq N$ independent one- or two-particle problems.
Figure~\ref{fig:GFRD-Principle} illustrates this principle.
When---after a domain-specific time $\tau_{\mathcal D}$ that can be sampled from the Green's functions---one of 
the particles hits a domain boundary or experiences a reaction that changes its biochemical properties, 
the state of the involved particle(s) is updated, the old domain removed, and one or more new domains initialized.
The use of protective domains is the key innovation of \eGFRD compared to the original \GFRD \cite{vanZon2005-PRL,vanZon2005-JChemPhys,vanZon2006},
which was based on the same motivation, but had to operate with a maximal cut-off time for particle updates
in order to render particle interactions not captured by the used unbounded Green's functions sufficiently improbable.
Since in \eGFRD by construction all position updates remain confined to the respective domain,
any interference with the situation outside is not just improbable, but completely impossible.
\eGFRD therefore is an exact algorithm.

We will now describe how Green's functions can be used to generate next-event times and corresponding new particle states within the protective domains in more detail.
Let us first focus on the \domaintype{Single} domain and \CH{denote by $p_1(\myvec r,t|\myvec r_0)$} the probability density function (PDF)
for the diffusing particle being at position $\myvec r$ at time $t$, given that it started at position $\myvec r_0$.
Then $p_1(\myvec r,t|\myvec r_0)$ is the Green's function of the boundary value problem
\begin{align}
 \partial_t \, p_1(\myvec r,t|\myvec r_0) &= D \nabla^2_{\myvec r} \, p_1(\myvec r,t|\myvec r_0) + \delta(\myvec r - \myvec r_0)\delta(t - t_0)	\label{eq:Diffusion_Single}\\
 p_1(\myvec r,t|\myvec r_0) 	&= 0 \hspace{10ex}\text{for}\hspace{10ex} \myvec r \in \partial\mathcal{D}_1					\label{eq:BC_Single}
\end{align}
where the last equation imposes absorbing boundary conditions on the outer shell ($\partial\mathcal{D}_1$) of the domain.
Note that here we do not specify the Laplace operator $\nabla^2_{\myvec r}$ in detail yet;
its precise form depends on the dimensionality of the underlying diffusion process.

Similarly, the exact solution for the PDF $p_2(\myvec r_{\rm A}, \myvec r_{\rm B}, t|\myvec r_{A0}, \myvec r_{B0})$, describing the positions $\myvec r_{\rm A}$ and $\myvec r_{\rm B}$ of two particles A and B inside a \domaintype{Pair} domain after time $t$ given that they started at positions $\myvec r_{A0}$ and $\myvec r_{B0}$,
can be obtained by solving the \person{Smoluchowski} equation \cite{Smoluchowski1915,vanZon2005-JChemPhys}
\begin{align}
\partial_t \, p_2(\myvec r_{\rm A}, \myvec r_{\rm B}, t|\myvec r_{A0}, \myvec r_{B0})	
    &=	[ D_{\rm A}\nabla_{{\myvec r}_{\rm A}}^2 + D_{\rm B}\nabla_{{\myvec r}_{\rm B}}^2 ] \, p_2(\myvec r_{\rm A}, \myvec r_{\rm B}, t|\myvec r_{A0}, \myvec r_{B0})
\label{eq:Smoluchowski}
\end{align}
\TODO{MORE DETAIL HERE!} after separating it into two individual diffusion equations for the 
interparticle vector $\myvec{r}$ and a ``center of motion'' coordinate $\myvec{R}$ via a product ansatz 
$p_2(\myvec r_{\rm A}, \myvec r_{\rm B}, t) \equiv p_{\rm r}(\myvec r,t)p_{\rm R}(\myvec R,t)$ 
(as described in detail in sec.~\ref{sec:GFRD-GeneralCoordinateTransform} of the \SI), and imposing the following boundary conditions:
\begin{align}
q_\sigma(t) 		&\equiv - \int_{|\myvec r|=\sigma} D\nabla_{\myvec r} \, p_{\rm r}(\myvec r,t|\myvec r_0) d\myvec r 
				= k \, p_{\rm r}(|\myvec r| = \sigma, t)	\label{eq:BC_rad}\\
p_{\rm r}(\myvec r,t)	&= 0 \hspace{10ex}\text{for}\hspace{10ex} \myvec r \in \partial\mathcal{D}_{\rm r}	\label{eq:BC_abs_r}\\
p_{\rm R}(\myvec R,t) 	&= 0 \hspace{10ex}\text{for}\hspace{10ex} \myvec R \in \partial\mathcal{D}_{\rm R}	\label{eq:BC_abs_R}
\end{align}
Here, \Eqref{eq:BC_rad}, representing the particle reaction at intrinsic association rate $k$, 
imposes a radiating (flux) boundary condition to the interparticle vector $\myvec r$ 
at the particle contact radius $\sigma=R_{\rm A} + R_{\rm B}$,
while \Eqref{eq:BC_abs_r} imposes an absorbing boundary condition at the outer radius of the ``interparticle domain'' $\partial\mathcal{D}_{\rm r}$,
and \Eqref{eq:BC_abs_R} does the same to the center-of-motion vector $\myvec R$;
the subdomains $\partial\mathcal{D}_{\rm r}$ and $\partial\mathcal{D}_{\rm R}$ have to be chosen such that they fit inside the shell
$\partial\mathcal{D}_2$ of the original \domaintype{Pair} domain constructed around the two particles (this rule is exemplified in sec.~\ref{sec:GFRD-Pairs} / Fig.~\ref{fig:PairAndMulti}\subref{fig:PairSubdomains} of the \SI \TODO{MOVE TO MAIN TEXT}).
Importantly, since the form of the Laplacian and the precise form of the integral in the radiating condition
\Eqref{eq:BC_rad} vary with the dimensionality of the problem,
the Green's functions are different in 1D, 2D and 3D and have to be calculated
for each dimension separately.

Quantities that derive from the Green's function, the survival probability
and the boundary fluxes, can be used to generate tentative next-event times for each domain individually.
Most importantly, since $p(\myvec r,t|\myvec r_0)$ completely describes the transient dynamics within the domain, 
it enables exact sampling of new particle positions at {\it any} time $t$ after domain construction,
and at the next-event time $\tau_{\mathcal D}$ in particular, rendering incremental sampling of particle trajectories unnecessary.
The procedure of sampling next-event times and new positions from the Green's functions
is described in detail in sec.~\ref{sec:GFRD-General-Sampling-Procedures} of the \SI.
After each domain update the domain is removed, the new configuration
of particles is reanalysed and new domains are constructed around the displaced particles.
Then the newly calculated next-event times are inserted into the ordered scheduler
in the right place, and the domain with the foremost next-event time is updated next.
To enhance the formation of two-particle domains, recently updated particles can force
a ``premature'' update of domains in their proximity, called ``bursting'',
by which the domain is propagated towards a time prior to its originally scheduled update. \TODO{BURSTING COMIC}
If bursting causes particles to move close enough, creation of a two-particle domain will be attempted.
If there is not enough space to construct any \eGFRD domain due to nearby obstacles or particle crowding, 
the particles are propagated by a Brownian Dynamics fallback simulator, as explained in sec.~\ref{sec:BD} and supplementary sec.~\ref{sec:BD-SI}.
A compact overview of the basic \eGFRD algorithm in pseudo-code is given by Algorithm \ref{alg:eGFRD} in the \SI (p.~\pageref{alg:eGFRD}), 
while a detailed account of the bursting and domain construction rules is found in supplementary sec.~\ref{sec:GFRD-shellmaking}.

\subsection{Extension to lower dimensions}
\label{sec:NewDomains}
\begin{figure}[t]  
  \centering
  \subfigure[][]{
    \label{fig:Domain-PlanarSurfaceInteraction}
    \includegraphics[width=\DomainTypeSketchWidth]{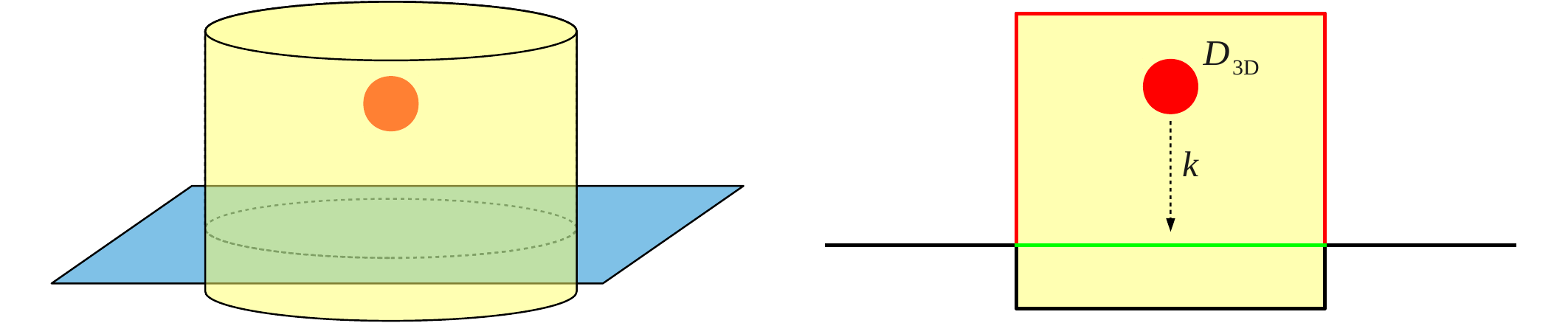}
  }
  \subfigure[][]{
    \label{fig:Domain-PlanarSurfaceSingle}
    \includegraphics[width=\DomainTypeSketchWidth]{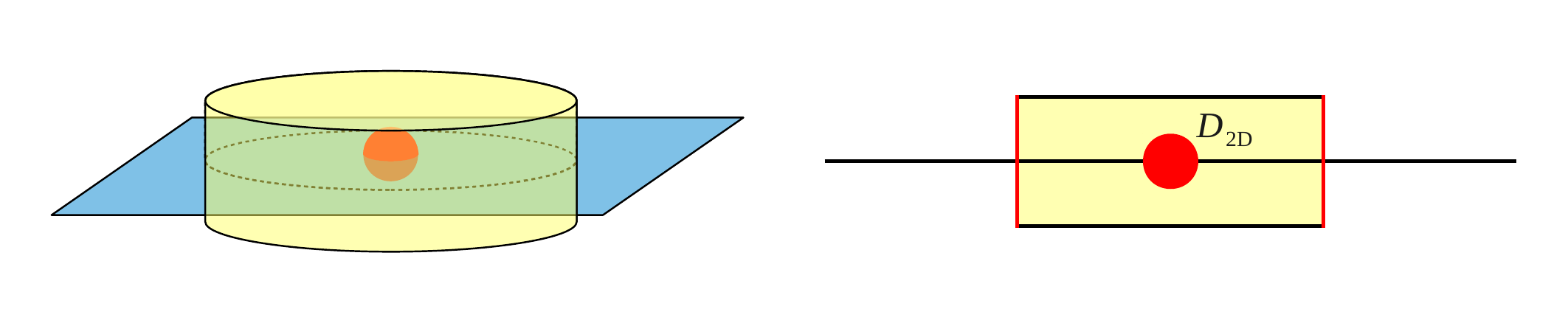}
  }
  \subfigure[][]{
    \label{fig:Domain-PlanarSurfacePair}
    \includegraphics[width=\DomainTypeSketchWidth]{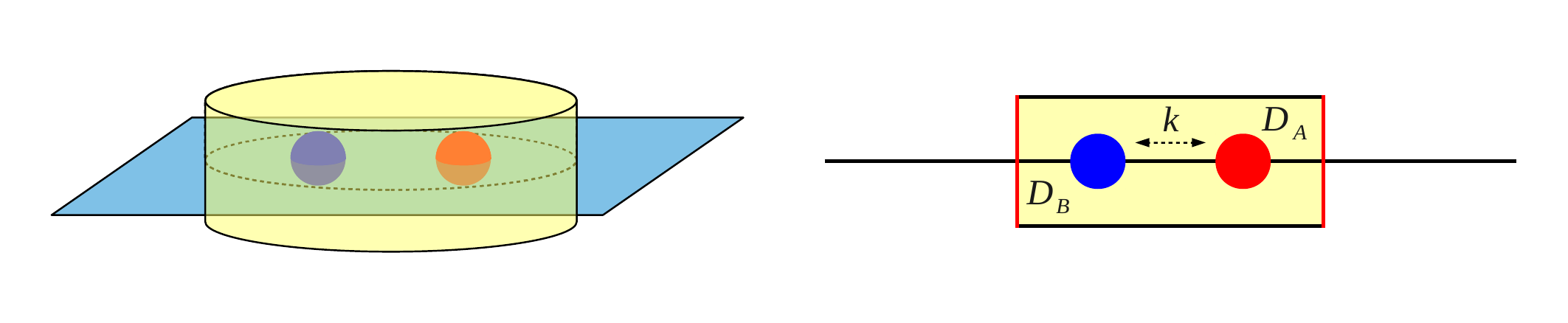}
  }
\caption{  
  \textbf{New protective domain types for interaction with and reaction-diffusion on planar surfaces.}  
  \subfigref{\subref{fig:Domain-PlanarSurfaceInteraction}} \domaintype{Planar Surface Interaction} domain;
  \subfigref{\subref{fig:Domain-PlanarSurfaceSingle}} \domaintype{Planar Surface Single} domain;
  \subfigref{\subref{fig:Domain-PlanarSurfacePair}} \domaintype{Planar Surface Pair} domain.
  Right panels show sections of 3D objects.
  Absorbing boundaries are highlighted by red, radiating boundaries by green. 
  \label{fig:NewDomains-2D}
}
\end{figure}

In order to port the \eGFRD principle to lower dimensions we introduced static reactive
surfaces capturing the essential geometric features of subcellular structures:
finite planes, which can be used to model membranes, and (thin) finite cylinders,
representative, e.g., of elongated DNA or cytoskeletal tracks (microtubules, actin filaments, etc.).
Based on this we defined a new set of protective domains for interactions of particles
{\it with} the new structures (\domaintype{``Interaction''} domains)
and new \domaintype{Single} and \domaintype{Pair} domains for diffusion and interparticle reactions {\it on} the structures,
and calculated the Green's functions for the associated reaction-diffusion problems within the respective geometry.
Figures~\ref{fig:NewDomains-2D} and \ref{fig:NewDomains-1D} contain an overview of the most important new domains in 2D and 1D, respectively;
most of them are cylindrical, reflecting the natural coordinate separation for the respective binding or transport process.

Below we motivate and explain the \CH{principal} new domain types in more detail,
and briefly sketch the derivation of the associated new Green's functions;
for the complete mathematical derivations, sampling and domain making rules we will refer the reader to the \SI.
\CH{Several domain types devised for special applications are described in sec.~\ref{sec:SpecialDomains}.}
\TODO{A hierarchic overview of all domain types used in our new \eGFRD version is shown in Fig.~TODO in the SI.}

\subsubsection{Binding to planes} \label{sec:2D-Binding}
Fig.~\ref{fig:NewDomains-2D}\subref{fig:Domain-PlanarSurfaceInteraction} schematically shows 
the \domaintype{``Planar Surface Interaction''} domain used for interactions of a bulk particle (undergoing 3D diffusion) with a reactive plane.
While not strictly necessary, we chose a cylindrical geometry for the domain, because it facilitates its scaling with respect to the other
cylindrical domains for plane-bound particles that we will introduce further below.
The height of the domain over the plane $h = \delta + \beta R_0$ is composed of the particle-plane distance $\delta$ plus $\beta R_0$, 
where $R_0$ is the particle radius and $\beta>1$ a safety factor (the ``single-shell factor'', defined in section \ref{sec:GFRD-shellmaking} of the \SI);
the domain radius is determined by the available space in the vicinity of the domain.
Association to the plane is modeled via a radiating boundary condition at particle-plane contact,
whereby the particle is defined to be at contact when its center touches the plane.
For that reason, the domain slightly extends behind the reactive plane by a length $h'=\beta R_0$,
to prevent the bound particle from overlapping with particles on the opposite side of the plane.
By default we allow for \CH{association from both sides of the plane}. 

The particle can exit the domain by either binding to the plane
or by hitting one of the absorbing domain boundaries, i.e. the cylinder tube or the (more distant) cylinder cap.
Let $p(r,\varphi,z,t|r_0,\varphi_0,z_0)$ be the probability density function for this problem,
written in cylindrical coordinates $\myvec r = (r,\varphi,z)$.
We can separate diffusion along the cylinder axis ($z$) from diffusion in the polar plane ($r,\varphi$) via the ansatz
\begin{align}
p(r,\varphi,z,t|r_0,\varphi_0,z_0) = p_{\rm r}^{\rm A}(r,\varphi,t|r_0,\varphi_0) \, p_{\rm z}^{\rm RA}(z,t|z_0)
\end{align}
which yields a one-dimensional diffusion equation for $p_{\rm z}^{\rm RA}$ and \textsc{Bessel}'s equation for $p_{\rm r}^{\rm A}$.
The 1D-problem for $p_{\rm z}^{\rm RA}$ has to be solved with a radiating boundary at $z=0$ and an absorbing boundary at $z=h$;
this is a special case of the 1D Green's function that we portray below in sec.~\ref{sec:eGFRD-on-cyl} and in supplementary sec.~\ref{sec:1D-GreensFunction-Rad-Abs}
(with $v=0$, $\sigma=0$ and $a=h$).
The equation for $p_{\rm r}^{\rm A}$ has perfect radial symmetry by construction and describes 2D diffusion
in polar coordinates for a particle starting at $r=0$ and a circular absorbing boundary at $r=R$.
The solution to this problem is well-known and presented in section \ref{sec:2D-GF-AbsSym} of the \SI.

From the two Green's functions $p_{\rm r}^{\rm A}$ and $p_{\rm z}^{\rm RA}$ we sample next-event times $\tau_{\rm r}$ and $\tau_{\rm z}$ in the usual way
and take their minimum as the next-event time \CH{$\tau_\nu = \min(\tau_{\rm r}, \tau_{\rm z})$} of the interaction domain.
In the case $\tau_\nu=\tau_{\rm z}$, i.e. when the particle exits the cylindrical domain through one of
its caps, we compare the fluxes through the opposite boundaries to
determine whether the particle left through the absorbing (\textit{``IV Escape''} event, particle remains in the cytoplasm)
or through the radiating boundary (\textit{``IV Interaction''}, particle associates with the plane).
For $\tau_\nu=\tau_{\rm r}$ we know with certainty that a (radial) escape through the cylinder tube occured.
In both cases the respective other coordinate is sampled from the corresponding Green's function
normalized by the respective survival probability, following the principle explained in sec.~\ref{sec:GFRD-Pairs} of the \SI.

\subsubsection{2D-diffusion and reactions on the plane}
\label{sec:2d-GF-plane}
\begin{figure}[t]
  \centering
  \subfigure[][]{
    \label{fig:Domain-CylindricalSurfaceInteraction}
    \includegraphics[width=\DomainTypeSketchWidth]{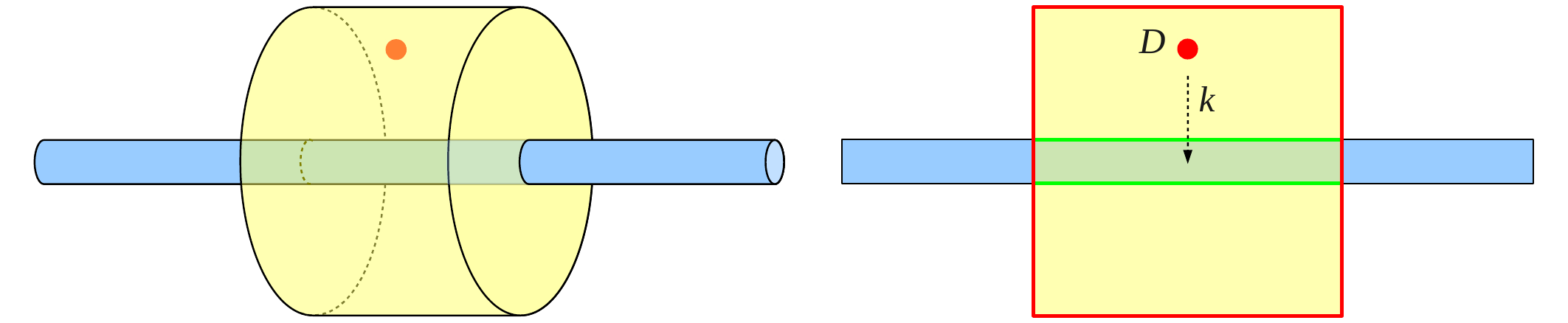}
  }
  \subfigure[][]{
    \label{fig:Domain-CylindricalSurfaceSingle}
    \includegraphics[width=\DomainTypeSketchWidth]{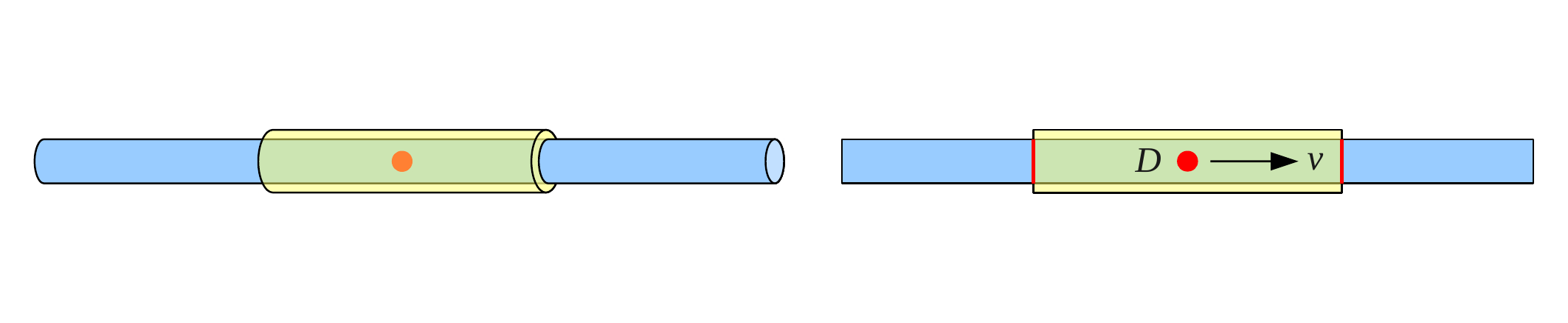}
  }
  \subfigure[][]{
    \label{fig:Domain-CylindricalSurfacePair}
    \includegraphics[width=\DomainTypeSketchWidth]{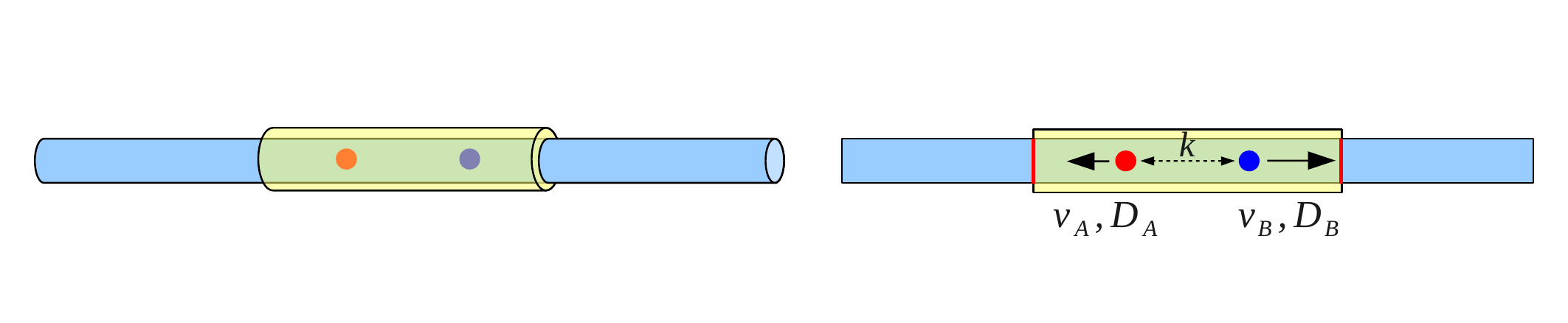}
  }
\caption{
  \textbf{New protective domain types for interactions with and transport and reactions on 1D structures.}
  \subfigref{\subref{fig:Domain-CylindricalSurfaceInteraction}} \domaintype{Cylindrical Surface Interaction} domain;
  \subfigref{\subref{fig:Domain-CylindricalSurfaceSingle}} \domaintype{Cylindrical Surface Single} domain;
  \subfigref{\subref{fig:Domain-CylindricalSurfacePair}} \domaintype{Cylindrical Surface Pair} domain.
  Right panels show sections of 3D objects along the common cylinder axis.
  Absorbing boundaries are highlighted by red, radiating boundaries by green.  
  Note that drift velocities ($v$, $v_{\rm A}$, $v_{\rm B}$) can be towards any cylinder end.  
  \label{fig:NewDomains-1D}
}
\end{figure}

\CH{
Diffusion and interaction of particles bound to the plane is simulated using
the \domaintype{Planar Surface Single} and \domaintype{Planar Surface Pair} domains, respectively,
shown in Fig.~\ref{fig:NewDomains-2D}\subref{fig:Domain-CylindricalSurfaceSingle} 
and Fig.~\ref{fig:NewDomains-2D}\subref{fig:Domain-CylindricalSurfacePair}.
In analogy to the spherical \domaintype{Single} and \domaintype{Pair} domains in 3D,
the 2D domains are cylidrical;
the height of their cylindrical shells is determined by the particle diameter (times safety factors),
while the radius again is dependent on the available space in its surroundings.}

In the \domaintype{Planar Surface Single} the particle starts out from the center of the domain,
and the only exit channel is the absorbing boundary at its outer radius.
The Green's function for this problem is precisely the one that describes
the polar movement in the \domaintype{Planar Surface Interaction} domain, $p_{\rm r}^{\rm A}$ (see supplementary sec.~\ref{sec:2D-GF-AbsSym}).

\CH{In the \domaintype{Planar Surface Pair} we perform the coordinate transform initially described in sec.~\ref{sec:eGFRD-Principle},
partitioning the available space among the center-of-motion coordinate $\myvec{R}$ and the interparticle vector $\myvec{r}$;
this way the reaction-diffusion process is once again separated into two independent diffusion processes,
while the particle interaction can be completely characterized by a radiating boundary condition to the $\myvec{r}$ coordinate.}
Next-event times and new positions for the polar diffusion in the $\myvec{R}$ coordinate 
\CH{then can be} sampled from the Green's function $p_{\rm r}^{\rm A}(R,\Phi,t|R_0,\Phi_0)$.
\CH{For the interparticle coordinate $\myvec{r}$} we use the Green's function $p_{\rm r}^{\rm RA}(r,\varphi,t|r_0,\varphi_0)$,
which solves the following boundary value problem:
\begin{align}
 \partial_t \, p_{\rm r}^{\rm RA}(r,\varphi,t|r_0,\varphi_0)	&=	D_{\rm r}\nabla_{\myvec r}^2 \, p_{\rm r}^{\rm RA}(r,\varphi,t|r_0,\varphi_0)			 \nonumber\\
 &= D_{\rm r} \left[ \partial^2_r + \frac{1}{r}\partial_r + \frac{1}{r^2}\partial^2_\varphi \right] \, p_{\rm r}^{\rm RA}(r,\varphi,t|r_0,\varphi_0)			 \label{eq:PDE_2D}\\
 &																 					 \nonumber\\
 2\pi\sigma D_{\rm r} ~\partial_r \, p_{\rm r}^{\rm RA}(r,\varphi,t|r_0,\varphi_0)\takenat_{r=\sigma}	&=	k \, p_{\rm r}^{\rm RA}(|\myvec r|=\sigma|r_0,\varphi_0) \label{eq:BC_rad_2D}\\
 p_{\rm r}^{\rm RA}(r,\varphi,t|r_0,\varphi_0)\takenat_{r=a}						&= 	0			 				 \label{eq:BC_abs_2D}\\
 &																 					 \nonumber\\
 p_{\rm r}^{\rm RA}(r,\varphi,t=0|r_0,\varphi_0)	&= \frac{1}{r} \delta(r-r_0)\delta(\varphi-\varphi_0)					 			 \label{eq:IC_2D}
\end{align}
Herein, \Eqref{eq:PDE_2D} is the diffusion equation in polar coordinates, \Eqref{eq:BC_rad_2D} the radiating boundary condition
modeling reactions at a contact radius $\sigma = R_{\rm A} + R_{\rm B}$ with intrinsic reaction rate $k$,
\Eqref{eq:BC_abs_2D} an absorbing boundary condition at the outer radius $a$ of the interparticle subdomain,
and \Eqref{eq:IC_2D} the initial condition of the interparticle vector properly transformed into polar coordinates.
We solve this problem \CH{explicitly} in sec.~\ref{sec:2D-GreensFunction} of the \SI.

To determine the next event for the \domaintype{Pair} domain, we first sample next-event times
$\tau_{\rm r}$ and $\tau_{\rm R}$ for the interparticle and center-of-motion coordinates, respectively;
the smaller of the two is taken to be the next-event time for the whole domain.
If $\tau_{\rm R}<\tau_{\rm r}$, the corresponding event is an exit of the $\myvec{R}$ coordinate from its subdomain,
such that its new length $|\myvec{R}|$ is fixed (equal to the outer radius of the subdomain),
but we still have to sample \CH{the corresponding new angle $\Phi$, and moreover a new length and a new angle for the interparticle vector $\myvec{r}$;}
the latter is achieved by plugging the time $\tau_{\rm R}$ into $p_{\rm r}^{\rm RA}(r,\varphi,t|r_0,\varphi_0)$.
If, conversely, $\tau_{\rm r}<\tau_{\rm R}$, we can sample a new $\myvec{R}$ vector in a similar way
from $p_{\rm r}^{\rm A}(R,\Phi,\tau_{\rm r}|R_0,\Phi_0)$.
\CH{However, now concerning the dynamics of $\myvec{r}$, two events are possible:}
either the $\myvec{r}$ coordinate hit the inner boundary and reacted at interparticle contact (\textit{``IV Reaction''}), 
or it left the $\myvec{r}$-subdomain at its outer radius $a$ (\textit{IV Escape});
which of the two events occurs is determined by comparing the magnitude of the probability fluxes through
the two subdomain boundaries at time $\tau_{\rm r}$ \TODO{MORE DETAIL / SI?}.
If the event was an \textit{IV Reaction} we directly replace the two particles by a particle of the product species
at the new position of the center-of-motion $\myvec{R}$;
if, instead, an \textit{IV Escape} occured, we know that $|\myvec{r}|=a$ and still have to sample a new angle $\varphi$
to construct the new interparticle vector $\myvec{r}$.
Finally, we transform the coordinates $\myvec{R}$ and $\myvec{r}$ back to new particle positions $\myvec{r}_{\rm A}$
and $\myvec{r}_{\rm B}$.
The detailed procedure of sampling the event type and new positions is described in supplementary sec.~\ref{sec:GFRD-Pairs}.

\subsubsection{Binding to cylinders}
\label{sec:eGFRD-binding-to-cyl}
We handle the binding of bulk particles to reactive cylindrical structures via the \domaintype{Cylindrical Surface Interaction} domain,
shown in Fig.~\ref{fig:NewDomains-1D}\subref{fig:Domain-CylindricalSurfaceInteraction}.
Since only the radial distance of the particle from the cylinder is relevant to the binding problem,
the natural geometry of this problem is again cylindrical, only that the cylindrical domain now has
four boundaries: an inner boundary that wraps around the reactive cylinder at contact radius $\sigma=R_{\rm part} + R_{\rm cyl}$,
an opposite absorbing outer boundary at a distant radius $a$, and two absorbing boundaries in axial ($z$) direction.
As before, we can separate the polar and axial movement and determine two next-event times $\tau_{\rm r}$ and $\tau_{\rm z}$,
the smaller of which determines which coordinate hit the corresponding boundary first.

In the two polar coordinates, cylinder binding is akin to the two-particle reaction-diffusion problem formulated in the interparticle coordinate $\myvec{r}$ in sec.~\ref{sec:2d-GF-plane},
and indeed we employ the same Green's function $p_{\rm r}^{\rm RA}(r,\varphi,t|r_0,\varphi_0)$, obtained by solving the problem defined by Eqs.~(\ref{eq:PDE_2D})--(\ref{eq:IC_2D}), 
to sample a next-event time $\tau_{\rm r}$ for the polar motion;
for the movement in $z$-direction, we use the Green's function for one-dimensional free diffusion with two absorbing boundaries, which is a special case (for $v=0$) of Green's function $p_{\rm X}^{\rm A}$, introduced in the following section~\ref{sec:eGFRD-on-cyl}.
As before, in the case $\tau_{\rm r}>\tau_{\rm z}$, \CH{i.e., when the particle first hits one of the radial boundaries,} 
we compare the probability fluxes at the opposite boundaries of the radial coordinate to determine whether
the event is a binding reaction to the cylinder (\textit{IV Interaction}) or exit through the distant boundary (\textit{IV Escape});
upon binding, the particle is placed onto the axis of the cylinder.
\CH{In the case $\tau_{\rm z}>\tau_{\rm r}$, i.e. when the particle exits the cylindrical domain through one of its caps,
the respective probability fluxes through the corresponding absorbing boundaries are compared to determine through which $z$-boundary the particle exits;
the latter can be omitted when a symmetric cylinder centered at the initial particle position is used---then the boundary of exit can be chosen randomly with probability $1/2$.}
In either case \CH{($\tau_{\rm r}>\tau_{\rm z}$ or $\tau_{\rm z}>\tau_{\rm r}$)} the new value of the respective other coordinate(s) is sampled from the corresponding Green's function
evaluated at the next-event time.

\subsubsection{Movement and reactions on cylinders}
\label{sec:eGFRD-on-cyl}
In biological cells, motion confined to one-dimensional structures is widespread.
Perhaps the most prominent examples are
the diffusive search, hopping and sliding of transcription factors on the DNA \CH{\cite{Wang2006,Bonnet2008,Tafvizi2011,Hammar2012,Nguyen2014,Loverdo2008,Loverdo2009-PRL,Loverdo2009-PRE,Benichou2010,Benichou2011,Paijmans2014}},
and active transport of proteins and vesicles by motor proteins
on cytoskeletal filaments, such as microtubules and actin filaments \CH{\cite{Ross2008,Hirokawa2009,Verhey2009,Tischer2010,Akhmanova2011,Hammer2012,Hancock2014,Khaitlina2014,Huber2015}}.
In the latter case, processive motor proteins, such as members of the dynein and kinesin families,
undergo an ATP-fueled periodic cycle of reactions, this way breaking detailed balance 
and creating a random walk with a clear directional bias on the filament, markedly different from simple diffusion.

In our new \eGFRD implementation, particles that are bound to cylinders can move both via 1D diffusion and/or active transport, 
and engage in interparticle reactions.
We model active transport by supplementing the PDE governing the time evolution of the PDF $p_{\rm x}(x,t|x_0)$,
describing the probability to find the particle at position $x$ at time $t$ given an initial position $x_0$,
by an additional convection term with a constant ``drift'' velocity $v$,
representative of the unidirectional motion component of active transport:
\begin{align}
\partial_t \, p_{\rm x}(x,t|x_0) = \left[ D \partial_x^2 - v \partial_x \right] \, p_{\rm x}(x,t|x_0)
\label{eq:PDE_1D}
\end{align}
This equation, together with the respective initial and boundary conditions,
defines the boundary value problems that yield the Green's functions for the
\domaintype{Cylindrical Surface Single} and \domaintype{Cylindrical Surface Pair} domains,
shown in Fig.~\ref{fig:NewDomains-1D}\subref{fig:Domain-CylindricalSurfaceSingle} and \ref{fig:NewDomains-1D}\subref{fig:Domain-CylindricalSurfacePair}, 
\CH{used, respectively, to simulate the diffusion-drift process of a single particle on the cylinder and the interaction of two particles moving by diffusion and drift.}
Naturally, both domains are cylindrical, with a radius correponding to the (largest) particle radius,
and a length depending on the available free space in the surroundings of the cylinder-bound particle.
The calculation of the necessary Green's functions and the procedure of sampling the
next-event time and type, and new positions of the particle(s), is analogous
to the \domaintype{Planar Surface Pair}, with minor additional precautions owed to the presence of the drift, \CH{as described further below.}
The Green's function $p_{\rm x}^{\rm A}(x,t|x_0)$ used by the \domaintype{Cylindrical Surface Single}
is obtained by solving \Eqref{eq:PDE_1D} subject to absorbing boundary conditions
on both sides of the domain.

For the \domaintype{Cylindrical Surface Pair}, the original equation governing the time-evolution
of the two-particle PDF, $p_{\rm 2,x}(x_{\rm A},x_{\rm B},t|x_{0,A},x_{0,B})$, reads
\begin{align}
\partial_t \, p_{\rm 2,x}  &=	[ D_{\rm A}\partial_{x_{\rm A}}^2 + D_{\rm B}\partial_{x_{\rm B}}^2
				      -v_{\rm A}\partial_{x_{\rm A}} -v_{\rm B}\partial_{x_{\rm B}} ] \, p_{\rm 2,x}
\label{eq:Smoluchowski-Drift}
\end{align}
where each of the two particles has its own diffusion constant $D_{\rm A}$ ($D_{\rm B}$)
and drift velocity $v_{\rm A}$ ($v_{\rm B}$).
In the \SI, sec.~\ref{sec:1D-CoordinateTransform}, we show that the same coordinate transform as introduced in sec.~\ref{sec:eGFRD-Principle}
can be carried out also with the convection terms, again separating the above equation
into one for the center-of-motion $X$ and one for the interparticle separation $x = x_{\rm B} - x_{\rm A}$,
which both have the form of \Eqref{eq:PDE_1D},
with the following diffusion coefficients and drift velocities for the transformed coordinates:
\begin{align}
 D_x &= D_{\rm A} + D_{\rm B}					\quad,		&v_x &= v_{\rm B} - v_{\rm A}							\quad,	\\
 D_X &= \frac{D_{\rm A} D_{\rm B}}{D_{\rm A} + D_{\rm B}}	\quad, 		&v_X &= \frac{D_{\rm B} v_{\rm A} + D_{\rm A} v_{\rm B}}{D_{\rm A} + D_{\rm B}}	\quad,
\end{align}
For the center-of-motion coordinate $X$, the Green's function is identical 
to the one used in the \domaintype{Cylindrical Surface Single}, i.e. $p_{\rm x}^{\rm A}(X,t|X_0)$.
For the interparticle Green's function $p_{\rm x}^{\rm RA}(x,t|x_0)$ once more
a radiating boundary condition has to be imposed at particle contact, $x=\sigma=R_{\rm A}+R_{\rm B}$,
in order to model interparticle reactions.
Here it is important to take into account that the inclusion of convective motion also
changes the definition of the probability flux at the boundary; 
the radiating boundary condition at contact with diffusion and drift therefore reads:
\begin{align}
-D \partial_x \, p(x,t)\takenat_{x=\sigma} + vp(\sigma,t) &= -k p(\sigma,t) 			\label{eq:BC_rad_1D_def}\\
  \Leftrightarrow \qquad \partial_x \, p(x,t)\takenat_{x=\sigma} &= \frac{v+k}{D} p(\sigma,t) 	\label{eq:BC_rad_1D}
\end{align}
where $k$ stands for the intrinsic reaction rate at contact which in 1D has the same unit as the drift velocity $v$.
The minus sign on the right side of \Eqref{eq:BC_rad_1D_def} reflects the flux direction within the chosen coordinate system,
which at the inner boundary for the interparticle separation $x$, by convention, is negative with respect to the $x$-axis.

We present the full derivation of the Green's function $p_{\rm x}^{\rm RA}$
in sec.~\ref{sec:1D-GreensFunctions} of the \SI,
obtaining Green's function $p_{\rm X}^{\rm A}$ as a special case in sec.~\ref{sec:1D-GreensFunction-Abs-Abs}.

\subsubsection{Connected reactive structures and transitions between them}
While isolated planar and cylindrical reactive structures already are well-suited to conceptually study
their effect on biochemical reactions, a more faithful representation of \textit{in vivo} conditions
requries the possiblity to create closed 3D compartments bounded by reactive surfaces,
and let particles transit between them.
\CH{\eGFRDii allows the creation of} such compartments from interconnected orthogonal planes,
while particles can diffuse and react accross their connection seams via special \domaintype{``Transition''} domains.
Moreover, cylindrical surfaces can be connected to planes via an ``interface disk'' structure,
allowing for transitions from cylinder to plane and vice versa;
the same structure can be also used to ``cap'' a finite cylinder such that particles can accumulate at its end 
and unbind into the bulk.
The different \domaintype{Transition} domains and their working principle are explained in more detail
in sections \ref{sec:Plane-Transitions}--\ref{sec:GFRD-FiniteCylinders} of the \SI.

\subsubsection{Further domains for special applications}
\label{sec:SpecialDomains}
We devised two further domains for special applications:
The \domaintype{Mixed Pair 2D-3D} domain, in which a bulk particle diffusing in 3D
can, \CH{upon contact,} directly react with a plane-bound particle diffusing in 2D (see supplementary sec.~\ref{sec:DirectBinding-MixedPair2D3D}),
and the \domaintype{Cylindrical Surface Sink} domain, in which a 1D particle moving on a cylinder
can interact with a static reactive ``sink'' (binding site) while diffusing over it (see supplementary sec.~\ref{sec:1D-GreensFunction-Sink});
\CH{this makes it possible to model, for example, the binding of transcription factors to their promoter, as we showed recently \cite{Paijmans2014}.}
While for the former we make use of a special coordinate transform that allows us to employ Green's functions already implemented for simpler binding scenarios,
for the latter we derived a new Green's function with specialized boundary conditions;
the detailed calculations are found in the respective sections of the \SI listed above.

\subsection{Brownian Dynamics based on the Reaction Volume Method provides a fallback propagation mode} 
\label{sec:BD}
\CH{While at low particle densities \eGFRD can be orders of magnitude more efficient than Brownian Dynamics,
the process of sampling next-event times and new positions inside \eGFRD domains is computationally expensive,
such that the usage of Brownian Dynamics is advantageous again when the domain size becomes comparable to the particle radius.
The crux of GFRD is to construct domains that do not overlap with each other---indeed, this is what turns GFRD into an exact algorithm. 
However, even at low densities it can happen that more than two particles come so close to each other that only very small non-overlapping domains could be constructed; 
this can also occur when one single particle comes close to static structures (planes or cylinders) with which it cannot react.
In such situations, we resort to Brownian Dynamics when \eGFRD domains larger than a predefined minimal size cannot be made any more.}

Therefore, our \eGFRD simulator comprises a fully-featured BD simulator capable of simulating all modes of particle motion,
particle reactions and particle-surface interactions, 
and is equipped with a set of rules that \CH{makes it possible} to seamlessly shuffle particles between the two simulator types.

To guarantee that interparticle and particle-surface reactions fulfill detailed balance, 
we devised a new BD algorithm based on the ``reaction volume method'' (rvm-BD);
rvm-BD is akin in spirit to the Reaction Brownian Dynamics scheme \cite{Morelli2008-JCP}, but more versatile.
At its heart, the new algorithm 
\CH{introduces} \TODO{RETHINK}
a ``reaction volume'' around each reactive object from which forward reactions can occur; 
detailed balance is maintained by placing the unbinding particle inside the reaction volume
with a properly rescaled rate directly derived from the detailed balance condition.
We implemented the scheme such that particles that can potentially interact with each other are automatically grouped
into special \domaintype{``Multi''} domains, each of which constitutes an independent rvm-BD simulator instance
with a specific optimal reaction volume size and time step jointly determined from the set of rates involved;
this way the performance of the propagation in rvm-BD mode is enhanced.
We give a detailed description of the scheme, including a step-by-step derivation of the detailed balance condition,
in sec.~\ref{sec:BD-SI} in the \SI.

\subsection{Performance}
In order to assess the performance of \CH{\eGFRDii},
we profiled our new \eGFRD implementation for representative simulation scenarios (both with and without reactions involved)
by recording the CPU time per real (simulated) time as a function of the particle number,
always comparing to simulations in which only rvm-BD is used to propagate the particles as a reference.
Since in the lower dimensions particle crowding builds up much faster than in 3d,
the profiling was carried out for each dimension separately,
in order to avoid that the lowest dimension becomes the limiting factor and obstructs the performance gains in higher dimensions.
The detailed profiling protocol and results are described in sec.~\ref{sec:Profiling} of the \SI.
In brief, the profiling results demonstrate that our new \eGFRD implementation outperforms rvm-BD by up to 3 orders of magnitude
for particle densities $\lesssim 100/\mu m^d$ ($\lesssim 1000/\mu m^3$ in 3d, which translates to $\lesssim \mu M$ concentrations).
Here it should be emphasized that rvm-BD is a smart BD scheme which optimally adapts the choice of the reaction volumes and propagation time steps to the set of rates involved;
this typically results in average time steps (more than) 3 orders of magnitude larger than classical time step settings ensuring sufficiently fine resolution of particle collisions,
such as $\Delta t \sim 10^{-6}~\sigma^2/D$, where $\sigma^2$ is the particle contact radius and $D$ the (interparticle) diffusion constant; 
brute-force schemes based on such ``safe'' (but comparably naive) choices of the time step thus are outperformed by up to 6 orders of magnitude by \eGFRD.
\CH{Moreover, we expect significant performance gains from a currently ongoing code optimization, as detailed in the subsequent section~\ref{sec:Code}.}

\subsection{Code availability}
\label{sec:Code}
A fully-functional prototype code of \CH{\eGFRDii} is available online at GitHub\footnote{\CH{\texttt{https://github.com/gfrd/egfrd/tree/develop}}}.
Most parts of the code, especially core-functions such as the scheduler system, the reaction-networks implementation and basic geometric objects, 
are written in C++, while Python has been used for the more top-level routines. The open-source {\it boost::python} libraries are used as an interface between C++ and Python.
As a benefit of this, our implementation offers a user-friendly Python interface which can be used for fast and easy scripting of simulations and associated measurement routines.
It also contains a visualization module based on {\it VPython}, and a module that exports the simulator output into a format readable by {\it Paraview}. 
While the use of Python comes with many user- and developer-friendly conveniences, we also found that it inflicts a significant overhead of computational cost.
In the future, we will present a more efficient, fully overhauled code-version which outsources all remaining simulator parts into C++, 
only retaining the scripting interface in Python, thus minimizing its overhead.

\newpage
\section{Results: Simulation of Pom1 gradient formation}
\label{sec:Example-sims}
In order to apply the \CH{newly implemented \eGFRDii framework} to a real biological problem and illustrate its capabilities
we sought to study a simple but nontrivial reaction mechanism within which spatial features and different modes of biochemical transport play a prominent role.
The reaction mechanism underlying the formation of the intracellular Pom1 gradient in bacteria, introduced below, 
ideally fulfills these criteria.

\subsection{The Pom1 gradient}
Protein gradients play a crucial role in cell biology;
they map protein concentration levels to the distance from the gradient source, creating positional cues for downstream targets.
The establishment of local protein accumulations acting as gradient sources oftentimes involves the cytoskeleton and active transport \cite{Bastiaens2006,Niethammer2004,Siegrist2007,Martin2009,LoPresti2011,Howard2012,Kruse2012,Schmick2014,Recouvreux2016}.
A representative example is the Pom1 gradient:
Pom1 is a strongly membrane-associated auto-kinase that marks the division site in elongated fission yeast cells
via concentration gradients decreasing from the cell poles \cite{Bahler1998,Moseley2009,Martin2009b,Tostevin2011,Vilela2010};
the required source-accumulations of membrane-bound Pom1 are established by microtubules that direct cytoplasmic Pom1 towards the opposite poles.
Recent experiments revealed that the membrane-associated gradient is shaped via a phosphorylation-dephosphorylation cycle of Pom1 \cite{Hachet2011}:
Pom1 has its highest affinity in its dephosphorylated form;
after membrane binding, dephosphorylated Pom1 starts self-phosphorylating, 
successively reducing its own membrane affinity as it diffuses away from the cell tip.
Upon reaching higher phosphorylation levels it is recycled to the cytoplasm.
Dephosphorylation of Pom1 by the phosphatase Dis2 is catalyzed by the polarity marker protein Tea4, which is transported towards the
cell tips via active transport on microtubules and itself accumulates at the cell tips.

While in \cite{Hachet2011} the basic principle of this intricate gradient formation mechanism was uncovered, 
some important details remained unknown,
in particular where and how precisely Dis2 dephosphorylates cytoplasmic Pom1,
and whether the autophosphorylation occurs in a intramolecular or intermolecular fashion (cis- vs. trans-autophosphorylation).
While other kinases from the DYRK family, to which Pom1 belongs, have been shown to undergo cis-autophosphorylation \CH{\cite{Lochhead2005,Lochhead2009,Saul2013}},
a more recent study \cite{Hersch2015} based on new experiments and an ODE model concluded that membrane-bound Pom1 phosphorylates in trans-fashion;
the resulting feedback of (local) Pom1 concentration on the (local) phosphorylation activity anti-correlates 
the gradient amplitude and length scale, thus implementing a buffering mechanism that makes the Pom1 density
away from the gradient origin insensitive to Pom1 abundance and the rate of (dephosphorylated) Pom1 delivery to the membrane \cite{Hersch2015}. \TODO{DISCUSS PROS AND CONS}
It was also shown that the underlying multi-step trans-phosphorylation on the membrane effectively is equivalent 
to a nonlinear membrane desorption rate with close-to-quadratic dependence on the Pom1 concentration.
Notwithstanding the benefit of such mechanism for buffering against different initial conditions between cells (extrinsic noise), 
it remains unclear how exactly it can be successfully implemented under the highly stochastic conditions encountered within each single yeast cell (intrinsic noise);
in particular, the nonlinear trans-phosphorylation at the heart of the scheme could amplify local density inhomogeneities,
and thus introduce additional fluctuations in the gradient profile, especially at low Pom1 abundance or delivery rate.


\subsection{Model}
\begin{figure}[h]
  \centering
  \subfigure[][]{
    \label{fig:Model-Box}
    \includegraphics[width=0.65\textwidth]{\SketchesDir/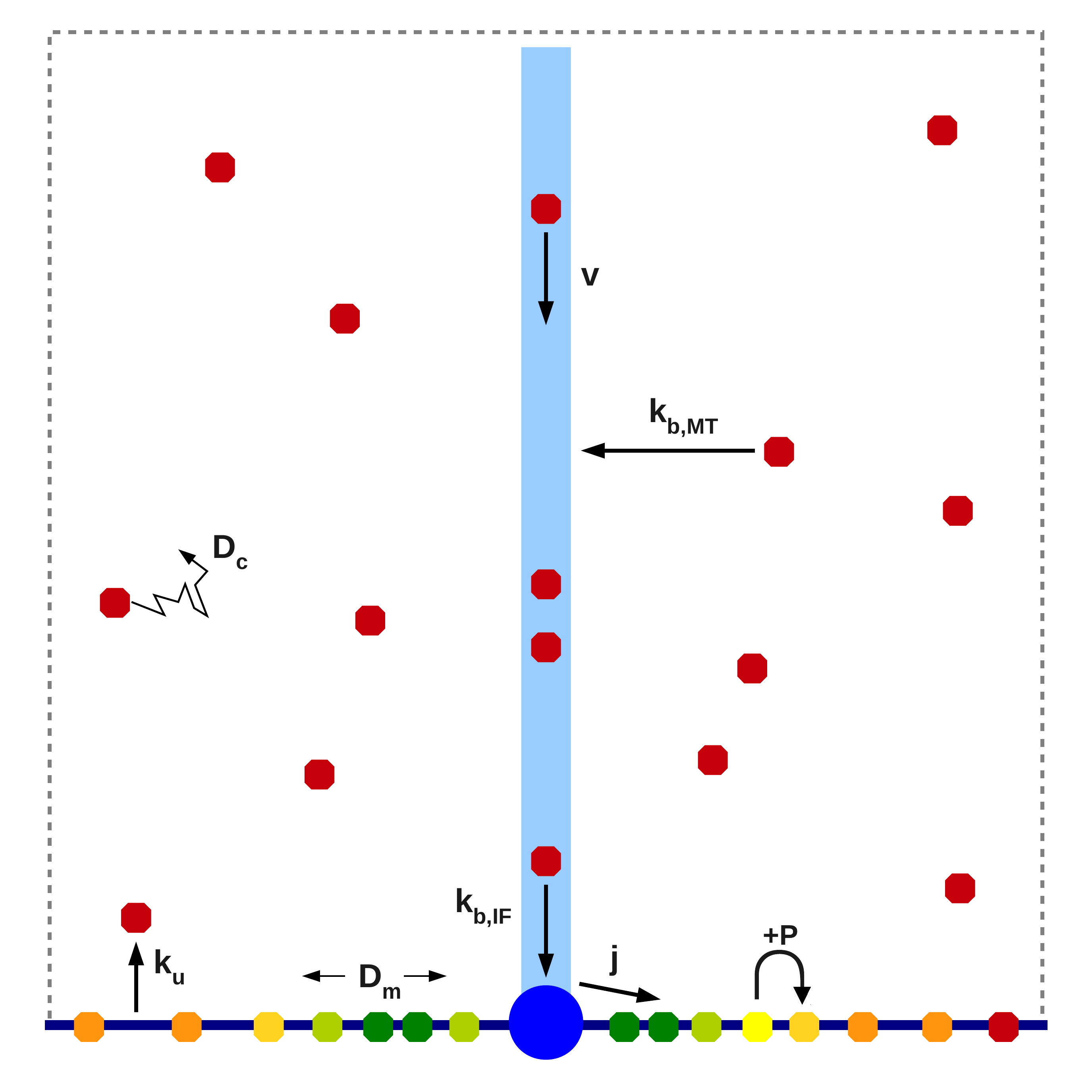}
  }
  \subfigure[][]{
    \label{fig:Model-cis-vs-trans}
    \includegraphics[width=0.3\textwidth]{\SketchesDir/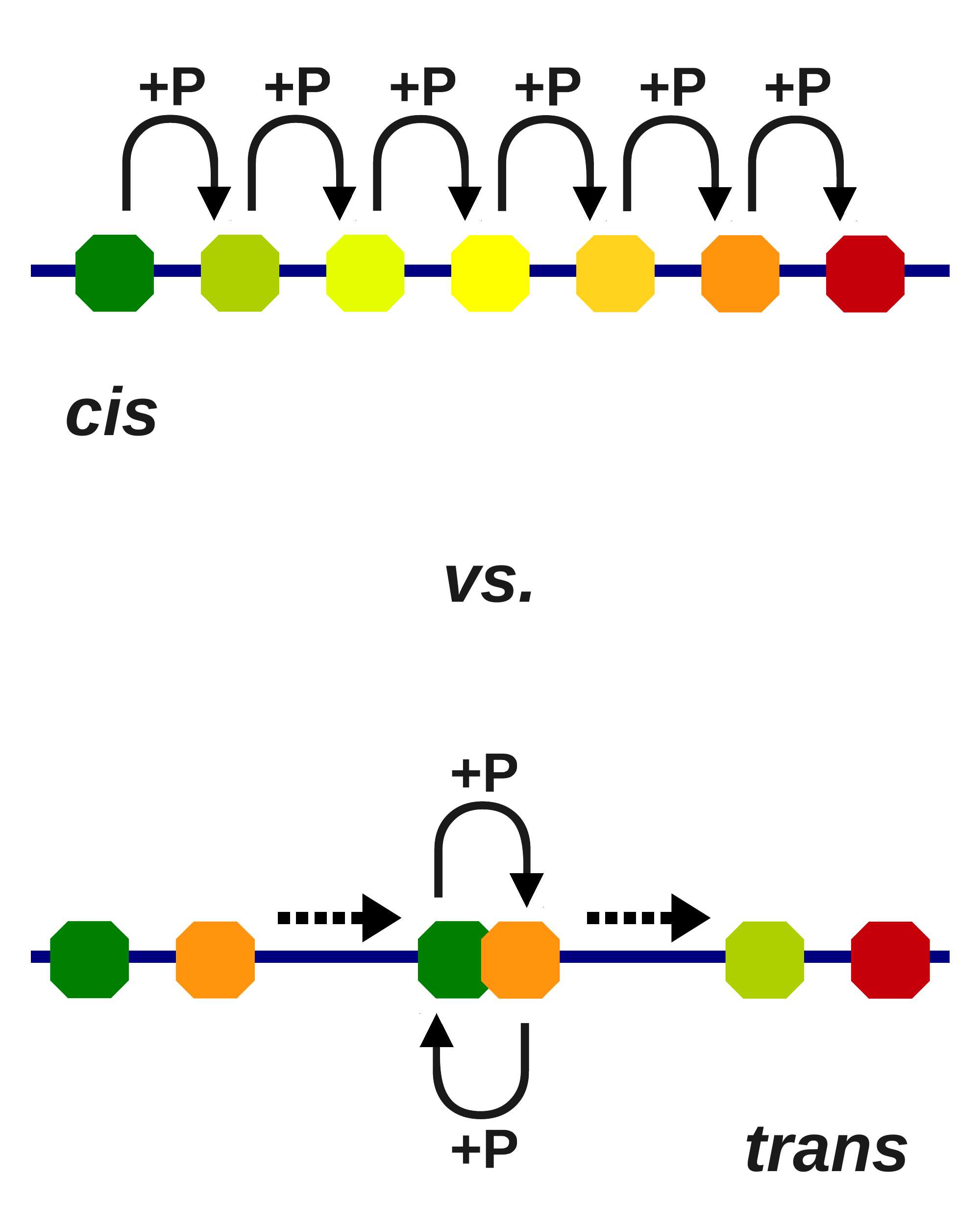}
  }
\caption{  \label{fig:Model}
  \textbf{Pom1 gradient formation model.}
  \subfigref{\subref{fig:Model-Box}}
    Simplified 2D schematic of our Pom1 gradient formation model: A reactive plane (``membrane'', dark-blue line) is placed at the bottom of a 3D simulation box (grey-dotted lines)
    with uniform side length and periodic boundary conditions and connected to a reactive cylinder (``microtubule'', light-blue rectangle).
    Pom1 molecules (hexagons) can exist in 7 different phosphoforms ($n=0$ / unphosphorylated to $n=6$ / 6-fold phosphorylated, indicated by different colors);
    they diffuse in the cytoplasm with diffusion constant $D_{\rm c}$ and bind to the cylinder with diffusion-limited rate \CH{(reaction rate $k_{\rm b,MT}$ at contact is very high)}.
    Cylinder-bound Pom1 undergoes a biased random walk towards the plane with drift velocity $v$ and binds to the cylinder-plane interface with diffusion-limited rate 
    \CH{(very high $k_{\rm b,IF}$ at contact)}.
    The interface can absorb $N\geq1$ Pom1 particles from the cylinder into a ``conversion cluster'' (blue circle)
    assumed to be responsible for Pom1 dephosphorylation.
    From the conversion cluster, single fully dephosphorylated Pom1 particles are unbound directly onto the membrane at random angle with injection rate $j$,
    where they diffuse with diffusion constant $D_{\rm m}$ and undergo successive autophosphorylation, using either the {\it cis} or {\it trans} scheme (explained below).
    The fully phosphorylated Pom1 ($n=6$) can unbind back into the cytoplasm with rate $k_{\rm u}$, thus closing the reaction cycle.
  \subfigref{\subref{fig:Model-cis-vs-trans}} 
    We consider two alternative autophosphorylation schemes: In {\it cis}-autophosphorylation (top) each particle can spontaneously self-phosphorylate 
    towards the respective higher phosphoform with rate $k_{\rm p,n}$. In the {\it trans}-scheme (bottom) two Pom1 particles first have to meet via diffusion
    and form a complex, within which they phosphorylate each other and dissociate again into two separate particles with increased phosphorylation level.
    \CH{For simplicity, here we assume that during this process the phosphorylation level of both particles is always incremented by one,
    and that the complex dissociates instantly afterwards; this combined autophosphorylation-dissociation process is governed by the 
    same zero-order ``in-complex phosphorylation-dissociation rate'' $k_{\rm pt}$ for all combinations of phosphoforms.
    Moreover, we assume that the process of complex formation is diffusion-limited, meaning that the complex formation rate at contact is very high.}
}
\end{figure}

To address this question, we considered a stochastic particle-based model with a simplified geometry 
containing the minimal set of components necessary for Pom1 gradient formation on the membrane at one of the cell poles, shown in Fig.~\ref{fig:Model}.
The unfolded cell cortex is represented by a single (xy-) plane.
A single static cylinder orthogonal to the plane (i.e. pointing in z-direction), representing a microtubule, intersects with the plane at its center.
\TODO{While this minimalistic geometry neglects membrane curvature and represents the cytoplasmic volume in a very idealized fashion,
it is representative of the real system if cytoplasmic diffusion is significantly faster than the other timescales involved;
then the cytoplasmic concentrations quickly become well-mixed, rendering the precise distance function between membrane and microtubule less important.
For the Pom1 system this assumption is likely to hold \cite{Saunders2012}.}

As a further simplification, we do not explicitly include the Tea4/Dis2 dimers responsible for Pom1 dephosphorylation.
Instead, we assume a static ``conversion cluster'' at the interface between membrane and microtubule,
represented by a single static particle with a large radius. 
Dephosphorylation of Pom1 via Tea4/Dis2 is assumed to occur 
with Poissonian statistics for the complete conversion from the fully phosphorylated to the fully dephosphorylated state of Pom1.

Cytoplasmic Pom1 particles bind to the microtubule with a diffusion-limited rate.
The microtubule-bound Pom1 then are actively transported towards the microtubule-membrane interface.
There they are absorbed to the conversion cluster and unbind from it directly onto the membrane upon (complete) dephosphorylation.
\CH{Membrane-bound Pom1 can autophosphorylate 6 times, as determined in experiments \cite{Hachet2011},
thereby decreasing its affinity to the membrane;
since it is unknown how exactly the Pom1 membrane unbinding rate increases with progressing phosphorylation level $n$,
we decided for the arguably simplest choice, setting the unbinding rate to zero for $n\leq5$ and to a finite value $k_{\rm u}>0$ for $n=6$.}
We studied both cis- and trans-authophosphorylation, but mainly focused on the latter.
Trans-phosphorylation is assumed to occur in a distributive fashion;
however, in our simulations we find that fast rebindings tend to convert the distributive to a pseudo-processive scheme, as observed earlier in other systems \cite{Takahashi2010}.
\CH{While cis-autophosphorylation is a simple zero-order reaction process, the process of trans-autophosphorylation consists of three steps: 
first, the two involved Pom1 particles have to encounter each other via membrane diffusion and react to form a complex; 
secondly, the two particles bound in complex can carry out the actual phosphorylation of each other;
finally, the complex dissociates into two monomers again.
In our model, we make two simplifying assumptions:
(1.) We assume that the rate of forming the complex is very high, such that the first (encounter) step is diffusion-limited;
(2.) we treat the second and third steps as one process with Poissonian statistics described by a single ``in-complex phosphorylation-dissociation rate'' $k_{\rm pt}$,
assuming that the complex instantly dissociates after both monomers have increased their phosphorylation level by one;
$\tau_{\rm pt} \equiv 1/k_{\rm pt}$ thus is the average waiting time between complex formation and dissociation into monomers with (one-fold) incremented phospholevels.
}

We used experimentally determined parameters if available, and reasonable estimates otherwise. 
Details of parameter choice are described in sec.~\ref{sec:Pom1-Pars} and Table~\ref{tab:Pom1-Pars} in the \SI.

\FloatBarrier

\begin{figure}[t!]
  \centering
  \subfigure[][]{
    \label{fig:ExampleGradient-3D}
    \includegraphics[width=\GradientPlotWidth]{\PlotsDir/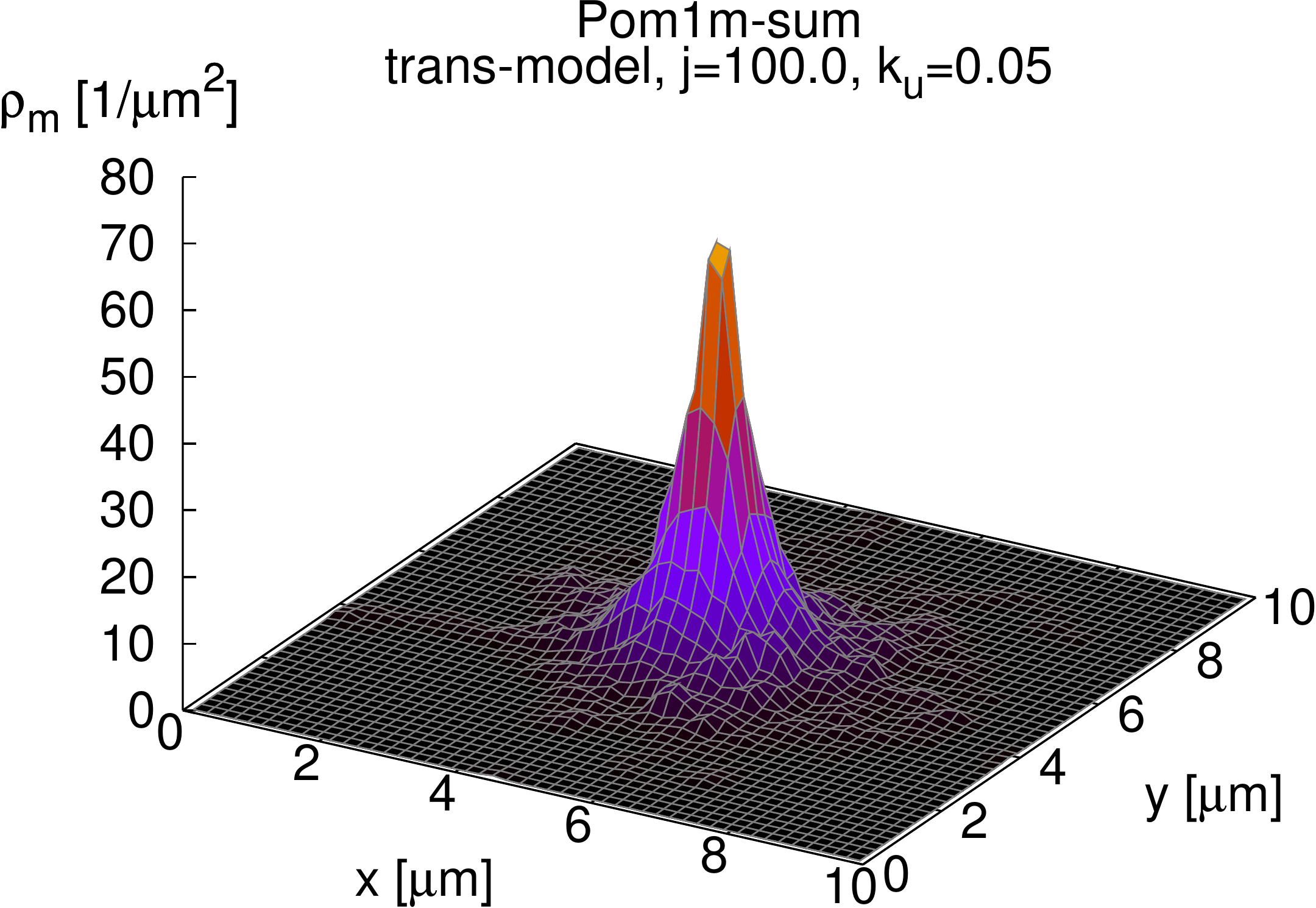}
  }
  \subfigure[][]{
    \label{fig:ExampleGradient-Section}
    \includegraphics[width=\GradientPlotWidth]{\PlotsDir/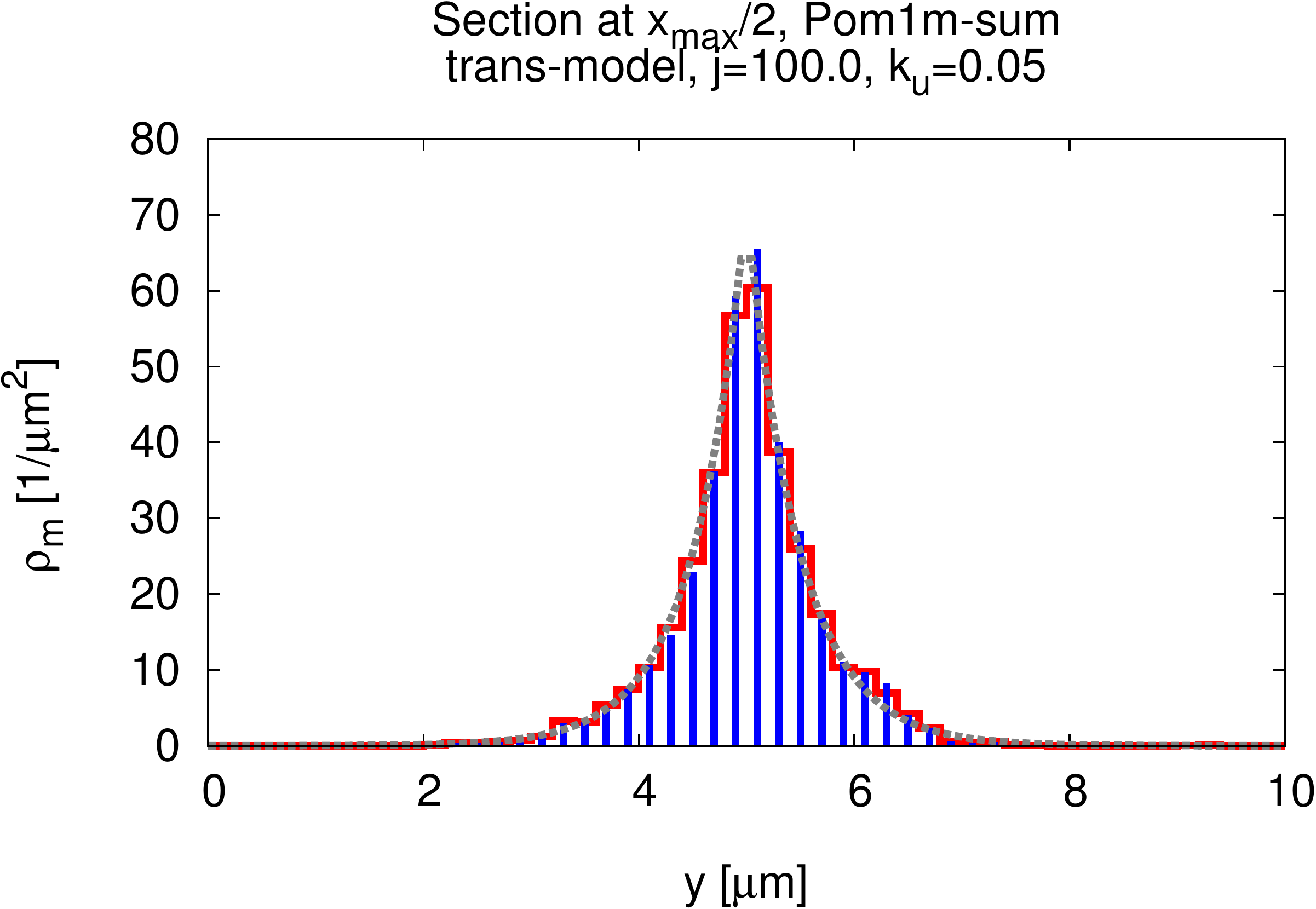}
  }\\
  \vspace{1EM}
  \subfigure[][]{
    \label{fig:ExampleGradient-RadialFit}
    \includegraphics[width=\GradientPlotWidth]{\PlotsDir/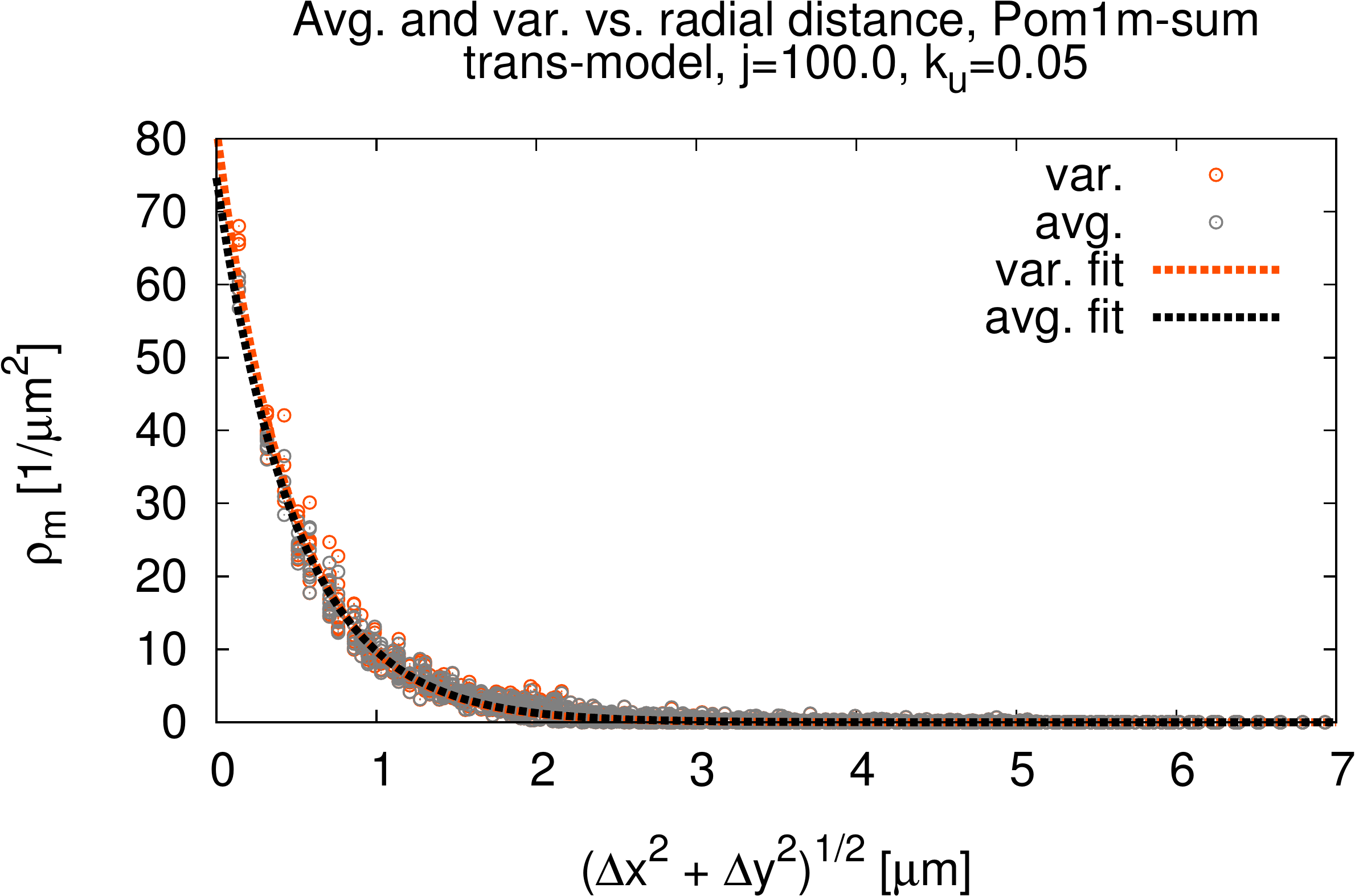}
  }
  \parbox{\GradientPlotWidth}{~~~}
\caption{  \label{fig:ExampleGradient}
  \textbf{Example gradient (trans-model).}
  \subfigref{\subref{fig:ExampleGradient-3D}}
    Histogram of total membrane-bound Pom1 density (comprising all phosphorylation states) on the membrane plane.
  \subfigref{\subref{fig:ExampleGradient-Section}} 
    Section through histogram in (\subref{fig:ExampleGradient-3D}) at $x=5~\mu m$; red steps show the local density, blue impulses its variance, grey lines exponential fits to the data.    
  \subfigref{\subref{fig:ExampleGradient-RadialFit}} 
    The same data plotted against the radial distance from the gradient origin, with exponential fits to the local mean density and its variance.
    Note that at all positions the variance essentially is in the Poisson limit.        
   \TODO{4th PANEL: SIM. SNAPSHOT}
}
\end{figure} 

\subsection{Simulation results}
We performed particle-based stochastic simulations of the model, using \eGFRD extended by the new functionalities introduced in section~\ref{sec:Methods}.
In order to be able to observe the buffering effect we varied---at constant particle number---the key parameters affecting the gradient amplitude
and length scale: $j$, the ``membrane injection rate'' (rate of unbinding from the microtubule-membrane interface towards the membrane),
and $k_{\rm u}$, the rate of unbinding of the fully phosphorylated Pom1.
Starting from random intial conditions with cytoplasmic, unphosphorylated Pom1 only, for each parameter set we first propagated the system
for (at least) $30~s$ of simulated time, in order to allow it to form and equilibrate the gradient;
we then recorded particle positions and species with a fixed data acquisition interval ($0.1~s$) for (at least) $30~s$ more of simulated time.
Subsequently, we binned the local number of membrane-bound Pom1 species in a 2D histogram and thus obtained the stationary gradient profiles
and its local variance, which were further characterized by fitting exponential functions to the data.
The details of the protocol are explained in sec.~\ref{sec:Pom1-sim-protocol} of the \SI.

In Fig.~\ref{fig:ExampleGradient} we show typical output from our simulations for a representative set of parameters (trans-model, $j=100/s$, $k_{\rm u}=0.05/s$).
Fig.~\ref{fig:ExampleGradient}\subref{fig:ExampleGradient-3D} displays a 2D histogram of the total (membrane-bound) Pom1m density 
(summing all phosphoforms) on the plane, Fig.~\ref{fig:ExampleGradient}\subref{fig:ExampleGradient-Section} is a $y$-section
through the 3D profile at $x=5~\mu m$; the variance of the local mean density in each histogram bin is plotted using blue impulses.
In Fig.~\ref{fig:ExampleGradient}\subref{fig:ExampleGradient-RadialFit} we present the same data as a function of the radial distance 
from the gradient origin (at $x=5~\mu m$, $y=5~\mu m$) in a scatter plot (local mean density = grey circles, local density variance = orange circles)
and the exponential functions fitted to both the density and variance data.
Figs.~\ref{fig:ExampleGradient}\subref{fig:ExampleGradient-Section} and \ref{fig:ExampleGradient}\subref{fig:ExampleGradient-RadialFit}
demonstrate that (1.) the shape of the stationary gradient profile is well approximated by an exponential function
\CH{(as already observed in \cite{Hersch2015}),} \TODO{CROSS-CHECK FOR PRECISE ARGUMENT}
and (2.) that the local density fluctuations are essentially in the Poissonian limit (variance $\simeq$ mean).
These findings are representative for all parameter sets that we present in the following.

\begin{figure}[h!]
  \centering
  \subfigure[][]{
    \label{fig:trans-Gradients-Ampl}
    \includegraphics[width=\GradientPlotWidth]{\PlotsDir/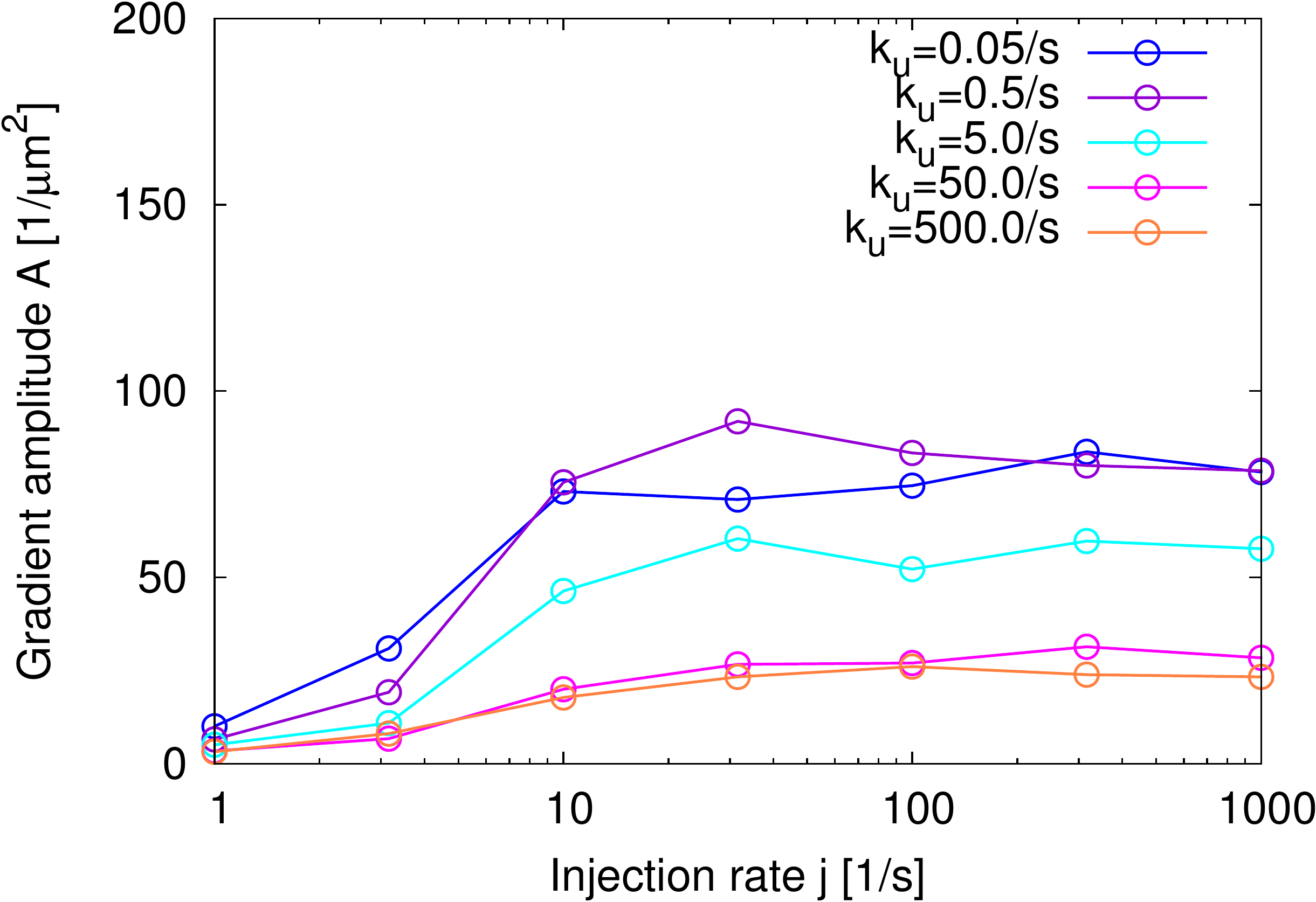}
  }
  \subfigure[][]{
    \label{fig:trans-Gradients-Length}
    \includegraphics[width=\GradientPlotWidth]{\PlotsDir/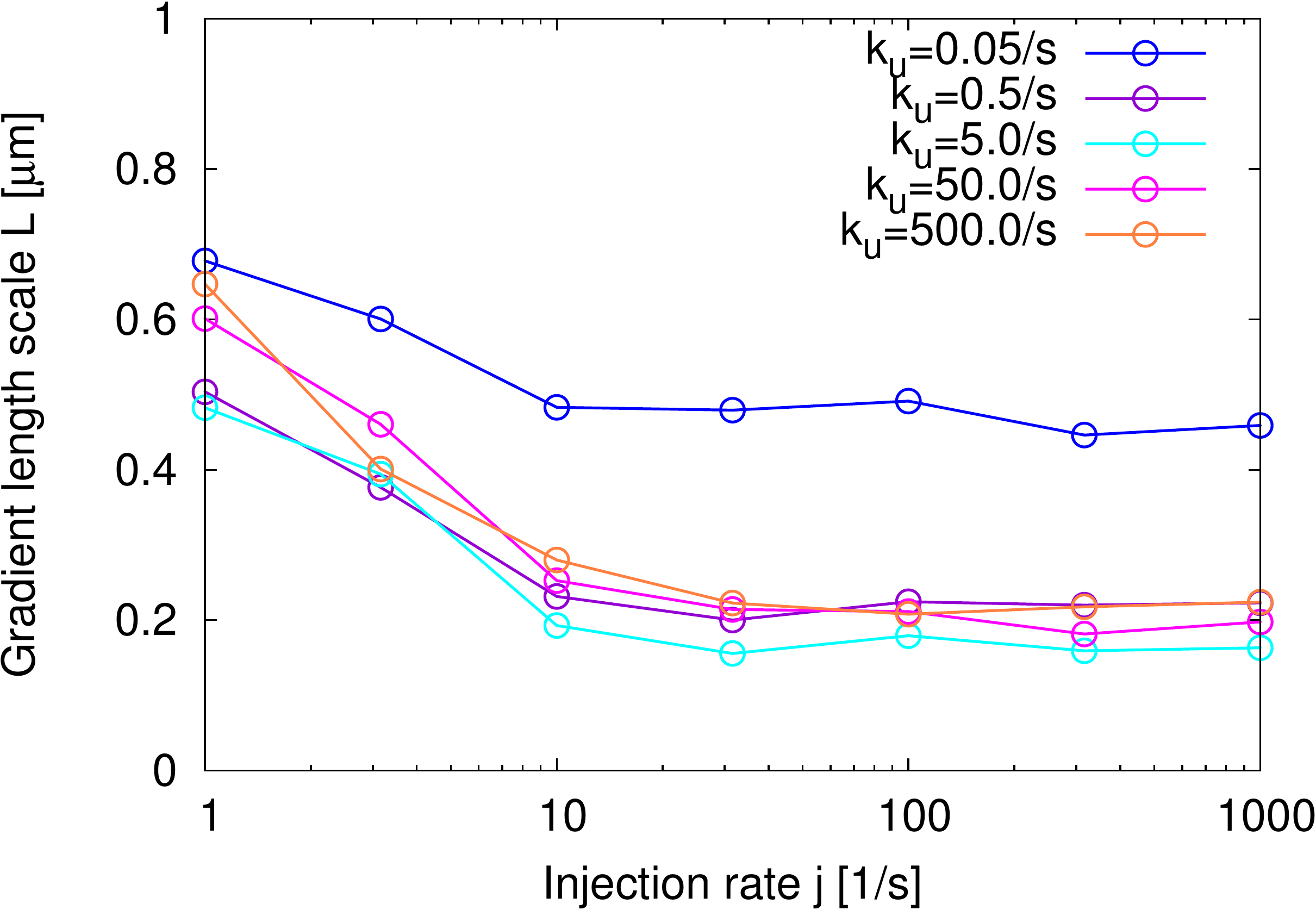}
  }\\  
  \vspace{1EM}
  \subfigure[][]{
    \label{fig:trans-Gradients-Corr}
    \includegraphics[width=\GradientPlotWidth]{\PlotsDir/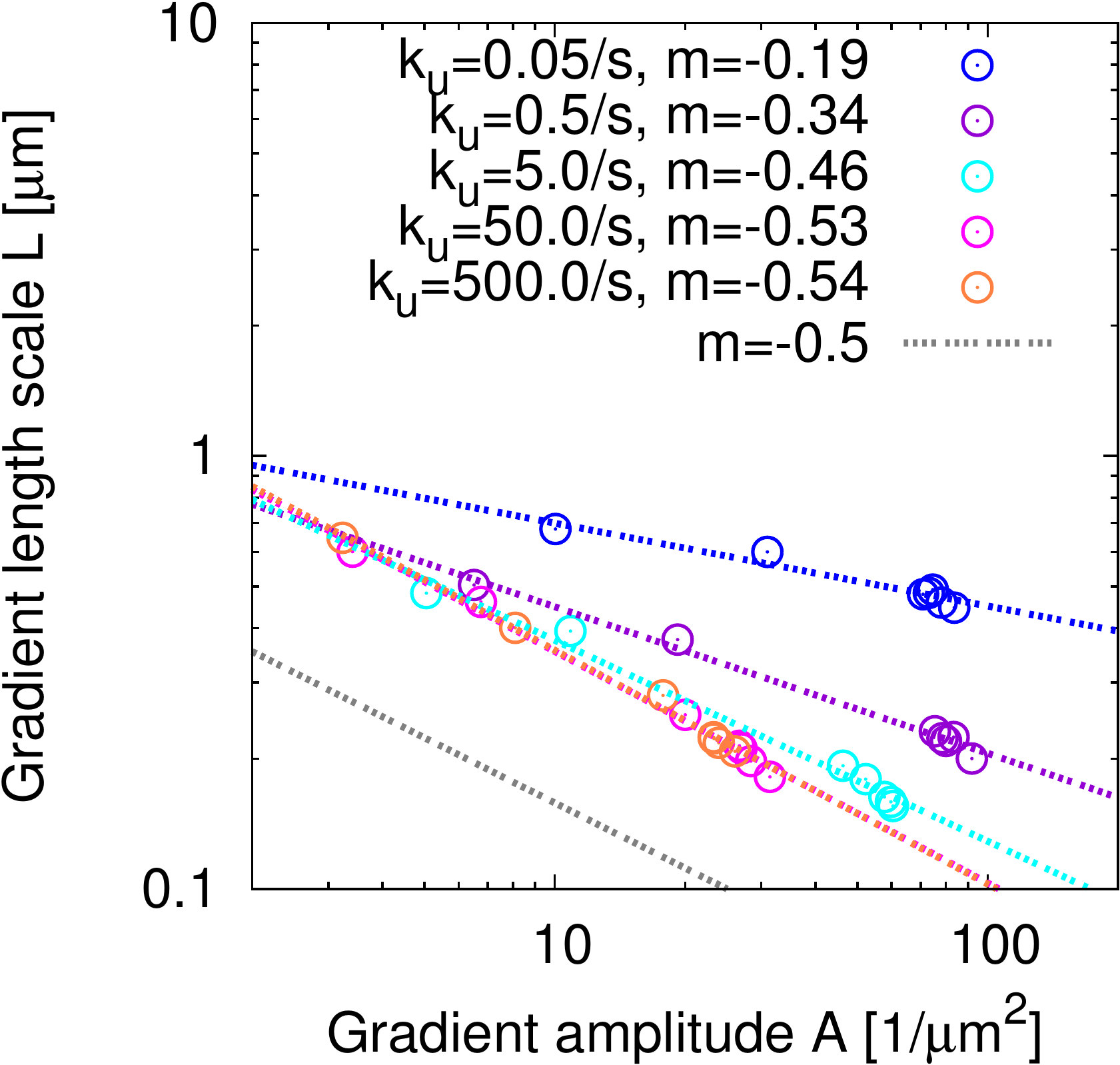}
  }
    \subfigure[][]{
    \label{fig:trans-Gradients-Noise}
    \includegraphics[width=\GradientPlotWidth]{\PlotsDir/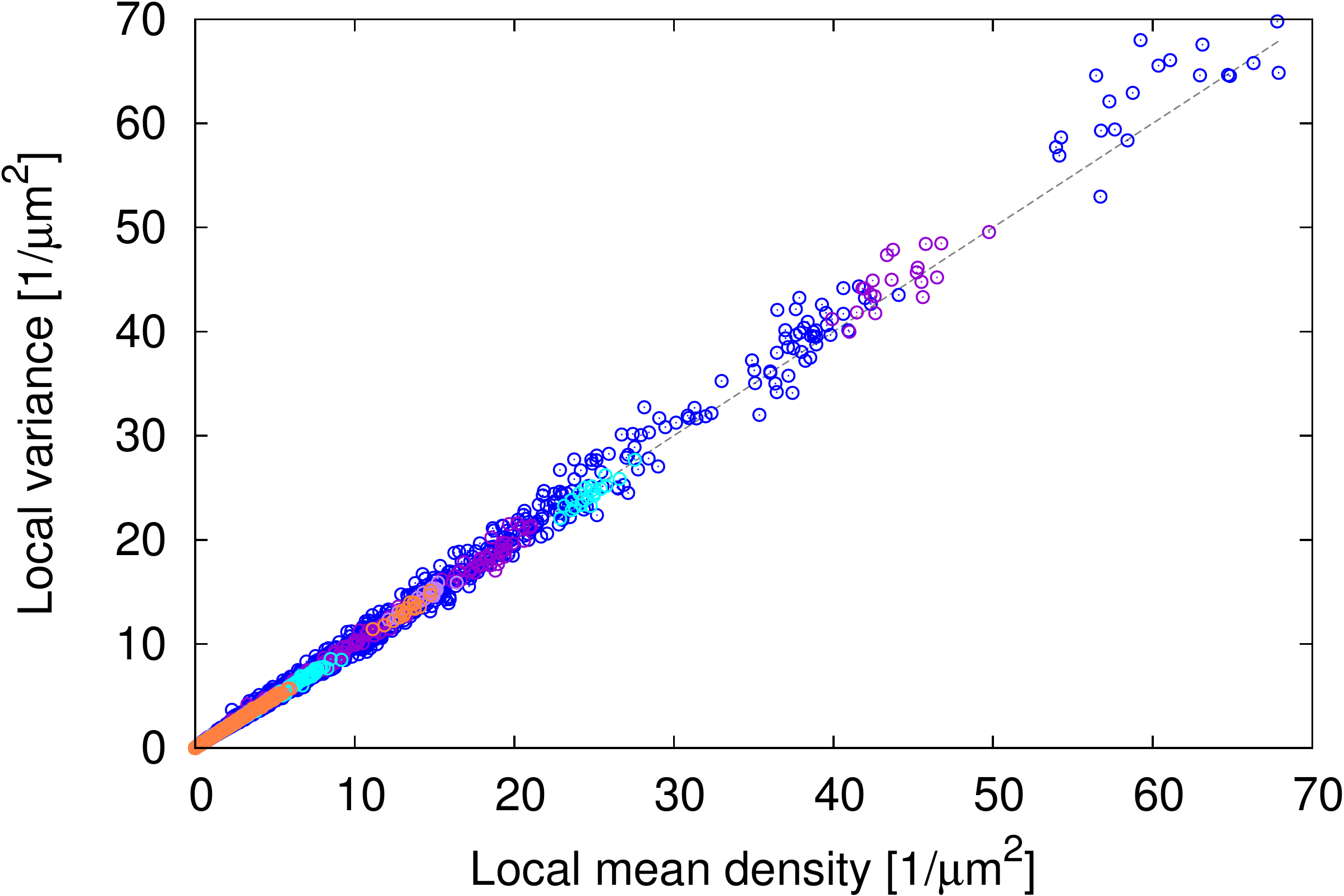}
  }  
\caption{  \label{fig:trans-Gradients}
  \textbf{Gradient characteristics as a function of key parameters (trans-model).}
  Gradient amplitude \subfigref{\subref{fig:trans-Gradients-Ampl}} and length scale $L$ \subfigref{\subref{fig:trans-Gradients-Length}},
  obtained by fitting an exponential function to stationary radial profiles (cf.~Fig.~\ref{fig:ExampleGradient}\subref{fig:ExampleGradient-RadialFit}),
  plotted against the injection rate $j$ at the gradient origin for different unbinding rates $k_{\rm u}$ of the fully phosphorylated Pom1.
  Lines are guidelines for the eye.
  Note that $L$ becomes shorter for increasing $j$ due to \CH{increased concentration at the membrane}, resulting in higher probability
  of autophosphorylation and faster unbinding to the cytoplasm.
  \subfigref{\subref{fig:trans-Gradients-Corr}} Log-log plot of the gradient length $L$ vs. gradient amplitude $A$ for the same data.
  The grey line shows the theoretically predicted \cite{Hersch2015} scaling (slope $m=-0.5$), only reached in the limit of fast unbinding 
  of fully phosphorylated Pom1.
  \subfigref{\subref{fig:trans-Gradients-Noise}} Variance vs. mean for the local total membrane-bound Pom1 density,
  combining the values of all histogram bins into one scatter plot;
  point colors correspond to the colors in \subfigref{\subref{fig:trans-Gradients-Ampl}} - \subfigref{\subref{fig:trans-Gradients-Corr}} (different values of $k_{\rm u}$).
  The data shown was obtained in the limit of very fast \CH{in-complex phosphorylation-dissociation rate} ($k_{\rm pt}=1000/s$),
  such that the overall rate of autophosphorylation is (membrane-) diffusion limited.
}
\end{figure} 

Fig.~\ref{fig:trans-Gradients} shows a more systematic analysis of the key gradient properties, comparing different parameter sets.
In Fig.~\ref{fig:trans-Gradients}\subref{fig:trans-Gradients-Ampl} and Fig.~\ref{fig:trans-Gradients}\subref{fig:trans-Gradients-Length} we plot the fitted gradient amplitude $A$
and length scale $L$, respectively, as a function of the injection rate $j$.
Not surprisingly, the amplitude $A$ is an increasing function of the Pom1 injection rate at the gradient origin, but saturates for $j\gtrsim30/s$
when the overall abundance of Pom1 in the system and the flux of microtubule-bound Pom1 towards the membrane become the limiting factors;
this holds for all considered unbinding rates $k_{\rm u}$, but the maximal amplitude slightly reduces with increasing $k_{\rm u}$.
In contrast, the gradient length scale decreases with increasing Pom1 injection at the origin;
this is precisely the effect of enhancing the effective unbinding rate and thus shortening the gradient via 
trans-autophosphorylation and its nonlinear dependence on the Pom1 concentration.
For the faster unbinding rates $k_{\rm u}\geq 0.5/s$ the gradient length scale $L$ is essentially independent of $k_{\rm u}$,
because in this regime only the average time required for full phosphorylation matters;
remarkably, $L$ becomes insensitive to further increases of $k_{\rm u}$ much earlier than the amplitude $A$.

In Fig.~\ref{fig:trans-Gradients}\subref{fig:trans-Gradients-Corr} we plot the fitted length scale against the fitted amplitude on a double-logarithmic scale.
The figure shows that gradient length and amplitude indeed are anti-correlated, 
but the degree of anticorrelation depends on the unbinding rate, as demonstrated by the fitted slopes:
for slow unbinding ($k_{\rm u}=0.05/s$), the amplitude decreases less strongly with increasing length scale (slope $m=-0.19$),
while the theoretically predicted \cite{Hersch2015} slope $m=-0.5$ (grey line) is only approached for the faster unbinding rates ($k_{\rm u}\geq 5/s$).

Fig.~\ref{fig:trans-Gradients}\subref{fig:trans-Gradients-Noise} summerizes the noise properties of the gradients.
Here we plot the local variance (density) against the local mean density in a scatter plot,
whereby each point corresponds to a single bin of a two-dimensional histogram as shown in Fig.~\ref{fig:ExampleGradient-3D},
for all parameter sets considered.
The colors correspond to the ones used in Figs.~\ref{fig:trans-Gradients}\subref{fig:trans-Gradients-Ampl}-\ref{fig:trans-Gradients}\subref{fig:trans-Gradients-Corr}
and distinguish points from parameter sets with a certain value of the unbinding rate $k_{\rm u}$;
the point clouds of one particular color thus include values for different positions on the plane and different injection rates $j$.
The plot shows that the gradient profiles for all parameter values considered are, in essence, in the Poissonian limit (variance = mean, grey line in the plot),
apart from a small deviation towards slightly higher variances at the highest Pom1 densities, corresponding to low unbinding rate ($k_{\rm u}=0.05/s$) 
and positions close to the gradient origin, where unphosphorylated Pom1 is injected onto the membrane.

\clearpage
\subsection{Influence of the trans-phosphorylation rate}
In order to test how much our results depend on our choice of very fast \CH{in-complex phosphorylation-dissociation rate} ($k_{\rm pt}=1000/s$),
we varied this parameter at a constant, intermediately fast unbinding rate $k_{\rm u}=5/s$.
Here we only summarize the principal observations and comment on them in more detail in sec.~\ref{sec:trans-slowerPhos} of the \SI.
We observe that while the gradient amplitude markedly increases with increasing rate $k_{\rm pt}$, 
the scaling of the anticorrelation between gradient length and amplitude is preserved as long as $k_{\rm pt}\gtrsim 10/s$. 
Moreover, for all values of $k_{\rm pt}$, the noise remains Poissonian.
The buffering effect observed in the trans-model therefore turns out to be very robust with respect to \CH{$k_{\rm pt}$, the rate governing autophosphorylation in complex and subsequent dissociation}.

\subsection{\CH{The cis-phosphorylation model}}
In order to compare the trans- and cis-autophosphorylation schemes, we repeated the simulations above also for the cis-scheme.
The results are presented in Fig.~\ref{fig:cis-Gradients} in the \SI and only briefly summarized here.
In accordance to previous results and our initial expectations, the cis-scheme does not provide any buffering mechanism
that compensates increases in amplitude with decreases in length scale.
However, both the amplitude and length scale are remarkably invariant to most of the relevant system parameters,
in particular to the membrane unbinding rate $k_u$, highlighting that the successive Poissonian phosphorylation steps
of the cis-scheme effectively provide a ``timer'' function for the unbinding from the membrane.
\TODO{DISCUSS REGIME OF $k_u$}

\section{Conclusion and Outlook}
Due to limited tractability of analytical models, and thanks to advances both in biophysical theory and computational power during the last decades, 
spatial-stochastic simulations have become an important tool for exploring the mechanistic behavior of complex biochemical systems.
Driven by the continuous desire to make simulations more detailed and realistic, and by the recognition of the fact that in biological cells a myriad of chemical species coexists at predominantly low copy numbers, particle-based simulation techniques, mostly based on the principle of Brownian Dynamics, recently have taken a prominent role among these efforts.
While substatial achievements have been made in making Brownian Dynamics more realistic and biochemically accurate,
even with the recent advances they remain computationally demanding;
with time steps typically in the $\unit{\mu s}$ regime and below, the run time required to simulate $1~s$ of real time for a system of $\sim 1000$ particles can easily reach the order of $\sim 10^6~s$ ($\sim 10~d$).

To overcome these limitations, the community has recently begun to pursue two distinct approaches: the parallelization of particle-based simulation algorithms, 
in order to allow them to exploit the full power of large CPU or GPU clusters \CH{\cite{Miyauchi2016,Drawert2016,Dematte2010,Gladkov2011,Dematte2012}}, 
and the development of hybrid techniques that shuffle particles between simulators with coarser or finer spatial resolution, depending on the local density \CH{\cite{Erban2013,Franz2013,Flegg2014,Flegg2015,Robinson2014,Robinson2015}}.
\eGFRD puts forward a different approach.
The key idea of \eGFRD is to get rid of ``unnecessary'' detail, namely the microscopic trajectories of diffusive motion between particle encounters,
while retaining all other, ``informative'' details, without sacrificing access to individual particle-positions at any time, and without compromising accuracy.
To this end, \eGFRD partitions the original $N$-particle system into $M<N$ smaller, analytically tractable systems for which an exact solution (probability distribution) 
for the time evolution of the underlying stochastic dynamics can be obtained.
This allows to implement an event-driven algorithm in which the subsystems are updated locally and asynchronously,
while the overall behavior of the full system is correctly preserved,
and individual particle positions can be sampled with exact statistics at any desired time.

While this approach endows \eGFRD with an extraordinarily high computational efficiency, reducing the CPU time per simulated time up to $10^6$-fold compared to brute-force Brownian Dynamics,
until now \eGFRD has been limited to a classical three-dimenional setting, such that an adequate representation of the intracellular architecture, oftentimes featuring lower-dimensional structures like membranes and elongated macropolymers, was not possible.
In this work, we presented \CH{\eGFRDii, a new version of \eGFRD which incorporates finite low-dimensional reactive structures,}
and derived the Green's functions describing the reaction-diffusion processes of particles on these structures, and their interactions with them.
Using these exact analytical solutions, we implemented a diverse set of new protective domains that seamlessly integrates with the spherical protective domains of the original \eGFRD, 
and supplemented the new \eGFRD with an efficient and accurate Brownian Dynamics scheme (rvm-BD) capable of propagating particles in all dimensions with situation-adaptive time steps.
A benchmarking carried out for biologically representative parameters reveals that \CH{the \eGFRDii simulator}---for sufficiently low particle densities---is up to 3 orders of magnitude more efficient than rvm-BD, and even more so compared to more naive BD schemes; however, in the lower dimensions the crossover point at which BD becomes more efficient than \eGFRD is approached faster, because the effects of crowding build up more quickly as the dimension decreases.
While our framework is very sophisticated and efficient in computing next-event times and new particle positions,
the methods used for domain (re)creation are still comparably simplistic; yet, these processes make up a considerable amount of the computational effort required, especially when the simulation space gets more crowded. Using more sophisticated domain making schemes therefore is expected to further improve the performance of \eGFRDii.
Moreover, we find that the use of hybrid code (C++ for core routines, Python for more upstream routines and the user interface) comes at the cost of sacrificing efficiency.
Forthcoming releases of our simulator will reduce the amount of Python code to a minimum required for user-friendliness, 
and is expected to result in significant further increases in simulation efficiency.
\CH{In fact, as a first step we have recently rewritten the \eGFRD code for simulating 3D systems, such that all routines are now in C++. 
This code is up to 6 orders of magnitude faster than brute-force BD\footnote{See: \texttt{gfrd.org}}.}

As an example application, we used the \CH{new \eGFRDii framework} to carry out particle-based simulations of Pom1 gradient formation; this process is driven by a reaction cycle in which fully phosphorylated Pom1 is collected from the cytoplasm by a microtubule, afterwards directed towards the membrane via active transport, and injected onto the membrane in its fully dephosphorylated state; on the membrane, the gradient then is shaped by diffusion and a multi-step phosphorylation cascade tuning the Pom1 unbinding rate.
Comparing a trans-autophosphorylation mechanism to cis-autophosphorylation, we varied the crucial parameters of a minimal model capturing all essential processes involved in Pom1 gradient formation, and recorded how they affect the stationary gradient profiles. Our results confirm the buffering effect arising from an anticorrelation between the gradient amplitude and length scale in the trans-phosphorylation model, found by earlier studies \CH{\cite{Hachet2011,Hersch2015}}; in addition, we find that even at low copy numbers the fluctuations in the gradient concentration are Poissonian at any distance from the source.
Our results also suggest an important role for the trans-phosphorylation rate: the predicted scaling is only achieved for \CH{sufficiently fast trans-phosphorylation and complex dissociation}.
Finally, while---as expected---no similar buffering effect can be observed in the cis-phosphorylation model, 
we find that here the emerging gradients are remarkably insensitive to most of the varied parameters,
highlighting diametral benefits of the two different authophosphorylation schemes (buffering or ``elasticity'' in the trans-model vs. insusceptibility in the cis-model).

Our ongoing efforts pursue three directions:
(1.) Since the tractability of Green's function derivations demands working with simple, abstracted geometries, the level of detail with which real cell environments can be represented in \eGFRD remains limited. However, since \eGFRD necessarily must be integrated with a BD fallback simulator, this offers the opportunity to resort to BD simulations on triangulated structures if desired. We currently work on integrating particle motion and reactions on triangulated meshes with \eGFRD.
(2.) Also the fact that all particles are treated as spheres with uniform surface reactivity limits the level of detail in \eGFRD. In reality, large molecules in particular have a markedly non-spherical structure with well-defined reactive spots, and upon coming close first have to engage in rotational diffusion within interaction potentials before being able to form a complex. In order to equip GFRD with the capability of resolving the reaction process with such detail, recently some of us developed MD-GFRD (``Molecular Dynamics GFRD'') \cite{Vijaykumar2015,Vijaykumar2017}, which allows to propagate spherical particles with well-defined reactive patches via \person{Langevin} dynamics once they come close together, with the option to switch to molecular dynamics at even closer particle proximity if desired. This principle will be fully integrated into \eGFRD in the future.
(3.) While the parallelization of event-driven spatial simulators is a daunting task, because different parts of the simulated space may quickly desynchronize, it has been recently achieved for a simple (3d only) \eGFRD variant, named pGFRD, as part of the e-cell project \cite{Kaizu2017}. Future versions of our \eGFRD implementation will borrow from the techniques developed in pGFRD.
Until these further extensions of \eGFRD are fully elaborated, the framework presented here provides the community with an efficient quantitative tool for studying the behavior of biochemical systems in geometries representing the essential architecture of real cells, perfectly suited to explore the impact of principal geometric constraints on biochemical reactions at low concentrations in a particle-based, genuinely stochastic setting.

\section{Acknowlegdements}
\CH{This work is part of the research programme of the Netherlands Organisation for Scientific Research (NWO) and was performed at the research institute AMOLF.}

\CH{The authors thank Filipe Tostevin, Wiet H. de Ronde, Thomas E. Ouldridge, Andrew Mugler and Sorin T\u{a}nase-Nicola for many useful discussions about the principles that \eGFRD is based on.}

\newpage
\bibliographystyle{styleTomek-v2}
\bibliography{General,GFRD,Intro,StochSim,SpatialEffects,SingleMolecule,ActiveTransport,Polarization,TF-Sliding,GapGenes}


\setcounter{page}{0}
\setcounter{section}{0}
\setcounter{figure}{0}
\setcounter{table}{0}
\setcounter{equation}{0}

\renewcommand{\thesection}{S\arabic{section}}
\renewcommand{\thefigure}{S\arabic{figure}}
\renewcommand{\thetable}{S\arabic{table}}
\renewcommand{\theequation}{S-\arabic{equation}}

\clearpage
\pagestyle{empty}
~\\
\vfill
\begin{center}
\begin{Huge}
\textsc{Supplementary Information}
\end{Huge}
\end{center}
\vfill

\clearpage
\pagestyle{plain}

\section{Algorithmic details}
\subsection{The eGFRD algorithm}
The box entitled ``Algorithm~\ref{alg:eGFRD}'' shows a pseudo-code listing of the basic \eGFRD algorithm.
Note that for brevity and clarity details of event (time) sampling and domain construction have been omitted here.
A similar pseudo-code listing for the domain and shell construction procedure used in our new \eGFRD implementation
is found in box ``Algorithm~\ref{alg:Shellmaking}'' in Section~\ref{sec:GFRD-shellmaking}.

\begin{algorithm}
\begin{algorithmic}[p!]
 \vspace{1EM}
 \State Initialize:
 \State $t_{\rm sim} \gets 0$, scheduler $S \gets \lbrace\rbrace$
 \ForAll{ particles $p_i$ }
      \If{ \NOT $p_i$ already in domain }
	\State $\mathcal{D}_j \gets$ create domain for $p_i$
	\State $\tau_j \gets$ draw next-event time for $\mathcal{D}_j$
	\State insert $\tau_j$ into $S$ ordered by increasing time
      \EndIf
 \EndFor
 \State
 \State Main loop:
 \While{ $S \neq \lbrace\rbrace$ \AND $t_{\rm sim} < t_{\rm end}$}
    \State $t_{\rm sim} \gets \tau_n =$ topmost element in $S$
    \State remove $\tau_n$ from $S$
    \State propagate $\mathcal{D}_n$ to $\tau_n$ and remove $\mathcal{D}_n$
    \State reset particle update list: $U \gets \lbrace\rbrace$
    \State $U \gets U \cup \lbrace p_{n_i} \rbrace$ \textbf{for all} particles $p_{n_i} \in \mathcal{D}_n$
    \While{ $U \neq \lbrace\rbrace$ }
	\State $p_u \gets$ next particle in $U$
	\ForAll{domains $\mathcal{D}_{u_j}$ close to $p_u$ }
	    \State burst: propagate $\mathcal{D}_{u_j}$ to $\tau_n$ and remove $\mathcal{D}_{u_j}$
	    \State remove $\tau_{u_j}$ from $S$	    
	    \State $U \gets U \cup \lbrace p_{u_{jk}} \rbrace$ \textbf{for all} particles $p_{u_{jk}} \in \mathcal{D}_{u_j}$
	\EndFor
    \EndWhile
    \ForAll{ $p_u \in U$ }
	\If{ \NOT $p_u$ already in domain }
	    \State $\mathcal{D}_u \gets$ create domain for $p_u$
	    \State $\tau_u \gets$ draw next-event time for $\mathcal{D}_u$
	    \State insert $\tau_u$ into $S$
	\EndIf
    \EndFor    
 \EndWhile
 \vspace{1EM}
\end{algorithmic}
\caption{
Basic outline of the \eGFRD algorithm.
Symbols $\mathcal{D}_x$ denote domains, $\tau_x$ next-event times.
The scheduler $S$ is the list of all next-event times in the system, ordered by increasing time.
List $U$ collects all particles that have been updated at a given time $\tau_x$ and require construction of a new domain.
$t_{\rm sim}$ is the time that passed since simulation start.
\label{alg:eGFRD}
}
\end{algorithm}

\subsection{Sampling of next-event information}
\label{sec:GFRD-General-Sampling-Procedures}
In the main text, section~\ref{sec:eGFRD-Principle}, we introduce the \domaintype{Single} and \domaintype{Pair} domains featured throughout eGFRD.
Here we describe in more detail how next-event times and types can be sampled from the analytical probability distributions
linked to the domains.

\subsubsection{Single domains}
\label{sec:GFRD-Singles}
Two types of next-events can happen within a \domaintype{Single} domain $\mathcal{D}_1$:
either the particle undergoes a monomolecular reaction, which can mean
decay or species change, or it hits the boundary of the domain by diffusion.
In \eGFRD, the former is called a \textit{Single Reaction}, the latter
a \textit{Single Escape}.
\index{monomolecular reaction} \index{Single domain} \index{Single} \index{Single Reaction} \index{Single Escape}

If we assume that monomolecular reactions are a Poissonian process independent
of particle diffusion, the propensity function for the next reaction
is simply an exponential distribution
\begin{align}
 q_m(t) = k_m e^{-k_m t}
\end{align}
where $k_m$ is the rate of the specific monomolecular reaction.
A tentative next-event time $\tau_m$ for a monomolecular reaction then 
can be sampled via the inversion method as
\begin{align}
 \tau_m = - k_m \ln(\mathcal{R}_m)
\end{align}
where $\mathcal{R}_m\in[0,1]$ is a uniformely distributed random number.

Let $p(\myvec r,t|\myvec r_0)$ be the density function for the probability that
a diffusing particle will be located at $\myvec r$ at time $t$ given 
that it started at $\myvec r_0$ at time $t_0=0$.
Within an unbounded volume, the evolution of $p(\myvec r,t|\myvec r_0)$ is well-described by the diffusion equation
\begin{align}
 \pd_t p(\myvec r,t|\myvec r_0) = D \nabla^2 p(\myvec r,t|\myvec r_0) + \delta(\myvec r - \myvec r_0)\delta(t - t_0)	\quad .
\end{align}
Note that due to the delta-peak inhomogeneity that represents the initial condition,
the solution $p(\myvec r,t|\myvec r_0)$ technically is a Green's function.

To sample a first-passage time for the particle to reach the outer shell $\pd\mathcal{D}_1$
of a domain $\mathcal{D}_1$ constructed around $\myvec r_0$,
additionally an absorbing boundary condition may be imposed as follows:
\begin{align}
 p(\myvec r,t|\myvec r_0) = 0 \quad\text{for}\quad \myvec r \in \pd\mathcal{D}_1
\end{align}
In the simplest case, for a spherical domain with radius $R$:
\begin{align}
 p(|\myvec r - \myvec r_0|=R,t|\myvec r_0) = 0
\end{align}
For more complicated domain geometries, e.g. cylinders, the problem has
to be transformed into a coordinate system that captures specific symmetries,
and boundary conditions have to be imposed for each coordinate separately.

Given that the Green's function $p(\myvec r,t|\myvec r_0)$ for the above boundary-value problem can
be found, integration of $p(\myvec r,t|\myvec r_0)$ over the whole domain yields its survival probability $S(t)$,
i.e. the probability for the particle to still remain within $\mathcal{D}_1$ at time~$t$.
Note that $S(t_0)=1$.
The survival probability is linked to the propensity function $q(t)$, which is the
probability for hitting $\pd\mathcal{D}_1$ within the time interval $[t,t+dt]$, via:
\begin{align}
  q(t) = -\pd_t S(t) = -\pd_t \int_{\mathcal{D}_1} p(\myvec r, t|\myvec r_0) d\myvec r
\end{align}
In other words, $1-S(t) = \int_{t_0}^t q(t')dt' = Q(t)$ is equal to the 
cumulative distribution function of $q(t)$ and may be used to sample 
a next-event time $\tau_e$ for exiting the domain 
via the inversion method as follows:
\begin{align}
 \tau_e = Q^{-1}(\mathcal{R}_e) = S^{-1}(1-\mathcal{R}_e)
\end{align}
Here $\mathcal{R}_e\in[0,1]$ again is a uniformely distributed random number.
In general, it can be difficult to calculate $S^{-1}$ analytically.
Then $\tau_e$ can be obtained by solving the equation $S(\tau_e) - \mathcal{R}_e =0$
with a numerical rootfinder\footnote{As a matter of course, using $1-\mathcal{R}_e$ and $\mathcal{R}_e$ is
equivalent if both are uniform random numbers from $[0,1]$.}.
\index{survival probability}	\index{propensity function}	\index{next-event time}

After construction of an \eGFRD\xspace \domaintype{Single} domain, first both
$\tau_e$ and $\tau_m$ are sampled as described.
Since we presuppose that diffusion and monomolecular reactions are occuring independently,
the next-event time for the domain is set as:
\begin{align}
 \tau_{\mathcal{D}_1} = \min(\tau_e, \tau_m)
\end{align}
This automatically determines the event type to be either a \textit{Single Escape}
or a \textit{Single Reaction}.
For reasons discussed in section~\ref{sec:eGFRD-Principle} of the main text and in section \ref{sec:GFRD-shellmaking}, 
\eGFRD also allows for ``bursting'', i.e. a ``premature'' update of the domain at times $\tau_b<\tau_{\mathcal{D}_1}$. \index{bursting}
Here we make use of the fact that \eGFRD is capable to generate a new particle position $\myvec r_\nu$
from the Green's function for an arbitrary update time $\tau_\nu\leq\tau_{\mathcal{D}_1}$.
Since in these cases the PDF $p(\myvec r,\tau_\nu|\myvec r_0)$ is not normalized within $\mathcal{D}_1$,
precisely because probability leaked out through $\pd \mathcal{D}_1$ during the time $\tau_\nu - t_0$,
it is important to sample $\myvec r_\nu$ from the conditional PDF $p_{S(\tau_\nu)}(\myvec r,\tau_\nu|\myvec r_0)\equiv \frac{1}{S(\tau_\nu)} p(\myvec r,\tau_\nu|\myvec r_0)$.
How this is done in detail depends on the geometry of the domain.
For a spherical domain with radius $R$ the angles $\theta\in[0,\pi]$ and $\phi\in[0,2\pi]$ 
in $\myvec r_\nu=(r_\nu,\theta,\phi)$ are
sampled from uniform distributions over the respective intervals, employing existing symmetry.
If the next-event type is a \textit{Single Escape}, $r_\nu=R$ with certainty and no further
steps are required.
For the other event types, i.e. \textit{Single Reaction} or bursting, the new radial distance $r_\nu$ is sampled from
\begin{align} 
 r_\nu = P^{-1}_{S(\tau_\nu)}(\mathcal{R}_S)
\end{align}
with a uniform random number $\mathcal{R}_S\in[0,1]$, using the cumulative conditional PDF
\begin{align}
 P_{S(\tau_\nu)}(r_\nu) = \frac{1}{S(\tau_\nu)} \int\limits_0^{r_\nu} \int\limits_0^{2\pi} \int\limits_0^{\pi} p(r,\theta,\phi,\tau_\nu|\myvec r_0) r^2 \sin(\theta)d\theta d\phi dr	\quad .
\end{align}

If a \textit{Single Reaction} event produces two particles from one, these are put at contact
at the sampled position $\myvec r_\nu$ with random spatial orientation.
If it is a true decay event the particle is removed from the system together with its domain.
Finally, when a \textit{Single Reaction} induces a change the particle species from $s_0$ to $s$, 
the $s_0$ particle is removed and a new $s$-particle is created at $\myvec r_\nu$.
In any other event the particle is simply moved to $\myvec r_\nu$.

\subsubsection{Pair domains}
\label{sec:GFRD-Pairs} \index{Pair domain} \index{Pair}
Sampling of next-event times for a \domaintype{Pair} domain $\mathcal{D}_2$ follows the
same principles as for \domaintype{Singles}.
However, here the two particles can react at contact,
which creates an additional channel of exit from the domain and a new next-event type.

\begin{figure}[ht!]
   \centering
   \subfigure[][]{
    \label{fig:PairSubdomains}
    \includegraphics[width=0.4\textwidth]{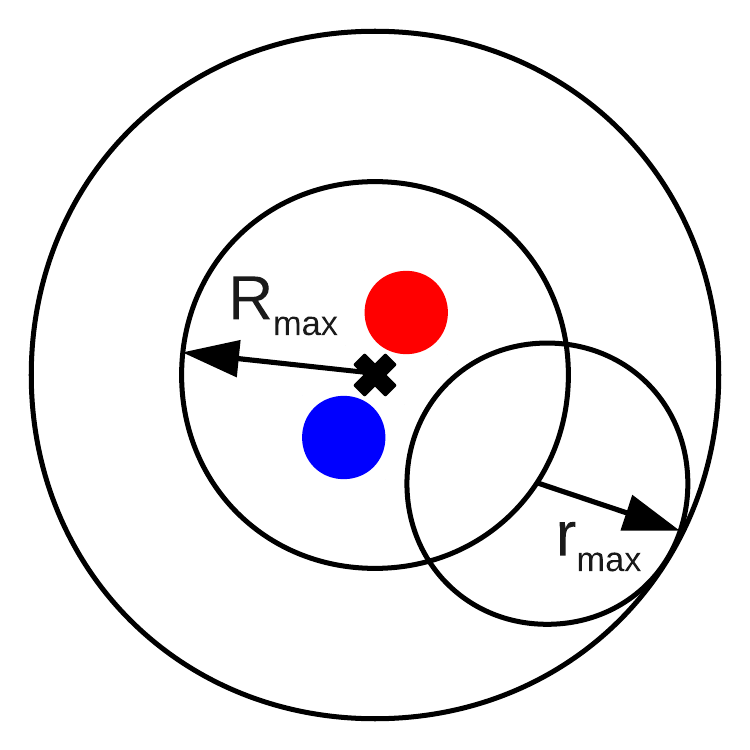}
  }\hfill
   \subfigure[][]{
    \label{fig:MultiExample}
    \includegraphics[width=0.5\textwidth]{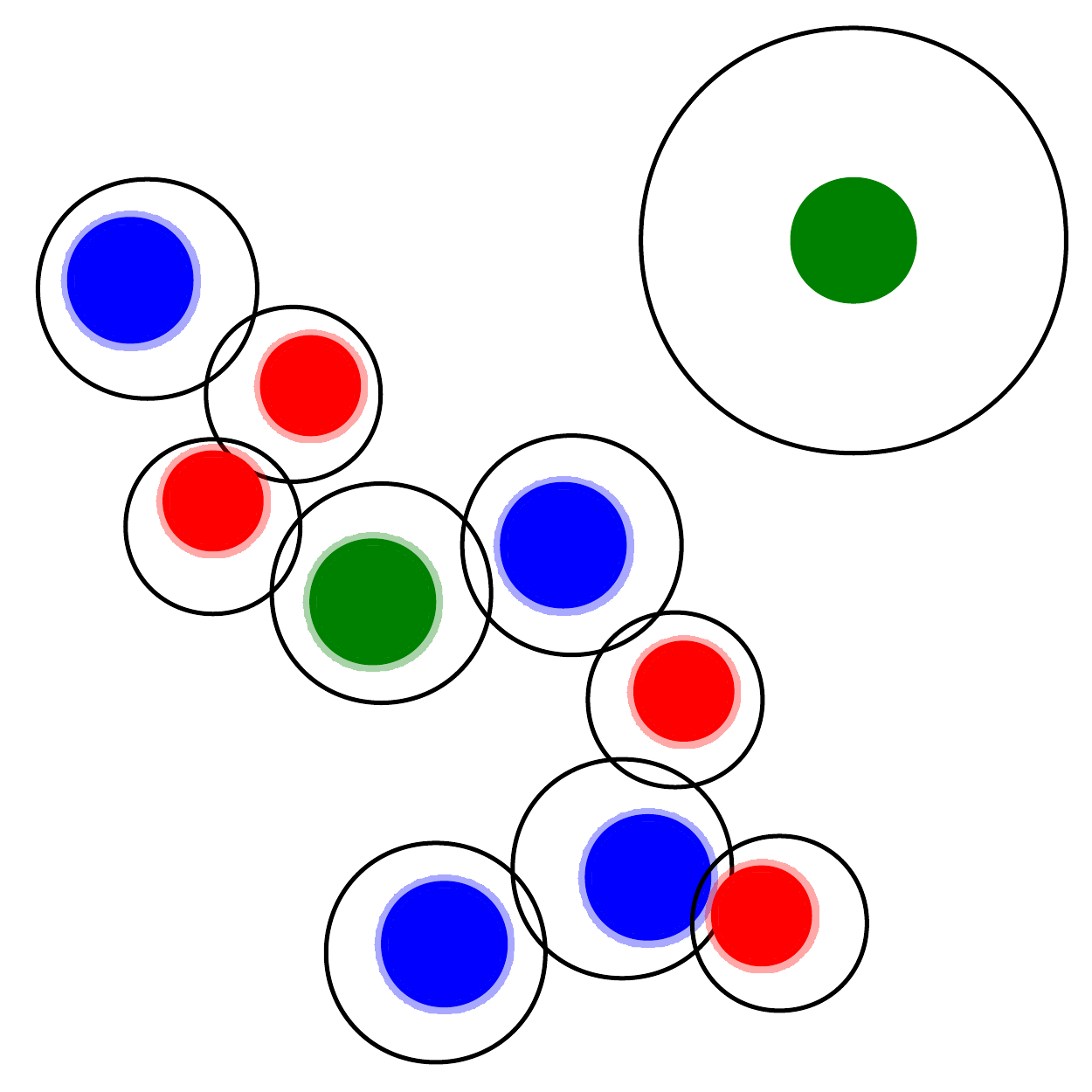}
  }
\caption{ \label{fig:PairAndMulti}
  \textbf{Pair and Multi domains in \eGFRD.}
  \subfigref{\subref{fig:PairSubdomains}} Decomposition of a (projected) spherical \domaintype{Pair} domain into subdomains for
    the center-of-mass vector $\myvec R$ and the interparticle vector $\myvec r$.
  \subfigref{\subref{fig:MultiExample}} An exemplary \domaintype{Multi} domain.
  Here we show the situation in which particles inside the \domaintype{Multi} have been already
  propagated by earlier updates (and thus are offset with respect to the shells), but none of them reached its outer shell yet.
  Thin faint-color rings around the particles indicate their reaction volume.
  A blue and a red particle in the lower-right part of the Multi overlap with their reaction volumes and
  will attempt a reaction.
  Since enough space was available, the top-right green particle formed a regular \domaintype{Single} domain.
}
\end{figure}

Let us denote by $p_2(\myvec r_A, \myvec r_B, t|\myvec r_{A0}, \myvec r_{B0})$ the PDF for the likelihood
of finding two diffusing particles $A$ and $B$,
initially located at positions $\myvec r_{A0}$ and $\myvec r_{B0}$ at $t=t_0$, 
at positions $\myvec r_{A}$ and $\myvec r_{B}$ at a later time $t$.
The time-evolution of $p_2$ is governed by the \textsc{Smoluchowski} equation:
\begin{align}
 \label{eq-1D-Smoluchowski}
 \pd_t \, p_2		&=	[ D_A\nabla_A^2 + D_B\nabla_B^2 ] \, p_2
\end{align}
Here $D_A$ and $D_B$ are the diffusion constants of particles $A$ and $B$.
As demonstrated in sec.~\ref{sec:GFRD-GeneralCoordinateTransform} for a more general case, this problem can be simplified 
by transforming coordinates $\myvec r_A$ and $\myvec r_B$ to $\myvec r$ and $\myvec R$,
where $\myvec r$ is the interparticle vector and $\myvec R$ the (weighted) center-of-mass of the particles.
A separation ansatz $p_2=p_r(\myvec r)p_R(\myvec R)$
then yields two separate, uncoupled diffusion equations for $\myvec r$ and $\myvec R$,
which are equivalent to (\ref{eq-1D-Smoluchowski}):
\begin{align}
 \pd_t \, p_r = D_r \nabla_r^2 \, p_r\;, & & \pd_t \, p_R = D_R \nabla_R^2 \, p_R \quad .
\end{align}
The constants $D_r$ and $D_R$ depend only on $D_A$ and $D_B$.
The uncoupling allows for the calculation of two Green's function solutions
$p_r(\myvec r,t|\myvec r_0)$ and $p_R(\myvec R,t|\myvec R_0)$ on two subdomains
$\mathcal{D}_r$ and $\mathcal{D}_R$ of $\mathcal{D}_2$, respectively, with
boundary conditions adapted to the problem as described further below.
$\mathcal{D}_r$ and $\mathcal{D}_R$ must be defined in a way that 
all possible positions constructed from sampled values of $\myvec r$ and $\myvec R$
remain within the protective domain $\mathcal{D}_2$.
Figure \ref{fig:PairAndMulti}\subref{fig:PairSubdomains} shows a valid 
definition of the subdomains for a (projected) spherical pair domain.

The Green's function $p_R$ for the $\myvec R$ diffusion is calculated in precisely the same way as the Green's function
for the one-particle problem in \domaintype{Single} domains, with an absorbing boundary condition $p_R(\myvec R, t) = 0$
for $\myvec R \in \pd\mathcal{D}_R$.
This yields a next-event time $\tau_R$ for first-arrival of $\myvec R$ to $\pd\mathcal{D}_R$,
called \textit{Center of Mass Escape} or \textit{CoM Escape}.
\index{Center of Mass Escape} \index{CoM Escape}

Reactions between $A$ and $B$ are modelled via
a radiating boundary condition to $p_r$ at the particle contact radius $\sigma = R_A + R_B$:
\begin{align}
 q_\sigma(t) \equiv \int_{\pd\mathcal{D}_r^\sigma} -D\nabla_{\myvec r}p_r(\myvec r,t|\myvec r_0) d\myvec r = k p_r(|\myvec r| = \sigma, t)
\end{align}
Here, $k$ is the intrinsic particle reaction rate, which is the rate at which the particles react
given that they are in contact, and $p_r(|\myvec r| = \sigma, t)$ is the probability that
the particles are indeed at contact at time $t$.
The integral on the left is the total probability (out)flux through the ``contact surface'' or
inner boundary of the $\myvec r$-subdomain, which is the set of all points at which $A$ and $B$ are in contact:
$\pd\mathcal{D}_r^\sigma = \left\lbrace \myvec r \big\vert |\myvec r| = \sigma \right\rbrace$.
At the outer boundary of the $\myvec r$-subdomain $\pd\mathcal{D}_r^a$ absorbing boundary conditions are imposed.
The initial condition for this boundary value problem is set by the inital separation of the two
particles, $p_r(\myvec r, t=0|\myvec r_0)=\delta(\myvec r - (\myvec r_{B0} - \myvec r_{A0}))$.
A tentative next-event time $\tau_r$ can be sampled from the survival probability 
$S_r(t)=\int_{\mathcal{D}_r} p_r(\myvec r,t|\myvec r_0) d\myvec r$ in the same way as before.
With this, however, it remains undetermined by which boundary the particle escaped.
To specify whether the exit from $\mathcal{D}_r$ happens through the radiating
(\textit{Pair Reaction} event)
or through the absorbing boundary (\textit{IV Escape}\footnote{\xspace IV = interparticle vector}),
the probability fluxes through the boundaries at time $\tau_r$ are compared:
If a uniformly distributed random number $\mathcal{R}_r\in[0,1]$ is smaller than the fractional propensity
\begin{align}
 q_{\rm frac}(\tau_r) = \frac{q_\sigma(\tau_r)}{q_\sigma(\tau_r) + q_a(\tau_r)} = \frac{q_\sigma(\tau_r)}{-\pd_t S_r(\tau_r)}
\end{align}
the next-event is a \textit{Pair Reaction}; otherwise it is a \textit{IV Escape}.
\index{Pair Reaction}\index{IV Escape}

In general, the particles $A$ and $B$ additionally can undergo monomolecular reactions,
for which next-event times $\tau_A$ and $\tau_B$ are calculated in the same manner as for \domaintype{Singles}.
Thus, during \domaintype{Pair} domain construction, altogether four next-event times $\tau_r$, $\tau_R$, $\tau_A$, $\tau_B$
with different next-event types are determined.
Since the four stochastic processes again are independent of each other, the tentative next-event time
for the \domaintype{Pair} domain is defined as:
\begin{align}
 \tau_{\mathcal{D}_2} = \min\left( \left\lbrace \tau_r, \tau_R, \tau_A, \tau_B \right\rbrace \right)
\end{align}

The precise procedure of sampling new positions for $A$ and $B$ at next-event time $\tau_{\mathcal{D}_2}$ depends on the
type of the next event and the coordinate system in which the problem is considered.
We present here the classical treatment for two interacting particles in 3D, which employs spherical coordinates,
for each possible event-type respectively:\\

\begin{itemize}
 \item \textit{Pair Reaction}: Here the CoM position $\myvec R_\nu(\tau_r)$ is sampled in the same manner as the new position $\myvec r_\nu(\tau<\tau_{\mathcal{D}_1})$ 
			     in \domaintype{Singles}.
			     A particle with the product species is created at $\myvec R_\nu(\tau_r)$.\\ \index{Pair Reaction}

 \item \textit{IV Escape}:   In this case $|\myvec r_\nu|=a$ with certainty, but the escape angle $\theta_\nu$ yet remains undetermined.
			     It is sampled from the propensity function for leaving the $\myvec r$-subdomain at the $\myvec r$-escape time $\tau_r$
			     through its outer boundary at an angle $\theta$, given by
			      $$q_a(\theta)=-\frac{a \sin(\theta)}{Q} D_r \int\limits_0^{2\pi} \left[ \pd_r p_r(r,\theta, \phi, \tau_r|\myvec r_0)\right]_{r=a} a d\phi\;.$$
			     Here the normalization factor is the total flux through the outer boundary $Q = \int_0^{\pi} q_a(\theta) d\theta$.
			     The second angle $\phi_\nu \in[0,2\pi]$ is drawn from a uniform distribution.\\ \index{IV Escape}

 \item \textit{CoM Escape}:  A new center-of-mass position $\myvec R_\nu(\tau_R)$ is sampled as $\myvec r_\nu(\tau_{\mathcal{D}_1})$ in the \domaintype{Single}.
			     To determine $\myvec r_\nu(\tau_R)$, first a new radius $r_\nu$ is sampled from the conditional probability
			     $$\tilde p_{r}(r,\tau_R)\equiv\frac{r^2}{S_r(\tau_R)} \int\limits_0^{\pi}\int\limits_0^{2\pi} p_r(r,\theta,\phi,\tau_R|\myvec r_0) \sin(\theta)d\theta d\phi\;.$$
			     Subsequently, a new angle $\theta_\nu$ is sampled from the density
			     $p_{r,\theta}(\theta,r_\nu,\tau_R) \break \equiv \frac{r_\nu \sin(\theta)}{\tilde p_{r}(r_\nu,\tau_R)} \int_0^{2\pi} p_r(r_\nu,\theta,\phi,\tau_R|\myvec r_0) d\phi$
			     and $\phi_\nu \in[0,2\pi]$ from the uniform distribution.\\ \index{CoM Escape}
			     
 \item \textit{Monomolecular reaction} (of $A$)\footnote{An identical procedure applies, with $A$ and $B$ interchanged, to the case in which $B$ undergoes a monomolecular reaction.}:
			The new CoM position $\myvec R_\nu(\tau_A)$ is sampled as in the case \textit{Pair Reaction},
			the new interparticle vector $\myvec r_\nu(\tau_A)$ as in the case \textit{CoM Escape}. 
			From this we obtain $\myvec r_{A,\nu}(\tau_A)$ and $\myvec r_{B,\nu}(\tau_A)$.
			Particle $B$ is simply moved to $\myvec r_{B,\nu}(\tau_A)$, while $A$ is treated as described for 
			the monomolecular reactions in \domaintype{Singles}.\\ \index{monomolecular reaction}
\end{itemize}

\TODO{We should still briefly describe how the procedures in 2D and 1D differ from the above.}

\clearpage
\section{Brownian Dynamics fallback-system and Multi domains}
\label{sec:BD-SI}
The strength of \eGFRD is that--thanks to the knowledge of the Green's function--detailed sampling 
of diffusive trajectories inside the domains can be omitted and particles are propagated with
large jumps in time and space.
This, however, comes at the cost of increased computational effort per update, because
drawing times and positions from Green's functions is significantly more expensive than
sampling of simple Gaussian displacements.
Therefore \eGFRD becomes more costly than Brownian Dynamics (BD) when particles get such crowded
that the maximal size of protective domains becomes comparable to particle radii.
This may be due to the presence of more than one other particle or other, static obstacles.
In such situations, \eGFRD seamlessly switches to a simulation mode in which particles
are propagated by Brownian Dynamics within specialized domains, called \domaintype{Multis}.

\subsection{Multi domains} \index{Multi domain}\index{Multi} \index{multi-shell factor}
Whenever particle distances fall below a predefined threshold and regular domain types
cannot be constructed, the algorithm prompts the construction of \domaintype{Multi}
domains, which can contain more than two particles.
An exemplary \domaintype{Multi} domain is shown in Figure~\ref{fig:PairAndMulti}\subref{fig:MultiExample}.
\domaintype{Multis} are composed of intersecting spherical shells with shell radii $\rho_n$ proportional
to particle radii $R_n$, i.e. $\rho_n=\mu R_n$, where the ``multi-shell factor'' $\mu>1$ is a simulation parameter.
Each \domaintype{Multi} constitutes an autonomous BD simulator isolated from its surroundings.
Within their shells, particles are propagated, one at a time, by sampling displacements $\Delta \myvec r$
from the free Gaussian propagator with a fixed, sufficiently small timestep $\Delta t$ 
that ensures $|\Delta \myvec r| \ll \rho_n$.
Particle propagation continues until either one of the particles hits its surrounding shell
or two overlapping particles react.
Then, the \domaintype{Multi} is broken apart and the new configuration is evaluated {\it de novo},
possibly resulting in \domaintype{Multi} reconstruction.
Particles that moved away sufficiently far from the particle crowd or obstacle at that moment
reform \domaintype{Single} domains and revert to propagation via Green's functions.

When and how \domaintype{Multis} are constructed is explained in more detail
in section \ref{sec:GFRD-shellmaking}.

\subsection{The Reaction-Volume Method ensures that reactions in BD fulfill detailed balance} \index{reaction volume} \index{reaction length} \index{detailed balance}
\label{sec:rvm-BD}
Particles that create overlaps within \domaintype{Multis} are tested for reactions.
Reaction events in BD mode are sampled such that detailed balance is obeyed.
Let $\myvec r_{12}$ be the interparticle vector of the two interacting particles.
Detailed balance demands that, for any $\myvec r_{12}$,
the probability of the unbound configuration at distance $|\myvec r_{12}|$
times the transition probability to move into the bound state from $\myvec r_{12}$ equals the
likelihood to be in the bound state times the probability of the inverse transition:
\begin{align}
  \label{eq-DetailedBalance}
  p_u(\myvec r_{12}) \pi_{u\rightarrow b}(\myvec r_{12})  = p_b \pi_{b\rightarrow u}(\myvec r_{12}) 
\end{align}
The occupancy ratio $p_b/p_u(\myvec r_{12})=K_{eq}$ is fixed by the equilibrium
constant of the reaction and $\pi_{u\rightarrow b}(\myvec r_{12})$ depends on algorithmic details of particle propagation.
This leaves us with the task to prescribe a backward move in a way that $\pi_{b\rightarrow u}(\myvec r_{12})$
obeys (\ref{eq-DetailedBalance}).
Originally, \eGFRD employed the Reaction Brownian Dynamics algorithm by \person{Morelli} and \person{ten Wolde} \cite{Morelli2008-JCP}.
While this scheme yields excellent results for diffusing spheres in 3D, it proved troublesome to extend it
to arbitrary dimensions and non-spherical objects.
In recent \eGFRD we therefore implemented a conceptually similar but more versatile scheme,
which we refer to as the ``reaction-volume method'', or ``rvm-BD''.
Its key assumption is that reactive objects, be it other particles or static structures, are surrounded by a small ``reaction volume'' $V^*$
within which the precise shape of the density $p_u(\myvec r_{12})$ may be ignored.
Reaction attempts only occur within $V^*$, and at the inverse reaction the particle is placed back uniformely into $V^*$.
The binding process is thus broken apart into a displacement and a reaction step.
As in \cite{Morelli2008-JCP}, it proves useful to rewrite the transition probabilities as a product
of a proposal (i.e. move generation) density and (reaction) acceptance probability:
\begin{align}
 \pi_{u\rightarrow b}(\myvec r_{12}) = P^{\rm gen}_{u\rightarrow V^*}(\Delta t)P^{\rm acc}_{V^* \rightarrow b}	\nn\\
 \pi_{b\rightarrow u}(\myvec r_{12}) = P^{\rm gen}_{V^*\rightarrow u}(\Delta t)P^{\rm acc}_{b \rightarrow V^*}
\end{align}
Herein $P^{\rm gen}_{u\rightarrow V^*}(\Delta t)$ is the probability to move diffusively into the reaction volume $V^*$ from a distance
$\myvec r_{12}$ in the unbound state within a time $\Delta t$, whereas $P^{\rm gen}_{V^*\rightarrow u}(\Delta t)$
is the probability of the inverse move.
It can be shown that these probabilities only differ by a factor $V^*$: $P^{\rm gen}_{V^*\rightarrow u}(\Delta t) = V^* P^{\rm gen}_{u\rightarrow V^*}(\Delta t)$.
Together with the assumption that unbinding occurs with Possonian statistics, i.e. $P^{\rm acc}_{b \rightarrow V^*}=k_u\Delta t$,
one finds that detailed balance is fulfilled when forward reaction attempts are accepted with a rate
\begin{align}
 P^{\rm acc}_{V^* \rightarrow b} = \frac{k_b\Delta t}{V^*}	\quad .
\end{align}
In practice, $V^*$ depends on the given situation and it is convenient to tune the magnitudes of
specifically occuring reaction volumes via a global ``reaction length'' parameter $\delta$.
For two spherical particles with contact radius $\sigma$ we have $V^*=\frac{4}{3}\pi\left( (\sigma+\delta)^2 - \sigma^2 \right)$.
For the particle-surface interactions that we introduce later
$V^*$ is calculated similarly, taking into account the particular geometry of the contact region.
This is described in more detail in \cite{Paijmans2012}.

The reaction length $\delta$ and the propagation time step $\Delta t$ are set for each \domaintype{Multi}
domain individually, subject to the following two constraints:
\begin{align}
 & D_{\rm max}\Delta t \leq (\phi R_{\rm min})^2		\\
 & \frac{k_{\rm b,max}\Delta t}{\delta} \leq P^{\rm acc}_{\rm max}
\end{align}
Here $D_{\rm max}$ is the maximal diffusion constant of a particle in the \domaintype{Multi},
$R_{\rm min}$ the minimal particle radius, $k_{\rm b,max}$ the fastest intrinsic forward rate (divided by dimension-specific contact-surface factors)
and $\phi\leq1$ a tuneable step size fraction.
The first requirement limits the maximal displacement within timestep $\Delta t$ to a fraction
of the smallest particle size; the second ensures that the acceptance probability remains
bounded by a value $\leq1$ also for fast reactions;
the standard choice in \eGFRD is $P^{\rm acc}_{\rm max}=0.01$ and $\delta=\phi R_{\rm min}$ with $\phi=0.05 - 0.10$.

Note that since each \domaintype{Multi} is locally constructed with a specific set of involved reaction rates and diffusion constants
that does not necessarily include the fastest of them in the system, $\delta$ and $\Delta t$ can be optimized in a situation-dependent manner, 
which is more flexible and efficient than initially setting a global time step based on the overall fastest rates.

\clearpage

\newcommand{\coeffC}{\gamma}
\newcommand{\coeffD}{\delta}
\section{Coordinate transforms}
\label{sec:GFRD-GeneralCoordinateTransform}
Here we describe how the \person{Smoluchowski} equation for two particles with independent position vectors $\myvec{r}_A$ and $\myvec{r}_B$
can be separated into two more tractable equations after transforming the original set of coordinates into new coordinates
(that have the interpretation of an interparticle vector and a weighted center-of-mass or center-of-diffusion vector),
and imposing a corresponding product ansatz.
We will first demonstrate the calculations for a general case that takes into account a distance-dependent force between the particles;
the case of constant forces acting on the two particles that we use for modeling 1d-transport in \eGFRD is derived as a special case in Sec.~\ref{sec:1D-CoordinateTransform} further below.

The \person{Smoluchowski} equation for the probability density function $p= \break p(\myvec{r}_A,\myvec{r}_B,t| \myvec{r}_{A0},\myvec{r}_{B0},t_0)$ 
of two diffusing particles A and B that can interact via a force $\myvec{F}$ depending on their distance and move with different
diffusion constants $D_A$ and $D_B$ is given by \cite{Smoluchowski1915,vanZon2005-JChemPhys}
\begin{align}
\label{eq-Smoluchowski-general}
\pd_t p		&=	[ D_A\nabla_A^2 + D_B\nabla_B^2 
		 +D_A\nabla_A\cdot\varphi\myvec{F}(\myvec{r}) -D_B\nabla_B\cdot\varphi\myvec{F}(\myvec{r}) ]\; p
\end{align}
where $\myvec{r}$ denotes the interparticle vector:
\begin{align}
\myvec{r}=\myvec{r}_B - \myvec{r}_A	\label{eq-r}
\end{align}
We define the weighted center-of-mass vector $\myvec{R}$ as follows:
\begin{align}
\myvec{R}=\coeffC\myvec{r}_A + \coeffD\myvec{r}_B	\label{eq-R}
\end{align}

Equations (\ref{eq-r}) and (\ref{eq-R}) define new coordinates $\myvec{r}(\myvec{r}_A,\myvec{r}_B)$ and $\myvec{R}(\myvec{r}_A,\myvec{r}_B)$.
Notice that this is not a coordinate transformation in the strict sense, as in general $\myvec{r}$ and $\myvec{R}$ will not be orthogonal.

Moreover, we define the operators:
\begin{align}
\label{eq-r-R-operators}
\nabla_r \equiv \frac{\pd}{\pd \myvec{r}} = \begin{pmatrix}\pd_{r_1}\\ \pd_{r_2}\\ \pd_{r_3}\end{pmatrix}\;,
\quadquad \nabla_R \equiv \frac{\pd}{\pd \myvec{R}} = \begin{pmatrix}\pd_{R_1}\\ \pd_{R_2}\\ \pd_{R_3}\end{pmatrix}\;.
\end{align}

If the differential operator on the right side of (\ref{eq-Smoluchowski-general}) equation can be written as a sum of $\nabla_r^2$, $\nabla_r$, $\nabla_R^2$ and $\nabla_R$,
we may separate (\ref{eq-Smoluchowski-general}) into two independent PDEs for $\myvec r$ and $\myvec R$ by a product ansatz for $p$.
In the following we will calculate different options for the choice of coefficients $\coeffC$ and $\coeffD$ with which the above objective is reached.

\subsection{Rewriting $\nabla_A$ and $\nabla_B$}
\label{sec:GFRD-CT-CoeffCondition}
First we rewrite $\nabla_A$ and $\nabla_B$ in terms of $\nabla_r$ and $\nabla_R$. Let $r_{A,i}=r_{A,i}(\myvec{r},\myvec{R})$ 
denote the $i$-th component of the vector $\myvec{r_A}$, and $r_j$, $R_k$ components of $\myvec{r}$ and $\myvec{R}$ respectively.
Then the derivative of $p$ with respect to $r_{A,i}$ is
\begin{align}
\frac{\pd p}{\pd r_{A,i}}	&= \sum_j \frac{\pd p}{\pd r_j}\frac{\pd r_j}{\pd r_{A,i}} + \sum_k \frac{\pd p}{\pd R_k}\frac{\pd R_k}{\pd r_{A,i}} \nn\\
				&= \sum_j (-1)\delta_{ij}\frac{\pd p}{\pd r_j} + \sum_k \coeffC\delta_{ik}\frac{\pd p}{\pd R_k} \nn\\
				&= \coeffC \frac{\pd p}{\pd R_i} - \frac{\pd p}{\pd r_i} = \left( \coeffC\frac{\pd}{\pd \myvec{R} } - \frac{\pd}{\pd \myvec{r}} \right)_i p
\end{align}
because $r_i$ and $R_i$ only depend on the component $r_{A,i}$ with the same index $i$.
Since this holds for every $i$, we have:
\begin{align}
\label{}
\nabla_A = \coeffC\nabla_R - \nabla_r
\end{align}
Analogously, one obtains:
\begin{align}
\nabla_B = \coeffD\nabla_R + \nabla_r
\end{align}
From this it follows that:
\begin{align}
\nabla_A^2 &= \coeffC^2\nabla_R^2 + \nabla_r^2 - 2\coeffC\nabla_r\nabla_R \nn\\
\nabla_B^2 &= \coeffD^2\nabla_R^2 + \nabla_r^2 + 2\coeffD\nabla_r\nabla_R
\end{align}

Here we use $\nabla_r\nabla_R=\nabla_R\nabla_r$, assuming the 2nd derivative of $p$ with respect to any of 
its variables to be a continuous function in $\mathbb{R}^3$.
The partial derivatives then may be interchanged by the theorem of \person{Clairaut \& Schwarz}.

Now that we have expressed $\nabla_A$ and $\nabla_B$ in terms of $\nabla_r$ and $\nabla_R$,
we can also rewrite the right side of the \person{Smoluchowski} equation.
First, for the case $\varphi=0$, we get:
\begin{align}
D_A\nabla_A^2 + D_B\nabla_B^2	&= \left( D_A+D_B \right) \nabla_r^2 \nn\\
				&\quad +\left( \coeffC^2 D_A+\coeffD^2 D_B \right) \nabla_R^2 \nn\\
				&\quad +2\left( \coeffD D_B-\coeffC D_A \right) \nabla_r\nabla_R
\end{align}
Rewriting the force term separately yields:
\begin{align}
\left( D_A\nabla_A-D_B\nabla_B \right) \cdot\varphi\myvec{F} = 
[\left( \coeffC D_A - \coeffD D_B \right) \nabla_R - \left( D_A+D_B \right)\nabla_r ]\cdot\varphi\myvec{F} \label{eq-force}
\end{align}
To get rid of the mixed term containing $\nabla_r\nabla_R$ we can make any choice for $\coeffC$ and $\coeffD$ that fulfills
\begin{align}
\label{eq-CF-MixedDerVanishingCond}
\coeffD &= \frac{D_A}{D_B}\coeffC	\quad.
\end{align}
Note that with this choice the force contribution (\ref{eq-force}) only depends on the derivative with respect to the interparticle vector ($\nabla_r$).

\subsection{Coefficient choice as in original GFRD}
\label{sec:CoefficientChoice-originalGFRD}
One possible choice for $\coeffC$ and $\coeffD$, which is the the same as in the original
version of GFRD, is the following:
\begin{align}
\coeffC \equiv	\sqrt{\frac{D_B}{D_A}}, &\qquad \coeffD \equiv	\sqrt{\frac{D_A}{D_B}}
\end{align}
This yields
\begin{align}
\myvec{R} &= \sqrt{\frac{D_B}{D_A}}\cdot\myvec{r}_A + \sqrt{\frac{D_A}{D_B}}\cdot\myvec{r}_B 
\end{align}
and
\begin{align}
D_A\nabla_A^2 + D_B\nabla_B^2	= \left( D_A+D_B \right) \left( \nabla_r^2 + \nabla_R^2 \right)	\;.
\end{align}
The same prefactor also appears in the force term
\begin{align}
\left( D_A\nabla_A-D_B\nabla_B \right) \cdot\varphi\myvec{F} = - \left( D_A+D_B \right) \nabla_r\cdot\varphi\myvec{F}
\end{align}
so that equation (\ref{eq-Smoluchowski-general}) simplifies as follows:
\begin{align}
\pd_t p		=	( D_A + D_B ) \left(\nabla_r^2 + \nabla_R^2 - \nabla_r\cdot\varphi\myvec{F} \right) p
\end{align}
We can now separate the equation by the product ansatz
\begin{align}
p(\myvec{r},\myvec{R},t|\myvec{r}_0,\myvec{R}_0,t_0)=p_{r}(\myvec{r},t|\myvec{r}_0,t_0)p_R(\myvec{R},t|\myvec{R}_0,t_0)
\end{align}
into two equations describing two independent diffusion processes (here with the same diffusion constant $D_A+D_B$):
\begin{align}
\pd_t p_r		&=	\braceunderset{D_r}{( D_A + D_B )} \left(\nabla_r^2 - \nabla_r\cdot\varphi\myvec{F} \right) p_r 	\\
\pd_t p_R		&=	\braceunderset{D_R}{( D_A + D_B )} \nabla_R^2 \text{\space}p_R
\end{align}
Note that, as expected, the force contribution is present only in the equation for the interparticle vector $\myvec r$.

\subsection{Coefficient choice as in eGFRD}
\label{sec:CoefficientChoice-eGFRD}
The following, slightly different choice for $\coeffC$ and $\coeffD$ is used in \eGFRD:
\begin{align}
\coeffC \equiv \frac{D_B}{D_A+D_B},	&\qquad \coeffD \equiv \frac{D_A}{D_A+D_B}
\end{align}
This leads to:
\begin{align}
D_A\nabla_A^2 + D_B\nabla_B^2 & =\left[\frac{D_A D^2_B + D_B D^2_A}{\left( D_A + D_B \right)^2} \right] \nabla_R^2 + \left( D_A+D_B \right) \nabla_r^2 	\nn\\
&=\left(\frac{D_AD_B}{D_A+D_B}\right) \nabla_R^2 + \left( D_A+D_B \right) \nabla_r^2
\end{align}

Everything that has been said for the previous choice of $\coeffC$ and $\coeffD$ also applies to this case,
except for the fact that $\myvec R(\myvec r_A, \myvec r_B)$ now has a different weighting as before:
\begin{align}
\myvec{R}	&= \frac{ D_B \myvec{r}_A + D_A \myvec{r}_B }{ D_A + D_B }
\end{align}
Therefore, also the diffusion constant $D_R$ now is different from $D_r$.
Using the same separation ansatz as before we arrive at:
\begin{align}
\pd_t p_r		&=	( D_A + D_B ) \left(\nabla_r^2 - \nabla_r\cdot\varphi\myvec{F} \right) p_r \nn\\
\pd_t p_R		&=	\left( \frac{D_AD_B}{D_A+D_B} \right) \nabla_R^2 \text{\space}p_R
\end{align}

\pagebreak[4]
\subsection{Special case: Diffusion-drift equation in 1d}
\label{sec:1D-CoordinateTransform}
If there is no distance-dependent force interaction between the particles, but still constant forces that--in addition to diffusion--cause the two particles
to move with average drift velocities $\myvec{v}_A$ and $\myvec{v}_B$, the general \person{Smoluchowski} equation, \Eqref{eq-Smoluchowski-general}, becomes:
\begin{align}
\label{eq-Smoluchowski-Drift}
\pd_t p_2	&=	[ D_A\nabla_A^2 + D_B\nabla_B^2
			  -\myvec{v}_A\nabla_A -\myvec{v}_B\nabla_B ] p_2
\end{align}
Defining the new coordinates $\myvec r$ and $\myvec R$ and operators $\nabla_r$ and $\nabla_R$ as before [Eqs.~(\ref{eq-r}), (\ref{eq-R}) and (\ref{eq-r-R-operators})] 
and once again imposing \Eqref{eq-CF-MixedDerVanishingCond}, after some intermediate steps one arrives at:
\begin{align}
\pd_t p_2	=&	\big[ (D_A+D_B)\nabla_r^2 + (\coeffC^2D_A+\coeffD^2D_B)\nabla_R^2 \nn\\
		 &		\quadquad -\myvec{v}_A\left(\coeffC\nabla_R-\nabla_r\right) -\myvec{v}_B\left(\coeffD\nabla_R+\nabla_r\right) \big] p_2 \nn\\
		=&	\big[ (D_A+D_B)\nabla_r^2 + (\coeffC^2D_A+\coeffD^2D_B)\nabla_R^2 \nn\\
		 &		\quadquad +(\myvec{v}_A-\myvec{v}_B)\nabla_r -(\coeffC\myvec{v}_A+\coeffD\myvec{v}_B)\nabla_R \big] p_2
\end{align}

Via the separation ansatz $p_2 \equiv p_r p_R$ we can rewrite the above equation in terms of two
diffusion-drift equations, one for $\myvec r$ and one for $\myvec R$, with diffusion and drift
constants made up from the corresponding constants of the individual particles:
\begin{align}
\pd_t p_r	=&	\big[ \braceunderset{D_r}{(D_A+D_B)}\nabla_r^2 - \braceunderset{\myvec{v}_r}{(\myvec{v}_B-\myvec{v}_A)}\nabla_r \big]
															\text{\space}p_r \nn\\
\pd_t p_R	=&	\big[ \braceunderset{D_R}{(\coeffC^2 D_A + \coeffD^2 D_B)}\nabla_R^2 
						- \braceunderset{\myvec{v}_R}{(\coeffC\myvec{v}_A+\coeffD\myvec{v}_B)} \nabla_R \big] \text{\space}p_R
\end{align}
The interpretation of the new drift constants is straightforward: $\myvec{v}_r$ describes the relative velocity of the particles 
(as in the case without diffusive motion) while $\myvec{v}_R$ is an effective weighted center-of-mass drift.

Choosing ``eGFRD style'' coefficients $\coeffC$ and $\coeffD$ as in Sec.~\ref{sec:CoefficientChoice-eGFRD} we obtain:
\begin{align}
 D_R = \frac{D_A D_B}{D_A+D_B},	&\qquad\qquad \myvec{v}_R = \frac{D_B\myvec{v}_A + D_A\myvec{v}_B}{D_A+D_B}
\end{align}

With the definition of 1D structures and domains introduced in Sec.~\ref{sec:NewDomains} of the main text,
movement of the particles is restricted to a straight line.
Then the vectors $\myvec{r}$, $\myvec{R}$, $\myvec{v}_r$ and $\myvec{v}_R$ are collinear,
and we can pass from the vector equation to a scalar equation.

\pagebreak[4]
\subsection{Prescribing an arbitrary centre-of-mass diffusion constant}
For completeness we briefly describe how to choose $\coeffC$ and $\coeffD$ in order to ensure that $D_R$
is equal to an arbitrary prescribed diffusion constant $D_C$, if desired.
$D_C$ might be, for example, the diffusion constant of the product of the $A+B\rightarrow C$ reaction.

In this case $\coeffC$ and $\coeffD$ have to obey the two equations:
\begin{align}
D_R = \coeffC^2 D_A+\coeffD^2 D_B, &\qquad \coeffD = \frac{D_A}{D_B}\coeffC
\end{align}
Combining these we obtain:
\begin{align}
& D_R	= \coeffC^2 D_A \left[ 1 + \frac{D_A}{D_B} \right] \overset{!}{=} D_C
\end{align}
Since all involved quantities are positive real numbers, it follows that:
\begin{align}
 \coeffC = \sqrt[+]{\frac{D_C}{D_A\left(1+\frac{D_A}{D_B} \right)}} &\qquad \Rightarrow \qquad \coeffD = \frac{D_A}{D_B}\coeffC = \sqrt[+]{\frac{D_C}{D_B\left(1+\frac{D_B}{D_A} \right)}}
\end{align}
This combination of $\coeffC$ and $\coeffD$ indeed leads to:
\begin{align}
D_R	= D_A\coeffC^2 + D_B\coeffD^2 &= \frac{D_C}{1+\frac{D_A}{D_B}} + \frac{D_C}{1+\frac{D_B}{D_A}} \nn\\
	&= \frac{D_B D_C + D_A D_C}{D_A + D_B} = D_C
\end{align}
Since $\coeffC$ and $\coeffD$ are always real and positive, except for the (usually uninteresting) cases $D_A=0$ or $D_B=0$,
one can indeed always find a coordinate transform for which $D_R$ matches an arbitrary diffusion coefficient, while $D_r=D_A+D_B$.

\subsection{Other coordinate transforms}
As we describe later in section~\ref{sec:Special-Domains}, we have also devised domains for special applications, some of which feature 
specialized coordinate transforms; 
these are not described here, but introduced in the respective sections (\ref{sec:Plane-Transitions} and \ref{sec:DirectBinding-MixedPair2D3D}).

\clearpage
\section{Green's function derivations and dimension-specific sampling procedures}
\subsection{Green's functions for the 1D-diffusion-reaction problem with drift under different boundary conditions}
\label{sec:1D-GreensFunctions}
After performing the coordinate transform explained in Sec.~\ref{sec:1D-CoordinateTransform},
the 1D diffusion-drift equation used for computing the Green's functions for the \domaintype{Cylindrical Surface Single} and \domaintype{Cylindrical Surface Pair},
introduced in sec.~\ref{sec:eGFRD-on-cyl} of the main text \TODO{check whether notations are consistent w. main text}, takes the common form
\begin{align*}
\pd_t p_x = \left[ D_x\pd_x^2 - v_x\pd_x \right] \; p_x
\tag{PDE}
\label{eq-PDE-1D}
\end{align*}
where $x$ stands for either the inter-particle distance $r$ or the center-of-mass position $R$.
In both cases the initial condition is $p_x(x,t_0=0)=\delta(x-x_0)$.
In the following we will drop the index and simply use $p=p(x,t|x_0,t_0)$
to denote the Green's function.
Following the standard treatment in eGFRD, we model chemical reactions between
particles $A$ and $B$ on the cylinder by imposing a radiating boundary condition 
to $r$ at particle contact,
while the $R$-equation has to be solved subject to absorbing boundary conditions.
For completeness, we will also give the solutions for the half-bounded problems.

To summarize, in the following we will derive the Green's function for (\ref{eq-PDE-1D})
on an interval $[\sigma,a]$ of length $L=a-\sigma$ or on a one-sided interval $[\sigma,\infty)$,
and the following boundary conditions, respecitvely:
\begin{itemize}
	\item \underline{Rad-Abs}: radiating left boundary at $\sigma$, absorbing right boundary at $a>\sigma$.
	\item \underline{Abs-Abs}: absorbing left boundary at $\sigma$, absorbing right boundary at $a>\sigma$.
	\item \underline{Rad-Inf}: radiating left boundary at $\sigma$, no right boundary.
	\item \underline{Abs-Inf}: absorbing left boundary at $\sigma$, no right boundary.
\end{itemize}

The Green's Functions are used to derive the resulting survival probability 
$S(t|x_0,t_0)=\int_a^{b\text{ or }\infty} p(x,t|x_0,t_0) dx$, 
the corresponding propensity function $\pi(t|x_0,t_0)=-\pd_t S(t|x_0,t_0)$
and expressions for the transient boundary fluxes.

\subsubsection{Free solution}
It is easily verified that the ``free'', i.e. unbounded, diffusion-drift equation (\ref{eq-PDE-1D}) 
with initial condition $p(x,t=0)=\delta(x-x_0)$ is solved by
\begin{align}
 p_{\rm free}(x,t|x_0) = \frac{1}{\sqrt{4\pi Dt}} e^{-\frac{1}{4Dt}\left[ (x-x_0) - vt \right]^2}
\end{align}
which describes a Gaussian distribution with a width that increases in time, centered around a 
mean value that moves with the drift velocity $v$.

\subsubsection{Green's function for 1D-diffusion with drift, Rad-Abs case}
\label{sec:1D-GreensFunction-Rad-Abs}
We start with the most general of the four cases, with the perspective of deriving other cases as special limits.

The radiation boundary condition relates the probability flux $j(x=\sigma,t)$ at the radiating boundary 
to the intrinsic reaction rate $k$ via:
\begin{align}
j(x=\sigma,t) = -k p(x=\sigma,t)
\end{align}
Here the flux contains a contribution from diffusion and a contribution from the drift:
\begin{align}
j(x,t) &= -D\pd_x p(x,t) + vp(x,t)
\end{align}
The correct boundary condition for the boundary at $\sigma<a$ therefore is:
\begin{align*}
-D \pd_x p(x,t)\takenat_{x=\sigma} + vp(\sigma,t) &= -k p(\sigma,t) \\
\Leftrightarrow \quadquad \pd_x p(x,t)\takenat_{x=\sigma} &= \frac{v+k}{D} p(\sigma,t)	\tag{BCr}\label{eq-PDE-1D-BCr}
\end{align*}
The minus sign on the right side of the equation is due to the fact that at the left boundary the flux out 
of the system is negative with respect to the $x$-axis.

The absorbing boundary at $x=a$ requires:
\begin{align}
p(a,t) &= 0 \quad\quad \forall t
\tag{BCa}
\end{align}

\subsubsection*{Dedimensionalization}
Before we attempt to solve the PDE with these boundary conditions it is convenient to perform a dedimensionalization. 
The natural length scale is given by the length $L=a-\sigma$ of the interval $[\sigma,a]$, 
while $T\equiv L^2/D$ defines a corresponding natural time scale.
We thus rescale our variables via
\begin{align}
\xi\equiv\frac{x-\sigma}{a-\sigma}=\frac{x-\sigma}{L},\quad		&\Rightarrow \pd_\xi\equiv L\pd_x \nn\\
\tau\equiv\frac{t}{T}=\frac{Dt}{L^2},\quad				&\Rightarrow \pd_\tau\equiv T\pd_t=\frac{L^2}{D}\pd_t
\end{align}
to obtain the following boundary value problem to solve:
\begin{align}
\pd_\tau p(\xi,\tau)			&= \left[ \pd_\xi^2 - \frac{vL}{D}\pd_\xi \right] \text{\space}p(\xi,\tau) \tag{PDE}\label{PDE-dedim}\\
\pd_\xi p(\xi,\tau)\takenat_{\xi=0}	&= \frac{(v+k)L}{D} p(0,\tau) \tag{BCr}\\
p(1,\tau)				&= 0 \tag{BCa} \\
p(\xi,\tau=0)				&= \frac{1}{L}\delta\left( \xi - \xi_0 \right) \tag{IC}
\label{eq-PDE-dedim}
\end{align}
The last equation represents the starting condition for a particle initially located at position $L\xi_0$
(or a pair having an initial separation $L\xi_0$), where for convenience we set $t_0=0=\tau_0$.
Note that we have to scale the delta function by $1/L$ because the integration norm scales as $d\xi = Ldx$.

Unfortunately the linear operator ${\Lambda\equiv\left[ \pd_\xi^2 - \frac{vL}{D}\pd_\xi \right]}$ is non-Hermitian.
Therefore we can not apply straightforward techniques like eigenfunction expansion to calculate the solution.
As we will see, a simple transform can resolve this issue.

\subsubsection*{Simplifying the problem with the help of an integrating factor}
The difficulties imposed by the non-Hermiticity of the operator can be overcome by introducing an integrating factor
$\phi(\xi)\equiv e^{\frac{vL}{2D}\xi}=e^{\frac{\nu}{2}\xi}$
(depending explicitly on $\xi$)\footnote{$\Lambda$ is non-Hermitian with respect to the usual Cartesian integration norm $d\xi$. 
However it is Hermitian with respect to the integration norm $d\phi=\frac{\nu}{2}e^{\frac{\nu}{2}\xi}d\xi$.}.
This technique was already used by \textsc{Smoluchowski} himself \cite{Smoluchowski1916}.
Multiplying (\ref{PDE-dedim}) with ${1/\phi(\xi)=e^{-\frac{\nu}{2}\xi}}$ and completing the square yields ($\nu\equiv\frac{vL}{D}$):
\newcommand{\IFp}{e^{\frac{\nu}{2}\xi}}
\newcommand{\IFm}{e^{-\frac{\nu}{2}\xi}}
\begin{align}
\pd_\tau \left[ \IFm p(\xi,\tau) \right] &= \IFm \left[ \pd_\xi^2 - \nu\pd_\xi \right] \text{\space}p(\xi,\tau)		\nn \\
& = 	\left[ \IFm\pd_\xi^2 - 2\frac{\nu}{2}\IFm\pd_\xi + \frac{\nu^2}{4}\IFm	\right.	\nn\\
&\hspace{3.8cm}	\left. - \frac{\nu^2}{4}\IFm \right] \text{\space}p(\xi,\tau) 	\nn \\
& = \pd_\xi^2\left[ \IFm p(\xi,\tau) \right] - \frac{\nu^2}{4} \IFm p(\xi,\tau)
\end{align}
Thus, by writing the solution with an ansatz
\begin{align}
p(\xi,\tau)=\phi(\xi) \pi(\xi,\tau)=\IFp \pi(\xi,\tau)
\label{eq-intfac-ansatz}
\end{align}
which means defining a new function
\begin{align}
\pi(\xi,\tau)\equiv\IFm p(\xi,\tau)
\label{eq-def-q}
\end{align}
equation (\ref{PDE-dedim}) is equivalent to:
\begin{align}
\pd_\tau \pi(\xi,\tau) = \pd_\xi^2 \pi(\xi,\tau) - \frac{\nu^2}{4} \pi(\xi,\tau)
\label{PDE-q}
\end{align}
Alternatively, this can be seen by plugging the new ansatz for $p$ into (\ref{PDE-dedim})
and applying the operators accordingly.

As a next step, also the boundary conditions and the initial condition have to be transformed analogously
to yield an equivalent of the whole problem.
Clearly, $\pi(\xi,\tau) = 0$ whenever $p(\xi,\tau) = 0$.
Thus, $\pi(\xi,\tau)$ fulfills the boundary condition at $\xi=1$ trivially if $p(\xi,\tau)$ does so.
Because of
\begin{align}
\pd_\xi \pi(\xi,\tau)	&= \left[\pd_\xi \IFm \right] p(\xi,\tau) + \IFm \pd_\xi  p(\xi,\tau) \nn
\end{align}
we have
\begin{align}
 \pd_\xi \pi(\xi,\tau)\takenat_{\xi=0}	&= -\frac{\nu}{2} p(0,\tau) + \left[ \IFm \pd_\xi  p(\xi,\tau)\right]_{\xi=0} \nn\\
					&= -\frac{\nu}{2} p(0,\tau) + \frac{(v+k)L}{D} p(0,\tau) \nn\\
					&= \left[ \frac{\nu}{2} + \frac{kL}{D} \right] \pi(0,\tau) \nn
\end{align}
where in the last step we use ${p(0,\tau)=\pi(0,\tau)}$ and $\nu=\frac{vL}{D}$.

The initial condition becomes:
\begin{align}
 \pi(\xi,0)	&= \IFm p(\xi,0) = \IFm \frac{1}{L}\delta\left( \xi - \xi_0 \right) \nn
\end{align}
In the prefactor of the delta function $\xi$ only takes values other than $\xi_0$ 
when the delta function is zero, so we can set $\xi=\xi_0$ here.
This facilitates further calculations.

In summary, after multiplication with the integrating factor $\phi$ the initial problem 
for $p(\xi,\tau)$ is equivalent to the following problem for $\pi(\xi,\tau)$:
\begin{align}
\pd_\tau \pi(\xi,\tau)			&= \left[ \pd_\xi^2 - \frac{\nu^2}{4} \right] \pi(\xi,\tau)	\tag{PDE}\label{PDE-intfac}\\
\pd_\xi \pi(\xi,\tau)\takenat_{\xi=0}	&= \left[\frac{\nu}{2} + \frac{kL}{D}\right] \pi(0,\tau) 	\tag{BCr}\\
\pi(1,\tau)				&= 0 \tag{BCa} \\
\pi(\xi,0)				&= e^{-\frac{\nu}{2}\xi_0} \frac{1}{L} \delta\left( \xi - \xi_0 \right)	\tag{IC}
\label{eq-PDE-intfac}
\end{align}
This result reveals that the diffusion-drift problem is mathematically equivalent 
to a diffusion-decay problem with a slightly modified radiating boundary condition.
The strategy now is to solve (\ref{PDE-intfac}) for $\pi(\xi,t)$ 
and reconstruct the solution $p(\xi,t)$ afterwards using (\ref{eq-intfac-ansatz}).

\subsubsection*{Solving the PDE via Laplace transform}
Applying the \person{Laplace} transform by integrating $\int_0^\infty \pi(\xi,\tau) e^{-s\tau}d\tau \equiv \hat \pi(\xi, s)$ 
on both sides of (\ref{PDE-intfac}) yields:
\begin{align}
s\hat \pi(\xi, s) - e^{-\frac{\nu}{2}\xi_0}\frac{1}{L}\delta\left( \xi - \xi_0 \right)	&=	\left[ \pd^2_\xi - \frac{\nu^2}{4} \right] \hat \pi(\xi, s) \nn\\
\Leftrightarrow\quadquad \left[ \pd^2_\xi - \kappa^2 \right] \hat \pi(\xi, s)	&= -\phi_0 \delta\left( \xi - \xi_0 \right)	\label{PDE-laplace} 
\end{align}
where we abbreviate $\kappa^2 \equiv \frac{\nu^2}{4} + s \geq 0$ and $\phi_0 \equiv \frac{1}{L} e^{-\frac{\nu}{2}\xi_0} > 0$.
According to transformation rules, 
the time derivative $\pd_\tau \pi(\xi,\tau)$ converts to $\hat \pi(\xi,s) - \pi(\xi,\tau=0)$ in \person{Laplace} space.

To solve the transformed equation we first calculate the solution of the homogenous problem.
This will be used to obtain two different specific solutions on the two parts 
of the underlying space separated by the delta peak, i.e. $[0,\xi_0]$ and $[\xi_0, 1]$,
employing the boundary conditions and a continuity/discontinuity condition at $\xi=\xi_0$.
The general solution to the homogenous problem $\pd^2_\xi = \kappa^2 \hat \pi(\xi, s)$ 
can be written as $\hat \pi_h(\xi,s) = \alpha \sinh(\kappa x) + \beta \cosh(\kappa x)$.
We thus make an ansatz for each part of the interval $[0,1]$ as follows:
\begin{align}
 \hat \pi(\xi,s) &= \hat \pi_{-}(\xi,s) \equiv \alpha_{-} \sinh(\kappa \xi) + \beta_{-} \cosh(\kappa \xi)	&\text{for}\quad \xi < \xi_0	\label{ansatzL} \\
 \hat \pi(\xi,s) &= \hat \pi_{+}(\xi,s) \equiv \alpha_{+} \sinh(\kappa \xi) + \beta_{+} \cosh(\kappa \xi)	&\text{for}\quad \xi > \xi_0	\label{ansatzR}
\end{align}
with constant, yet arbitrary, real coefficients $\alpha_{+}$, $\beta_{+}$ and $\alpha_{-}$, $\beta_{-}$.
Let us first apply the absorbing boundary condition at $\xi=1$ to (\ref{ansatzR}):
\begin{align}
 \hat \pi_{+}(1) = 0 \quad&\Rightarrow& \alpha_{+} \sinh(\kappa) &= - \beta_{+} \cosh(\kappa)
\end{align}
where we neglect the unphysical solution $\alpha_{+}=0,\; \beta_{+}=0$.

Applying the transformed radiating boundary condition at $\xi=0$ to (\ref{ansatzL}) yields:
\begin{align}
 & \pd_\xi \hat \pi_{-}(\xi,s)\takenat_{\xi=0} = \braceunderset{\Omega}{\left[\frac{\nu}{2} + \frac{kL}{D}\right]} \hat \pi_{-}(0,s)	\nn\\
 & \Leftrightarrow	\left[\alpha_{-} \kappa \cosh(\kappa \xi) + \beta_{-} \kappa\sinh(\kappa \xi) \right]_{\xi=0}			\nn\\
 & \quadquad\quadquad\quad = \Omega \left[\alpha_{-} \sinh(\kappa \xi) + \beta_{-} \cosh(\kappa \xi) \right]_{\xi=0}						\nn\\
 & \Leftrightarrow	\kappa\alpha{-} = \Omega \beta_{-}
\end{align}
Reinsertion into (\ref{ansatzL}) and (\ref{ansatzR}) leads to:
\begin{align}
 \hat \pi_{-}(\xi,s)	&= \alpha_{-} \left( \sinh(\kappa \xi) + \frac{\kappa}{\Omega} \cosh(\kappa \xi) \right)	\label{reinsertL}\\
 \hat \pi_{+}(\xi,s)	&= \alpha_{+} \left( \sinh(\kappa \xi) - \tanh(\kappa) \cosh(\kappa \xi) \right)		\label{reinsertR}
\end{align}

In order to determine coefficients $\alpha_{+}$ and $\alpha_{-}$ we, firstly,
impose continuity of $\hat \pi(\xi,s)$ at $\xi=\xi_0$,
i.e. $\hat \pi_{-}(\xi_0,s) = \hat \pi_{+}(\xi_0,s)$ for all $s$.
Secondly, by integrating equation (\ref{PDE-laplace}) over $[\xi_0-\epsilon, \xi_0+\epsilon]$
and taking the limit $\epsilon\rightarrow0$, we obtain the following 
discontinuity condition for the left- and right-hand derivative $\pd_\xi \hat \pi(\xi,s)\takenat_{\xi=\xi_0}$:
\begin{align}
 & \int_{\xi_0-\epsilon}^{\xi_0+\epsilon} \left[ \pd^2_\xi \hat \pi(\xi, s)- \kappa^2 \hat \pi(\xi, s)\right] d\xi	= - \int_{\xi_0-\epsilon}^{\xi_0+\epsilon} \phi_0 \delta(\xi-\xi_0) d\xi	\quad\quad \Leftrightarrow \nn\\
 & \left[ \pd_\xi \hat \pi(\xi, s)\takenat_{\xi_0+\epsilon} - \pd_\xi \hat \pi(\xi, s)\takenat_{\xi_0-\epsilon} \right] -\kappa^2 \left[ \hat \Pi(\xi_0+\epsilon, s) - \hat \Pi(\xi_0-\epsilon, s) \right] = -\phi_0	\nn\\
 & \overset{\epsilon\rightarrow 0}{\Rightarrow} \quad\quad \pd_\xi \hat \pi_{+}(\xi, s)\takenat_{\xi_0} - \pd_\xi \hat \pi_{-}(\xi, s)\takenat_{\xi_0} = -\phi_0
\end{align}
The term $[ \hat \Pi(\xi_0+\epsilon, s) - \hat \Pi(\xi_0-\epsilon, s) ]$ vanishes for $\epsilon\rightarrow0$ 
because continuity of $\hat \pi(\xi,s)$ at $\xi=\xi_0$ implies continuity of the stem function 
$\hat \Pi(\xi,s)=\int \hat \pi(\xi,s) d\xi$ at this point.

Applying the two additional constraints to (\ref{reinsertL}) and (\ref{reinsertR}) determines, after some algebraic steps,
the coefficients $\alpha_{-}$ and $\alpha_{+}$:
\begin{align}
 \alpha_{-} &= \frac{-\phi_0}{\kappa \left( \frac{\kappa}{\Omega} + \tanh(\kappa) \right)}	\left( \sinh(\kappa \xi_0) - \tanh(\kappa) \cosh(\kappa \xi_0) \right)		\nn\\
 \alpha_{+} &= \frac{-\phi_0}{\kappa \left( \frac{\kappa}{\Omega} + \tanh(\kappa) \right)}	\left( \sinh(\kappa \xi_0) + \frac{\kappa}{\Omega}\cosh(\kappa \xi_0) \right)
\end{align}
Hence,
\begin{align}
 \hat \pi_{-}(\xi,s)	&= \frac{-\phi_0}{\kappa \left( \frac{\kappa}{\Omega} + \tanh(\kappa) \right)}	\times		\nn\\
			&\quad\quad	\left( \sinh(\kappa \xi) + \frac{\kappa}{\Omega} \cosh(\kappa \xi) \right)
				\left( \sinh(\kappa \xi_0) - \tanh(\kappa) \cosh(\kappa \xi_0) \right) 			\nn\\
 \hat \pi_{+}(\xi,s)	&= \frac{-\phi_0}{\kappa \left( \frac{\kappa}{\Omega} + \tanh(\kappa) \right)}	\times		\nn\\
			&\quad \quad	\left( \sinh(\kappa \xi_0) + \frac{\kappa}{\Omega}\cosh(\kappa \xi_0) \right)
				\left( \sinh(\kappa \xi) - \tanh(\kappa) \cosh(\kappa \xi) \right)			\nn\\	
\end{align}
or, after multiplying numerator and denominator by $\cosh(\kappa)$:
\begin{align}
 & \hat \pi_{-}(\xi,s) =	\nn\\
 &\quad\quad \frac{\phi_0}{\kappa}
	\frac{\left( \sinh(\kappa \xi) + \frac{\kappa}{\Omega} \cosh(\kappa \xi) \right) \left( \sinh(\kappa)\cosh(\kappa \xi_0) - \cosh(\kappa)\sinh(\kappa \xi_0) \right)}
	{\sinh(\kappa) + \frac{\kappa}{\Omega} \cosh(\kappa)}	\nn\\
 & \hat \pi_{+}(\xi,s) = 	\nn\\
 &\quad\quad \frac{\phi_0}{\kappa}
	\frac{\left( \sinh(\kappa \xi_0) + \frac{\kappa}{\Omega} \cosh(\kappa \xi_0) \right) \left( \sinh(\kappa)\cosh(\kappa \xi) - \cosh(\kappa)\sinh(\kappa \xi) \right)}
	{\sinh(\kappa) + \frac{\kappa}{\Omega} \cosh(\kappa)}
\end{align}
Here we shall not forget that $\kappa = \kappa(s) = \sqrt{s+\frac{\nu^2}{4}}$.

With this we have determined unique solutions to the diffusion-drift-reaction problem for the left 
($\xi\leq\xi_0$) and right ($\xi\geq\xi_0$) part of the spatial domain in \person{Laplace} space.
Now we can attempt the back transform to the time domain,
where we will find that the solution becomes symmetric in $\xi$ and $\xi_0$ again.

\subsubsection*{Inverse Laplace transform via residue formula}
Having $\hat \pi(\xi,s)$ we can obtain the corresponding function in the time domain via
the \textsc{Bromwich} / \textsc{Fourier}-\textsc{Mellin} integral: \index{Bromwich integral}
\begin{align}
 \pi(\xi,t) &= \mathcal{L}^{-1}\left[ \hat \pi(\xi,z) \right] = \frac{1}{2\pi i} \lim_{T\rightarrow\infty} \int_{\gamma-iT}^{\gamma+iT} \hat \pi(\xi,z) e^{zt} dz
\end{align}
Herein $\hat \pi(\xi,z)$ is the extension of $\hat \pi(\xi,s)$ to the complex plane.
The integration has to be performed on a line perpendicular to the real axis at the positive real value $\gamma$,
which must be greater than the real part of any singularity of the complex function $\hat \pi(\xi,z)$.
Usually this is a daunting task.
It is simplified a lot if $\hat \pi(\xi,z)$ is a holomorphic function.
In that case we can apply residue calculus to compute the line integral via a contour integral.
To that purpose we close the line path from $\gamma-iT$ to $\gamma+iT$ by a half-circle in the space left 
to it ($\lbrace z|\text{Re}(z)\leq \gamma\rbrace$) to obtain contour $\gamma'(T)$.
In the limit $T\rightarrow\infty$ the half-circle contribution vanishes, i.e.
$\lim_{T\rightarrow\infty} \oint_{\gamma'(T)} \hat \pi(\xi,z) dz = \lim_{T\rightarrow\infty} \int_{\gamma-iT}^{\gamma+iT} \hat \pi(\xi,z) dz$.
The residue formula states that the integral of $\hat \pi(\xi,z)$ along the contour is equal to the
sum of residues at the singularities of $\hat \pi(\xi,z)$ enclosed by the contour times $2\pi i$. \index{residue formula}
Thus, if all singularities of $\hat \pi(\xi,z)$ are to the left of $\gamma$, 
the inverse \person{Laplace} transform can be calculated as:
\begin{align}
 \pi(\xi,t) = \frac{1}{2\pi i} \lim_{T\rightarrow\infty} \oint_{\gamma'(T)} \hat \pi(\xi,z) e^{zt} dz = \sum_n \text{Res}_{\hat \pi(\xi,z)e^{zt}}(z_n)
\end{align}

Since $\hat \pi_{-}(\xi,s)$ and $\hat \pi_{+}(\xi,s)$ only differ by the fact that $\xi$ and $\xi_0$ are interchanged
it is sufficient to carry out the inverse transform for $\hat \pi_{+}(\xi,s)$ and obtain $\pi_{-}(\xi,t)$ by
substituting $\xi\leftrightarrow\xi_0$.

As a first step we reinsert $\kappa=\sqrt{s+\frac{\nu^2}{4}}$ and substitute
$s=z-\frac{\nu^2}{4}$ in $\hat \pi_{+}(\xi,s)$ where $z$ is a complex variable now:
\begin{align}
 \hat \pi_{+}(\xi,z) &= \frac{\phi_0}{\sqrt{z}}
	\frac{\left( \sinh(\xi_0 \sqrt{z}) + \frac{1}{\Omega} \sqrt{z} \cosh(\xi_0 \sqrt{z}) \right)}
	{\sinh(\sqrt{z}) + \frac{1}{\Omega} \sqrt{z} \cosh(\sqrt{z})}	\times \nn\\
 &\quadquad\triquad \left( \sinh(\sqrt{z})\cosh(\xi \sqrt{z}) - \cosh(\sqrt{z})\sinh(\xi \sqrt{z}) \right)
\end{align}
Later, instead of reinserting $z=s+\frac{\nu^2}{4}$, we will apply 
the standard \person{Laplace}-inversion rule $\mathcal{L}^{-1}[\hat f(s+c)] = e^{-ct}f(t)$ for $c=const$.

The fact that $\hat \pi_{+}(\xi,z)$ contains $\sqrt{z}$ which is non-holomorphic on the negative
real branch recommends testing holomorphicity of the function.
By \person{Taylor}-expanding the $\sinh(const\cdot\sqrt{z})$ and $\cosh(const\cdot\sqrt{z})$ functions
one can show that $\hat \pi_{+}(\xi,z)$ can be written as a sum over purely integer powers $z^n$ 
and thus is indeed a holomorphic (even entire) function.
Its complex roots $z_n$ are found by setting the denominator to zero, which yields:
\begin{align}
 & \tanh(\sqrt{z_n}) = -\frac{1}{\Omega}\sqrt{z_n}
\end{align}
It can be shown that all $z_n$ lie on the negative real axis and that there is no
singularity at $z=0$.
Since $\hat \pi_{+}(\xi,z)$ can be written as $\hat \pi_{+}(\xi,z)=g(z)/h(z)$ with
functions $g(z)$ and $h(z)$ that are holomorphic in the neighborhood of each $z_n$,
we may calculate the residue of $\hat \pi_{+}(\xi,z)$ at $z_n$ via the well-known formula
$\Res_{\hat \pi_{+}(\xi,z)e^{zt}}(z_n) = g(z_n)/h'(z_n)$.
Here we find, with $h(z)=\sinh(\sqrt{z}) + \frac{1}{\Omega} \sqrt{z} \cosh(\sqrt{z})$:
\begin{align} 
h'(z_n)	= \frac{1}{2\sqrt{z_n}} \left[ \left( 1 + \frac{1}{\Omega} - \frac{z_n}{\Omega^2} \right) \cosh(\sqrt{z_n}) \right]
\end{align}
In the next step we additionally substitute the square roots via $\sqrt{z_n}\equiv \pm i\zeta_n$
with $\zeta_n\in\mathbb{R}^+$.
Taking into account $\sinh(\pm ix)=\pm\sin(x)$ and $\cosh(\pm ix)=\cos(x)$ and cancelling multiplicative minus signs,
the solution in the time domain as a sum of residues reads:
\begin{align}
 & \pi_{+} (\xi,\xi_0,t) = \sum_n \Res_{\hat \pi_{+}(\xi,z)e^{zt}}(z_n)	\nn\\
 & = e^{-\frac{\nu^2}{4}t} \cdot 2\phi_0 \sum_n e^{z_nt} \frac{ \sinh(\xi z^\frac{1}{2}_n) + \frac{z^\frac{1}{2}_n}{\Omega} \cosh(\xi z^\frac{1}{2}_n) }
			 { \left( 1+\frac{1}{\Omega}-\frac{z_n}{\Omega^2} \right) \cosh(z^\frac{1}{2}_n) }		\nn\\
 & \quadquad\quadquad\quad \times \left( \sinh(z^\frac{1}{2}_n)\cosh(\xi_0 z^\frac{1}{2}_n) - \cosh(z^\frac{1}{2}_n)\sinh(\xi_0 z^\frac{1}{2}_n) \right)	\nn\\
 & = -\frac{2}{L} e^{-\frac{\nu^2}{4}t-\frac{\nu}{2}\xi_0} \cdot \sum_n e^{-\zeta_n^2 t} \frac{ \sin(\xi \zeta_n) + \frac{\zeta_n}{\Omega} \cos(\xi \zeta_n) }
			 { \left( 1+\frac{1}{\Omega}+\frac{\zeta_n^2}{\Omega^2} \right) \cos(\zeta_n) }		\nn\\
 & \quadquad\quadquad\quadquad \times \left( \sin(\zeta_n)\cos(\xi_0 \zeta_n) - \cos(\zeta_n)\sin(\xi_0 \zeta_n) \right)	\nn\\
\end{align}
where the $n$-summation goes over the (positive) roots of the implicit equation
\begin{align}
  \label{eq-GF-1D-drift-roots}
  \tanh(\sqrt{z_n}) = -\frac{1}{\Omega} \sqrt{z_n} \quad\Leftrightarrow\quad \tan(\zeta_n) = -\frac{1}{\Omega} \zeta_n
\end{align}
with $\Omega=\frac{\nu}{2}+\frac{kL}{D}=\left(\frac{v}{2}+k\right)\frac{L}{D}$.
Using the root equation (\ref{eq-GF-1D-drift-roots}) we can further simplify
\begin{align}
 \frac{\left( \sin(\zeta_n)\cos(\xi_0 \zeta_n) - \cos(\zeta_n)\sin(\xi_0 \zeta_n) \right) }{\cos(\zeta_n)} &=	\nn\\
 \tan(\zeta_n)\cos(\xi_0 \zeta_n) - \sin(\xi_0 \zeta_n) &= -\left( \sin(\xi_0 \zeta_n) + \frac{\zeta_n}{\Omega}\cos(\xi_0 \zeta_n) \right)
\end{align}
and realize that the denominator of the summation terms in the time domain is completely symmetric in $\xi$ and $\xi_0$.
Hence, the solution in the time domain is invariant to interchanging $\xi$ and $\xi_0$.
After taking into account the integrating factor via $p(\xi,\xi_0,t) = e^{\frac{\nu}{2}\xi}\pi(\xi,\xi_0,t)$
and reverting dedimensionalization we can write the final solution for both sides of the spatial domain as:\\
\\
\fbox{ \begin{minipage}{0.96\textwidth}
\vspace{-1em}
\begingroup
\begin{align}
 p_{\rm RA}(x,x_0,t) & \equiv  			\nn\\
 p(x,x_0,t) & = \frac{2}{L} e^{\frac{v}{2D}(x-x_0)-\frac{v^2}{4D}t}
  \sum_n e^{\frac{-\zeta_n^2 Dt}{L^2}} 
  \frac{
      F_n(x) F_n(x_0)
  }
  {
      \Omega^2+\Omega+\zeta_n^2
  }
\label{eq-GF-1D-drift}
\end{align}
with
\begin{align}
 F_n(x) & \equiv \Omega \sin(\zeta_n \frac{x-\sigma}{L}) + \zeta_n \cos(\zeta_n \frac{x-\sigma}{L}) \\
 \Omega  &= \left(\frac{v}{2}+k\right)\frac{L}{D}		\quadquad \text{and}	\label{eq-GF-1D-drift-Omega} \\
 \zeta_n &\quad\text{positive roots of}\quad\quad	\tan(\zeta_n) = -\frac{1}{\Omega} \zeta_n	\label{eq-GF-1D-drift-root-eq}
\end{align}
\endgroup
\end{minipage} }\\

It can be easily verified that this function fulfills the imposed boundary conditions.
Also the initial condition at $t=0$ is recovered, which can be seen by expanding
the delta function into the orthogonal functions $F_n(x)$
and utilizing the straightforwardly proven orthogonality relation:
\begin{align}
\int_\sigma^a F_n(x)F_m(x) dx = \left\lbrace \begin{array}{cr} \frac{L}{2}\left( \zeta_n^2+\Omega^2 + \Omega \right) &\text{for}\quad n=m \\
						       0						     &\text{for}\quad n\neq m
                                     \end{array}\right.
\end{align}

In the limit $v\rightarrow0$ the solution reproduces the well-known solution for the case without drift,
which can be found in \cite[14.3II, p. 360]{CarslawJaeger}.
To verify this, set $\zeta_n=\alpha_n L$ in (\ref{eq-GF-1D-drift}) and $k_1=1$, $k_2=0$, $h_1=\frac{k}{D}$, $h_2=1$
in the reference formula.

Exemplary time evolution plots of $p_{\rm RA}(x,x_0,t)$ are shown in Figure \ref{fig:GF1DwDrift}
for different values of diffusion coefficient $D$, drift velocity $v$ and intrinsic reaction rate $k$.

\newcommand{\GFGraphHeight}{0.3\textwidth}
\begin{figure}[p!]
  \centering
  \subfigure[][]{    
    \includegraphics[height=\GFGraphHeight]{\PlotsSIDir/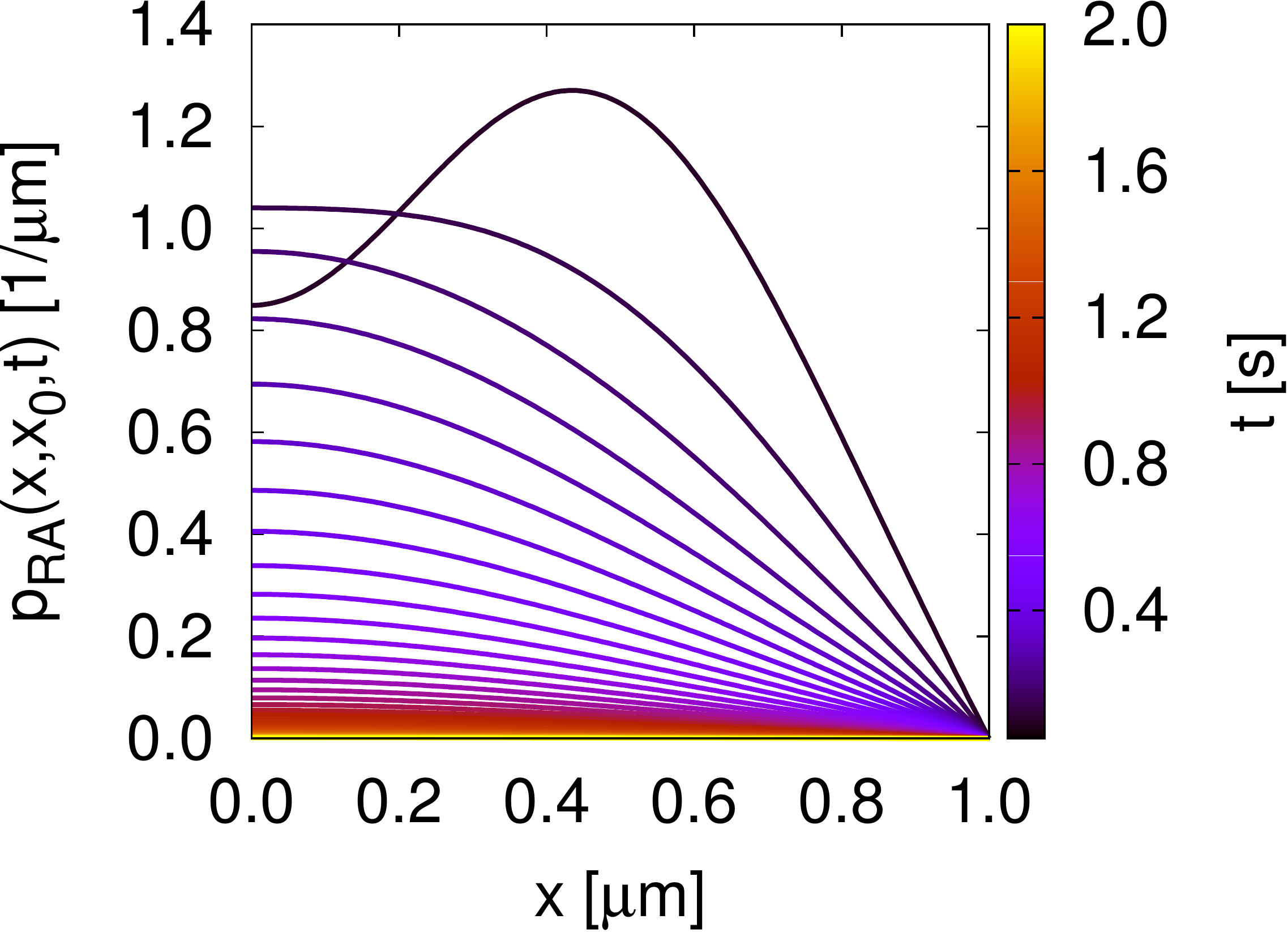}
    \label{fig:GF1DwDrift-1}
  } \hspace{0.05\textwidth}
  \subfigure[][]{    
    \includegraphics[height=\GFGraphHeight]{\PlotsSIDir/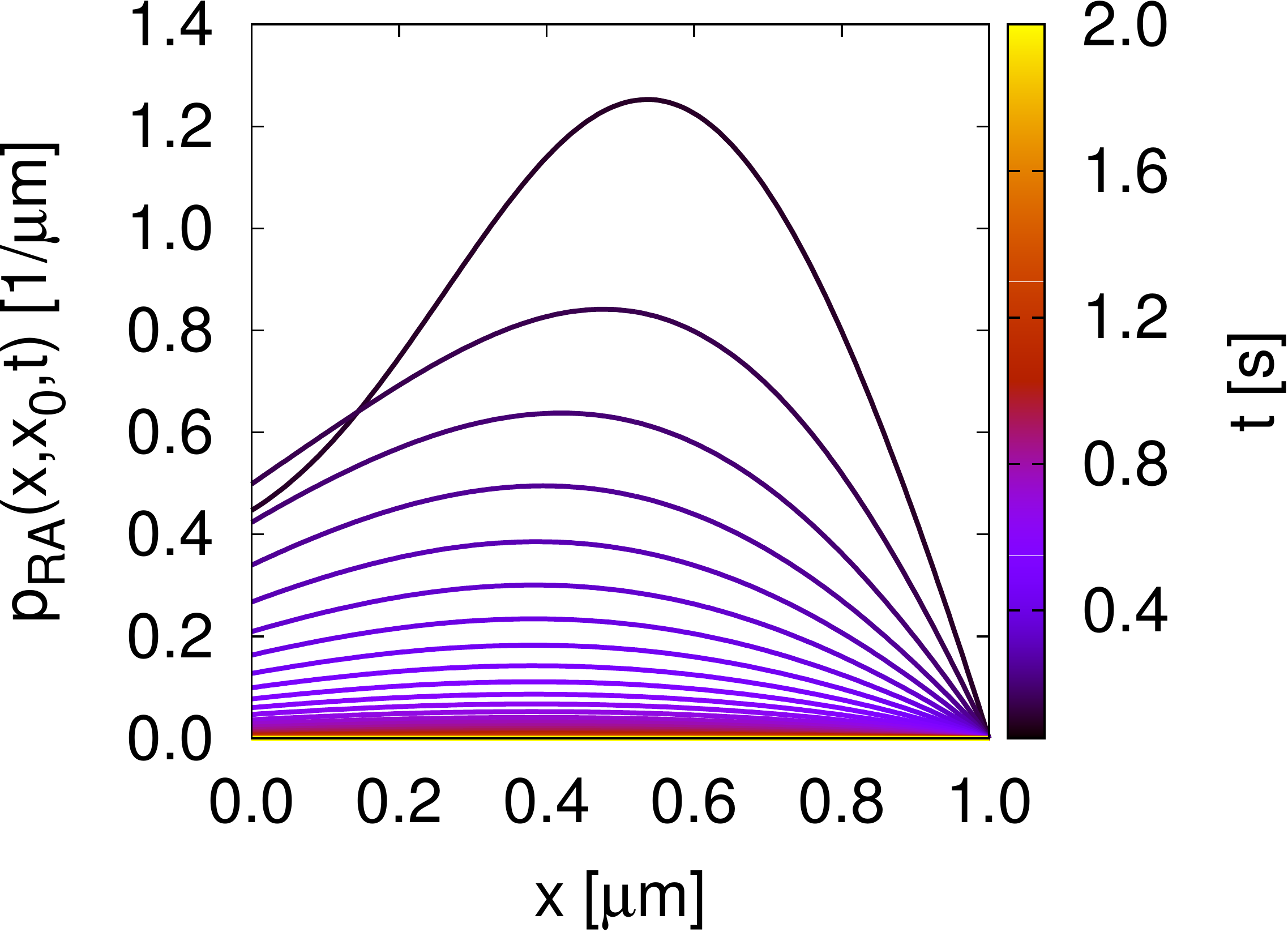}
    \label{fig:GF1DwDrift-2}
  }\\
  \subfigure[][]{    
    \includegraphics[height=\GFGraphHeight]{\PlotsSIDir/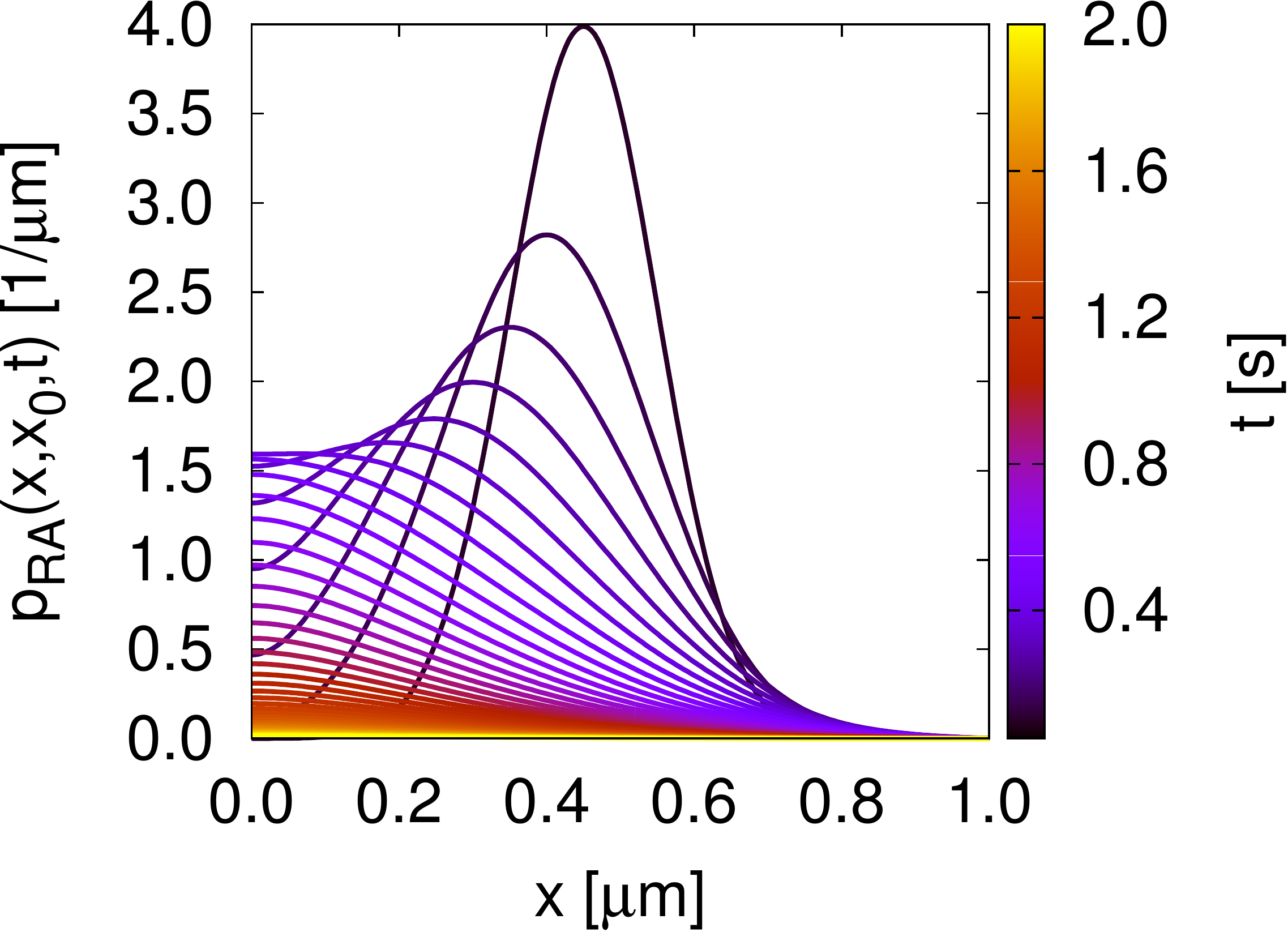}
    \label{fig:GF1DwDrift-3}
  } \hspace{0.05\textwidth}
  \subfigure[][]{    
    \includegraphics[height=\GFGraphHeight]{\PlotsSIDir/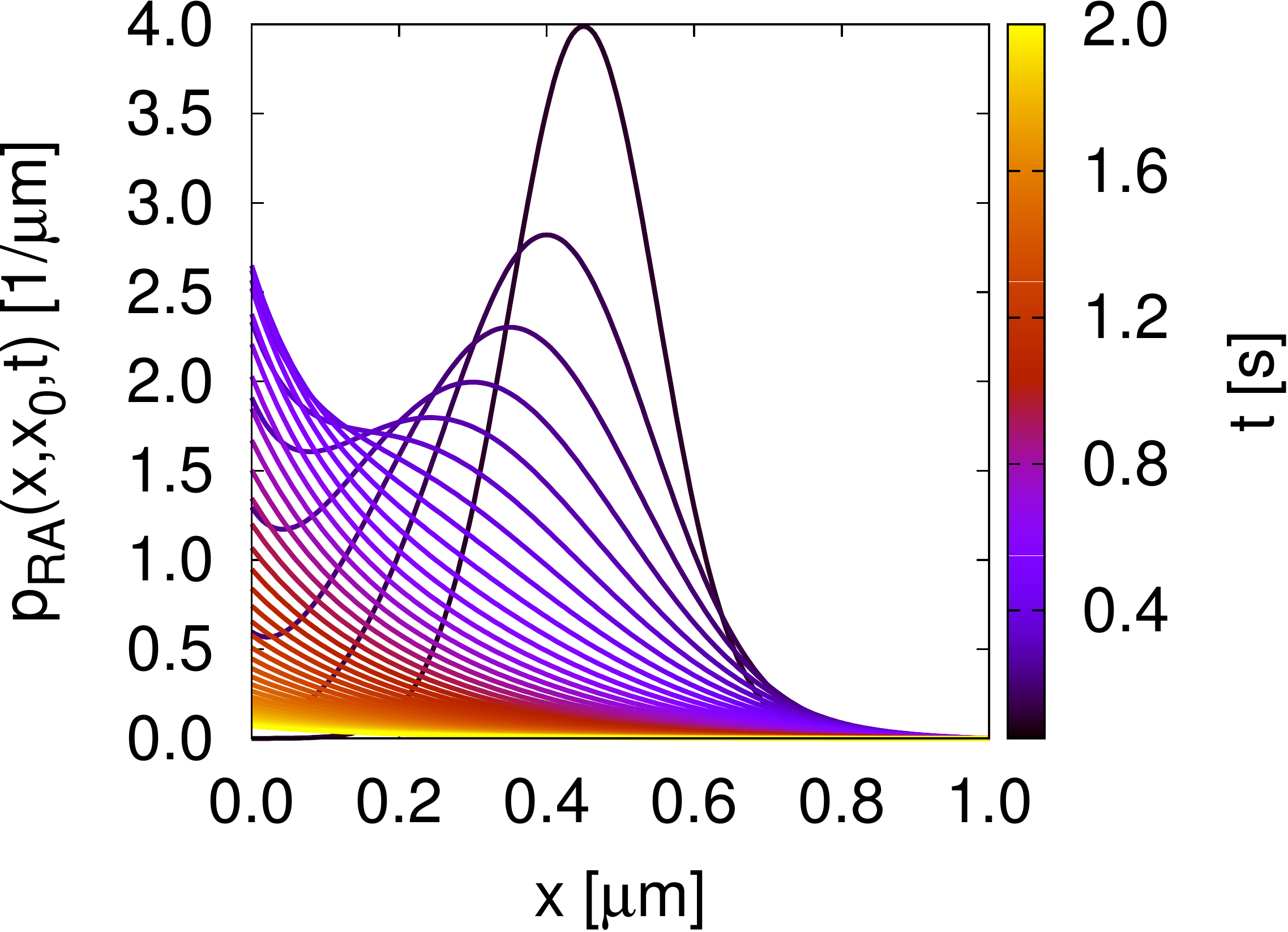}
    \label{fig:GF1DwDrift-4}
  }\\
  \subfigure[][]{    
    \includegraphics[height=\GFGraphHeight]{\PlotsSIDir/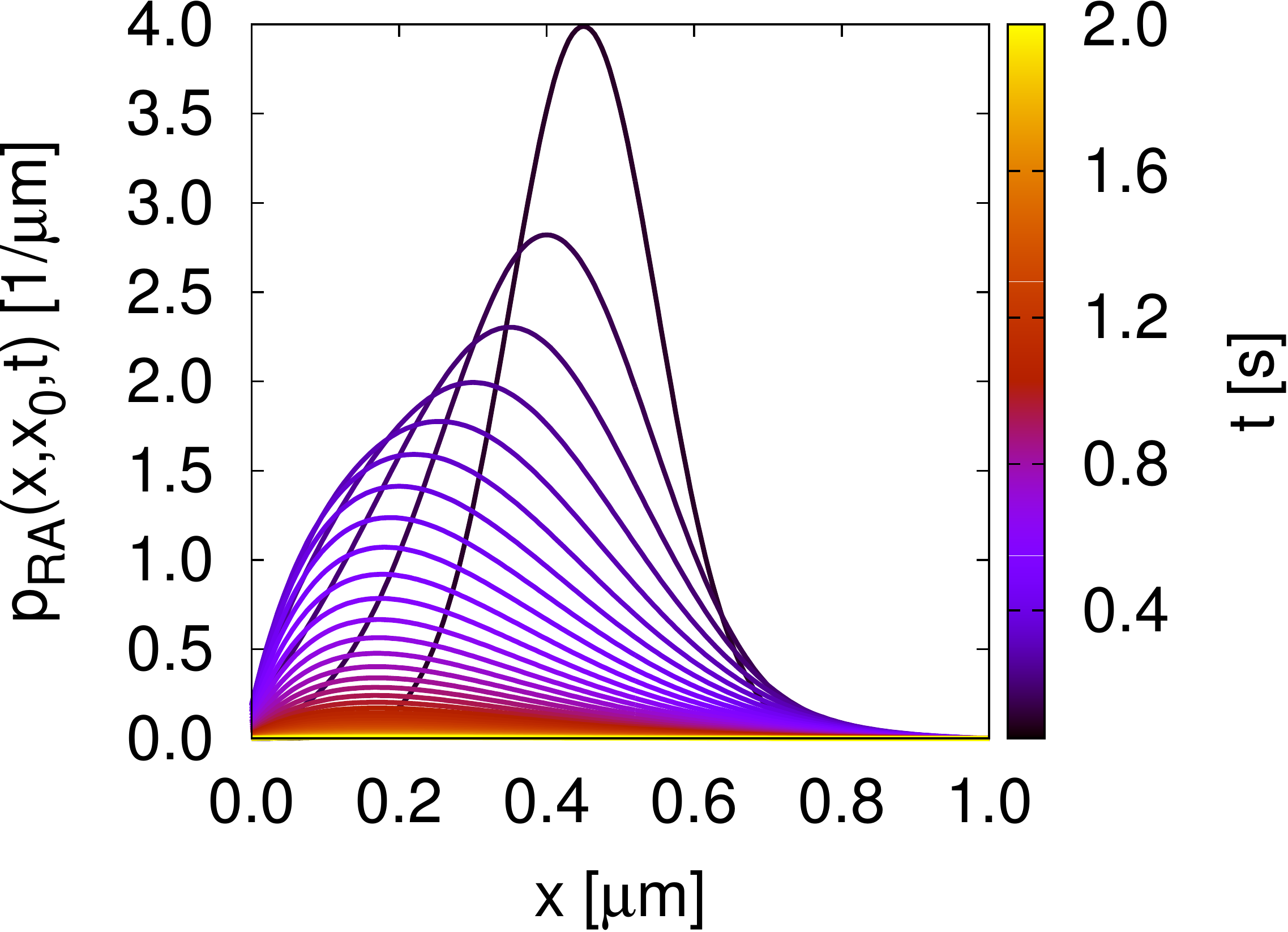}
    \label{fig:GF1DwDrift-5}
  } \hspace{0.05\textwidth}
  \subfigure[][]{    
    \includegraphics[height=\GFGraphHeight]{\PlotsSIDir/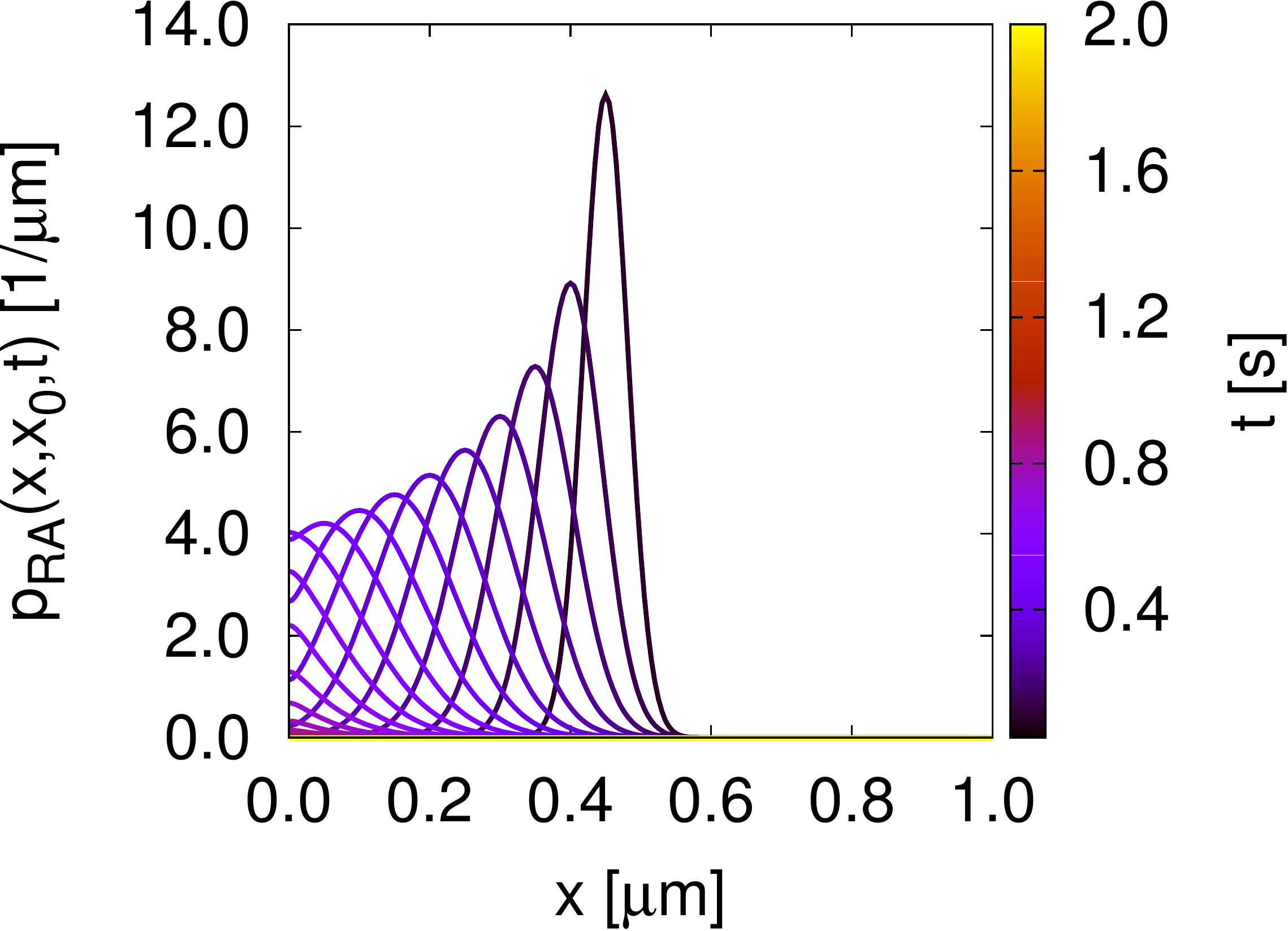}
    \label{fig:GF1DwDrift-6}
  }
\caption{ \label{fig:GF1DwDrift}
  \textbf{Green's function for the 1D diffusion-drift problem with radiating ($x=0$) and absorbing ($x=1$) boundary.}
  \subfigref{\subref{fig:GF1DwDrift-1}} $D=1.0$, $v=-1.0$, $k=|v|$;
  \subfigref{\subref{fig:GF1DwDrift-2}} $D=1.0$, $v=+1.0$, $k=v$;
  \subfigref{\subref{fig:GF1DwDrift-3}} $D=0.1$, $v=-1.0$, $k=|v|$;
  \subfigref{\subref{fig:GF1DwDrift-4}} $D=0.1$, $v=-1.0$, $k=\frac{|v|}{2}$;
  \subfigref{\subref{fig:GF1DwDrift-5}} $D=0.1$, $v=-1.0$, $k=10|v|$;
  \subfigref{\subref{fig:GF1DwDrift-6}} $D=0.01$, $v=-1.0$, $k=|v|$.
  Values are in $[\unit{\frac{\mu m^{(2)}}{s}}]$.
  $x_0=0.5~\unit{\mu m}$.
}
\end{figure}

\subsubsection*{Green's function for 1D-diffusion with drift, Abs-Abs case}
\label{sec:1D-GreensFunction-Abs-Abs}
\newcommand{\convergesto}{\;\underset{k\rightarrow\infty}{\longrightarrow}\;}

From (\ref{eq-GF-1D-drift}) we can easily obtain the Green's function for 1D-diffusion with drift
on a finite domain with two absorbing boundaries by taking the limit $k\rightarrow\infty$.
The originally radiating boundary condition at $x=\sigma$ (\ref{eq-PDE-1D-BCr}) then becomes
\begin{align}
p(\sigma,t) = \frac{D \pd_x p(x,t)\takenat_{x=\sigma} - vp(\sigma,t)}{k} \convergesto 0	\quad .
\end{align}

First notice that because of $\Omega = \left(\frac{v}{2}+k\right)\frac{L}{D} \convergesto 0$
the root equation (\ref{eq-GF-1D-drift-root-eq}) in the limit $k\rightarrow\infty$ reads:
\begin{align}
 &\tan(\zeta_n) = -\frac{1}{\Omega} \zeta_n \convergesto 0	\nn\\
 &\Rightarrow \quad \zeta_n = n\pi,\quad n\in\mathbb{N} \quad\text{for}\quad k\rightarrow\infty
\end{align}

To obtain the limit of (\ref{eq-GF-1D-drift}) we multiply the numerator
and denominator of the summation terms by $1/k^2$.
Because of
\begin{align}
 \frac{\Omega}{k} = \left(1+\frac{v}{2k}\right)\frac{L}{D} &\convergesto \frac{L}{D}
\end{align}
we have
\begin{align}
 \frac{\Omega}{k} \sin\left(\zeta_n \frac{x-\sigma}{L}\right) + \frac{\zeta_n}{k} \cos\left(\zeta_n \frac{x-\sigma}{L}\right) &\convergesto \frac{L}{D} \sin\left(\zeta_n \frac{x-\sigma}{L}\right)	\\
 \frac{\Omega^2}{k^2} + \braceunderset{\rightarrow 0}{\frac{\Omega}{k^2}+\frac{\zeta_n^2}{k^2}} &\convergesto \frac{L^2}{D^2}
\end{align}
and therefore:\\
\\
\fbox{ \begin{minipage}{0.96\textwidth}
\vspace{-1em}
\begin{align}
 & p_{\rm AA}(x,x_0,t) = p_{\rm RA}(x,x_0,t)\takenat_{k\rightarrow\infty} 	\nn\\
 & \quad = \frac{2}{L} e^{\frac{v}{2D}(x-x_0)-\frac{v^2}{4D}t} \sum_n e^{-\left(\frac{n\pi}{L^2}\right)^2 Dt} \sin\left(n\pi \frac{x-\sigma}{L}\right) \sin\left(n\pi \frac{x_0-\sigma}{L}\right)
\end{align}
\end{minipage} }

\subsubsection*{Green's function for 1D-diffusion with drift, Rad-Inf case}
\label{sec:1D-GreensFunction-Rad-Inf}
For completeness, we also mention here the Green's functions for the corresponding half-bounded problems.
A solution for the case with only one radiating boundary at $x=0$ and constant drift was already published 
by \textsc{Lamm} and \textsc{Schulten} \cite{Lamm1983} and reads:
\begin{align}
  & p_{R\infty}(x,t|x_0,t=0) \nn\\
  & = \frac{1}{\sqrt{4\pi D t}} \left( e^{ -\frac{(x-x_0-vt)^2}{4Dt} } + e^{ -\frac{vx_0}{D} } e^{ -\frac{(x+x_0-vt)^2}{4Dt} } \right)	\nn\\
  & \quad\quad -\frac{v/2+k}{D} e^{\frac{vx_0}{D}} e^{\frac{k}{D} \left[ (x+x_0) + (k + v) t \right]} \,\erfc\left(\frac{x+x_0}{\sqrt{4Dt}} + \frac{v/2+k}{D}\sqrt{Dt} \right)
\end{align}

\subsubsection*{Green's function for 1D-diffusion with drift, Abs-Inf case}
\label{sec:1D-GreensFunction-Abs-Inf}
For the situation with only one absorbing boundary at $x=a\geq0$ the Green's function for
the 1D-diffusion-drift problem can be straightforwardly obtained via the method
of images:
Since any linear combination of the free solution $p_{\rm free}(x,t|x_0)$ fulfills the
diffusion-drift equation,
we can easily construct a solution $p_{A\infty}(x,t|x_0)$
that will obey $p_{A\infty}(a,t|x_0)=0\;\forall t$ by subtracting from the free solution
the antisymmetric solution for a particle starting from a distance $x_0 - a$ 
to the left of the boundary and with inverted drift $v\rightarrow-v$:
\begin{align}
 p_{A\infty}(x,t|x_0) &= p_{\rm free,v_+}(x,t|x_0) - p_{\rm free,v_-}(x,t|a-(x_0-a))	\nn\\
  & = \frac{1}{\sqrt{4\pi Dt}} \left( e^{-\frac{1}{4Dt}\left[ (x-x_0)-vt \right]^2} - e^{-\frac{1}{4Dt}\left[ (x-2a+x_0)+vt \right]^2} \right)
\end{align}
We verified that the above solution is equivalent
to the solution for this problem calculated explicitly by applying the boundary
conditions in \person{Laplace} space and inverting via the Residue formula,
following the workflow described in section \ref{sec:1D-GreensFunction-Abs-Abs}.

\subsubsection*{Survival probabilities}
The survival probability is calculated by integration of the Green's function over the whole
interval on which it is defined:
\begin{align}
 S(t) = \int\limits_a^b p(x,t|x_0) dx
\end{align}
For the half-bounded solutions $b=\infty$.
The cumulative distribution function $P(x,t|x_0)=\int_a^x p(x',t|x_0) dx'$ is needed
besides $S(t)$ to sample positions at next-event times $\tau_\nu$.
It is therefore convenient to first calculate $P(x,t|x_0)$ and then $S(t)=P(b,t|x_0)$ as a special case.

The Green's functions presented in this section all have the the form
\begin{align}
 & p(x,t|x_0) = C_0 e^{\frac{v(x-x_0)}{2D}} \nicesum{n}{} c_n f_n(x)
\end{align}
where $C_0$ and $c_n$ do not depend on $x$ and $f_n(x)$ are either 
trigonometric or \textsc{Gauss} functions.
To calculate $P(x,t|x_0)$ the integration is most conveniently performed term-wise,
i.e. by computing
$\int_a^x e^{\frac{v(x'-x_0)}{2D}} f_n(x') dx'$
with the help of partial integration and reassembling the sum.
Differentiation of the survival probability gives the propensity function
$q(t)\equiv -\partial_t S(t)$.
These are all straightforward calculations and therefore omitted here.

\subsubsection*{Boundary fluxes}
With drift $v\neq 0$ the probability flux at position $x=x'$
is calculated from the Green's function $p(x,t|x_0)$ as follows:
\begin{align}
 \label{eq-1D-flux-with-drift}
 & q_{x'}(t) = -D\pd_x p(x,t|x_0)\takenat_{x=x'} + vp(x',t|x_0)
\end{align}
Note that for an absorbing boundary at $x'=a$ we have $p(a,t|x_0)=0$
and the drift-dependent term vanishes.

\pagebreak[4]
For a radiating boundary at $x'=\sigma$ with intrinsic reaction rate $k$
it is more convenient to calculate the flux directly from
\begin{align}
 & q_{\sigma}(t) = k p(\sigma,t|x_0)
\end{align}
which equals (\ref{eq-1D-flux-with-drift}) with $x'=\sigma$ by construction of the problem.

For the Green's functions introduced in this chapter these expressions
are again easily calculated and therefore not shown here.
A complete collection of the surivival probabilities, cumulative distribution functions and boundary fluxes
for all Green's functions presented in this chapter is available as part of the eGFRD technical documentation.

\newcommand{\coeffA}{\alpha}
\newcommand{\coeffB}{\beta}
\renewcommand{\coeffC}{\gamma}
\renewcommand{\coeffD}{\delta}
\newcommand{\costerm}{\cos\left(m(\phi-\phi_0)\right)}

\subsection{Green's function for the 2D diffusion-reaction problem}
\label{sec:2D-GreensFunction}
In this section we describe the derivation of the Green's function in polar coordinates $\myvec r=(r,\phi)$ 
for a particle starting at an arbitrary position $\myvec r_0=(r_0,\phi_0)$ within an annular region
bounded by a radiating inner and absorbing outer boundary.
This Green's function is required for next-event sampling in the \domaintype{Planar Surface Pair} domain (section \ref{sec:2d-GF-plane} of the main text),
but also for the \domaintype{Cylindrical Surface Interaction} domain (section \ref{sec:eGFRD-binding-to-cyl}).

We assume here that the problem of two particles that interact on a plane has been transformed correctly
into a diffusion problem for their center-of-mass vector $\myvec{R}$ and a diffusion-reaction problem for their
interparticle vector $\myvec{r}=(r,\phi)$, with a radiating boundary at particle contact, i.e. $r=|\myvec{r}|=\sigma=R_A+R_B$,
and an absorbing boundary at $r=a$.
While the problem for $\myvec R$ is solved by the Green's function presented in section \ref{sec:2D-GF-AbsSym},
the spatio-temporal evolution of $\myvec{r}$ is goverened by the following diffusion equation in polar coordinates
\begin{align}
 \pd_t p_{\myvec r}(r,\phi,t|r_0,\phi_0) &= D_r\nabla_{\myvec r}^2 p_{\myvec r}(r,\phi,t|r_0,\phi_0)			\nn\\
 &= D_r \left[ \pd^2_r + \frac{1}{r}\pd_r + \frac{1}{r^2}\pd^2_\phi \right] p_{\myvec r}(r,\phi,t|r_0,\phi_0)	\tag{PDE} \label{PDE-2D}
\end{align}
subject to boundary conditions
\begin{align}
 2\pi\sigma D_r \pd_r p_{\myvec r}(r,\phi,t|r_0,\phi_0)\takenat_{r=\sigma} 	&= k p_{\myvec r}(|\myvec r|=\sigma|r_0,\phi_0)	\tag{BCr} \label{BC-Rad}\\
 p_{\myvec r}(r,\phi,t|r_0,\phi_0)\takenat_{r=a}				&= 0							\tag{BCa} \label{BC-Abs}
\end{align}
and initial condition
\begin{align}
 p_{\myvec r}(r,\phi,t=0|r_0,\phi_0)	&= \frac{1}{r} \delta(r-r_0)\delta(\phi-\phi_0)						\tag{IC} \label{IC}
\end{align}
where $k$ is the intrinsic particle reaction rate.

\subsubsection*{Solution in Laplace space}
The above boundary value problem is again solved most conveniently in \person{Laplace} space.
Applying the \person{Laplace} transform $\hat p_{\myvec r}(r,\phi,s|r_0,\phi_0) \equiv \int_{-\infty}^{\infty} p_{\myvec r}(r,\phi,s|r_0,\phi_0) e^{-st} dt$
on both sides of the equations yields:
\begin{align}
& s \hat p_{\myvec r}(r,\phi,s|r_0,\phi_0) - p_{\myvec r}(r,\phi,t=0|r_0,\phi_0) = D_r \nabla_{\myvec r}^2 \hat p_{\myvec r}(r,\phi,s|r_0,\phi_0)	\quad \Leftrightarrow \nn\\
& \left[ \pd^2_r + \frac{1}{r}\pd_r + \frac{1}{r^2}\pd^2_\phi - \frac{s}{D_r}\right] \hat p_{\myvec r}(r,\phi,s|r_0,\phi_0) 
	     = \frac{-1}{D_r r} \delta(r-r_0)\delta(\phi-\phi_0)	\tag{PDE} \label{PDE-2D-LS}\\
& 2\pi\sigma D_r \pd_r \hat p_{\myvec r}(r,\phi,s|r_0,\phi_0)\takenat_{r=\sigma} 	= k \hat p_{\myvec r}(\myvec r =\sigma|r_0,\phi_0)	\tag{BCr}\\
& \hat p_{\myvec r}(r,\phi,s|r_0,\phi_0)\takenat_{r=a}		= 0								\tag{BCa}
\end{align}

As usual we first attempt to find a general solution to the homogenous problem corresponding to (\ref{PDE-2D-LS}) and 
specialize it afterwards by applying the initial and boundary conditions.
Let us set $s/D_r\equiv q^2 \geq 0$. The homogenous problem then reads:
\begin{align}
 \left[ \pd^2_r + \frac{1}{r}\pd_r + \frac{1}{r^2}\pd^2_\phi - q^2\right] \hat p_{\myvec{r},h}(r,\phi,q) = 0	\label{PDE-2D-LS-hom}
\end{align}
Via the separation ansatz $\hat p_{\myvec{r},h}(r,\phi,q) = R(r)\Phi(\phi)$ one can show that the above PDE is equivalent
to the following two differential equations coupled by a positive parameter $m^2$:
\begin{align}
 \pd^2_\phi \Phi(\phi) = -m^2 \Phi(\phi)	\label{PDE-2D-LS-Phi}	\\
 \left[ r^2\pd^2_r + r\pd_r - (r^2q^2 - m^2)\right] R(r)	\label{PDE-2D-LS-R}
\end{align}
The solution to (\ref{PDE-2D-LS-Phi}) is readily obtained as $\Phi(\phi)=\coeffA \cos(m(\phi-\phi_0))$
where we exploit that $\Phi(\phi)$ must be an even function because the operator $\pd^2_\phi$ conserves
the symmetry of $\delta(\phi-\phi_0)$. $\coeffA$ is a yet undetermined real constant.
With $rq\equiv\rho$ equation (\ref{PDE-2D-LS-R}) is equivalent to the modified \textsc{Bessel} equation:	\index{Bessel equation} 
\begin{align}
 \left[ \rho^2\pd^2_\rho + r\pd_\rho - (\rho^2 - m^2)\right] R(\rho)
\end{align}
which is solved by any linear combination of the modified \textsc{Bessel} functions				\index{Bessel function} \index{modified Bessel function}
$R(\rho)=\coeffB I_m(\rho) + \coeffC K_m(\rho)$.
The solution to (\ref{PDE-2D-LS-hom}) thus reads
\begin{align}
& \hat p_{\myvec{r},h}(r,\phi,q)	\nn\\
& \qquad = (\coeffA\coeffB) \costerm I_m(\rho) + (\coeffA\coeffC) \costerm K_m(\rho)	\nn\\
& \qquad \equiv A \costerm I_m(qr) + B \costerm K_m(qr)	\label{PDE-2D-LS-hom-sol}
\end{align}
with constants $A$ and $B$.

We now can construct an ansatz for the inhomogenous problem.
For further calculation it is convenient to write the ansatz as
\begin{align}
  \hat p_{\myvec{r}} = \hat p_{f} + \hat p_{c}
\end{align}
where $\hat p_{f}(\myvec{r},q|\myvec{r}_0)=\frac{1}{2\pi D_r}K_0(q(\myvec{r}\cdot\myvec{r}_0))$
is the ``free'' solution to the unbounded 2D diffusion problem for a point particle starting
at $(r_0,\phi_0)$, written in \person{Laplace} space, and $\hat p_{c}$ a correction resulting from the boundaries.
Although $\hat p_f$ fulfills the initial condition by construction, this does not automatically apply
to the entire ansatz and must be separately proven later on.
Since until now $m^2$ is an arbitrary constant, we shall construct the ansatz as a sum over all possible $m$.
For $\hat p_c$ we thus write
\begin{align}
 \hat p_c(r,\phi,q|r_0,\phi_0) = \nicesum{m=-\infty}{\infty} \costerm \left[ A_m I_m(qr) + B_m K_m(qr) \right]
\end{align}
with real coefficients $A_m$ and $B_m$.
Now it is also convenient to expand $\hat p_f$ into functions that resemble (\ref{PDE-2D-LS-hom-sol}),
using a formula from \cite[p. 365]{CarslawJaeger, Watson1962}:
\begin{align}
\hat p_{f}(\myvec{r},q|\myvec{r}_0) = \left\lbrace \begin{array}{lr}
                                      \frac{1}{2\pi D_r} \nicesum{m=-\infty}{\infty} \costerm I_m(qr)K_m(qr_0)\;,	& r < r_0	\\
				      \frac{1}{2\pi D_r} \nicesum{m=-\infty}{\infty} \costerm I_m(qr_0)K_m(qr)\;,	& r > r_0
                                     \end{array} \right.
\end{align}
This yields the combined ansatz:
\begin{align}
& \hat p_{\myvec{r}}(r,\phi,q|r_0,\phi_0)		\nn\\
& \qquad = \left\lbrace
   \begin{array}{ll}
     \frac{1}{2\pi D_r} \nicesum{m=-\infty}{\infty}& \costerm \times\\
						   & \quad \left[ I_m(qr)K_m(qr_0) + A_m I_m(qr) + B_m K_m(qr) \right],\\
     &\\
     \multicolumn{2}{r}{\text{for}\; r < r_0} \\
     \frac{1}{2\pi D_r} \nicesum{m=-\infty}{\infty}& \costerm \times\\
						   & \quad \left[ I_m(qr_0)K_m(qr) + A_m I_m(qr) + B_m K_m(qr) \right],\\
     &\\
     \multicolumn{2}{r}{\text{for}\; r > r_0} \\
   \end{array}
   \right.
   \nn\\
\end{align}
By applying the boundary conditions at $r=\sigma$ and $r=a$ term-wise for each $m$ we find, after some algebraic steps
\begin{align}
 A_m =&\; K_m(qa) 	 \frac{\mathcal{I}_m(q)K_m(qr_0) - \mathcal{K}_m(q)I_m(qr_0)}  {\mathcal{K}_m(q)I_m(qa) - \mathcal{I}_m(q)K_m(qa)}	\nn\\
 B_m =&\; \mathcal{I}_m(q) \frac{I_m(qr_0)K_m(qa) - K_m(qr_0)I_m(qa)}  {\mathcal{K}_m(q)I_m(qa) - \mathcal{I}_m(q)K_m(qa)}
\end{align}
where we abbreviated:
\begin{align}
 \mathcal{I}_m(q) &= \kappa I_m(q\sigma) + qI'_m(q\sigma) \nn\\
 \mathcal{K}_m(q) &= \kappa K_m(q\sigma) + qK'_m(q\sigma)
\end{align}
With these coefficients the particular solution to the initial problem is completely determined in \person{Laplace} space.

\subsubsection*{Inverse Laplace transform}
We may transform the solution back into the time domain as usual by calculating the \textsc{Bromwich} integral \index{Bromwich integral}
\begin{align} 
 p_{\myvec{r}}(r,\phi,t|r_0,\phi_0) &= \lim\limits_{T\rightarrow\infty} \int\limits_{\gamma-iT}^{\gamma+iT} \hat p_{\myvec{r}}(r,\phi,s|r_0,\phi_0) e^{st} ds	\nn\\
				  &= \lim\limits_{T\rightarrow\infty} \int\limits_{\gamma-iT}^{\gamma+iT} \hat p_{\myvec{r}}(r,\phi,q|r_0,\phi_0) e^{q^2D_rt} (2D_rq)dq
\end{align}
where the integration as usual occurs on a line through a real constant $\gamma$ located 
to the right of all singularities of
function $\hat p_{\myvec{r}}(r,\phi,q|r_0,\phi_0)$ extended to the complex plane.
Assuming convergence of the sum, we can perform the integration term-wise.
This somewhat lengthy calculation shall be omitted here, and we only state the final result on the next page. \TODO{rework}
\\
\fbox{ \begin{minipage}{0.96\textwidth}
\vspace{-1em}
\begingroup
\begin{align}
& p_{\myvec r}(r,\phi,t|r_0,\phi_0) =	\nn\\
& \triquad \frac{\pi}{4} \nicesum{m=-\infty}{\infty} \costerm \nicesum{n=1}{\infty} \frac{ \zeta^2_{mn} \mathcal{R}^2_{m}(\zeta_{mn}) \mathcal{J}_{mn}(r)\mathcal{J}_{mn}(r_0) }
										      { \mathcal{R}^2_{m}(\zeta_{mn}) - \zeta^2_{mn} - \kappa^2 - \frac{m^2}{\sigma^2} } e^{-\zeta^2_{mn}D_rt}
\label{PDE-2D-sol}
\end{align}
where $\zeta_{mn}$ are the roots of the implicit equation
\begin{align}
  \frac{\kappa J_m(\sigma\zeta_{mn}) - \zeta_{mn}J'_m(\sigma\zeta_{mn})}{J_m(a\zeta_{mn})} 
  &= \frac{\kappa Y_m(\sigma\zeta_{mn}) - \zeta_{mn}Y'_m(\sigma\zeta_{mn})}{Y_m(a\zeta_{mn})}	\nn\\
  &\equiv \mathcal{R}_{m}(\zeta_{mn})
\end{align}
with $\kappa=\frac{k}{D_r}$ and
\begin{align}
 \mathcal{J}_{mn}(r) \equiv J_m(r\zeta_{mn})Y_m(a\zeta_{mn}) - Y_m(r\zeta_{mn})J_m(a\zeta_{mn})
\end{align}
\endgroup
\end{minipage} }
\\

As demonstrated in \cite{Bossen2010}, the above function fulfills the diffusion equation (\ref{PDE-2D}),
the boundary conditions (\ref{BC-Rad}) and (\ref{BC-Abs}) and the initial condition (\ref{IC}).

By separating out the $m=0$ term from the sum
and unifying the summation over $m<0$ and $m>0$ with the help of cosine and \textsc{Bessel} function
(anti)symmetry relations, the Green's function (\ref{PDE-2D-sol}) can be rewritten into a form that
proves more convenient for further usage:
\begin{align}
& p_{\myvec r}(r,\phi,t|r_0,\phi_0) =	\nn\\
& \quad \frac{\pi}{4} \nicesum{n=0}{\infty} \frac{ \zeta^2_{0n} \mathcal{R}^2_{0}(\zeta_{0n}) \mathcal{J}_{0n}(r)\mathcal{J}_{0n}(r_0) }
						 { \mathcal{R}^2_{0}(\zeta_{0n}) - \zeta^2_{0n} - \kappa^2} e^{-\zeta^2_{0n}D_rt} 		\nn\\
& \quad + \frac{\pi}{2} \nicesum{m=1}{\infty} \nicesum{n=1}{\infty} \costerm \frac{ \zeta^2_{mn} \mathcal{R}^2_{m}(\zeta_{mn}) \mathcal{J}_{mn}(r)\mathcal{J}_{mn}(r_0) }
										  { \mathcal{R}^2_{m}(\zeta_{mn}) - \zeta^2_{mn} - \kappa^2 - \frac{m^2}{\sigma^2} } e^{-\zeta^2_{mn}D_rt}
\label{PDE-2D-sol-simp}
\end{align}
The advantage of this form of the Green's function is that here the double-sum term
vanishes under the $\int_0^{2\pi} d\phi$ integral,
which significantly facilitates the calculation of the survival probability and the boundary fluxes.

\subsection*{Survival probability}
The survival probability $S(t)$ here is obtained by integrating the Green's function over
the entire circular domain on which the 2D diffusion takes place:
\begin{align}
 S_{\myvec r}(t) = \int\limits_\sigma^a \int\limits_0^{2\pi} p_{\myvec r}(r,\phi,t|r_0,\phi_0) r d\phi dr
\end{align}
This is done most conveniently starting from (\ref{PDE-2D-sol-simp}), where the part with the sum over $m>0$
disappears under the $\phi$-integral because of $\int_0^{2\pi} \costerm d\phi = 0$.
Hence, after interchanging the order of summation and integration:
\begin{align}
\label{PDE-2D-Jder}
& S_{\myvec r}(t) = 2\pi \cdot \frac{\pi}{4} \nicesum{n=0}{\infty} e^{-\zeta^2_{0n}D_rt}
					    \frac{ \zeta^2_{0n} \mathcal{R}^2_{0}(\zeta_{0n}) \mathcal{J}_{0n}(r_0) }
						 { \mathcal{R}^2_{0}(\zeta_{0n}) - \zeta^2_{0n} - \kappa^2} \int\limits_\sigma^a r\mathcal{J}_{0n}(r) dr	\nn\\
& = \frac{\pi^2}{2} \nicesum{n=0}{\infty} e^{-\zeta^2_{0n}D_rt} 
					  \frac{ \mathcal{R}^2_{0}(\zeta_{0n}) \mathcal{J}_{0n}(r_0) }{ \mathcal{R}^2_{0}(\zeta_{0n}) - \zeta^2_{0n} - \kappa^2}
					  \left[ \frac{2}{\pi} + \sigma \mathcal{J}'_{0n}(\sigma) \right]	\\
& \text{with}\;\; \mathcal{J}'_{0n}(r) = \frac{d}{dr} \mathcal{J}_{0n}(r) = \zeta_{0n} \left[ J_1(r\zeta_{0n})Y_0(a\zeta_{0n}) - Y_1(r\zeta_{0n})J_0(a\zeta_{0n}) \right]	\nn
\end{align}
To calculate the integral we have employed the well-known relations
 $\int rJ_0(r) dr = rJ_1(r)$ and $J'_1(r)=-J_0(r)$ which analogously apply to $Y_0(r)$.
The $\frac{2}{\pi}$ term within the brackets originates from $\mathcal{J}'_{0n}(a) = -\frac{2}{\pi a}$
which can be shown with the help of the boundary conditions.

\subsection*{Boundary fluxes}
We can calculate the probability fluxes through the radiating ($q_{\myvec r,\sigma}$) and absorbing ($q_{\myvec r,a}$)
domain boundaries at time $t$ by integrating the probability density gradient over the two
circular contours that constitute the boundaries:
\begin{align}
 q_{\myvec r,\sigma}(t) 	&= \int_0^{2\pi} +D_r \pd_r p_{\myvec r}(r,\phi,t|r_0,\phi_0)\takenat_{r=\sigma} rd\phi \nn\\
 q_{\myvec r,a}(t) 	&= \int_0^{2\pi} -D_r \pd_r p_{\myvec r}(r,\phi,t|r_0,\phi_0)\takenat_{r=a} rd\phi 
\end{align}
Here the signs account for opposite flux directions.
In the modified form of the Green's function (\ref{PDE-2D-sol-simp}) again the $m\neq0$ terms vanish
under the $\phi$-integral.
The only $r$-dependent part of the $m=0$ term is $\mathcal{J}_{0n}(r)$,
the derivative of which we have calculated in (\ref{PDE-2D-Jder}).
With this we arrive at:
\begin{align}
& q_{\myvec r,\sigma}(t) 	= 2\pi D_r \cdot \frac{\pi}{4} \nicesum{n=0}{\infty} e^{-\zeta^2_{0n}D_rt}
					    \frac{ \zeta^2_{0n} \mathcal{R}^2_{0}(\zeta_{0n}) \mathcal{J}_{0n}(r_0) }
						 { \mathcal{R}^2_{0}(\zeta_{0n}) - \zeta^2_{0n} - \kappa^2} \sigma\mathcal{J}'_{0n}(\sigma)	\nn\\
& q_{\myvec r,a}(t) 	= 2\pi D_r \cdot \frac{\pi}{4} \nicesum{n=0}{\infty} e^{-\zeta^2_{0n}D_rt}
					    \frac{ \zeta^2_{0n} \mathcal{R}^2_{0}(\zeta_{0n}) \mathcal{J}_{0n}(r_0) }
						 { \mathcal{R}^2_{0}(\zeta_{0n}) - \zeta^2_{0n} - \kappa^2} \braceunderset{\frac{2}{\pi}}{(-a\mathcal{J}'_{0n}(a))}	\nn\\
\end{align}


\subsection{Green's function with absorbing outer boundary in polar coordinates}
\label{sec:2D-GF-AbsSym}
Here we sketch the calculation of the Green's function in polar coordinates $\myvec r=(r,\phi)$
for a diffusing particle starting at radius $r=0$ with a symmetric absorbing boundary at a radial distance $r=a$;
this function is used to sample next-event times and new positions in the 
\domaintype{Planar Surface Interaction} domain (section \ref{sec:2D-Binding}), 
\domaintype{Planar Surface Single} and \domaintype{Planar Surface Pair} domains (section \ref{sec:2d-GF-plane}),
and the transition domains derived from the latter two (section \ref{sec:Plane-Transitions}).

Since we assume perfect radial symmetry here, the Green's function $p_{\rm s}(r,t)$ does not depend on the
angular coordinate $\phi$, so that the corresponding boundary value problem can be written as follows:
\begin{align}
 & \pd_t p_{\rm s}(r,t) = D \left[ \frac{1}{r}\pd_r \left( r \pd_r\right) \right] p_{\rm s}(r,t)	\\
 & p_{\rm s}(r,t_0=0) = \frac{1}{2\pi r}\delta(r)	\\
 & p_{\rm s}(a,t)=0
\end{align}

\pagebreak[4]
The solution to this problem is well-known \cite[p. 368f]{CarslawJaeger} 
and, for the above initial condition, reads\\
\fbox{
\begin{minipage}{0.96\textwidth}
\vspace{-1em}
\begin{align}
 p_{\rm s}(r,t|r_0=0) = \frac{1}{\pi a^2}\nicesum{n=1}{\infty} e^{-\rho_n^2 D t} \frac{J_0(r\rho_n)}{J_1^2(a\rho_n)}
\end{align}
where $J_0$ and $J_1$ are regular \textsc{Bessel} functions and $\rho_n$ the roots of the equation:
\begin{align}
  J_0(a\rho_n)=0
\end{align}
\end{minipage} }\\
\\
The corresponding survival probability $S_{\rm s}(t)$ and (radial) cumulative PDF $P_{\rm s}(r,t)$ 
follow by integration of $p_{\rm s}(r,t)$ over the considered circular domain:
\begin{align}
 S_{\rm s}(t) &= \int\limits_0^a \int\limits_0^{2\pi} p(r,t|r_0) rd\phi dr
	= \frac{2}{a} \nicesum{n=1}{\infty} e^{-\rho_n^2 D t} \frac{1}{\rho_nJ_1(a\rho_n)}	\\
 P_{\rm s}(r,t) &= \int\limits_0^r \int\limits_0^{2\pi} p(r',t|r_0) r'd\phi dr'
	= \frac{2}{a^2} \nicesum{n=1}{\infty} e^{-\rho_n^2 D t} \frac{rJ_1(r\rho_n)}{\rho_n J_1^2(a\rho_n)}
\end{align}
Here the standard formula $\int_0^r r' J_0(r') dr' = r J_1(r)$ is used.

Next-event times $\tau_\nu$ are sampled from $p_{\rm s}(r,t)$ in the usual way by comparing a uniform
random number from $[0,1]$ with $S_{\rm s}(t)$ via the inversion method.
For an arbitrary time $\tau$, a new radius $r_\nu(\tau)$ is obtained from $\frac{1}{S_{\rm s}(\tau)}P_{\rm s}(r,\tau)$,
whereas a new angle $\phi_\nu(\tau)$ is sampled from the uniform distribution on $[0,2\pi]$.
If $\tau=\tau_\nu$, we directly set $r_\nu(\tau_\nu) = a$.


\TODO{SUBSECTION: Do we want to state the 3d GFs here once more, for completeness?}

\clearpage
\section{Special domains}
\label{sec:Special-Domains}
\subsection{Plane-plane transition domains} 
\begin{figure}[hp!]
  \centering
  \subfigure[][]{
    \label{fig:Domain-PlanarSurfaceTransition}
    \includegraphics[width=\DomainTypeSketchWidth]{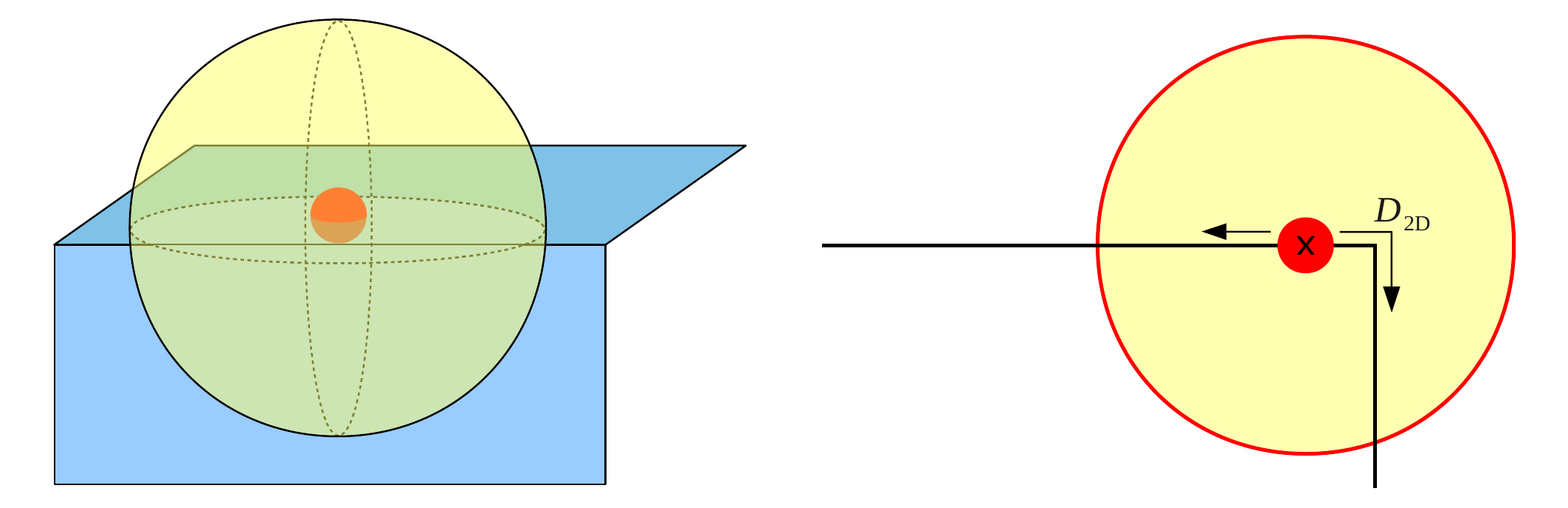}
  }
  \subfigure[][]{
    \label{fig:Domain-PlanarSurfaceTransitionPair}
    \includegraphics[width=\DomainTypeSketchWidth]{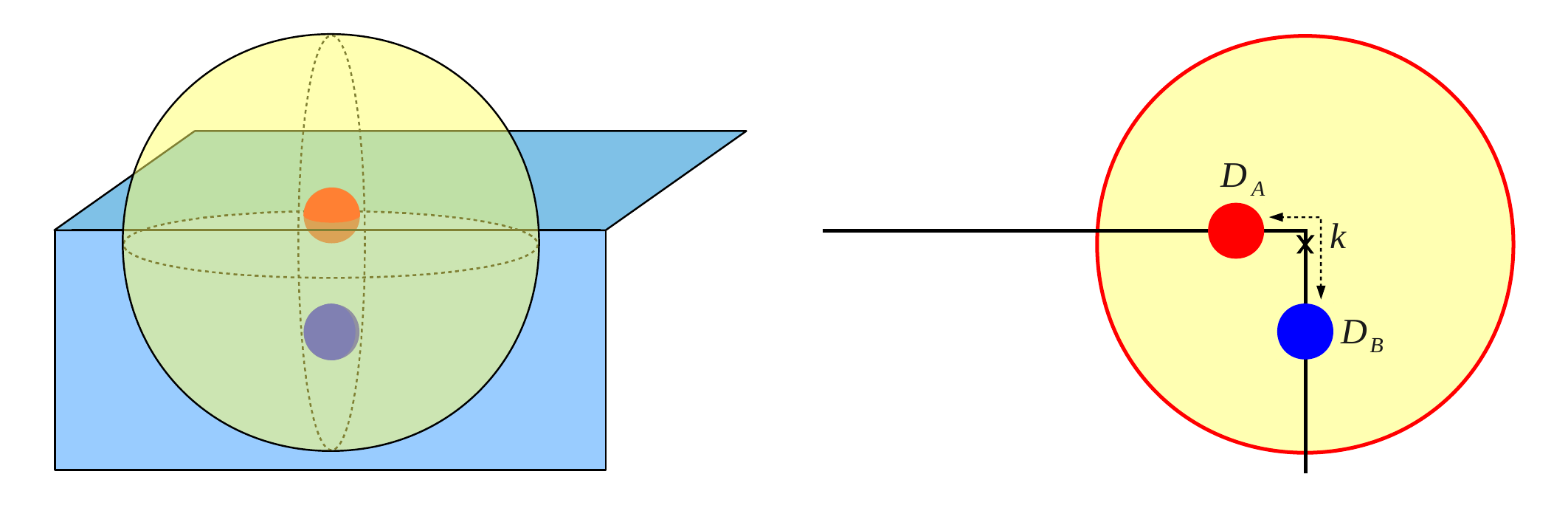}
  } 
  \subfigure[][]{
    \label{fig:Domain-MixedPair2D3D}
    \includegraphics[width=\DomainTypeSketchWidth]{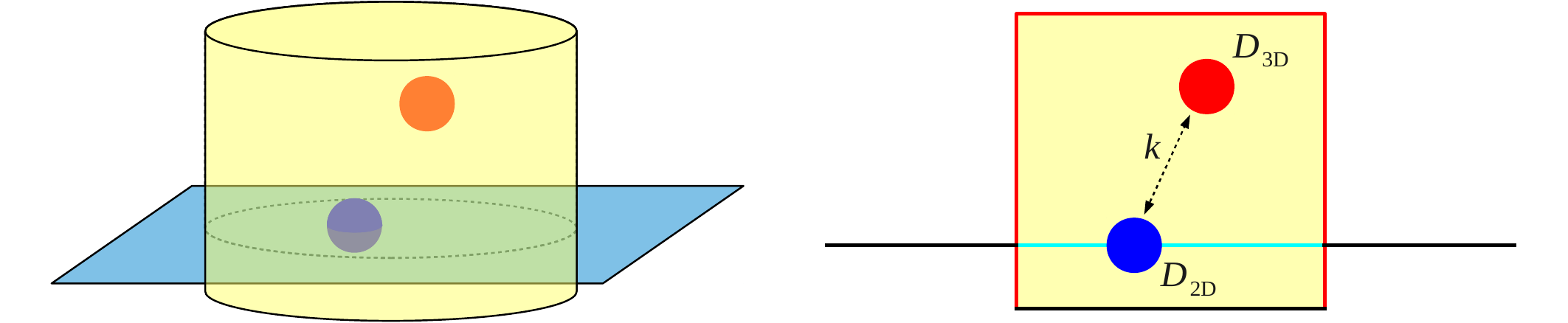}
  }  
\caption{ \label{fig:NewDomains-2D-Special}
  \textbf{Special protective domains for particles on planar surfaces.}  
  \subfigref{\subref{fig:Domain-PlanarSurfaceTransition}} \domaintype{Planar Surface Transition} domain;
  \subfigref{\subref{fig:Domain-PlanarSurfaceTransitionPair}} \domaintype{Planar Surface Transition Pair} domain;
  \subfigref{\subref{fig:Domain-MixedPair2D3D}} \domaintype{Mixed Pair 2D-3D} domain. \TODO{Make this an own figure}
  Right panels show sections of 3D objects.
  Here, absorbing boundaries are highlighted by red, reflective boundaries by cyan color.
  The center of spherical domains is indicated by a black cross.
}
\end{figure}

\index{Planar Surface Transition}
\label{sec:Plane-Transitions}
When introducing the new domain types for membrane-interaction and -diffusion
we implicitly assumed that particles interact with a single unbounded planar surface.
An extension towards bounded planes does not require significant
changes: domains associated to the plane must simply be constructed such 
that they do not reach out of it.
In contrast, the implementation of particle transitions between two bordering orthogonal planes
within the box-arrangement that we describe in the main text \TODO{where?} requires new domain types.
We imagine that the two connected planes--in an abstract fashion--represent a continuous part of
the membrane.
This means that the edge does not constitute an obstacle for the diffusing particle;
when it reaches the edge its movement is instantly redirected into the orthogonal 
direction imposed by the bordering plane.
Under the assumption that this holds for each diffusive trajectory,
we devised the following procedure for transitions of a particle between orthogonal planes:
First we construct a spherical \domaintype{``Planar Surface Transition''} domain
in a way that it contains the particle on the surface of origin and an empty region on 
the target surface. The domain is centered around the original particle position $\myvec r_0$.
This construction is shown in Figure \ref{fig:NewDomains-2D-Special}\subref{fig:Domain-PlanarSurfaceTransition}.
As a second step a next-event time $\tau_\nu$ and a new position $\myvec r_\nu$ is sampled
in the same way as for the \domaintype{Planar Surface Single}, where the
radius of the absorbing outer circle is equal to the radius of the spherical transition domain.
If $\myvec r_\nu$ is inside the finite plane of origin, 
the particle is moved to that point at $\tau_\nu$.
If, in contrast, $\myvec r_\nu$ lies beyond the boundaries of the original plane,
the new position is deflected onto the orthogonal target plane;
this is done by rotating the part of the displacement vector $\Delta \myvec r = \myvec r_\nu - \myvec r_0$ 
that reaches out of the original plane about the line that marks the edge
between the planes by an angle of $\pi/2$.
Details of this simple geometrical transform are described in the subsequent section~\ref{sec:GFRD-Deflection}.

The above principle can be straightforwardly extended to the case of a pair of particles that reside
on different neighboring planes and interact ``around the edge''.
Let us assume that particle A is located at position $\myvec r_A$ on plane A and
and particle B at $\myvec r_B$ on plane B.
Here first position $\myvec r_B$ is transformed into plane A via the inverse of the deflection transform (see sec.~\ref{sec:GFRD-Deflection}).
Then a next-event time and new particle positions are determined in plane A,
following the procedure for the \domaintype{Planar Surface Pair}.
Finally, new positions that lie beyond the boundaries of plane A are transformed into plane B.
Also the construction of the protective domain is slightly different as compared to the case with one particle:
We encapsulate the pair constellation with a spherical \domaintype{``Planar Surface Transition Pair''} domain 
centered around the weighted center-of-mass $\myvec R$ of the particles, 
as shown in Figure \ref{fig:NewDomains-2D-Special}\subref{fig:Domain-PlanarSurfaceTransitionPair}.
Note that here vector $\myvec R$ is calculated as follows:
First $\myvec r_B$ is transformed into plane A, yielding $\myvec r'_B$,
and the weighted center-of-mass $\myvec R'$ is computed in plane A from $\myvec r'_B$ and $\myvec r_A$.
If $\myvec R'$ is within the bounds of plane A, we set $\myvec R=\myvec R'$;
otherwise $\myvec R$ is obtained by deflecting $\myvec R'$ back into plane B.
The latter case is shown in the example situation in Figure \ref{fig:NewDomains-2D-Special}\subref{fig:Domain-PlanarSurfaceTransitionPair}.
\index{Planar Surface Transition Pair}

Special treatment is required in the rare event that two particles end up very close to each other
in the proximity of an edge between two planes.
This may happen due to a single reaction, in which the products are put at contact with a random angle,
but also when a \domaintype{Planar Surface Transition Pair} is bursted and its two particles happen
to end up close to each other.
In these cases configurations are possible in which one of the particles reaches out of the plane,
but application of the deflection transform would lead to particle overlap, 
because the transform shortens the effective distance between the two particles.
Therefore particles are slightly moved apart in such situations, introducing a minor error.

\subsubsection{Deflection of particle trajectories at the edge between orthogonal planes}
\label{sec:GFRD-Deflection}
We will now describe the mathematical procedure that ``deflects'' the new position $\myvec r_\nu$ 
of a plane-bound particle towards an adjacent, orthogonal ``target'' plane when $\myvec r_\nu$ reaches 
beyond the boundaries of the original plane.

Imagine that the particle originally was located at position $\myvec r_0$.
To transform the trajectory of the diffusing particle towards the target plane
first we calculate the point $\myvec S$ at which the edge between the two planes 
intersects with the line $\myvec r_0 + \lambda \Delta \myvec r$ that links $\myvec r_0$ and the new position $\myvec r_\nu$.
Let $\uvec u_x$ and $\uvec u_z$ be the unit vectors that define the orientation of the target plane
and $\uvec u_z \equiv \uvec u_x \times \uvec u_y$ the corresponding normal vector.
Since $\myvec S$ lies both on the line $\myvec r_0 + \lambda \Delta \myvec r$ and in the target plane,
it must obey
\begin{align}
 & \myvec S \cdot \uvec u_z = (\myvec r_0 + \lambda_S \Delta \myvec r) \cdot \uvec u_z = \myvec C \cdot \uvec u_z
\end{align}
where $\myvec C$ is the center point of the target plane\footnote{Note that instead of $C$, alternatively we could choose any point located in the target plane.}.
With this we find
\begin{align}
 & \myvec S = \myvec r_0 + \lambda_S \Delta\myvec r	\qquad\text{with}\qquad
   \lambda_S = \frac{(\myvec C - \myvec r_0)\cdot \uvec u_z}{\Delta\myvec r \cdot \uvec u_z}
\end{align}
and the protruding part of displacement vector $\Delta\myvec r$:
\begin{align}
 & \Delta\myvec r' = (1-\lambda_S) \Delta\myvec r
\end{align}
Instead of applying a rotation transform to $\Delta\myvec r'$,
here it is more convenient to construct the deflected position $\myvec r_\nu'$ directly via
\begin{align}
 & \myvec r_\nu' = \myvec S + \Delta r_\|' \uvec u_\| + \Delta r_{\bot}' \uvec u_\bot
\end{align}
where $\Delta r_\|' = \Delta\myvec r'\cdot \uvec u_\|$ is the component of $\Delta\myvec r'$ parallel to the edge
and $\Delta r_\bot' = \Delta\myvec r'\cdot \uvec u_z$ its component perpendicular to the edge.
$\uvec u_\|$ is the target plane's unit vector parallel to the edge, whereas $\uvec u_\bot$ is the target plane's second 
unit vector, which is perpendicular to both $\uvec u_\|$ and $\uvec u_z$.
How precisely $\uvec u_\|$ and $\uvec u_\bot$ map onto the two unit vectors $\uvec u_x$ and $\uvec u_y$ that define 
the plane depends on the direction from which the particle enters the target plane.
To avoid recalculation at each edge crossing, this information is stored in a neighborhood table 
when the box structure is constructed.
It is easily proven that the deflected position is ensured to stay within the circular domain.

For the inverse transform we note that $\myvec S$ is obtained by projecting
$\myvec r_\nu'$ onto the original plane.
With this $\myvec r_\nu$ is easily constructed via:
\begin{align}
 \myvec r = \myvec r_0 + |\myvec r_\nu' - \myvec S| \cdot \frac{\myvec S - \myvec r_0}{|\myvec S - \myvec r_0|}
\end{align}

\subsection{Special domains on finite cylindrical structures}
\label{sec:GFRD-FiniteCylinders}
Until here we have assumed an infinite length for the 
cylindrical structures that particles interact with and are transported on.
In many applications, however, we seek to study systems with finite 1D tracks,
such as microtubules that canalize transported cargo to one of their ends
preferentially, where it may unbind at a certain rate.
Moreover, particles can behave differently after reaching the ends of
microtubules by, for example, forming tip clusters \cite{Galjart2003,Sawin2000}.

In order to include these features into eGFRD we introduce a new structure
type, the ``disk structure'', which is used to mark a special interaction site
on a 1D cylindrical structure.
A disk may be placed at the ends of a cylindrical structure of finite length
to model reactive tip sites, or in any other place on the cylinder to model
a point of interaction, e.g. a transcription-factor binding pocket on DNA.
Within our framework a disk located at the end of a cylindrical structure
is called a ``cap'', a disk located in between the ends a ``sink''.
Particles on disks are immobilized.
We allow particles on the cylinder to bind with a certain
affinity both to disk structures and to particles already immobilized on caps.
Unbinding from a cap returns the particle into the bulk, whereby the particle
is moved in radial direction and placed at contact with the disk.
When unbinding from a sink, the particle transfers back to the cylindrical structure,
i.e. becomes mobile again.
Concerning implementation, particles bound to caps or sinks are treated as
individual species, which enables the definition of different parameters 
and reactions for cylinder-bound and disk-bound species.
This may be used, for example, to introduce cap-bound species representing particle clusters
that ``grow'' by successively absorbing particles from the cylinder via a cascade of reactions,
in order to model particle clustering at filament tips.
\index{disk} \index{cap}

Since the problem of a diffusing 1D particle that interacts with a disk
is mathematically equivalent to the problem of two interacting particles that move in 1D,
here we may re-use the 1D Green's function with drift (\ref{eq-GF-1D-drift}).

\begin{figure}[p!]
  \centering
  \subfigure[][]{
    \label{fig:Domain-CapInteraction}
    \includegraphics[width=\DomainTypeSketchWidth]{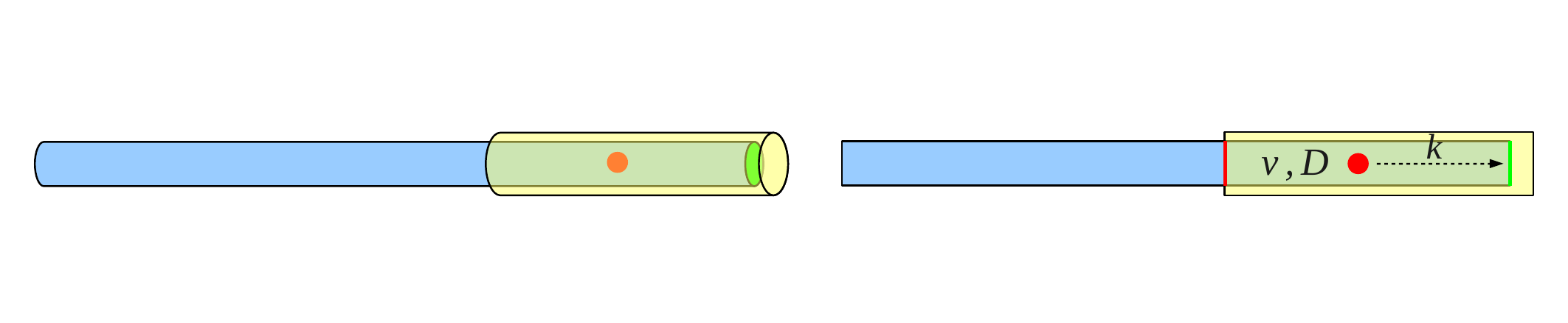}
  }
  \subfigure[][]{
    \label{fig:Domain-MixedPair1DCap}
    \includegraphics[width=\DomainTypeSketchWidth]{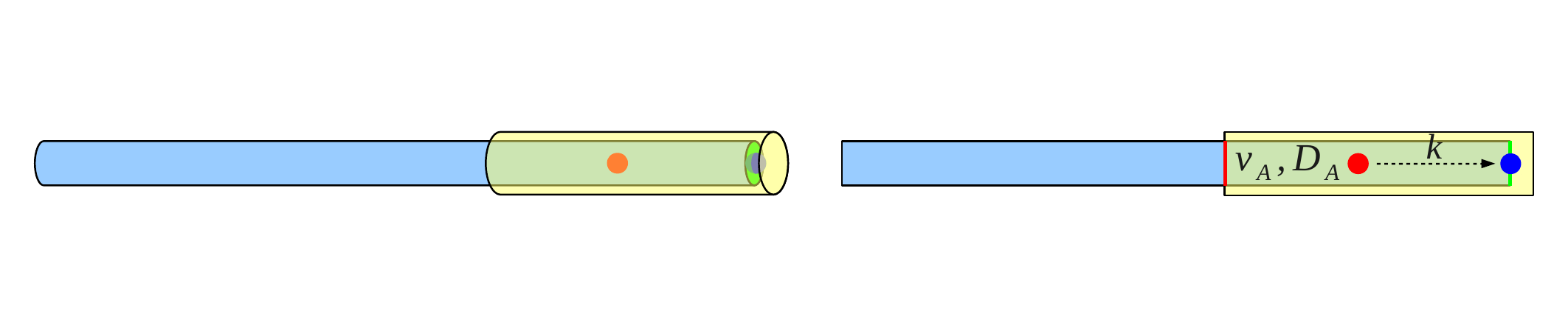}
  }
  \subfigure[][]{
    \label{fig:Domain-DiskSurfaceSingle}
    \includegraphics[width=\DomainTypeSketchWidth]{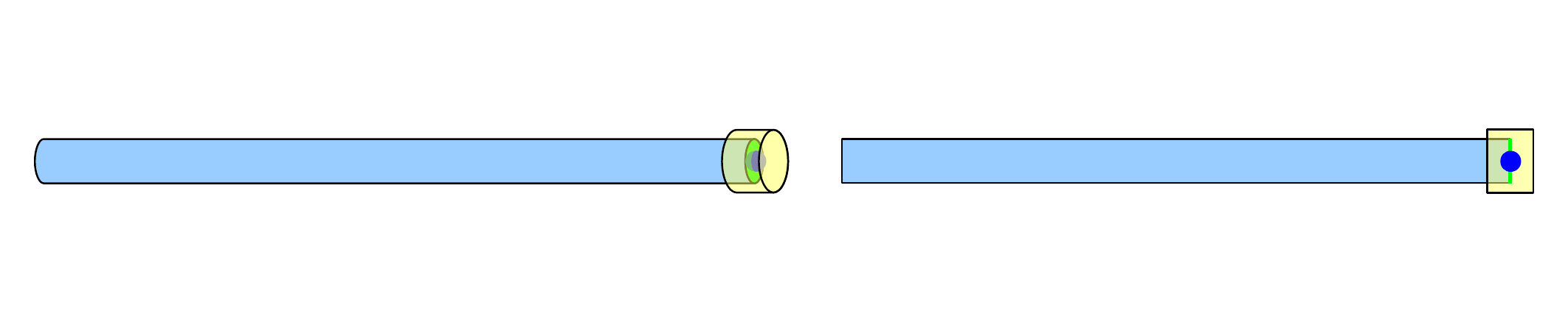}
  }
  \subfigure[][]{
    \label{fig:Domain-CylindricalSurfaceSink}
    \includegraphics[width=\DomainTypeSketchWidth]{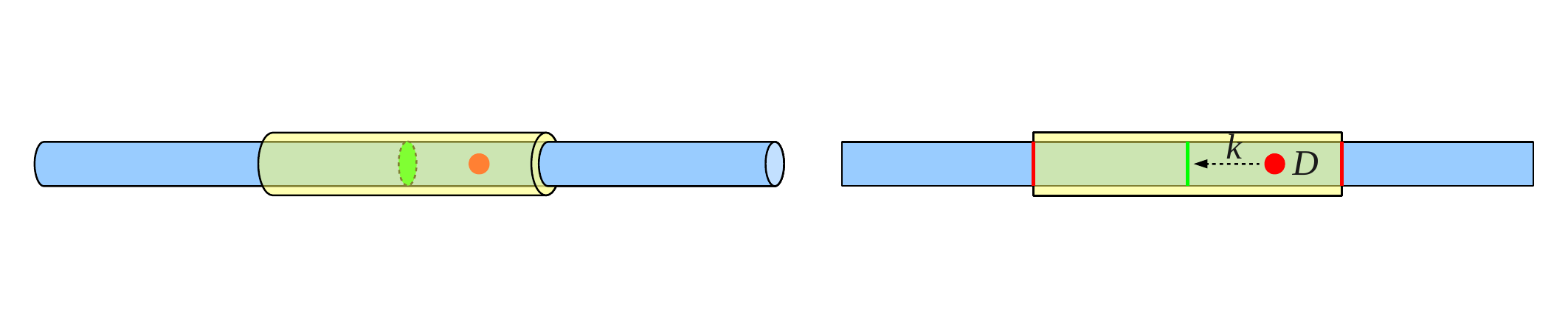}
  }  
\caption{ \label{fig:NewDomains-1D-Finite}
  \textbf{New protective domain types on finite 1D structures.}
  The cylindrical surface is limited by a reactive cap at its right end (green colour).
  \subfigref{\subref{fig:Domain-CapInteraction}} \domaintype{Cap Interaction} domain;
  \subfigref{\subref{fig:Domain-MixedPair1DCap}} \domaintype{Mixed Pair 1D-Cap} domain;
  \subfigref{\subref{fig:Domain-DiskSurfaceSingle}} \domaintype{Disk Surface Single} domain, here shown for a particle on a cap;
  \subfigref{\subref{fig:Domain-CylindricalSurfaceSink}} \domaintype{Cylindrical Surface Sink} domain.
  Right panels show sections of 3D objects along the common cylinder axis.
  Absorbing boundaries are highlighted by red, radiating boundaries by green. 
  Note that drift velocities ($v$, $v_A$, $v_B$) can be towards any cylinder end.
}
\end{figure}

To model interactions with disks we introduce the following new domain types:
\begin{itemize}
 \item \domaintype{Cap Interaction} domain: a cylindrical domain that encloses an empty cap and a nearby
	cylinder-bound 1D particle. Next-event times are calculated from the 
	1D Green's function with drift and Rad-Abs boundary conditions.
	\index{Cap Interaction}
\item \domaintype{Disk Surface Single} domain: a cylindrical domain that encloses
	a particle bound to a disk surface.
	The only possible next-events are unbinding reactions
	which are sampled from exponential distributions.
	\index{Disk Surface Single}
\item \domaintype{Mixed Pair 1D-Cap} domain: a cylindrical domain that encloses a cap with a bound particle
	and a nearby cylinder-bound 1D particle. Next-event times are calculated as for the
	\textit{Cylindrical Surface Pair},
	with drift and diffusion coefficient of the cap-bound partner set to zero.
	\index{Mixed Pair 1D-Cap}
\item \domaintype{Cylindrical Surface Sink} domain: a cylindrical domain that contains a sink and a proximate
	cylinder-bound particle. For this special case we calculate the Green's function explicitly in section \ref{sec:1D-GreensFunction-Sink}.
       \index{Cylindrical Surface Sink}
\end{itemize}
Sketches of these new domain types are shown in Figure \ref{fig:NewDomains-1D-Finite}.

%

\TODO{SUBSECTION: We should briefly describe here the new functionality for direct transitions from cylinder to plane and vice versa,
and also the variant with the ``interface disk'' in between, which were only implemented after the thesis.}

\subsection{Cylindrical Surface Sink domain}
\label{sec:1D-GreensFunction-Sink}
In transcription activation, transcription factors can perform a diffusive
search for their binding site on the DNA \cite{Halford2004,Shimamoto1999,vonHippel1989}.
To be able to model such and similar 1D random search processes in \eGFRD,
we introduced sink structures that mark point-sites at which particles
can react while diffusing over the cylindrical structure that they are bound to.
In order to isolate this interaction from interactions with other particles
on the cylinder we further introduced new domains (\domaintype{Cylindrical Surface Sink} domains)
that only contain a sink and the nearby particle.
Exits from these domains then can happen via two different events: either the particle
hits the (absorbing) boundaries of the domain, or it binds to the sink.
For the case without particle drift, we present here the Green's function for this problem.
One complication here is that the particle may diffuse over the sink without being absorbed.
It is, however, possible to incorporate this feature into the mathematical derivation
by imposing the correct probability flux balance at the sink position.

Assuming that the particle, starting from initial position $x=x_0$, 
can attach to the sink located at position $x_s$ with an intrinsic rate $k$,
the diffusive dynamics of the particle may be described via the modified diffusion equation:
\begin{align}
 \label{eq-1D-GreensFunction-Sink-PDE}
 & \pd_t p(x,t|x_0) = D\nabla^2 p(x,t|x_0) - k\delta(x-x_s)p(x,t|x_0)
\end{align}
with absorbing boundaries at $x=a$ and $x=b>x_s>a$
\begin{align}
 & p(a,t|x_0) = 0\; , \quadquad p(b,t|x_0) = 0
\end{align}
and initial condition
\begin{align}
 & p(x,t=0|x_0) = \delta(x-x_0)	\quad .
\end{align}

As with the 1D-Rad-Abs Green's function calculated in section \ref{sec:1D-GreensFunction-Rad-Abs}, 
this problem may be solved separately for the subintervals of $[a,b]$ separated 
by the delta peaks, imposing continuity of $p(x,t|x_0)$ and discontinuity
of its derivative at the junction points.
Here it is convenient to consider the intervals left and right to the sink
and to account for the initial condition directly by an adequate ansatz for the part that contains the starting point $x_0$.
Continuity-discontinuity relations then only have to be imposed
at $x_s$.

Let us denote the solution on interval $[a,x_s]$ by $p_{-}(x,t|x_0)$ and
the solution on $[x_s,b]$ by $p_{+}(x,t|x_0)$.
By integrating (\ref{eq-1D-GreensFunction-Sink-PDE}) over an $\epsilon$-interval
around the sink and taking the limit $\epsilon\rightarrow0$
we obtain the discontinuity condition for the probability flux at $x=x_s$:
\begin{align}
 \pd_x p_{-}(x,t|x_0)\takenat_{x=x_s} - \pd_x p_{+}(x,t|x_0)\takenat_{x=x_s} = -\frac{k}{D} p_{-}(x_s,t|x_0)
\end{align}
This equation simply states that the flux from/towards the left of the sink
equals the flux towards/from the region right of it, minus the reactive flux through the sink.
Moreover, continuity requires:
\begin{align}
 p_{-}(x_s,t|x_0) = p_{+}(x_s,t|x_0)
\end{align}

Once more, this problem is most conveniently solved in \person{Laplace} space.
The homogenous version of the \person{Laplace}-transformed PDE reads:
\begin{align}
 & s\hat p(x,s|x_0) = D\nabla^2 \hat p(x,s|x_0)
\end{align}
Let us without loss of generality assume $x_0\in[x_s,b]$ and that the sink is located at $x=x_s=0$,
implying $a=-|a|<0$.
Then we can make the following ansatz in \person{Laplace} space ($q\equiv\sqrt{\frac{s}{D}}$):
\begin{align}
 & \hat p_-(x,s|x_0) = A_- \sinh(qx) + B_- \cosh(qx),	&\; x < 0	\nn\\
 & \hat p_+(x,s|x_0) = A_+ \sinh(qx) + B_+ \cosh(qx) + \braceunderset{\hat p_{\rm free}(x,s|x_0)}{\frac{1}{2Dq}e^{-q|x-x_0|}} , &\; x > 0
\end{align}
Function $\hat p_{+}(x,s|x_0)$ contains the (\person{Laplace}-transformed) free solution for a point particle starting from $x=x_0$
and thus fulfills the initial condition by construction.

The coefficients $A_{\pm}$ and $B_{\pm}$ are calculated by applying the boundary and
continuity-discontinuity conditions.
Subsequently, the solution can be transformed back into the time-domain via the residue formula.
This procedure is precisely the same as in \ref{sec:1D-GreensFunction-Rad-Abs} and therefore
omitted here.

\pagebreak[4]
\noindent The final solution reads, with $L\equiv b-a=b+|a|$,
\\
\\
\fbox{ \begin{minipage}{0.96\textwidth}
\vspace{-1em}
\begin{align}
\label{eq-1D-GreensFunction-Sink-left}
 p_{-}(x,t|x_0) &= -2D \nicesum{n=1}{\infty}e^{-D\zeta^2_n t} \sin\left(\zeta_n (|a| + x)\right)
  \frac{ \zeta_n \sin\left(\zeta_n (b - x_0)\right) }{    \Xi_n	} \\
 \label{eq-1D-GreensFunction-Sink-right}
 p_{+}(x,t|x_0) &= -2D \nicesum{n=1}{\infty}e^{-D\zeta^2_n t} \sin\left(\zeta_n(b-\hat x )\right) \times	\nn\\
		& \quadquad\quad\quad \frac{
			      D \zeta_n \sin\left(\zeta_n (|a| + \check x )\right) + k \sin(\zeta_n (|a|) \sin(\zeta_n \check x )
			    }
			    {    \Xi_n	}
\end{align}
with $\hat x  \equiv \max(x,x_0)$, $\check x  \equiv \min(x,x_0)$ and a common denominator:
\begin{align}
  \Xi_n &= D \left[ L\zeta_n \cos(\zeta_n L) +\sin(\zeta_n L) \right] 	\nn\\
		   & \quad\qquad + k \left[ b \cos(\zeta_n b)\sin(\zeta_n |a|) + |a| \cos(\zeta_n |a|)\sin(\zeta_n b) \right]
\end{align}
\end{minipage} }\\
\\
The numbers $\zeta_n$ are all real and positive and the roots of the equation
\begin{align}
 & D \zeta_n \sin\left( \zeta_n L \right) + k \sin\left(\zeta_n |a|\right)\sin\left(\zeta_n b\right) = 0
\end{align}
which, with the help of trigonometric relations and setting $\Delta L\equiv b-|a|$,
may be written in the more convenient form:\\
\\
\fbox{ \begin{minipage}{0.96\textwidth}
\vspace{-1em}
\begin{align}
 & D \zeta_n \sin\left( \zeta_n L \right) = \frac{k}{2} \left[ \cos\left( \zeta_n L \right) - \cos\left( \zeta_n \Delta L \right) \right]
\end{align}
\end{minipage} }\\
\\
The necessity of interchanging $x$ and $x_0$ in (\ref{eq-1D-GreensFunction-Sink-right}) 
when the sign of $(x-x_0)$ changes arises from the presence of $|x-x_0|$ in the ansatz for $p_+(x,s|x_0)$.

The survival probability for the whole domain $[a,b]$ is given by
\begin{align}
 & S(t) = \int\limits_a^{x_s=0} p_-(x,t|x_0) dt + \int\limits_{x_s=0}^b p_+(x,t|x_0) dt
\end{align}
and easily obtained by simple integration.

The next-event time $\tau_\nu$ for the associated \domaintype{Cylindrical Surface Sink} 
domain is sampled from $1-S(t)$, as usual.
The next-event type, i.e. whether the particle exits the domain by being absorbed at the sink
or at one of the boundaries, is determined by comparing the probability fluxes through these
exit channels at $\tau_\nu$. These are most conveniently calculated via:
\begin{align}
 q_{s} &= k p_{-}(x_s,\tau_\nu|x_0) = k p_{+}(x_s,\tau_\nu|x_0)		\nn\\
 q_{a} &= -D \pd_x p_{-}(x,\tau_\nu|x_0)\takenat_{x=a}	\nn\\
 q_{b} &= -D \pd_x p_{+}(x,\tau_\nu|x_0)\takenat_{x=b}
\end{align}

\pagebreak
\subsection{Mixed Pair 2D-3D domain}
\label{sec:DirectBinding-MixedPair2D3D}
In order to be able to efficiently simulate a binding reaction of a particle diffusing in 3d
to a particle diffusing on a 2d plane in \eGFRD we devised a special domain type,
the \domaintype{Mixed Pair 2D-3D} domain, shown in Fig.~\ref{fig:NewDomains-2D-Special}\subref{fig:Domain-MixedPair2D3D};
it is based on a coordinate transform that takes into account the anisotropy in the diffusion behavior of the two reactants,
by essentially rescaling the interparticle vector in the direction perpendicular to the plane.
Once the particle positions have been transformed into the associated coordinate system,
Green's functions already derived for the more simple situations in 2D and 3D
can be used to sample next-event times and new positions for the reactants;
subsequently, these can be converted back to the original coordinate transform via a simple inverse transform.

In the following we first derive the generalized coordinate transform for the direct binding scenario,
and then describe how it can be used to implement the \domaintype{Mixed Pair 2D-3D} domain.

\newcommand{\Lcyr}{\mathlarger{\mathlarger{\lamed}}}
\renewcommand{\coeffA}{a}
\renewcommand{\coeffB}{b}
\renewcommand{\coeffC}{c}
\renewcommand{\coeffD}{d}
\subsubsection{Generalized linear coordinate transform for direct binding}
\label{sec:DirectBinding-Transform}
In this section we present a generalization of the linear coordinate
transform that maps two arbitrary particle positions $\myvec r_A$ and $\myvec r_B$
onto a (generalized) interparticle vector $\myvec r$ and weighted center-of-mass vector $\myvec R$
in a way that makes the diffusion anisotropy arising in the direct binding scenario
disappear in the transformed coordinates.
To this end we pursue the following approach:
First, we rewrite the linear operator (Laplacian) of the diffusion equation in
a generic matrix notation in order to account for anisotropic diffusion,
starting from the well-known form of the \textsc{Smoluchowski} equation.
As a second step, we write down the linear coordinate transform in its most general form
and apply it to both the linear operator and the original coordinates
$\myvec r_A$ and $\myvec r_B$ in order to decouple the equation.
Finally, the generic result will be used to specify a particular
transform for the direct binding scenario.
This involves some freedom in the choice of transformation coefficients.
We therefore postulate the following criteria to constrain the result:
First, the transformed Laplacian should not contain any mixed derivatives because
this complicates the solution of the transformed equation.
Second, the new coordinates should capture existing symmetries 
and, moreover, resemble the previously defined interparticle vector
and center of mass as much as possible,
so that we can use previously derived Green's functions.

\subsubsection*{Rewriting the Laplace operator in matrix notation}
Let us recall the \textsc{Smoluchowski} equation for the density of the probability \linebreak
$p\equiv p(\myvec{r}_A,\myvec{r}_B,t|\myvec{r}_{A0},\myvec{r}_{B0},t_0)$ 
to find two diffusing particles A and B with diffusion constants $D_A$ and $D_B$ at positions
$\myvec{r}_A$ and $\myvec{r}_B$,
given that they started from $\myvec{r}_{A0}$ and $\myvec{r}_{B0}$ (compare to section \ref{sec:GFRD-Pairs}):
\begin{align}
\pd_t p		&=	[ D_A\nabla_A^2 + D_B\nabla_B^2 ]\text{\space}p
\end{align}
Here for simplicity we neglect the force-interaction term, i.e. $F(\myvec r_A - \myvec r_B)=0$.

\noindent By introducing 
\begin{align}
\nablaX \equiv \begin{pmatrix} \nabla_A \\ \nabla_B \end{pmatrix}
\end{align}
and its transpose $\nablaX^T\equiv(\nabla_A, \nabla_B)$
the linear operator $\mat{L}^2 \equiv D_A\nabla_A^2 + D_B\nabla_B^2$ 
may be written in vector-matrix-notation as
\begin{align}
\mat{L}^2 = \nablaX^T \begin{pmatrix} \mat{D}_A &  \\  & \mat{D}_B \end{pmatrix} \nablaX \equiv \nablaX^T \mat{D} \nablaX
\end{align}
where the sub-matrices
\begin{align}
\mat{D}_{A(B)} \equiv \begin{pmatrix} D_{A(B)} & & \bigzero \\ & D_{A(B)} & \\ \bigzero & & D_{A(B)} \end{pmatrix}
\end{align}
define a diffusion matrix $\mat{D}$.
While here the entries of matrices $\mat{D}_A$ and $\mat{D}_B$ are equal along their diagonals,
in general they may differ if diffusion is anisotropic.
Importantly, the Laplacian $\mat{L}^2$ has no mixed derivatives only if $\mat{D}$ is diagonal.

\subsubsection*{Writing the linear coordinate transform in matrix notation}
A generic linear coordinate transform ${\mat{M}: \mathbb{R}^6 \rightarrow \mathbb{R}^6}$ 
for two arbitrary $\mathbb{R}^3$-vectors $\myvec{r}_A$ and $\myvec{r}_B$ is described via:
\begin{align}
\myvec{r}\equiv \coeffA\myvec{r}_A + \coeffB\myvec{r}_B\;,	\quadquad \myvec{R}\equiv \coeffC\myvec{r}_A + \coeffD\myvec{r}_B \;.
\end{align}
In matrix notation this reads
\begin{align}
\myvec{Y}\equiv\begin{pmatrix} \myvec{r} \\ \myvec{R} \end{pmatrix} = \mat{M}\begin{pmatrix} \myvec{r}_A \\ \myvec{r}_B \end{pmatrix} \equiv \mat{M}\myvec{X}
\end{align}
with
\begin{align}
\mat{M} = \begin{pmatrix}
		\coeffA & 0 & 0 & \coeffB & 0 & 0 \\
		0 & \coeffA & 0 & 0 & \coeffB & 0 \\
		0 & 0 & \coeffA & 0 & 0 & \coeffB \\
		\coeffC & 0 & 0 & \coeffD & 0 & 0 \\
		0 & \coeffC & 0 & 0 & \coeffD & 0 \\
		0 & 0 & \coeffC & 0 & 0 & \coeffD \\
\end{pmatrix}
\equiv \begin{pmatrix}
		\mat{A} & \mat{B} \\
		\mat{C} & \mat{D}
\end{pmatrix} \;.
\end{align}
We may generalize this transform further by allowing the nonzero coefficients 
to differ from each other as long as the full rank of the matrix is preserved.
In the following we therefore assume that $\mat{M}$ has the form:
\begin{align}
\mat{M} = \begin{pmatrix}
		\mat{A} & \mat{B} \\
		\mat{C} & \mat{D}
\end{pmatrix}
= \begin{pmatrix}
		\coeffA_1 & 0 & 0 & \coeffB_1 & 0 & 0 \\
		0 & \coeffA_2 & 0 & 0 & \coeffB_2 & 0 \\
		0 & 0 & \coeffA_3 & 0 & 0 & \coeffB_3 \\
		\coeffC_1 & 0 & 0 & \coeffD_1 & 0 & 0 \\
		0 & \coeffC_2 & 0 & 0 & \coeffD_2 & 0 \\
		0 & 0 & \coeffC_3 & 0 & 0 & \coeffD_3 \\
\end{pmatrix}
\end{align}

\subsubsection*{Transforming the Laplace operator}
Let us now apply the generalized transform to the linear operator $\mat{L}^2$.
This means expressing the derivatives $\partial_{X_i}$ of the original coordinates
in terms of derivatives $\partial_{Y_i}$ of the new, transformed coordinates.
The chain rule of differentiation yields:
\begin{align}
\frac{\partial}{\partial X_i} = \frac{\partial Y_j}{\partial X_i} \frac{\partial}{\partial Y_j} \equiv N_{ij} \frac{\partial}{\partial Y_j}
\end{align}
The coefficeints $N_{ij}=\frac{\partial Y_j}{\partial X_i}$ define a new matrix $\mat{N}$.
Since the considered transform is linear these coefficients must be constants and related to the entries of the matrix $\mat{M}$ via
\begin{align}
\frac{\partial Y_i}{\partial X_j}=(\mat{M})_{ij}=N_{ji}\;.
\end{align}
Since for a linear transform the Jacobian and the matrix of the transform are identical,
we have:
\begin{align}
\mat{N} = \mat{M}^T \nn\\
\nablaX = \mat{M}^T \nablaY
\end{align}
With this we may rewrite the linear operator as follows\footnote{We denote the new representation of the operator with a different sign, but formally $\Lcyr^2=\mat{L}^2$.}:
\begin{align}
\mat{L}^2 	&= \nablaX^T \left(\begin{array}{cc} \mat{D}_A & \\ & \mat{D}_B \end{array}\right) \nablaX \nn\\
		&= \nablaY^T \mat{M} \left(\begin{array}{cc} \mat{D}_A & \\ & \mat{D}_B \end{array}\right) \mat{M}^T\nablaY 
		\equiv \nablaY^T \mat{D}' \nablaY \equiv \Lcyr^2
\end{align}
Here $\mat{D}'$ is the transformed diffusion matrix in the new coordinates.
Recall that $\Lcyr^2$ will not contain mixed derivatives after the coordinate transform only if
$\mat{D}'$ is diagonal.
By carrying out explicitly the above calculation we arrive at:
\begin{align}
\mat{D}'=
\begin{pmatrix}
	\mat{D}_A\mat{A}^2 + \mat{D}_B\mat{B}^2 		& \mat{D}_A\mat{A}\mat{C} + \mat{D}_B\mat{B}\mat{D} \\
	\mat{D}_A\mat{A}\mat{C} + \mat{D}_B\mat{B}\mat{D}	& \mat{D}_A\mat{C}^2 + \mat{D}_B\mat{D}^2 \\
\end{pmatrix} \nn\\
\end{align}
Since all matrices involved in the above expression are diagonal by definition, the diagonality condition reduces to:
\begin{align}
\mat{D}_A\mat{A}\mat{C} + \mat{D}_B\mat{B}\mat{D} = \bigzero \nn\\
\Leftrightarrow \forall j:\quad (\mat{D}_A)_{jj}\coeffA_j\coeffC_j + (\mat{D}_B)_{jj}\coeffB_j\coeffD_j = 0 \label{DiagCond}
\end{align}
We have now established a condition for transforming the Laplacian in a way that mixed derivatives
disappear in the new coordinates. This is a generalization of the condition already mentioned in section \ref{sec:GFRD-CT-CoeffCondition}.
However note that we still can choose the transform coefficients freely as long as the above equation is fulfilled.
We will now determine a specific choice that is appropriate for the considered direct binding scenario.

\subsubsection*{Particular transform for the direct binding scenario}
In direct binding 
one of the particles (A) diffuses on a planar 2D submanifold of the $\mathbb{R}^3$  
while the other particle (B) performs a standard isotropic 3D diffusion in $\mathbb{R}^3$.
Let us assume that the 2D plane corresponds to the $xy$-plane of the Cartesian coordinate system, 
implying $D_{Az}\equiv(\mat{D}_A)_{33}=0$.
The diffusion matrix in the original coordinates $\lbrace \myvec{r}_A, \myvec{r}_B \rbrace$ then reads:
\begin{align}
 \mat{D}
= \begin{pmatrix}
		D_A & 0 & 0 & 0 & 0 & 0 \\
		0 & D_A & 0 & 0 & 0 & 0 \\
		0 & 0 & 0 & 0 & 0 & 0 \\
		0 & 0 & 0 & D_B & 0 & 0 \\
		0 & 0 & 0 & 0 & D_B & 0 \\
		0 & 0 & 0 & 0 & 0 & D_B \\
\end{pmatrix}
\end{align}
We now specify the coefficients of the transformation matrix $\mat{M}$
such that (\ref{DiagCond}) holds and the off-diagonal elements of $\mat{D}'$ vanish. 
The latter requires either $\coeffB_3=0$ or $\coeffD_3=0$ because of $D_{Az}=0$ and $D_{Bz}\neq 0$.
To ensure that the $z$-component of the generalized interparticle vector $\myvec{r}$ is nonzero,
as in previous definitions, we opt for $\coeffD_3=0$.
The transformed diffusion matrix $\mat{D}' = \mat{M}\mat{D}\mat{M}^T$ then becomes:
\begin{align}
\mat{D}' = \left(\begin{smallmatrix}
		D_A^2\coeffA_1^2+D_B^2\coeffB_1^2 & 0 & 0 & 0 & 0 & 0 \\
		0 & D_A^2\coeffA_2^2+D_B^2\coeffB_2^2 & 0 & 0 & 0 & 0 \\
		0 & 0 & D_B\coeffB_3^2 & 0 & 0 & 0 \\
		0 & 0 & 0 & D_A^2\coeffC_1^2+D_B^2\coeffD_1^2 & 0 & 0 \\
		0 & 0 & 0 & 0 & D_A^2\coeffC_2^2+D_B^2\coeffD_2^2 & 0 \\
		0 & 0 & 0 & 0 & 0 & 0 \\
\end{smallmatrix}\right)
\end{align}
With this we may rewrite the Laplacian as follows:
\begin{align}
 \Lcyr^2	&= \nablaY^T (\mat{M}^T\mat{D}\mat{M}) \nablaY \nn\\
 &= (D_A\coeffA_1^2 + D_B\coeffB_1^2)\partial_{r_1}^2 + (D_A\coeffA_2^2 + D_B\coeffB_2^2)\partial_{r_2}^2 + D_B\coeffB_3^2\partial_{r_3}^2 \nn\\
 &\quad + (D_A\coeffC_1^2 + D_B\coeffD_1^2)\partial_{R_1}^2 + (D_A\coeffC_2^2 + D_B\coeffD_2^2)\partial_{R_2}^2 \label{NewL2}\nn\\
\end{align}
The fact that the prefactor of $\partial_{r_3}^2$ is different from the prefactors of the other two components 
prevents us from regrouping the separate differential operators into a closed form.
This is precisely the signature of anisotropic diffusion.
Provided that we do not change the (full) rank of $\mat{M}$ we may choose the yet undetermined coefficients freely.
Here we set
\begin{align}
\coeffA_{1,2} = -1 \qquad
\coeffB_{1,2} = +1 
\end{align}
to ensure that the first two components of $\myvec{r}$ reproduce the ones of the standard interparticle vector.
Then from (\ref{NewL2}) it is evident that setting $\coeffB_3 = \pm \sqrt{\frac{D_A+D_B}{D_B}}$ yields equal 
prefactors for all three derivatives.
With this choice we get rid of diffusion anisotropy by adequately rescaling the $r_z$-coordinate.

\noindent The corresponding transformation matrix $\mat{M}_{DB}$ has the form
\begin{align}
 \mat{M}_{DB}
= \begin{pmatrix}
		-1 & 0 & 0 & 1 & 0 & 0 \\
		0 & -1 & 0 & 0 & 1 & 0 \\
		0 & 0 & \coeffA_3 & 0 & 0 & \pm \sqrt{1+\frac{D_A}{D_B}} \\
		\coeffC_1 & 0 & 0 & \coeffD_1 & 0 & 0 \\
		0 & \coeffC_2 & 0 & 0 & \coeffD_2 & 0 \\
		0 & 0 & \coeffC_3 & 0 & 0 & 0 \\
\end{pmatrix}
\end{align}
and in this particular case the diagonality condition (\ref{DiagCond}) reads:
\begin{align}
-D_A\coeffC_j + D_B\coeffD_j &= 0 \;,\quad\quad j=1,2 \nn\\
\undernull{(\mat{D}_A)_{33}}\coeffA_3\coeffC_3 + D_B\coeffB_3\undernull{\coeffD_3} &= 0 \label{DiagCondBosson}
\end{align}
Evidently, the second line is fulfilled for any choice of $\coeffA_3$, $\coeffB_3$ and $\coeffC_3$.
However, preserving full rank requires $\coeffC_3\neq 0$.
An adequate choice is $\coeffC_3=\frac{1}{D_A+D_B}$.
Moreover it is convenient to set $\coeffA_3 = -\coeffB_3$ and
\begin{align}
\coeffC_j = \frac{D_B}{D_A+D_B}, \quad \coeffD_j = \frac{D_A}{D_A+D_B}, \quad j=1,2	\quad .
\end{align}
With this the particular transform is completely determined,
and we finally arrive at the Laplacian in transformed coordinates:
\begin{align}
\Lcyr^2_{DB}	&= \underset{\equiv D_r}{\underbrace{(D_A+D_B)}}\nabla_r^2 + \underset{\equiv D_R}{\underbrace{\left(\frac{D_A D_B}{D_A+D_B}\right)}} \left(\partial_{R_1}^2+\partial_{R_2}^2\right)
\end{align}
This is structurally analogous to the operator yielded by the coordinate transform defined in section \ref{sec:CoefficientChoice-eGFRD}.
However, here the transformed coordinates are different.
The explicit (forward) transformation rules read:\\
\\
\fbox{ \begin{minipage}{0.96\textwidth}
\vspace{-1em}
\begin{align}
\myvec{r}		&= \begin{pmatrix} r_{B_1} - r_{A_1} \\ r_{B_2} - r_{A_2} \\ b_3 r_{B_3} + a_3 r_{A_3} \end{pmatrix} 
		  = \begin{pmatrix} r_{B_1} - r_{A_1} \\ r_{B_2} - r_{A_2} \\ \varepsilon(r_{B_3} - r_{A_3}) \end{pmatrix}, \qquad
		  \varepsilon \equiv \pm \sqrt{1+\frac{D_A}{D_B}} \nn\\
\myvec{R}		&= \frac{1}{D_A+D_B}\begin{pmatrix} D_B r_{A_1} + D_A r_{B_1} \\ D_B r_{A_2} + D_A r_{B_2} \\ r_{A_3} \end{pmatrix}
\end{align}
\end{minipage} }\\
\\
The formula for $\myvec{r}$ demonstrates that in the new coordinate system anisotropy is cancelled
by rescaling the $z$-component of the interparticle vector.
The sign of the scaling factor $\varepsilon$ may be chosen freely; in the following
we opt for the positive solution.
Note that $r_{A_3}=const$ because the 2D particle (A) is always in the plane by definition.

For completeness we once again explicitly state the final version of the transformation matrix $\mat{M}_{DB}$:
\begin{align}
\label{eq-Mat-specific-final}
\mat{M}_{\rm DB} = \begin{pmatrix}
	-1 & 0 & 0 & 1 & 0 & 0 \\
	0 & -1 & 0 & 0 & 1 & 0 \\
	0 & 0 & -\varepsilon & 0 & 0 & \varepsilon \\
	\varepsilon^{-2} & 0 & 0 & \frac{D_A}{D_B}\varepsilon^{-2} & 0 & 0 \\
	0 & \varepsilon^{-2} & 0 & 0 & \frac{D_A}{D_B}\varepsilon^{-2} & 0 \\
	0 & 0 & D_r^{-1} & 0 & 0 & 0 \\
\end{pmatrix}
\end{align}
The determinant of this matrix is
\begin{align}
\det(\mat{M}_{DB}) = \frac{- \varepsilon}{D_A+D_B} = \frac{\mp 1}{\sqrt{D_B(D_A+D_B)}} \neq 0
\end{align}
confirming that our specific coefficient choice preserves the full-rank property.

\subsubsection*{Inverse transform}
To obtain the inverse transformation rule we simply calculate the inverse of matrix $\mat{M}_{DB}$ (\ref{eq-Mat-specific-final}):
\begin{align}
\mat{M}_{DB}^{-1} =
\begin{pmatrix}
	-\frac{D_A}{D_B}\varepsilon^{-2} & 0 & 0 & 1 & 0 & 0 \\
	0 & -\frac{D_A}{D_B}\varepsilon^{-2} & 0 & 0 & 1 & 0 \\
	0 & 0 & 0 & 0 & 0 & D_r \\
	\varepsilon^{-2} & 0 & 0 & 1 & 0 & 0 \\
	0 & \varepsilon^{-2} & 0 & 0 & 1 & 0 \\
	0 & 0 & \varepsilon^{-1} & 0 & 0 & D_r\\
\end{pmatrix}
\end{align}
This results in the following back-transform rules:\\
\\
\fbox{ \begin{minipage}{0.96\textwidth}
\vspace{-1em}
\begin{align}
\myvec{r}_A &= 	\begin{pmatrix} R_1 - \frac{D_A}{D_A+D_B} r_1 \\ R_2 - \frac{D_A}{D_A+D_B} r_2 \\ (D_A+D_B) R_3 \end{pmatrix} \nn\\
\myvec{r}_B &= 	\begin{pmatrix} R_1 + \frac{D_B}{D_A+D_B} r_1 \\ R_2 + \frac{D_B}{D_A+D_B} r_2 \\ (D_A+D_B) R_3 +\sqrt{\frac{D_B}{D_A+ D_B}}r_3 			\end{pmatrix}
\end{align}
\end{minipage} }\\

We will now explain how the derived transform may be used to sample
next-event information for the direct binding scenario using some of the
Green's functions that we have already presented within this thesis.

\subsubsection*{Using known Green's Functions for the transformed problem}
By applying the coordinate transform in the way described we succeeded in transforming the two-particle problem in 3D 
into two separate diffusion problems, namely a 2D-diffusion of the center-of-mass vector $\myvec{R}$
and a 3D-diffusion of the interparticle vector $\myvec{r}$, with a $z$-axis rescaled by $\varepsilon\geq1$.
Since diffusion of $\myvec R$ in the plane is still isotropic after the transform, we may sample next-event times and new positions
for $\myvec R$ in the same way as for the \domaintype{Planar Surface Pair} (see sec.~\ref{sec:2d-GF-plane} in the main text), 
i.e. by imposing a circular absorbing boundary at $|\myvec R|=R_{\rm max}$
and reusing the Green's function presented in section \ref{sec:2D-GF-AbsSym}.

\begin{figure}[t!]
\centering
\includegraphics[width=8cm]{\SketchesDir/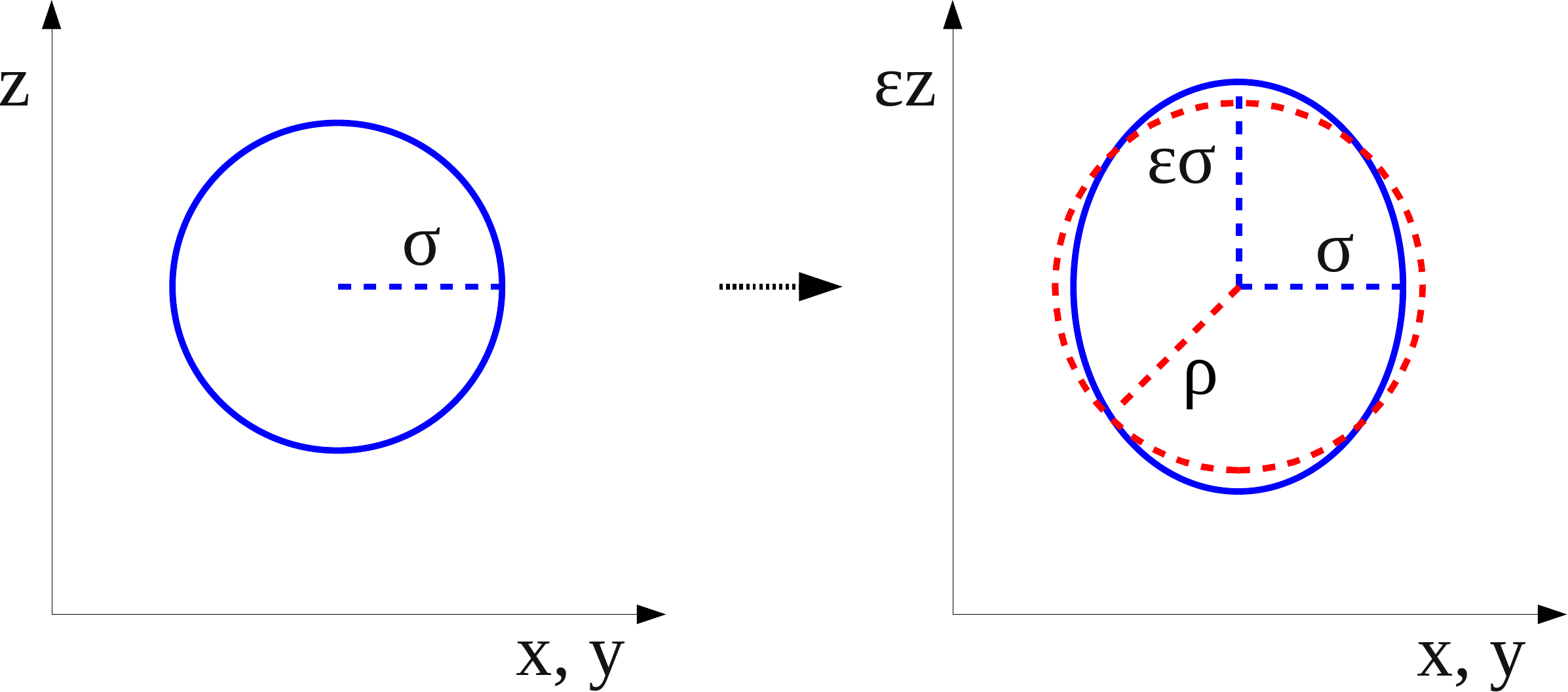}
\caption{Distorsion of a spherical boundary due to the anisotropic coordinate transform.
\label{fig:ProlateSpheroid}
}
\end{figure}

The situation is different for the rescaled interparticle vector $\myvec r$.
In 3D, where diffusion of the interparticle vector is isotropic, radiating or absorbing boundary conditions are defined on spheres.
Here, by rescaling the $z$-axis all lengths in $z$-direction become slightly longer with respect to the other directions in the new coordinate system,
meaning that boundaries originally represented by spheres now become prolate spheroids (Figure \ref{fig:ProlateSpheroid}).
Since it is technically challenging to compute the Green's function for such boundary conditions,
we opted for a simpler, approximative approach, in which the prolate spheroidal boundaries in the transformed coordinates
are substituted by spherical boundaries.
As an evident advantage, with spherical boundaries we may reuse the well-known 3D Green's function
for a radiating inner and absorbing outer boundary.
Given that the diffusion constant in the plane is significantly smaller than the diffusion constant in the bulk, 
e.g. $D_A\simeq D_B/10$, the scaling factor $\varepsilon=\sqrt{1+D_A/D_B}$ is rather small ($\varepsilon\simeq1.05$), 
implying only a minor error induced by the substitution.

As a further modification, we choose the radius of the inner, radiating sphere in such a way
that its surface area equals the surface area of the prolate spheroidal.
The rationale here is that equal surface areas will ensure that the total probability flux
through the new, spherical boundary will be approximately equal to the total flux through the original, prolate boundary.
Let the radius of the boundary sphere in the untransformed coordinate system be $\sigma$. 
The sphere transforms to a prolate spheroidal with semi-major axis length $A=\varepsilon\sigma$,
whereas the semi-minor axis $a$ is identical to the radius $\sigma$ (Figure \ref{fig:ProlateSpheroid}).
The surface area of a prolate spheroidal is given by:
\begin{align}
A_P(a,A)=2\pi\left[ a^2 + \frac{ aA\arccos\left(\frac{a}{A}\right) }{ \sin\left(\arccos\left(\frac{a}{A}\right)\right) }\right]
\end{align}

\noindent Setting this equal to the surface of a sphere with radius $\rho$, with the particular values for the half-axes from above, we obtain:
\begin{align}
4\pi\rho^2	&=	2\pi\left[ \sigma^2 + \frac{ \sigma(\varepsilon\sigma)\arccos\left(\frac{\sigma}{\varepsilon\sigma}\right) }{ \sin\left(\arccos\left(\frac{\sigma}{\varepsilon\sigma}\right)\right) }\right] \nn\\
\rho		&= 	\sqrt[+]{\frac{1}{2}
			\left( 1 + \frac{\varepsilon\arccos\left(\frac{1}{\varepsilon}\right)}{\sin\left(\arccos\left(\frac{1}{\varepsilon}\right)\right)} \right)} \cdot \sigma
\end{align}
To facilitate calculations we sample next-event information for the transformed interparticle vector
from the regular Green's function that assumes a spherical boundary, using $\rho$ as defined above for the contact radius.
Note that after back-transform sampled positions are located on oblate spheroids.
This may cause particle overlap when the sampled new distance between the particles is short (comparable to $\sigma$);
in this case particles are slightly moved apart, the error of which again is small.

\clearpage
\section{Domain and shell construction rules}
\label{sec:GFRD-shellmaking}
\subsection{Principal domain making strategy}
The governing principle in constructing and sizing of domains in eGFRD is to minimize the computational cost associated with this process.
Domain making consists of the following successive steps:
\begin{enumerate}
 \item Determine which type of domain to construct.
 \item Determine the available space for the domain (i.e. its shell).
 \item Construct the domain with an optimally sized shell.
 \item Draw the next-event time and type for the newly constructed domain.
 \item Re-schedule the domain in the central scheduler.
\end{enumerate}
While the cost of the first and last step is roughly the same for all domains, it may vary among different domain types for the other steps.
In general, cylindrical domains are more expensive to construct than spherical domains because of the increased computational effort
for scaling up cylinders within a specific configuration of other cylinders and spheres. Similarly, \domaintype{Pair} domains are more expensive than
\domaintype{Single} domains because they require an additional coordinate transform and employ Green's functions which are mathematically more complex.

It is unfeasible to foresee all possible configurations that may occur during eGFRD simulations as required for a real quantitative
optimization of the domain making rules. The strategy in defining a unique set of functional rules therefore is to minimize the likelihood of 
situations that lead to the waste of computational cost, such as repeated reconstruction of domains and construction of expensive domains when
it is not strictly advantageous.

\subsection{Social upsizing prevents premature and mutual bursting}
\index{social upsizing}
\index{premature bursting}
\index{mutual bursting}
Particular care must be put into determining the optimal size of the domain. In principle we want to construct domains as big as possible because
their next-event time directly correlates with their size. However, when we size up a large domain such that it will protrude into the
direct vicinity of a very small domain, the latter will most likely be updated long before the next event time of the freshly constructed large domain.
This may induce premature bursting of the large domain, which then has to be reconstructed from an almost identical situation as before after insignificant
time progress, wasting the initial investment of domain making cost. Therefore domains should not be sized up to the maximal available space
in any given situation but in a ``social'' manner, i.e. leave some space for their neighbour domains to avoid very small domains in their own
direct vicinity.

In particular, it is important to prevent repeated mutual bursting of two newly constructed adjacent \domaintype{Single} domains, which may even result
in an infinite cycle of domain (re)construction and bursting.
Repeated mutual bursting can occur when two particles are at a small distance, yet sufficiently far away to enable the construction of
two \domaintype{Single} domains, and the formation of a \domaintype{Pair} is disallowed for other reasons (e.g. presence of obstacles).
Since domains are sized up in a successive order, using the maximally available space for the first \domaintype{Single} domain (A) would result in
a very small size of the second (B), causing immediate update of B with negligible particle displacement. This in turn would force
bursting of domain A in order to size up the B domain again, which would restart the whole process all over from B.
This example demonstrates that maximizing domain size is not the same as optimizing it.

\subsection{A minimal Single domain size controls switching to Brownian Dynamics}
In section \ref{sec:BD-SI} we explain why it is necessary to switch from eGFRD to Brownian Dynamics
when propagating particles under crowded conditions.
Yet, it is not a priori clear when this switch should be performed.
In principle it should be done when the computational cost for \domaintype{Single} construction
divided by the maximal displacement within the domain, which is correlated to domain size,
becomes larger than the cost of sampling a trajectory covering the same distance with Brownian Dynamics.
Since in GFRD \domaintype{Single} construction cost is variable, it is hard to devise a general rule.
Nevertheless, it is clear that a minimal \domaintype{Single} has to be defined for proper working of
the algorithm.
We decided to make this a simulation parameter, the details of which will be described in more detail in
the following text.

\subsection{Summarized domain making objectives}
The abovementioned general reasoning can be summarized into the following set of simple objectives for efficient domain making rules:
\begin{itemize}
 \item Construct \domaintype{Pair} and \domaintype{Interaction} domains only when interaction is likely, i.e. when particles are close to other particles or reactive surfaces.
 \item Construct domains socially, i.e. reserve some space for neighbouring domains in order to prevent premature or mutual bursting.
 \item Construct \domaintype{Multi} domains (i.e. fall back into Brownian dynamics) if construction of a minimal-size \domaintype{Single} is impossible.
\end{itemize}

In order to transform these rules into an applicable algorithm we introduced two length factors which will determine when to start 
constructing \domaintype{Pairs} or \domaintype{Interactions} and when to go into the Brownian dynamics mode during runtime.

\subsection{Two length factors balance the domain making behaviour}
\label{Shellmaking-lengthscales}
Let us define the following two dimensionless constants:
\begin{itemize}
 \item $\beta \equiv$ ``single-shell factor''
 \item $\mu \equiv$ ``multi-shell factor''
\end{itemize}
\index{single-shell factor}
\index{multi-shell factor}

For a given particle radius $R$ these two factors define the half-size $\beta R$ of the minimal \domaintype{Single} shell
and the radius $\mu R$ of the (always spherical) \domaintype{Multi} shell of that particle, respectively.
Note that the minimal \domaintype{Single} shell can be either a sphere or a cylinder, depending on whether 
the particle is a 3D, 2D or 1D species.
The requirements $\beta \geq 1$ and $\mu \geq 1$ are obvious.
Since a \domaintype{Multi} shell shall never be constructed when there is enough space for a minimal \domaintype{Single} shell,
we also require $\beta > \mu$.

Let $R_0$ be the radius of a particle $P_0$ for which we want to construct a domain, $R_1$ the radius of
its nearest neighbour particle $P_1$ and $\sigma \equiv R_0 + R_1$.
Then, based on $\beta$ and $\mu$, we define the following lengths for $P_0$:
\begin{itemize}
 \item the ``reaction horizon'' $\equiv \beta R_0$
 \item the ``multi horizon'' $\equiv \mu R_0$
\end{itemize}
and, as specializations of the above:
\begin{itemize}
 \item the ``pair horizon'' $\equiv \beta \sigma$
 \item the ``surface horizon'' $\equiv$ ``reaction horizon'' $\equiv$ ``burst horizon/radius'' $= \beta R_0$
 \item the ``multi-partner horizon'' $\equiv \mu \sigma$
\end{itemize}
\index{reaction horizon}
\index{pair horizon}
\index{surface horizon}
\index{Multi horizon}
\index{Multi partner horizon}
\index{burst horizon}
\index{burst radius}
Different naming highlights different purpose for these of the above quantities that otherwise are equal to each other.

The reaction horizon is used to determine when a \domaintype{Pair} or \domaintype{Interaction} domain should be constructed instead of a \domaintype{Single} domain.
The multi horizon defines when \domaintype{Single} construction should be dropped in favor of \domaintype{Multi} construction.
The pair horizon and surface horizon are specifications of the reaction horizon for \domaintype{Pair} and \domaintype{Interaction Single} formation: while an
\domaintype{Interaction} is formed when a surface is within the reaction horizon, a \domaintype{Pair} construction is attempted only when the reaction
horizons of the two involved particles overlap, i.e. when the center of mass of $P_1$ is within the pair horizon of $P_0$.
Similarly, the algorithm will switch into Brownian Dynamics mode when a surface is within the multi horizon of $P_0$ or when
the multi horizons of $P_0$ and $P_1$ overlap, i.e. when $P_1$ is within the multi-partner horizon of $P_0$.

The burst horizon defines the volume within which a particle will burst neighbouring domains.
Since the objective of bursting is to generate space for at least a minimal \domaintype{Single} shell the burst horizon should
be at least as big as the reaction horizon. Since there is no evident necessity to make it bigger than
the reaction horizon, we conveniently set these lengths to be equal.

Practically $\beta$ and $\mu$ can be used to tune the overall behaviour and performance of domain making:
Increasing $\mu$ will prompt the simulation to switch to Brownian Dynamics earlier.
Whether this is advantageous or not depends the crossover radius at which construction of small \domaintype{Single} domains
yields a smaller average simulation time advance per computational cost unit than the construction of \domaintype{Multi} domains.
In a similar way, whether a larger or smaller $\beta$ is favorable depends on the average likelihood of reactions in the system.
Since the latter depend on the parameters, there is no obviously optimal choice for $\beta$ and $\mu$.
We find that $\beta \in [2, 3]$ and $\mu \in [\sqrt{3}, 2]$ gives reasonable performance.

\subsection{Domain making algorithm}
We can now compile the abovementioned postulations and definitions into a well-defined algorithm for domain making.
Let us imagine a particle which just exited from whatever domain type after an update.
An update can be either triggered by a next-event picked from the scheduler (i.e. a reaction, domain exit
or surface interaction) or by premature bursting of neighbouring particles induced in the aftermath of such
scheduler event.
The particle by default is put into the system as a \domaintype{Non-Interaction Single} with a shell that just envelopes
the particle. Note that this ``zero-shell'' is spherical for 3D particles and cylindrical for 2D and 1D particles.
We will call a \domaintype{Single} with a zero-shell a ``\domaintype{Zero-Single}''. Each \domaintype{Zero-Single} is put into the scheduler
with zero next-event time ($dt=0$) in order to reconstruct its domain immediately after it was produced.
\index{Zero-Single}
\index{zero-shell}

We then perform the following order of actions to construct a new domain:
\begin{enumerate}
 \item \textbf{Bursting}: Burst any neighbouring ``intruder'', i.e. a domain that intrudes into the burst radius of
	the particle, with the exception of \domaintype{Multi} domains and other \domaintype{Zero-Singles}, i.e. domains which are yet to 
	pass through the domain making procedure themselves. By default the burst radius is equal to the reaction horizon.
	Burst recursively, i.e. whenever a bursted intruder has intruders within its own reaction horizon, also burst these.
	The following steps then are repeated for each \domaintype{Zero-Single} present in the system after bursting.
	\index{intruder}
 \item \textbf{Reaction/interaction attempt}:
	Compile a list of all potential interaction partners (particles or reactive surfaces).
	Pick the closest interaction partner and try a reaction (with particles) or interaction (with surfaces) if the closest
	partner is within the specified reaction or interaction horizon. If a minimal reaction (\domaintype{Pair}) domain or \domaintype{Interaction} domain 
	can be constructed, size it up socially to the maximal available space and go directly to step (4.).
 \item \textbf{\domaintype{Single} domain upsizing attempt}: If a \domaintype{Pair} or \domaintype{Interaction} could not be constructed, yet the closest partner is within
	the multi horizon of the \domaintype{Non-Interaction Single}, then (recursively) construct a \domaintype{Multi} domain (as specified further below)
	and proceed directly to step (4.).
	Else, size up the \domaintype{Non-Interaction Single} domain socially to the maximally available space and continue.
 \item \textbf{Re-scheduling}: For the constructed domain type determine the next-event time and type and reinsert this information
	into the scheduler.
 \item \textbf{Repeat} the whole procedure for the next \domaintype{Zero-Single} until there are no more \domaintype{Zero-Singles} in the scheduler.
\end{enumerate}
The pseudo-code of the domain making algorithm is shown in box ``Algorithm \ref{alg:Shellmaking}''.

\begin{algorithm}[h!]
\begin{algorithmic}
 \vspace{1EM}
 \State $Z \gets$ \domaintype{Zero-Single}
 \While{$z \in Z$}
    \State
    \ForAll{ domain in burst radius of $z$ }
	\If{ domain is \NOT \domaintype{Multi} \AND dt(domain) $> 0$ }
	\State $Z_{\rm bursted} \gets$ burst domain recursively
	\State $Z \cup Z_{\rm bursted}$
	\EndIf
    \EndFor
    \State
    \State $S \gets \{ \text{neighbouring surfaces of $z$} \}$
    \State $P \gets \{ \text{neighbouring particles of $z$} \}$
    \State $c \gets \text{closest object } n \in S \cup P$
    \State
    \If{$c \in S$ \AND $c$ in surface horizon of $z$}
	\State successful $\gets$ try interaction of $z$ with $c$
    \ElsIf{\NOT successful \AND $c \in P$ \AND $c$ in pair horizon of $z$}
	\State successful $\gets$ try to form \domaintype{Pair} $(z, c)$
    \ElsIf{\NOT successful \AND $c$ out of multi horizon of $z$}
	\State successful $\gets$ try to scale up shell of $z$
    \ElsIf{\NOT successful}
	\State form \domaintype{Multi} from $z$ with $c$ recursively
    \EndIf
    \State
    \State re-schedule $z$
    \State remove $z$ from $Z$
    \State
  \EndWhile
  \vspace{1EM}
\end{algorithmic}
\caption{
The eGFRD domain making algorithm.
\label{alg:Shellmaking}
}
\end{algorithm}

\pagebreak[4]
\subsection{Multi construction} \index{Multi}
\label{Shellmaking-MultiConstruction}
In eGFRD \domaintype{Multis} are contingent three-dimensional objects made up from either one
spherical \domaintype{Multi} shell or a set of overlapping spherical \domaintype{Multi} shells.
The radius of a \domaintype{Multi} shell is equal to the particle radius plus the reaction length
multiplied by the multi-shell factor $\mu > 1$.
\domaintype{Multi} domains are constructed recursively:
When a \domaintype{Zero-Single} $z$ has been determined to form a \domaintype{Multi} object it checks for
objects within its surroundings.
Any other \domaintype{Zero-Single} $z'$ that is within the common multi horizon $\mu(R_z + R_{z'})$
will be added to the \domaintype{Multi}.
Then, for each $z'$ that was added, the same check is performed for its surroundings,
ignoring $z$.
This is repeated until no further \domaintype{Zero-Singles} can be added to the \domaintype{Multi} object.
Note that \domaintype{Multi} shells in such configurations in principle can overlap with more than
one other \domaintype{Multi} shells.
If there are only surfaces within the horizon the \domaintype{Multi} will consist of
only one \domaintype{Multi} shell, the one around $z$.

\subsection{The test-shell concept}
\label{Shellmaking-TestShells} \index{test shell}
In order to prevent wasted effort, in eGFRD the sizing of a domain shell upon domain (re)con\-struc\-tion
is decoupled from sampling of next-event information from the Green's function.
This is achieved by using ``test shells''.
In a particular situation in which a new domain has to be created,
the simulator first attempts to determine the maximal size of the tentative test shell
of the domain, taking into account the required shell geometry (cylinderical or spherical)
and particular scaling parameters (e.g. the scale aspect ratio of cylinders).
Starting from a (predefined) maximal shell size, the test shell then is scaled down 
successively with respect to each neighboring shell via collision detection.
During the collision detection step the maximal dimensions of the test shell 
that does not lead to an overlap with the particular other domain shell are calculated
(see subsequent section for more detail).
If at the end of the scaling procedure the dimensions of the test shell are not smaller
than the required minimal dimensions (determined by the factors $\beta$ and $\mu$ defined further above
and particle radii)
a new domain object is parametrized with the test shell,
and only after this step its next-event information is sampled.
In the opposite case the construction of the respective domain type is rejected
and the algorithm proceeds by attempting the construction of another domain type (e.g. \domaintype{Single} or \domaintype{Multi}).

\begin{figure}[t!]
  \centering
  \includegraphics[width=0.45\textwidth]{\SketchesDir/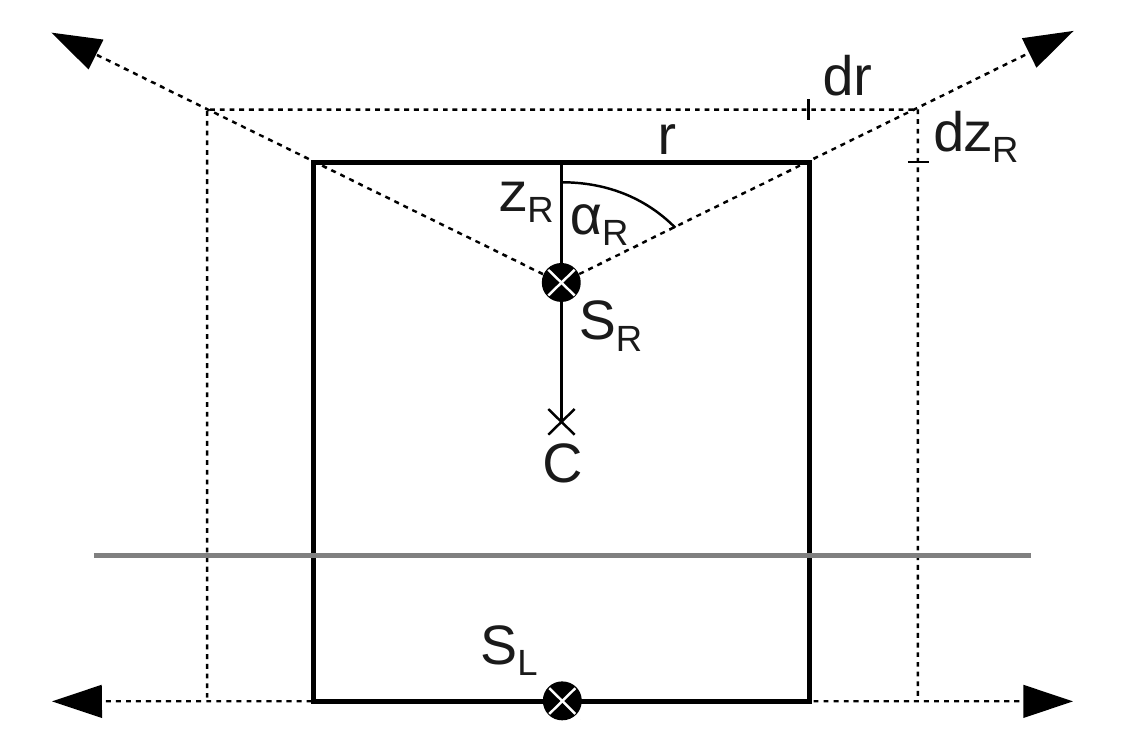}
  \caption{{\bf Cylinder scaling in eGFRD.}  
  The cylinder is scaled differently on its two sides (L and R) from
  two separate scale centers $S_L$ and $S_R$ which here do not coincide with the center point ($C$).
  For each side, the scale angle ($\alpha_L$, $\alpha_R$) defines the aspect ratio at scaling,
  i.e. how the respective height ($z_L$, $z_R$) scales with the cylinder radius ($r$).
  In the given example, $\alpha_L=\pi/2$, meaning that $z_L$ ($=0$) remains constant upon changing $r$.
  }
\label{fig:GFRD-CylinderScaling}
\end{figure}

\subsection{Shell collision detection}
\label{Shellmaking-CollisionDetection} \index{collision detection}
Considerable computational effort has to be put into detection of collisions
between a scaled test shell and another (static) shell.
While this problem is trivial when scaling spheres or cylinders against spheres
or parallel cylinders against each other, it is, maybe surprisingly, 
less straightforward for arbitrarily oriented and even orthogonal cylinders.
Note that in eGFRD cylinder scaling is performed subject to a fixed (but in principle arbitrary)
``scale aspect ratio'' (defining a certain ``scale angle''),
which links the change in cylinder height to the change in the radius.
The aspect ratio is usually set by the requirement of equalizing expected
first passage times of the enclosed particle towards the cap and the tube of the cylindrical shell;
as illustrated by Figure \ref{fig:GFRD-CylinderScaling}, the scale angle may differ for the two opposite sides of the cylinder\footnote{This is e.g. the case for the shell of the \domaintype{Planar Surface Interaction} domain (see section \ref{sec:2D-Binding}), 
the height of which is scaled only on the side of the planar surface facing the particle, while the height on the opposite side is kept fixed when the radius is scaled.
This is the example shown in Fig.~\ref{fig:GFRD-CylinderScaling}, where the gray line represents a planar surface.}.
In general, also the reference points of the scaling (``scale centers'') do not coincide with the midpoint
of the cylinder, particularly in cases in which scale angles are indeed different on both sides.
Thus, scaling a cylindrical shell in eGFRD conceptually consists of scaling the two sides of the cylinder separately, albeit linked via the common radius.
\index{scale aspect ratio} \index{scale angle} \index{scale center}

To scale orthogonal cylinders against each other in eGFRD we consider the two-cylinder problem
in a standardized cartesian coordinate system centered at the midpoint of the scaled cylinder,
with x-base-vector pointing towards the midpoint of the static cylinder and z-base-vector coinciding
with the axis of the scaled cylinder.
We then determine the specific type of collision that may occur upon scaling up the cylindrical test shell.
There are seven possible collision types:
\begin{enumerate}
 \item TF: the tube of the scaled cylinder hits the flat side of the static cylinder.
 \item TT: the tube of the scaled cylinder hits the tube of the static cylinder.
 \item TE: the tube of the scaled cylinder hits the edge\footnote{defined as the circular line that separates cylinder tube and flat side} of the static cylinder.
 \item FT: the flat side of the scaled cylinder hits the tube of the static cylinder.
 \item ET: the edge of the scaled cylinder hits the tube of the static cylinder.
 \item EE: the edge of the scaled cylinder hits the edge of the static cylinder.
 \item None: no collision possible in the given scenario\footnote{This e.g. may occur in cases in which only the radius or height are scaled.}.
\end{enumerate}
Identification of the collision type is facilitated by comparing the location
of the projected midpoint of the scaled cylinder to the projected edges of the static cylinder
in the xy-plane of the standardized coordinate system (in which the 
scaled cylinder appears circular and the static cylinder rectangular).
Certain respective locations exclude certain collision types;
for example, if the midpoint of the scaled cylinder is within the rectangular
projection of the static cylinder, the collision must be of type FT 
(given that the height is scaled).
Once the collision type is known, the maximal dimensions of the scaled cylinder
are determined taking into account its ``intrinsic'' scaling properties (scale angle, location of scale center, minimal size).
This is mostly achieved via straightforward geometric calculations.
For collision type EE a closed form for the new dimensions could not be obtained;
the values are therefore calculated from an implicit equation using a numerical rootfinder.
The detailed calculations are part of the eGFRD technical documentation
and beyond the scope of this thesis.

A concise scheme for scaling arbitrarily oriented eGFRD-type cylinders against each other
is yet to be devised.

\subsection{Convergence issues affecting domain construction}
The Green's functions, survival probabilities, cumulative PDFs
and the expressions for the boundary fluxes used in eGFRD
typically have the common form
\begin{align}
 C \cdot \nicesum{n=0}{\infty} e^{-\zeta^2_n Dt} X_n
\end{align}
where $C$ is constant and $X_n$ does not depend on $t$.
It can be shown that $\zeta^2_n$, to a good approximation, 
scales linearly with $n$ and inversely with the domain size $L$,
while typically $|X_n|\sim 1$.
Convergence of these sums thus is dominated by the exponential terms.
We found that it is severly hampered for evaluation times $t$ which are small 
on the typical timescale of the domain, i.e. the mean time required to traverse it by
diffusion with diffusion constant $D$.
This is the case when the distance $\Delta$ of the particle to
the closest boundary becomes very small.
Then evaluation times are of the order of $t_\Delta = \Delta^2 / 2dD$,
where $d$ is the dimensionality.
A fair estimate for the number $n$ of summation terms needed to reach a 
desired convergence threshold $\varepsilon$ follows from:
\begin{align}
 e^{- \left(\frac{cn}{L}\right)^2 Dt} < \varepsilon	\qquad\Leftrightarrow\qquad	n > \frac{L\sqrt{1/\varepsilon}}{c\sqrt{Dt}}
\end{align}
where we approximate $\zeta_n = c n / L$ with $c = const$.
Inserting $t_\Delta$ into the above equation yields
\begin{align}
 n_\Delta > \frac{L}{\Delta} \cdot const
\end{align}
showing that the required number of terms to reach a predefined convergence accuracy
scales inversely with the distance $\Delta$ to the closest boundary.
Therefore it should be avoided to construct domains in a way that $\Delta/L$ is small;
optimally domains should be constructed in a way that distances between the starting point
of the diffusing particle and the domain boundaries are approximately equal.

In practice this is handled in two ways:
Whenever particles start very close to radiating or absorbing boundaries 
we construct a domain that does not use all available space but is only sized up to
$L\simeq 2\Delta$.
While this requires the succesive creation of undersized domains (resulting in minor next-event times)
in order to elongate the distance between the reactive boundary and the particle,
it overcomes the abovementioned convergence issues by keeping $\Delta/L$ constant.
Alternatively, we scale domains up as much as possible and, where available,
employ Green's functions that are bounded unilaterally, i.e. neglect the distant second boundary.
These Green's functions typically are finite sums, which facilitates their implementation and computation,
and in the above case approximate the double-bounded solutions very well.

\clearpage
\section{Code profiling}
\label{sec:Profiling}
In order to assess the computational performance of our new eGFRD version we recorded the CPU time per simulated (``real'') time as a function of the (effective) number of simulated particles at constant volume, for the following scenarios:
\begin{itemize}
 \item ``3d, nR'': $N$ identical particles diffusing in a box (``cytoplasm'') with dimensions $L \times L \times L$ and periodic boundary conditions in $x$-, $y$- and $z$-directions.
 
 \item ``3d, wR'': The same setup, but with initially $N$ particles of species $A$ engaging in the simple reversible binding reaction $2A \rightleftharpoons B$, where $B$ is another cytoplasmic species that differs from $A$ only by its label and reactive behavior.
 
 \item ``2d, nR'': $N$ identical particles diffusing on a plane (``membrane'') with dimensions $L \times L$ and periodic boundary conditions in $x$- and $y$-directions.
 
 \item ``2d, wR'': The same setup, but with initially $N$ particles of species $A$ engaging in the simple reversible binding reaction $2A \rightleftharpoons B$, where $B$ is another plane-bound species that differs from $A$ only by its label and reactive behavior.
 
 \item ``1d, nR'': $N$ identical particles diffusing and drifting on a cylinder (``microtubule'') of length $L$ and periodic boundary conditions in $x$-direction.
 
 \item ``1d, wR'': The same setup, but with initially $N$ particles of species $A$ engaging in the simple reversible binding reaction $2A \rightleftharpoons B$, where $B$ is another cylinder-bound species that differs from $A$ only by its label and reactive behavior.
 
\end{itemize}
The parameters used for the respective cases are given in Table~\ref{tab:Prof-pars}. We chose the diffusion coefficients such that they correspond to realistic values in biological systems, and reaction parameters fast enough as to ensure that both the forward and backward reactions actually occur sufficiently often on the simulated time scale. To this end, we chose a fast dissociation rate $k_{\rm b}=100/s$ in all dimensions, while the forward rates were set to multiples of the predicted diffusion-limited rates in the respective dimensions; in the lower dimensions the latter are density-dependent.
Note that the drift in the 1d cases is only included in order to comprise all new features in the profiling; due to the periodic boundary conditions, it does not alter the expected spatial distribution on the cylinder.

\begin{table}
\centering
\begin{tabular}[b]{|l|l|c|c|l|}
 \hline
 {\bf Symbol} & {\bf Name} & {\bf Value / Formula} & {\bf Unit} \\
 \hline
 $L$ 			& System size				& 1.0 			& $\um$ \\
 $R_{\rm p}$ 		& Particle radius			& 5.0 			& $nm$ 	\\
 $\Rightarrow \sigma$ 	& Particle contact radius		& 10.0 			& $nm$ 	\\
 $N=N_{\rm A}$	& Number of $A$ particles 	& varied		&\\
 \hline
 $D_1$ 		& 1d-diffusion constant on cylinder	& 0.1 	& $\umsps$ 	\\
 $D_2$ 		& 2d-diffusion constant on plane 	& 0.1 	& $\umsps$ 	\\
 $D_3$ 		& 3d-iffusion constant in cytoplasm 	& 10.0 	& $\umsps$ 	\\
 \hline
 $v$		& drift velocity on microtubules	& 1.0		& $\umps$	\\
 \hline 
 $k_{\rm f,1}$	& intrinsic binding rate in 1d		& $2 D_1 N / L$				& $\unit{m/s}$	\\ 
 $k_{\rm f,2}$	& intrinsic binding in 2d		& $10 \cdot 4 \pi D_2 / \log(L^2/N\sigma^2)$	& $\unit{m^2/s}$	\\ 
 $k_{\rm f,3}$	& intrinsic binding in 3d		& $4 \pi \sigma D_1$			& $\unit{m^3/s}$	\\
 $k_{\rm b}$	& unbinding rate 			& 100					& $\unit{1/s}$	\\ 
 \hline
\end{tabular}
\caption{
\textbf{Overview of the standard parameters used in the profiling simulation runs.}
For the binding rates we use (multiples of) the diffusion-limited binding rates derived by \person{Kivenson} \& \person{Hagan} \cite{Kivenson2012}, which are different for each respective dimension and density-dependent in 1d and 2d (factors $N/L$ and $N/L^2$ in the formulas).}
\label{tab:Prof-pars}
\end{table}

For each setting and each set of parameters, we first carried out a calibration run by propagating 10 independent samples towards a stationary state in which the average copy numbers only fluctuate insignificantly between subsequent runs; this was followed by a profiling run in which the 10 samples were simulated further for a fixed amount of steps ($N_p=10^7$) while the CPU time and the (additional) simulated time were recorded.
In order to obtain a reference for comparison,
the profiling run was also carried out from the same initial conditions using our rvm-BD scheme (see sec.~\ref{sec:rvm-BD}) for particle propagation exclusively instead of \eGFRD (which occasionally also switches to rvm-BD, but only in very crowded situations).

The results for the respective cases above are summarized in Fig.~\ref{fig:Prof}.
The figure demonstrates that at low copy numbers ($\lesssim 100-1000$ for the given system size, translating to $\lesssim\mu M$ concentrations in 3d), 
the extended \eGFRD scheme is up to 3 orders of magnitude faster than our (optimized) rvm-BD scheme.
In 2d and 3d the advantage of \eGFRD over rvm-BD is less pronounced than in 3d, and the crossover at which rvm-BD becomes more efficient occurs at lower copy numbers, 
as expected from the fact that in lower dimensions the same amount of particles get crowded much faster than in 3d.
For the same reason, as explained in more detail in the following Section~\ref{sec:Profiling-Scaling}, the \eGFRD runtimes scale with the particle number with different exponents in the respective dimensions,
while the BD runtimes increase (approximately) linearly with the particle number in all dimensions;
the recorded data reproduces this scaling in parts of the scanned particle number ranges, with deviations at very low copy numbers and in the crossover regimes;
the latter can be attributed to crowding and an overall increased overhead originating from frequent particle interactions,
the former from the fact that this overhead is not present at particle numbers $\sim 1$.

\newcommand{\ProfGraphHeight}{0.3\textwidth}
\begin{figure}[p!]
  \centering
  \subfigure[][]{    
    \includegraphics[height=\ProfGraphHeight]{\PlotsSIDir/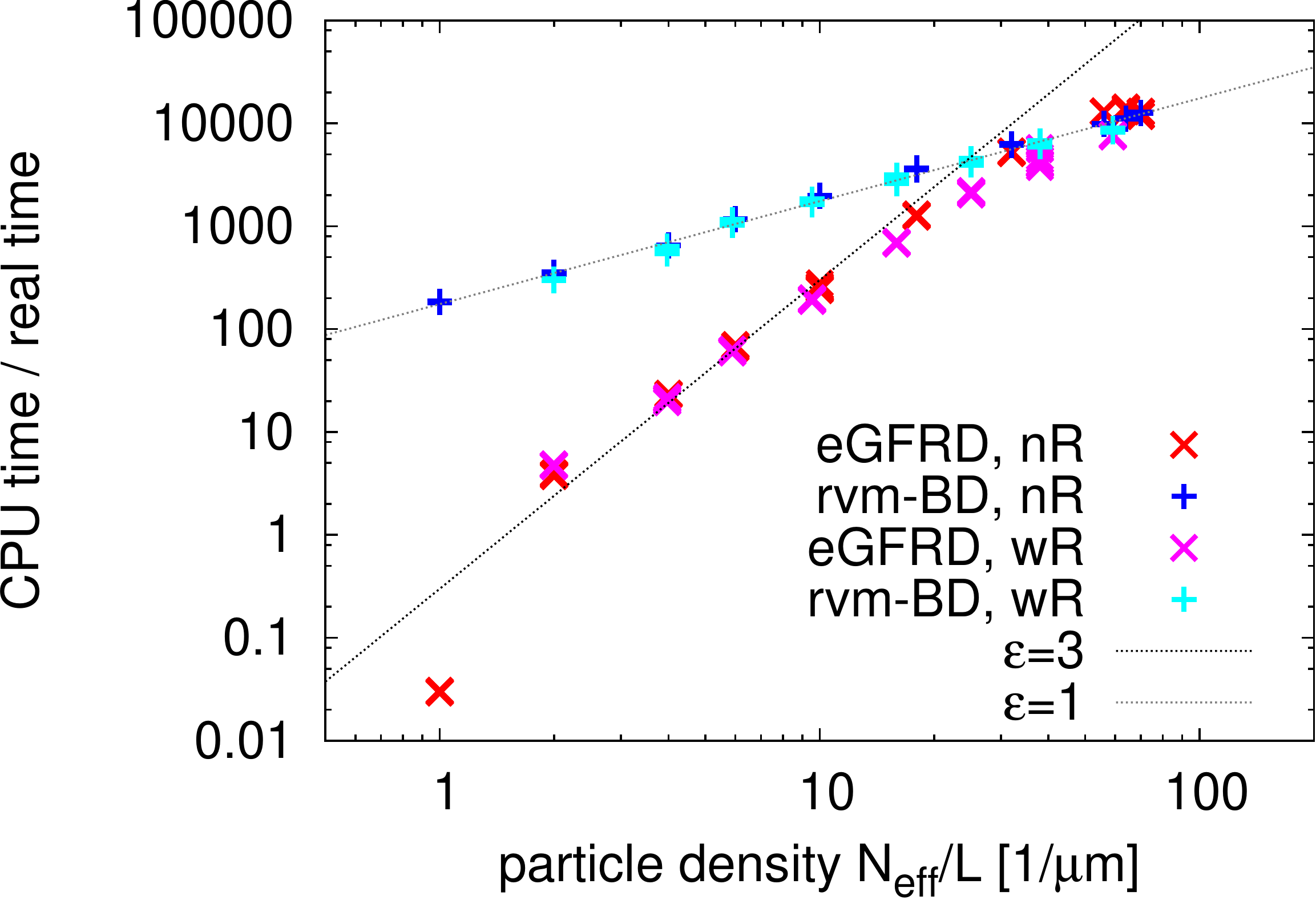}
    \label{fig:Prof-1d-nR+wR}
  } \hspace{0.05\textwidth}
  \subfigure[][]{    
    \includegraphics[height=\ProfGraphHeight]{\PlotsSIDir/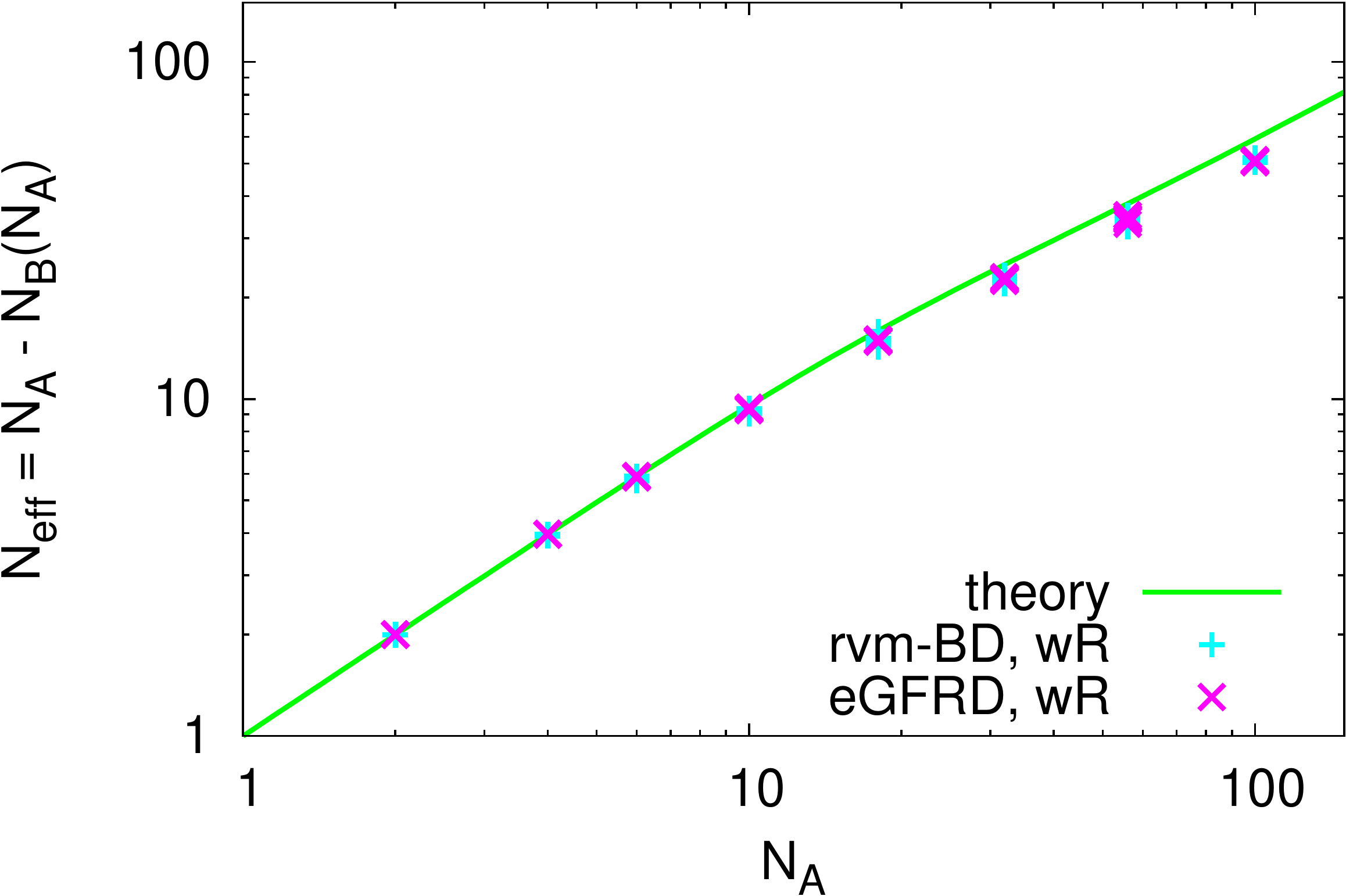}
    \label{fig:Prof-1d-CN}
  }\\
  \subfigure[][]{    
    \includegraphics[height=\ProfGraphHeight]{\PlotsSIDir/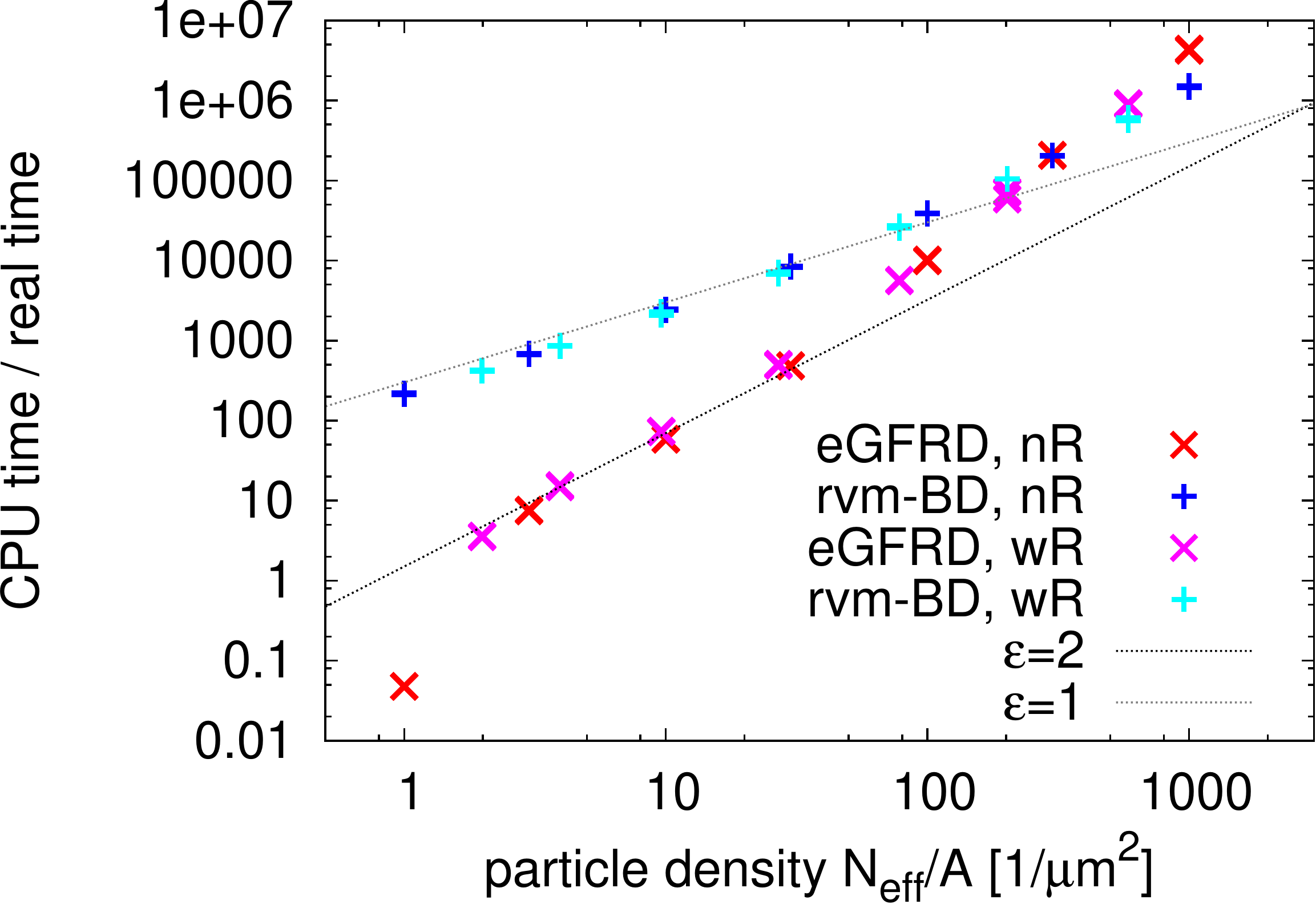}
    \label{fig:Prof-2d-nR+wR}
  } \hspace{0.05\textwidth}
  \subfigure[][]{    
    \includegraphics[height=\ProfGraphHeight]{\PlotsSIDir/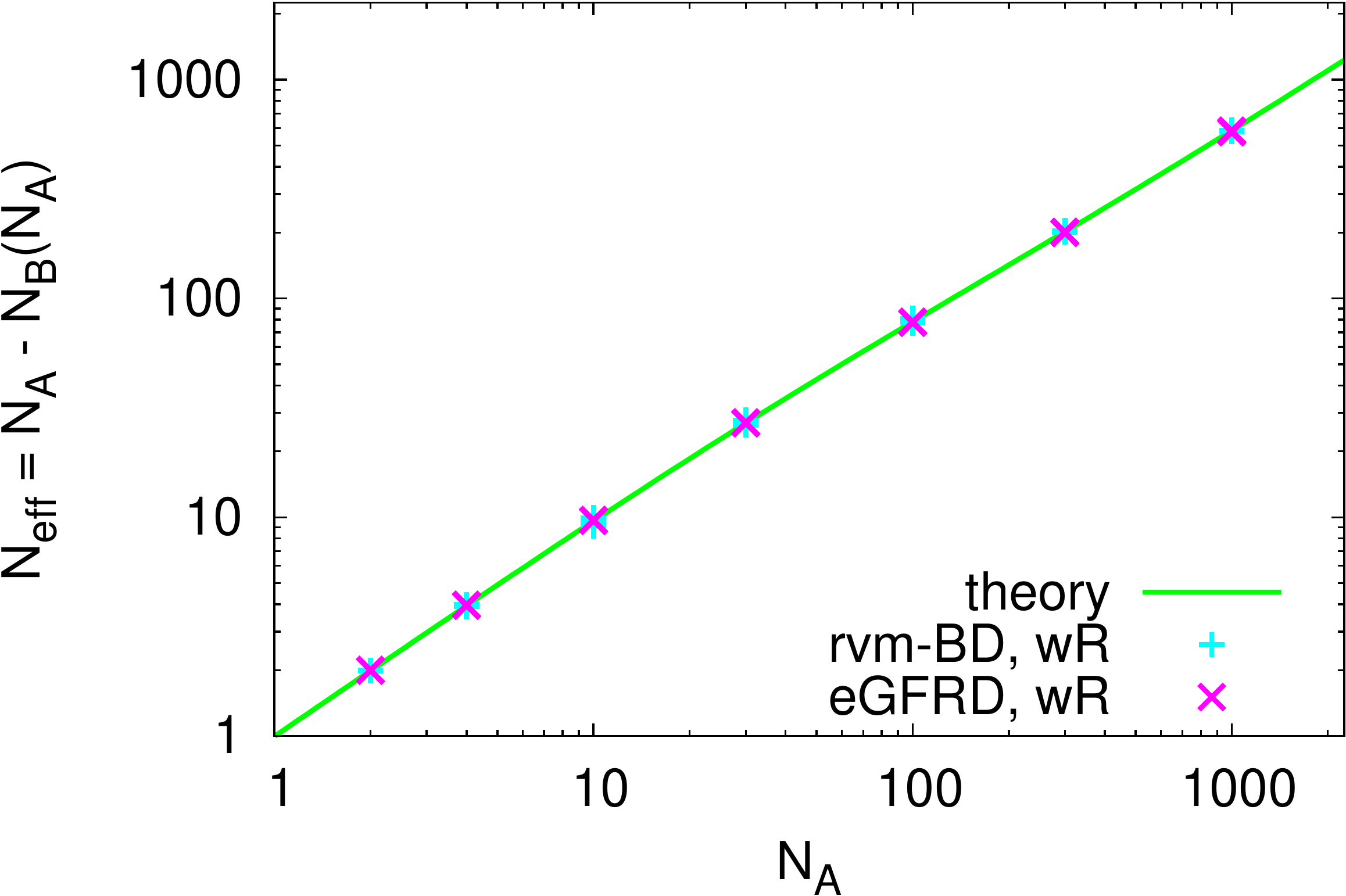}
    \label{fig:Prof-2d-CN}
  }\\  
  \subfigure[][]{    
    \includegraphics[height=\ProfGraphHeight]{\PlotsSIDir/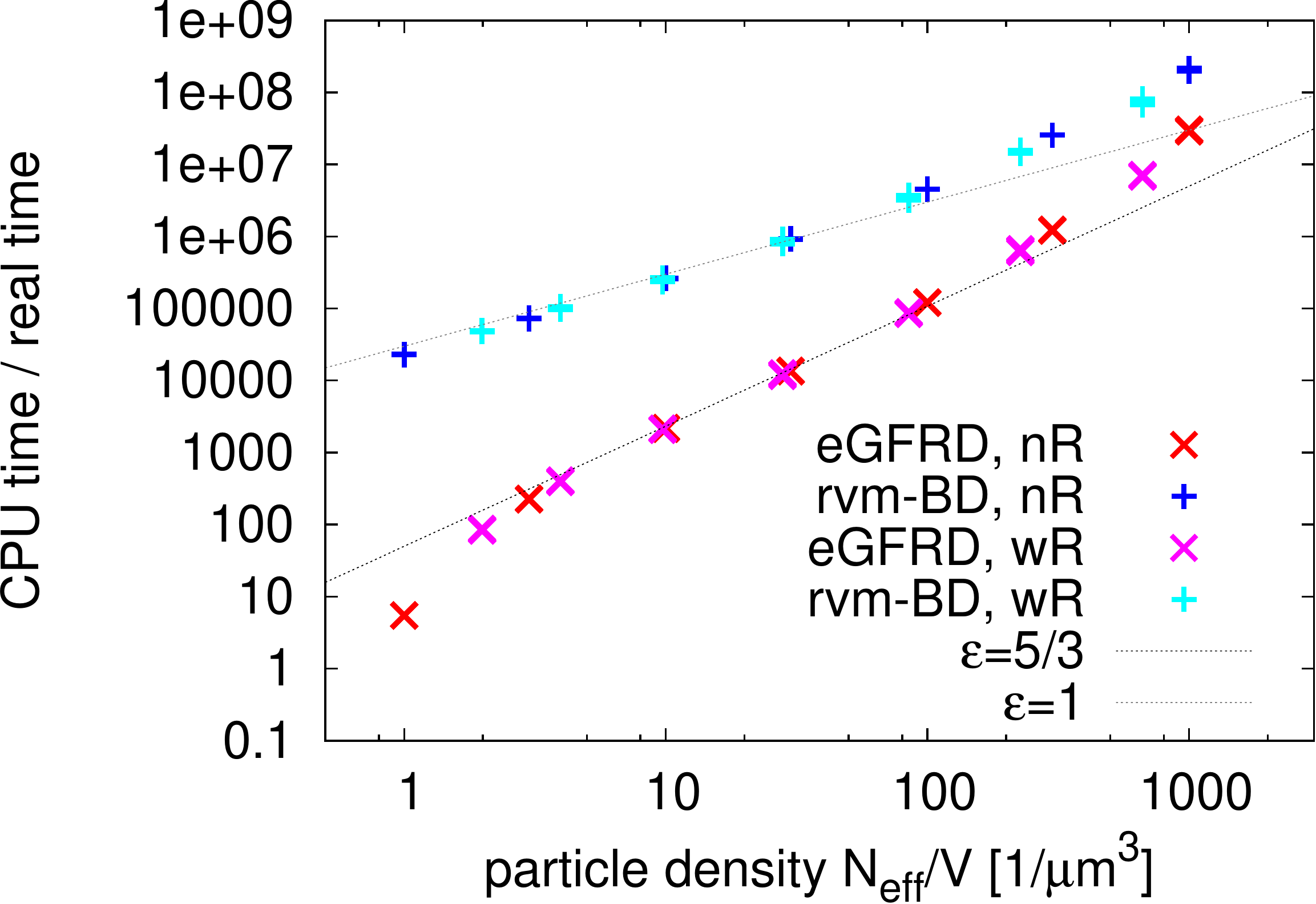}
    \label{fig:Prof-3d-nR+wR}
  } \hspace{0.05\textwidth}
  \subfigure[][]{    
    \includegraphics[height=\ProfGraphHeight]{\PlotsSIDir/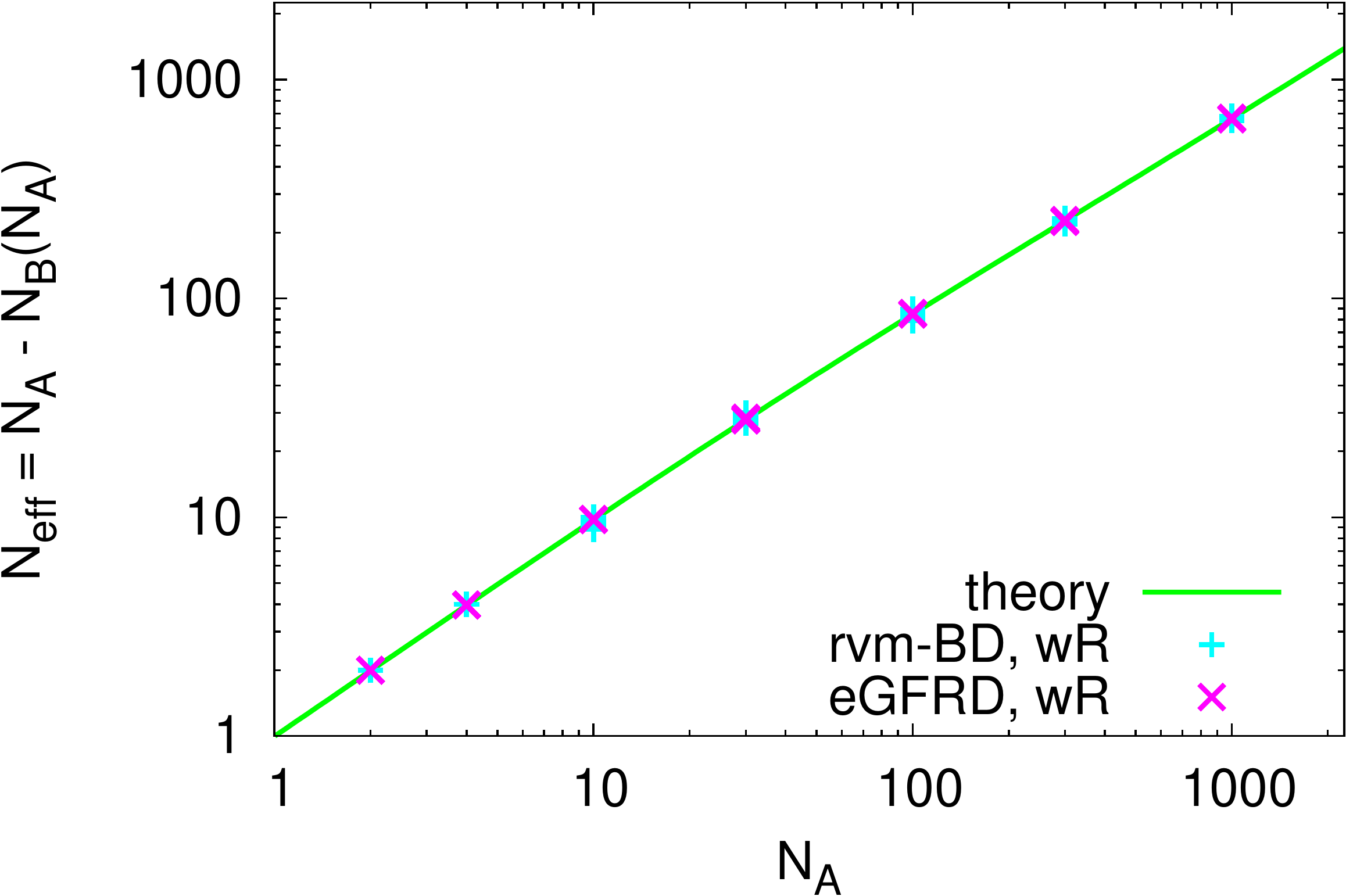}
    \label{fig:Prof-3d-CN}
  }\\  
\caption{
  \textbf{Code profiling results.}
  In the left column we show the CPU time per real (simulated) time as a function of the effective particle density $N_{\rm eff}/L^d$ ($d=1,2,3$),
  for the cases without reactions (nR) and with a reversible reaction $2A\rightleftharpoons B$ (wR),
  in 1d \subfigref{\subref{fig:Prof-1d-nR+wR}}, 2d \subfigref{\subref{fig:Prof-2d-nR+wR}} and 3d \subfigref{\subref{fig:Prof-3d-nR+wR}};
  in all cases, the eGFRD runs (red and magenta) are compared to runs carried out via our reaction-volume method BD scheme (rvm-BD, blue and cyan);
  the black and grey dashed lines show the expected scaling of the CPU time with $N_{\rm eff}$ for eGFRD ($\sim N^{\varepsilon>1}$, dimension-dependent, see Sec.~\ref{sec:Profiling-Scaling}) and BD (always $\sim N^1$), respectively.
  Note that for the ``nR'' runs $N_{\rm eff}=N$, while for the ``wR'' runs the total equilibrium copy number $N_{\rm eff}(N) \leq N$ is density-dependent;
  we plot the predicted and measured values of $N_{\rm eff}(N)$ in the right column, again for 1d \subfigref{\subref{fig:Prof-1d-CN}}, 2d \subfigref{\subref{fig:Prof-2d-CN}}, and 3d \subfigref{\subref{fig:Prof-3d-CN}}.
  Throughout all plots, each cross shows the data for one individual sample, with a sample size of 10 for each set of parameters.
  \TODO{Maybe we want to take out the points for N=1, because they are outliers (we know why)}
  \label{fig:Prof}
}
\end{figure}

\begin{figure}[p!]
  \centering
  \subfigure[][]{    
    \includegraphics[height=\ProfGraphHeight]{\PlotsSIDir/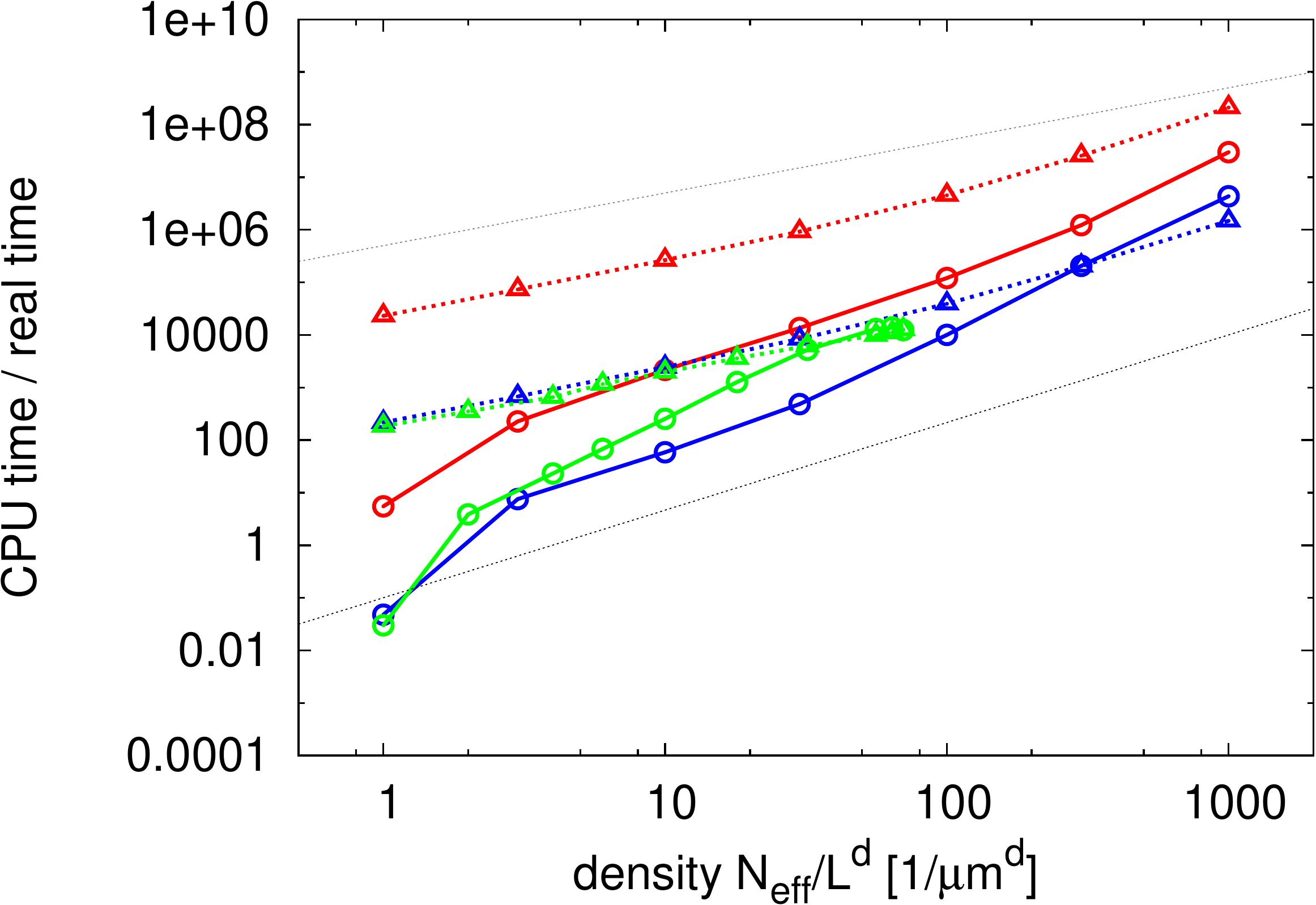}
    \label{fig:Prof-Comp-nR-N}
  } \hspace{0.05\textwidth}
  \subfigure[][]{    
    \includegraphics[height=\ProfGraphHeight]{\PlotsSIDir/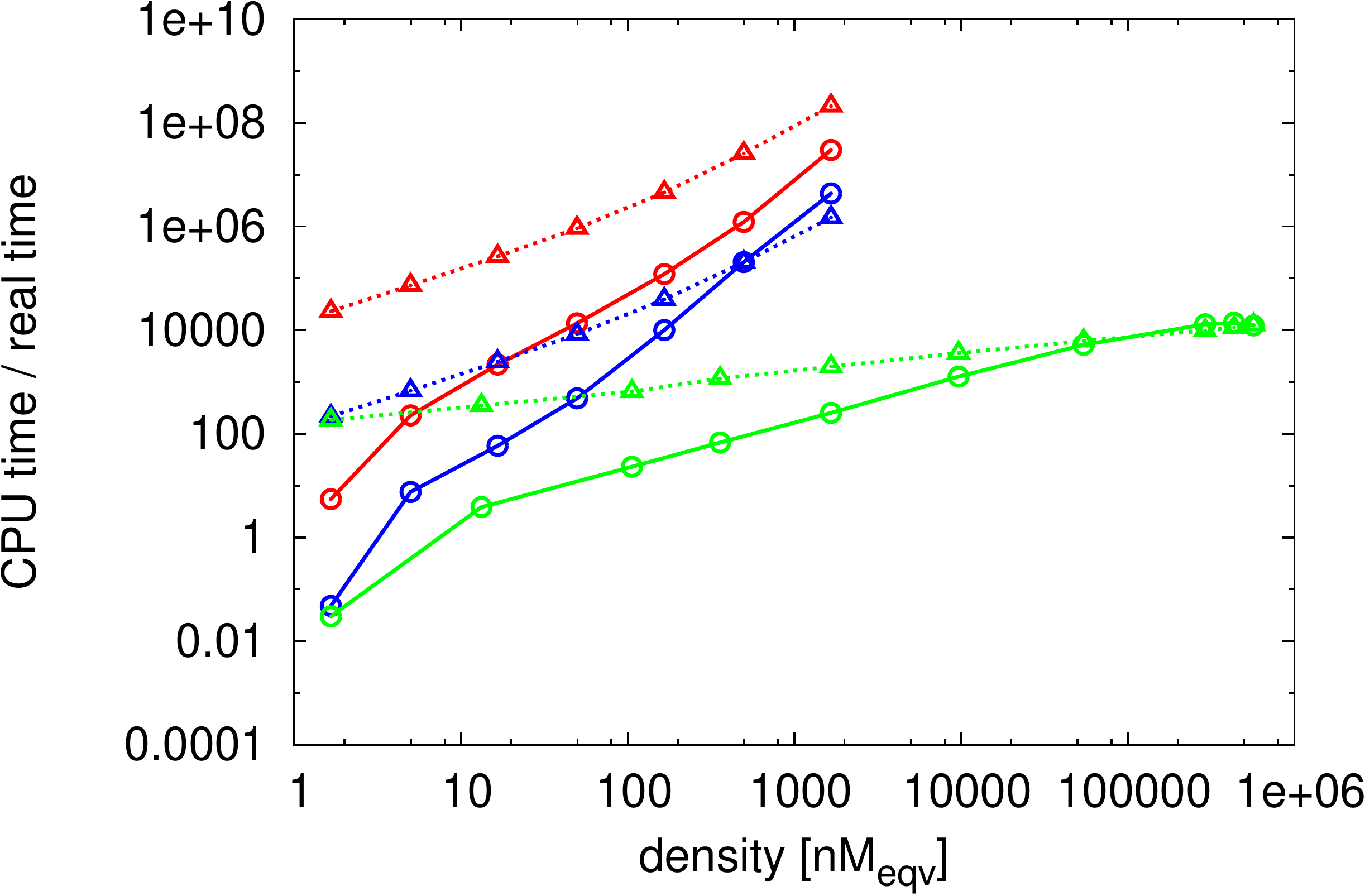}
    \label{fig:Prof-Comp-nR-nM}
  }\\
  \subfigure[][]{    
    \includegraphics[height=\ProfGraphHeight]{\PlotsSIDir/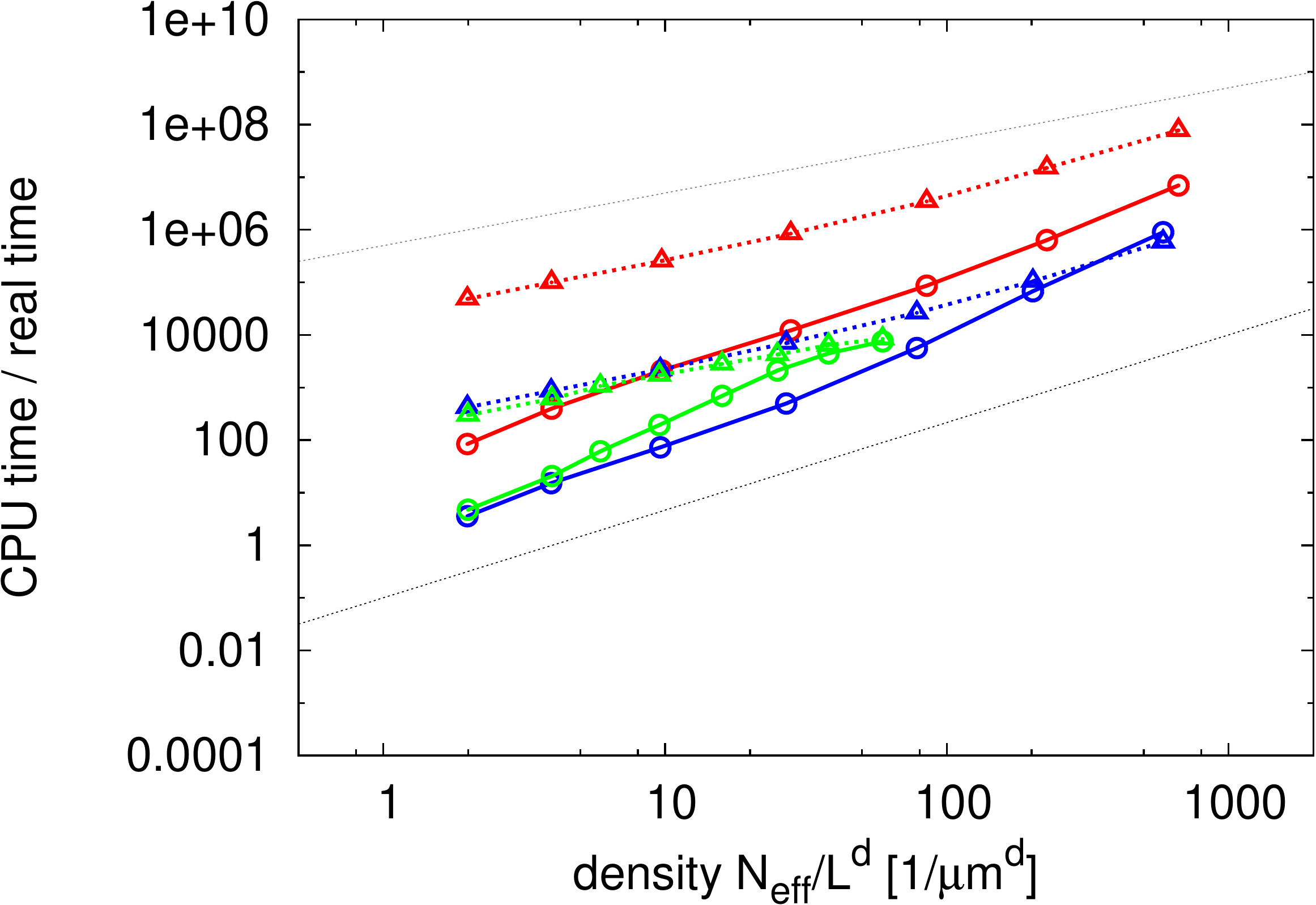}
    \label{fig:Prof-Comp-wR-N}
  } \hspace{0.05\textwidth}
  \subfigure[][]{    
    \includegraphics[height=\ProfGraphHeight]{\PlotsSIDir/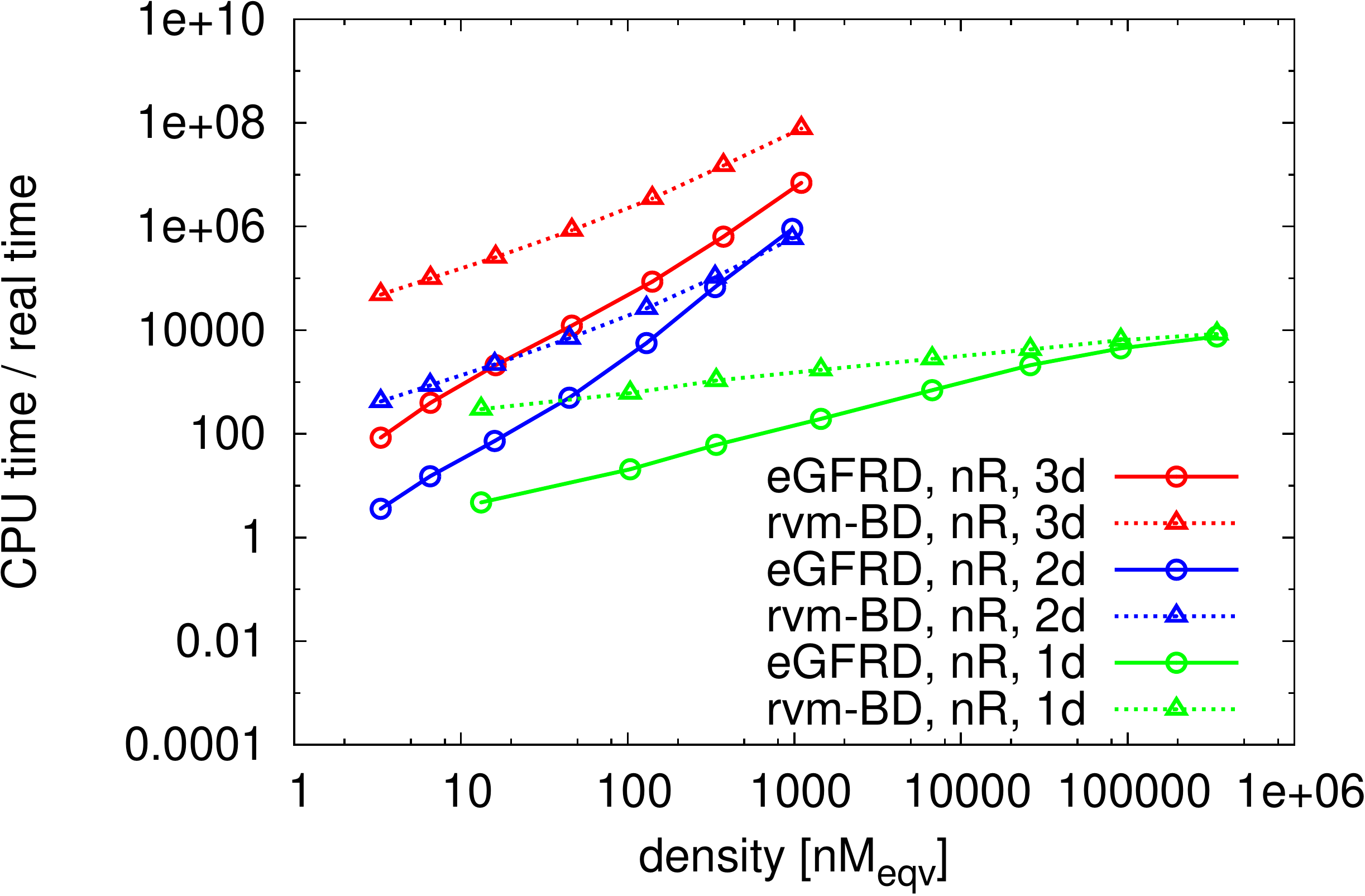}
    \label{fig:Prof-Comp-wR-nM}
  }\\  
\caption{
  \textbf{Code profiling comparison.}
  Here we show a comparison of the profiling results in the respective dimensions,
  for the case without (nR, top row) and with reversible reactions (wR, bottom row).
  They legend in panel \subfigref{\subref{fig:Prof-Comp-wR-nM}} is valid for all panels;
  the grey dashed line shows the predicted scaling for BD ($\sim N^1$),
  the black dashed line the predicted scaling for 3d-eGFRD ($\sim N^{5/3}$);
  the colored solid (\eGFRD) and dashed (rvm-BD) lines are guidelines for the eye.
  The plots in the left column display the CPU-time per simulated time as a function
  of the density $N_{\rm eff}/L^d$ in the respective dimension $d$, while the right column
  shows a ``fair comparison'' in which we convert the density to a ``nM equivalent'' ($\unit{nM_{eqv}}$)
  by rescaling the x-axis from $N_{\rm eff}/L^d$ to an effective 3d concentration $N_{\rm eff}^{3/d} / L^3$ measured in $\unit{nM}$;
  this unit allows a comparison at (a hypothetical) equivalent degree of crowding, or equivalent average interparticle distance.  
  Note that the 100-fold difference in CPU-time per real time in 3d as compared to 2d and 1d is expected,
  due to the 100-fold difference in the diffusion coefficients used in the respective dimensions.  
  \label{fig:Prof-Comp}
}
\end{figure}

Fig.~\ref{fig:Prof-Comp} shows a comparison between the dimensions, again for the cases without (upper panels) and with reversible reactions (lower panels).
In addition to replotting the data already shown in the separate panels of Fig.~\ref{fig:Prof} in the left column of Fig.~\ref{fig:Prof-Comp}, 
the respective right column panels display the data with an alternative quantity on the x-axis that allows for a more ``fair comparison'' at (a hypothetical) equivalent degree of crowding, captured by a ``3d density equivalent'' defined as
\begin{align}
 \rho_{\rm eqv}(N_{\rm eff}) \equiv \frac{N_{\rm eff}^{3/d}}{L^3}
\end{align}
for the respective dimension $d$.
For the lower dimensions, the ``3d density equivalent'' maps the lower-dimension density to a 3d concentration that would result in an average interparticle distance (roughly) equal to the one in the lower dimension; in 3d it is simply identical to the classical definition of the particle concentration. In the right column of Fig.~\ref{fig:Prof-Comp} we measure these concentrations in $\unit{nM}$ units, but we refer to it as ``$\unit{nM}$ equivalent'' ($\unit{nM}_{\rm eqv}$) to emphasize the hypothetical character of the quantity.
The plots in the left column demonstrate that, as expected, the crossover at which eGFRD becomes more expensive than BD happens at smaller particle numbers in 1d and 2d as compared to 3d;
however, the ``fair comparison'' in the right column also shows that the crossover occurs at lower equivalent degree of crowding in 2d than in 3d, but at a markedly higher degree of crowding in 1d.

Overall, although the profiling reveals that our prototype implementation still offers room for improvements, our \eGFRD scheme compares well against an optimized, smart version of BD;
it should be emphasized that in the rvm-BD scheme that we compare to, the average propagation time steps are more than 3 orders of magnitude higher than in a conventional brute-force BD scheme 
with a time step $\Delta t \sim 10^{-6}~\sigma^2/D$.

\subsection{Expected scaling of CPU time with the particle number}
\label{sec:Profiling-Scaling}
While in BD simulations with a fixed time step $\Delta t$ the computational cost is expected to scale linearly with the number of propagated particles $N$
(given that particle interactions are treated in a smart way, e.g. via cell lists, such that the cost for simulating particle reactions is not disproportionately higher than the cost of their propagation), the following scaling argument can be made for \eGFRD:
Generally, for a scheme with varying propagation time step $\Delta t$, 
we expect the runtime $T$ to scale in proportion to the particle number $N$, but inversely with the average time step $\la \Delta t \ra$.
In \eGFRD, $\la \Delta t \ra$ directly depends on the space available for diffusion without particle collisions, 
and therefore is roughly proportional to the square of the average distance between the particles $\la \lambda \ra$: $\la \Delta t \ra \sim \la \lambda \ra^2$.
The average interparticle distance $\la \lambda \ra$, however, scales differently with $N$ in the respective dimensions: $\la \lambda \ra \sim N^{-1/d}$.
Taken together this yields
\begin{align}
 T \sim \frac{N}{\la \Delta t \ra} \sim \frac{N}{\la \lambda \ra^2} \sim N^{1 + \frac{2}{d} }
\end{align}
and for the respective dimensions:
\begin{align}
 T &\sim N^{\frac{5}{3}}	&\text{in 3d}\hspace{3cm}	\\
 T &\sim N^{2}			&\text{in 2d}\hspace{3cm}	\\
 T &\sim N^{3}			&\text{in 1d}\hspace{3cm}
\end{align}

\subsection{Formulas for stationary equilibria in the simulations with reversible reactions}
For completeness, here we briefly derive the formulas for the effective copy number in our example reversible-binding reaction $A+A \rightleftharpoons B$, defined as
\begin{align}
 N_{\rm eff} \equiv \la N_A \ra + \la N_B \ra	\quad ,
\end{align}
where the brackets refer to the stationary mean of the copy numbers for the respective species.

The stationary copy numbers can be obtained by solving the following simple ODE for the particle densities $\rho_X \equiv N_X / S$ (where $S$ is the dimension-dependent, relevant system size)
\begin{align}
 \pd_t \rho_B = k_f \left(\frac{\rho_A}{2}\right)^2 - k_b~\rho_B
\end{align}
in steady state ($\pd_t \rho_B \equiv 0$),
under the side constraint that--at any moment--the initial total copy number $N\equiv N_A(0)$ is conserved as follows:
\begin{align}
 N_A(t) + 2 N_B(t) \equiv N
\end{align}
Note that this implies $N_{\rm eff} = N - \la N_B \ra$.
Furthermore, since in some of our simulations crowding effects are significant, it is important to use the effective system size reduced by the fraction taken up by the particles 
for calculating the densities.
The steady-state equation to solve then reads:
\begin{align}
 \pd_t \frac{\la N_B \ra}{S_0 - N_{\rm eff} S_p} &= \frac{k_f}{4} \left(\frac{N - 2\la N_B \ra}{S_0 - N_{\rm eff} S_p}\right)^2 - k_b~\frac{\la N_B \ra}{S_0 - N_{\rm eff} S_p}	= 0\\
 &\nn \\
 \Leftrightarrow \quad \left(N - 2\la N_B \ra\right)^2 &= 4 K_d \left[ S_0 - \left(N-\la N_B \ra\right) S_p \right] ~\la N_B \ra
\end{align}
where $K_d = k_b / k_f$ and $S_0$ denotes the total available space, while $S_p$ is the space taken up by a single particle; 
these quantities depend on the dimension as follows ($R_p$ being the particle radius): 
$S_0=L^3$ and $S_p=4/3~\pi R_p^3$ in 3d;
$S_0=L^2$ and $S_p=\pi R_p^2$ in 2d;
$S_0=L$ and $S_p=2 R_p$ in 1d.

This yields the stationary solution
\begin{align} 
 \la N_B \ra = \frac{ N + K_d (S_0 - N S_p) \pm \sqrt{N^2 S_p (K_d^2 S_p - 1) - 2 N S_0 (K_d^2 S_p - 1) + K_d^2 S_0^2} } {2 - 2 K_d S_p}
\end{align}
where only the solution with the negative sign matches the condition $\la N_B \ra = 0$ for $N=0$.
From this we obtain:
\begin{align}
 N_{\rm eff}(N) &= N - \langle N_B \rangle	\nonumber\\
		&= \frac{N - K_d(S_0 + N S_p) + \sqrt{N^2 S_p (K_d^2 S_p - 1) - 2 N S_0 (K_d^2 S_p - 1) + K_d^2 S_0^2}} {2 - 2 K_d S_p} 
 \label{eq:Neff}
\end{align}
This formula is valid in all dimensions, but note that in our profiling simulations $K_{\rm d} = k_{\rm b} / k_{\rm f}$ changes with $d$;
more specifically, $k_{\rm b}=100/s$ is held constant in all dimensions, while for the forward rate we chose multiples of the diffusion-limited rate in the respective dimensions (see Table~\ref{tab:Prof-pars}); for the diffusion-limited rates we use the expressions derived by \person{Kivenson} \& \person{Hagan} \cite{Kivenson2012}.

The average sampled copy numbers in our simulations agree very well the predictions of Eq.~(\ref{eq:Neff}), see Fig.~\ref{fig:Prof}, panels \subref{fig:Prof-1d-CN}, \subref{fig:Prof-2d-CN} and \subref{fig:Prof-3d-CN};
only in 1d (panel \subref{fig:Prof-1d-CN}) we observe that at higher copy numbers the stationary equilibrium is slightly biased towards the product state, resulting in an overall lower $N_{\rm eff}$; we attribute this to the fact that in crowded situations the dissociation reaction is more often rejected, due to lack of space;
the fact that both the \eGFRD and BD simulations deviate from the theory to the same degree support the notion
that the deviation is not an artifact of the simulation methods, but a limitation of the modelling assumptions (i.e., the naive handling of dissociation attempts with insufficient space).

\clearpage
\section{Pom1-gradient simulations}
\label{sec:Pom1-sim-protocol}

\subsection{Parameter choice}
\label{sec:Pom1-Pars}
Table~\ref{tab:Pom1-Pars} contains an overview of the parameters that we used for our exemplary simulations of Pom1 gradient formation,
described in detail in section~\ref{sec:Example-sims} of the main text.
Where available, we took parameter values from the literature (sources given);
otherwise we employed typical values or own estimates resulting in gradients that (roughly) reproduce the experimental observations,
in particular the gradient length scale; more detailed comments on this are found in the table and corresponding footnotes.

\subsection{Data acquisition and analysis}
Here we briefly describe how we recorded and analysed our Pom1 gradient data.

Starting from a situation in which all Pom1 particles are randomly placed in the cytoplasm in their fully phosphorylated state, 
we first propagated the simulations until all of them reached a simulated (real) time $t\geq 30~s$, to allow the Pom1 gradient on the membrane to be established and equilibrate.
We then continued the simulations for measurement runs for an additional simulated time of at least $30~s$, and recorded the complete particle trajectories (positions and species) with a data acquisition interval of $0.1~s$.

After the simulations had terminated, the copy numbers of the membrane-associated Pom1 in the respective phosphoforms, $N_n$, $n=1..n_{\rm p}$, and our main observable, the total number of membrane-bound Pom1 in all phosphoforms, $N_{\rm m,tot}\equiv \sum_{n=1}^{n_{\rm p}} N_{\rm n}$, was binned into 2d-histograms of particle density ($\rho_{\rm m}(x,y)\equiv$ local no. of particles / bin area) with $50 \times 50$ bins stretching over the whole membrane plane ($10~\um \times 10~\um$);
this was done both for individual time frames and with all measured data in one histogram, whereby in the latter case we also computed the (local) variance with respect to the time average in each histogram bin.
The same binning and averaging procedure was also carried out for a one-dimensional histogram in which the abscissa is the radial distance from the gradient origin (particle injection point) at $x_0=5~\um$, $y_0=5~\um$, i.e. a histogram of $\rho_{\rm m}(\Delta r)$ with $\Delta r \equiv \sqrt{(x-x_0)^2 + (y-y_0)^2}$.

We then fitted exponential functions of the form $f(\delta) = A \exp(-\delta/L)$ both to the x- and y-sections of the 2d-density histograms (thus with $\delta=(x-x_0)$ and $\delta=(y-x_0)$), and to the 1d-histogram (with $\delta = \Delta r$ as defined above) in order to obtain the gradient amplitude $A$ and the gradient decay length $L$ for the given set of parameters.
For the vast majority of gradients the fitted values of $A$ and $L$ do not depend much on which of the fits is used; however, we decided to proceed with the values from the fits to the radial 1d-histogram because they comprise all binned data in one graph.

\subsection{Results for the trans-model with slower phosphorylation rates}
\label{sec:trans-slowerPhos}
\begin{figure}[h!]
  \centering
  \subfigure[][]{
    \label{fig:trans-slowerPhos-Gradients-Ampl}
    \includegraphics[width=\GradientPlotWidth]{\PlotsDir/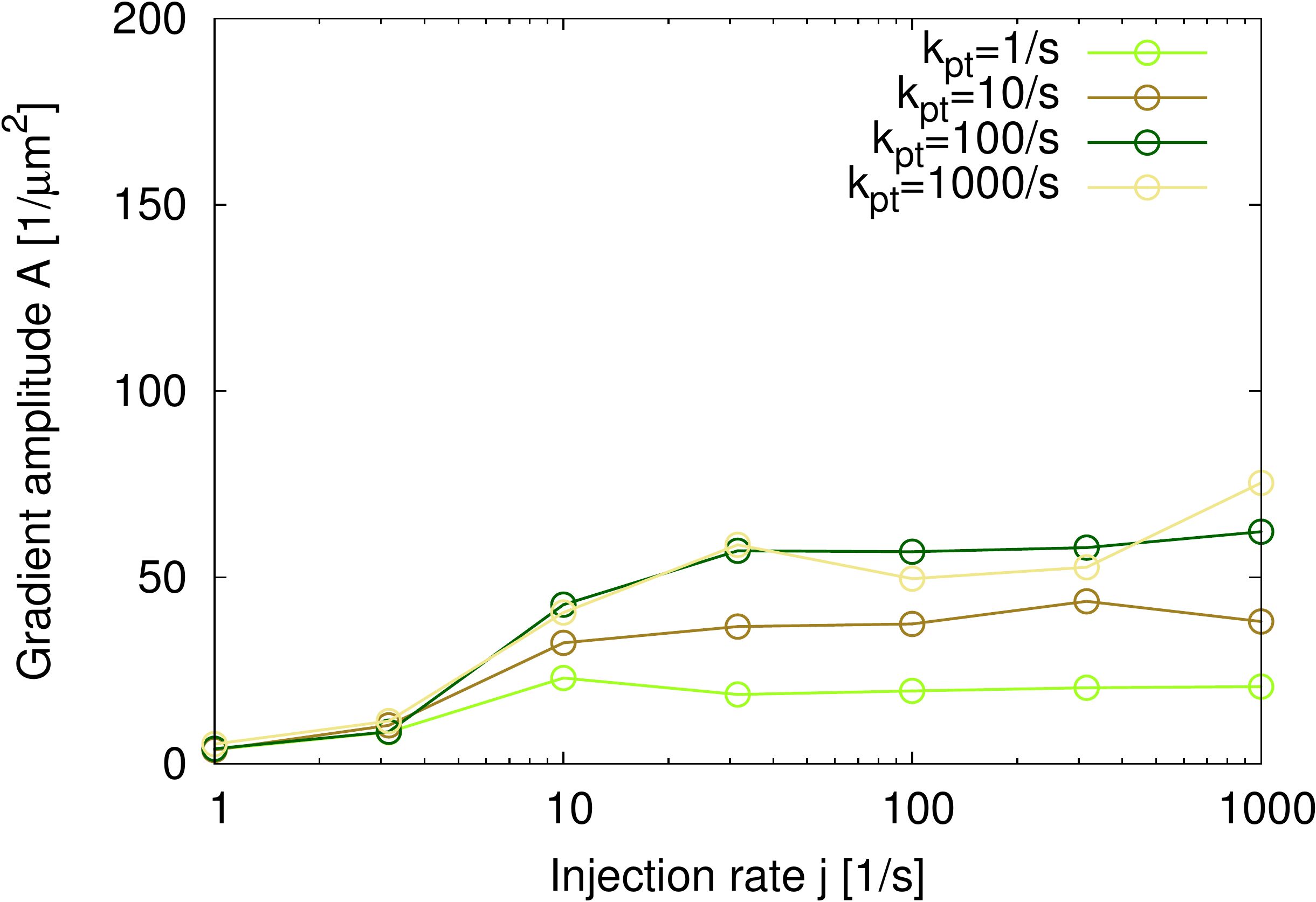}
  }
  \subfigure[][]{
    \label{fig:trans-slowerPhos-Gradients-Length}
    \includegraphics[width=\GradientPlotWidth]{\PlotsDir/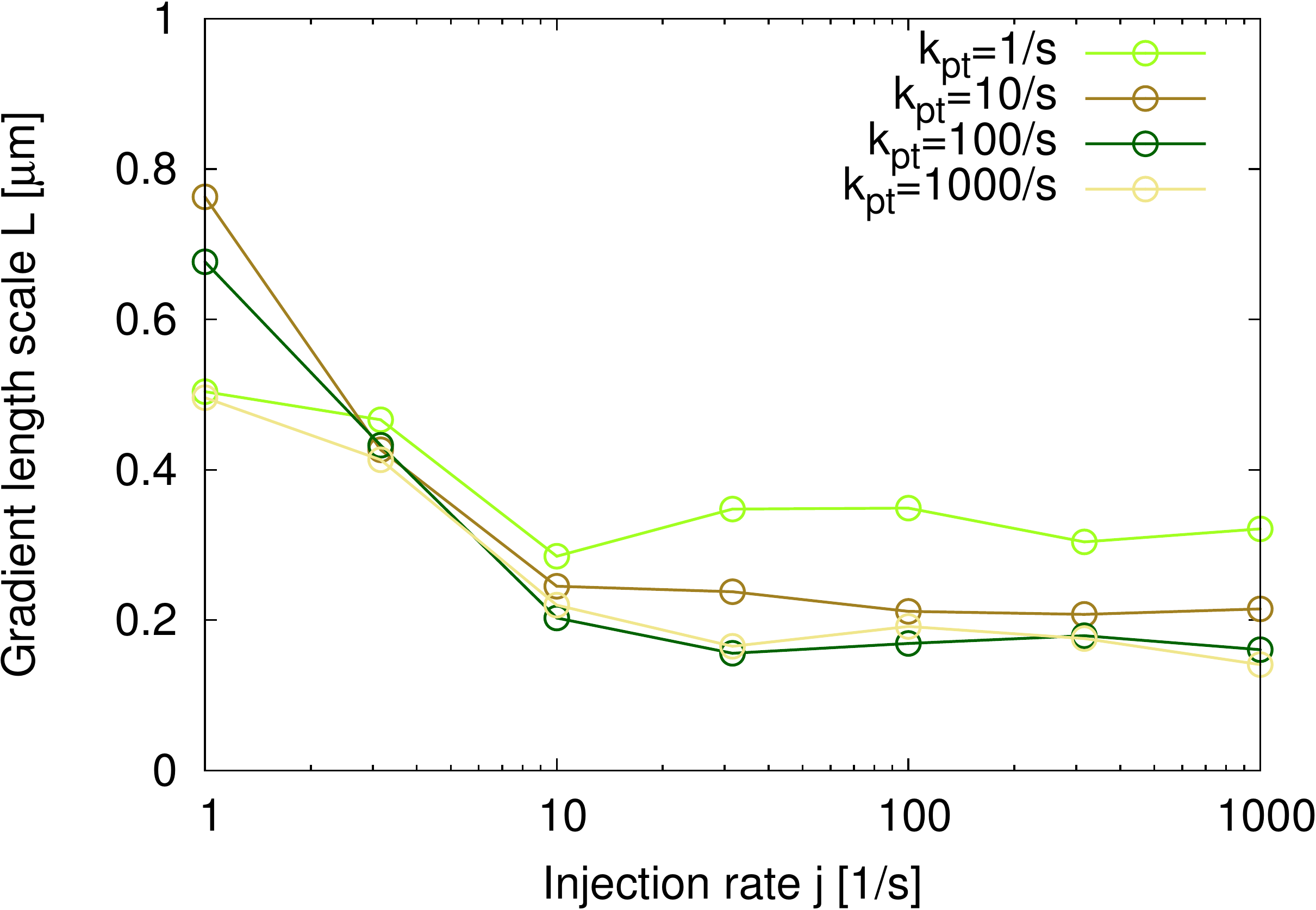}
  }\\  
  \vspace{1EM}
  \subfigure[][]{
    \label{fig:trans-slowerPhos-Gradients-Corr}
    \includegraphics[width=\GradientPlotWidth]{\PlotsDir/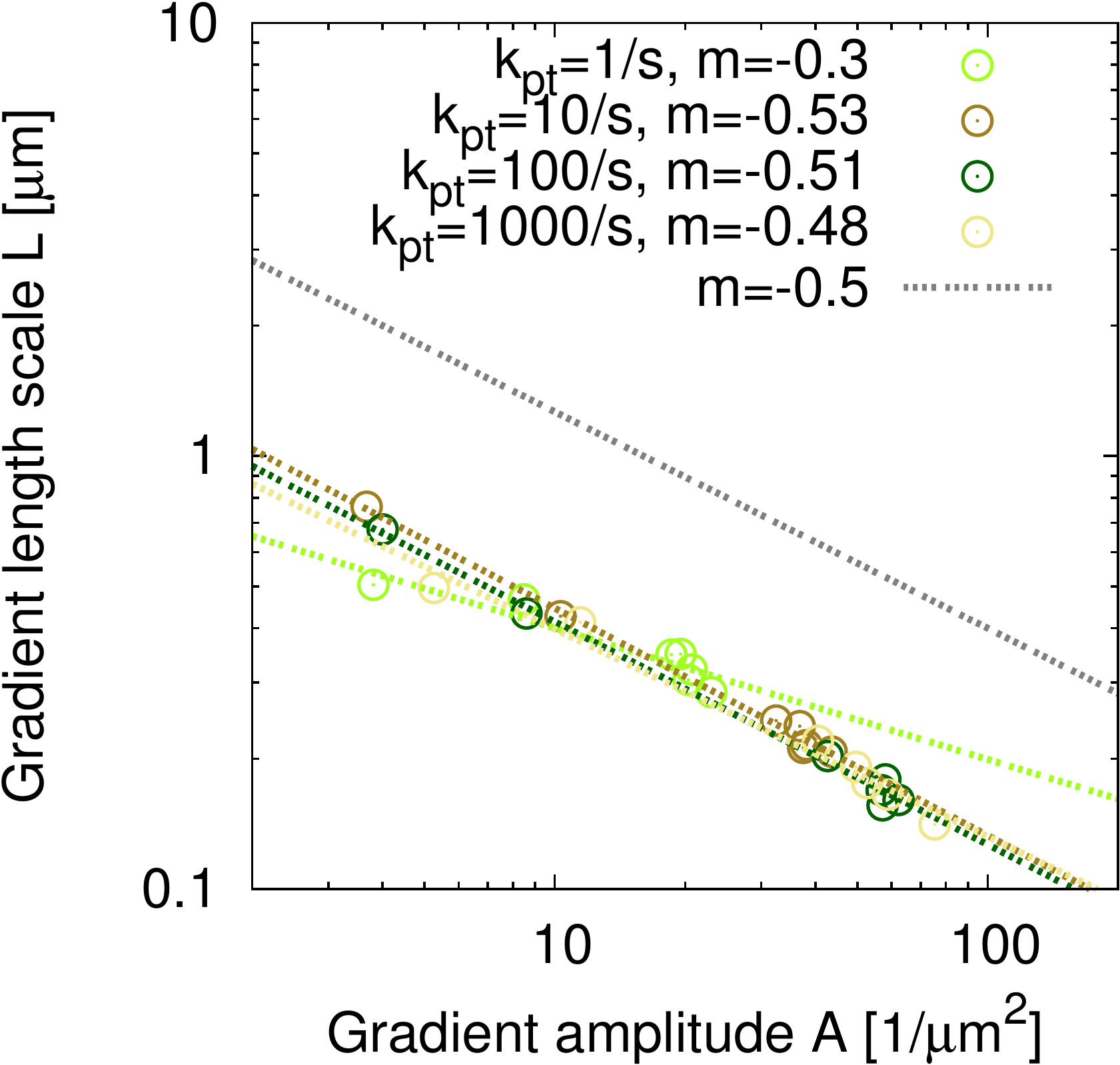}
  }
    \subfigure[][]{
    \label{fig:trans-slowerPhos-Gradients-Noise}
    \includegraphics[width=\GradientPlotWidth]{\PlotsDir/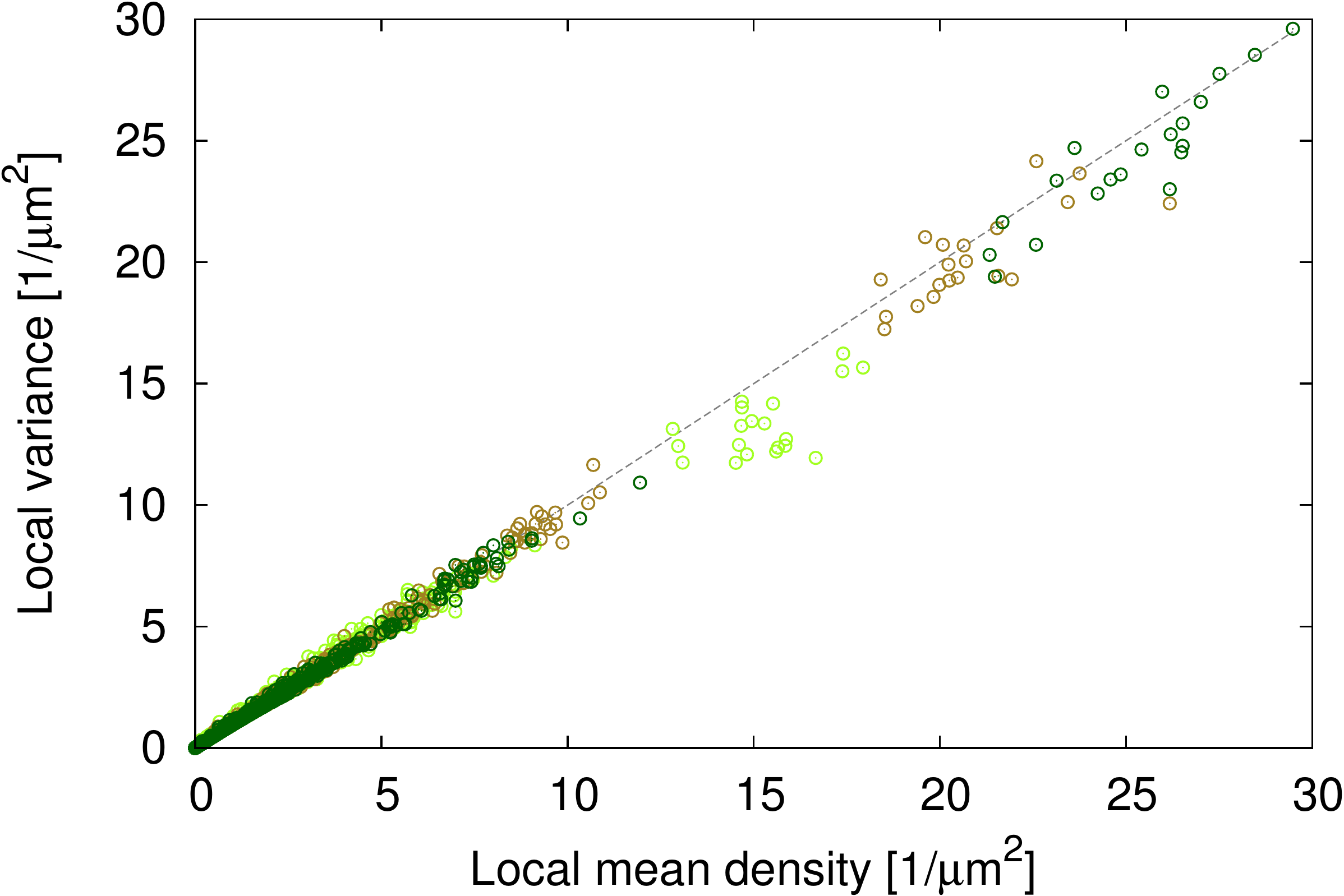}
  }  
\caption{
  \textbf{Gradient characteristics for different autophosphorylation rates (trans-model).}
  Gradient amplitude \subfigref{\subref{fig:trans-slowerPhos-Gradients-Ampl}} and length scale $L$ \subfigref{\subref{fig:trans-slowerPhos-Gradients-Length}},
  obtained by fitting an exponential function to stationary radial profiles,
  plotted against the injection rate $j$ at the gradient origin for different \CH{in-complex phosphorylation-dissociation rates $k_{\rm pt}$.}
  Lines are guidelines for the eye.
  \subfigref{\subref{fig:trans-slowerPhos-Gradients-Corr}} Log-log plot of the gradient length $L$ vs. gradient amplitude $A$ for the same data.  
  \subfigref{\subref{fig:trans-slowerPhos-Gradients-Noise}} Variance vs. mean for the local total membrane-bound Pom1 density,
  combining the values of all histogram bins into one scatter plot;
  point colors correspond to the colors in \subfigref{\subref{fig:trans-slowerPhos-Gradients-Ampl}} - \subfigref{\subref{fig:trans-slowerPhos-Gradients-Corr}} (different values of $k_{\rm pt}$).
  \label{fig:trans-slowerPhos-Gradients}
}
\end{figure}

In the main text we present our results for a trans-phosphorylation model in which $k_{\rm u}$, the unbinding rate of fully phosphorylated Pom1, is varied, while $k_{\rm pt}$, \CH{the rate of in-complex trans-phosphorylation and complex dissociation}, is set to a very fast value ($1000/s$).
This is an ad-hoc assumption, and the real phosphorylation rate is unknown.
In order to assess how our results depend on this, we also carried out simulations in which we vary $k_{\rm pt}$ at constant unbinding rate $k_{\rm u}=5/s$.
The results are summarized in Fig.~\ref{fig:trans-slowerPhos-Gradients}.
We observe that that slow trans-phosphorylation rates ($<100/s$) lower the maximal gradient amplitude (panel \subref{fig:trans-slowerPhos-Gradients-Ampl}),
while simultaneously the gradient length scale increases (panel \subref{fig:trans-slowerPhos-Gradients-Length});
this is consistent with the fact that slower release after Pom1-dimer formation allows the particles to diffuse away further from the injection point before they are fully phosphorylated and ready to unbind.

Yet, the observed dependencies of the amplitudes and length scales per se do not tell us how the buffering effect in the trans-phosphorylation model is affected by $k_{\rm pt}$.
This dependency is shown in panel \subref{fig:trans-slowerPhos-Gradients-Corr} of Fig.~\ref{fig:trans-slowerPhos-Gradients}.
Here we see that the anticorrelation between gradient length and amplitude is preserved as long as $k_{\rm pt}\geq 10/s$, although the maximal amplitudes for $k_{\rm pt}=10/s$ are already down to approximately half the values as for faster phosphorylation rates; only for very slow $k_{\rm pt}=1/s$ the slope in the anticorrelation plot is reduced from $\sim -0.5$ to $\sim -0.3$.
As evidenced by panel \subref{fig:trans-slowerPhos-Gradients-Noise} of Fig.~\ref{fig:trans-slowerPhos-Gradients}, altering $k_{\rm pt}$ does not affect the Poissonian character of the observed variances.

Overall, the buffering effect in the trans-model thus turns out to be very robust with respect to the value of the \CH{in-complex phosphorylation-dissociation rate $k_{\rm pt}$}.

\begin{table}
\centering
\begin{tabular}[h!]{|l|l|c|c|l|}
 \hline
 {\bf Symbol} & {\bf Name} & {\bf Value} & {\bf Unit} & {\bf R/C} \\
 \hline
 $L$	        & Side length of sim. box and membr. plane		& 10		& $\um$ &\\
 $L_{\rm cyl}$	& Length of microtubule cylinder				& 9.85		& $\um$ &\\
 $R_{\rm cyl}$	& Radius of microtubule cylinder				& 25		& $nm$  &\\
 $R_{\rm part}$	& Particle radius						& 10		& $nm$  &\\
 $N_{\rm p}$	& Total Pom1 particle number					& 480		&   &\\
 \hline
 $D_{\rm cyt}$ 	& Pom1 cytoplasmic diff. const. 	& 1.5 		& $\umsps$ 	& \cite{Saunders2012} \\
 $D_{\rm mem}$ 	& membrane-bound Pom1m diff. const. 	& 0.026 	& $\umsps$ 	& \cite{Saunders2012} \\
 $D_{\rm cyl}$	& diffusion const. on microtubules	& 0.05		& $\umsps$	& \\
 $v$		& drift velocity on microtubules	& 0.5		& $\umps$	& \\
 \hline
 $\lambda$ 	& Pom1 gradient decay length				& 1.5$\pm$0.4	& $\um$		& \cite{Saunders2012} \\
 $n_{\rm p}$	& no. of Pom1 phosphorylation sites\footnotemark	& 6 		&		& \cite{Hachet2011}\\
 \hline
 $k_{\rm p,n}, n=1..6$		& Pom1m cis-autophosphorylation rate 			& 0.05 		& $\unit{1/s}$   & OE\footnotemark\\
 $k_{\rm pt}$			& \CH{Pom1m in-complex phosphorylation-dissociation rate} & 1000 		& $\unit{1/s}$   & OE\\
 $k_{\rm u,n}, n=1..5$		& unbinding rate of n times phosph. Pom1m 		& 0.0		& $\unit{1/s}$	 & OE\\
 $k_{\rm u}\equiv k_{u,6}$	& unbinding rate of 6 times phosph. Pom1m 		& 0.05		& $\unit{1/s}$	 & OE\\ 
 $k_{\rm b,2d}$			& intrinsic bdg. rate on membr. and to cylinder 	& $10^{-11}$	& $\unit{m^2/s}$ & DL\footnotemark\\
 \hline 
\end{tabular}
\caption{
\textbf{Overview of the parameters used in the simulations of Pom1 gradient formation.}
In the right-most column we either list the source from which we obtained the given parameter value,
or alternatively comment on the motivation for our choice.
}
\label{tab:Pom1-Pars}
\end{table}
\addtocounter{footnote}{-2}
\footnotetext{that are relevant to lipid binding \stepcounter{footnote}}
\footnotetext{own estimate \stepcounter{footnote}}
\footnotetext{resulting in diffusion-limited binding; using the formulas from \cite{Kivenson2012} and the values of the other parameters in our system, the diffusion-limited rate of binding to the cylinder is estimated to be $\sim 10^{-12}~\unit{\frac{m^2}{s}}$, while the estimate for the interparticle reaction on the membrane is $\sim 10^{-14}~\unit{\frac{m^2}{s}}$. \stepcounter{footnote}}

\subsection{Results for the cis-model}
\begin{figure}[h!]
  \centering
  \subfigure[][]{
    \label{fig:cis-Gradients-Ampl}
    \includegraphics[width=\GradientPlotWidth]{\PlotsDir/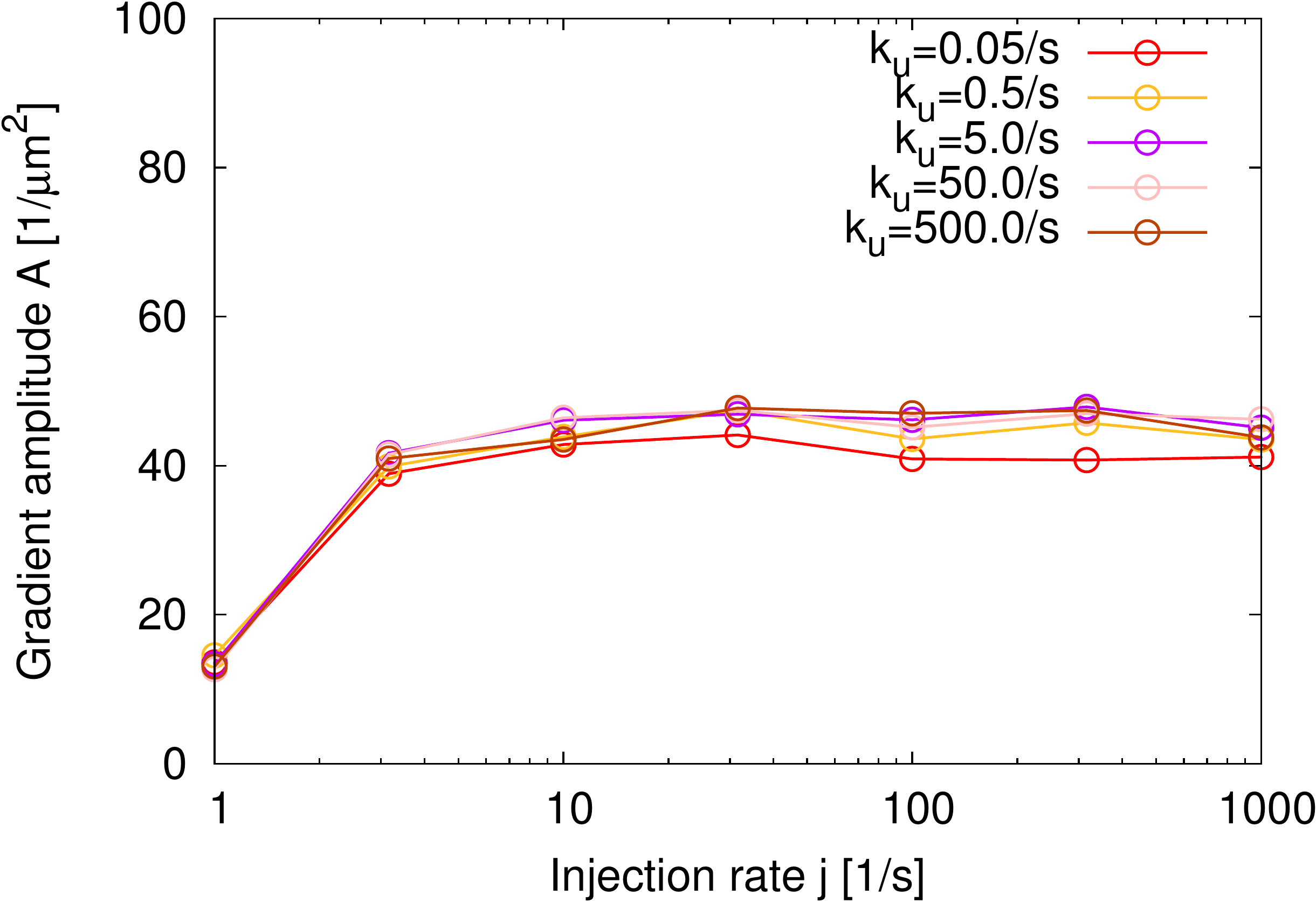}
  }
  \subfigure[][]{
    \label{fig:cis-Gradients-Length}
    \includegraphics[width=\GradientPlotWidth]{\PlotsDir/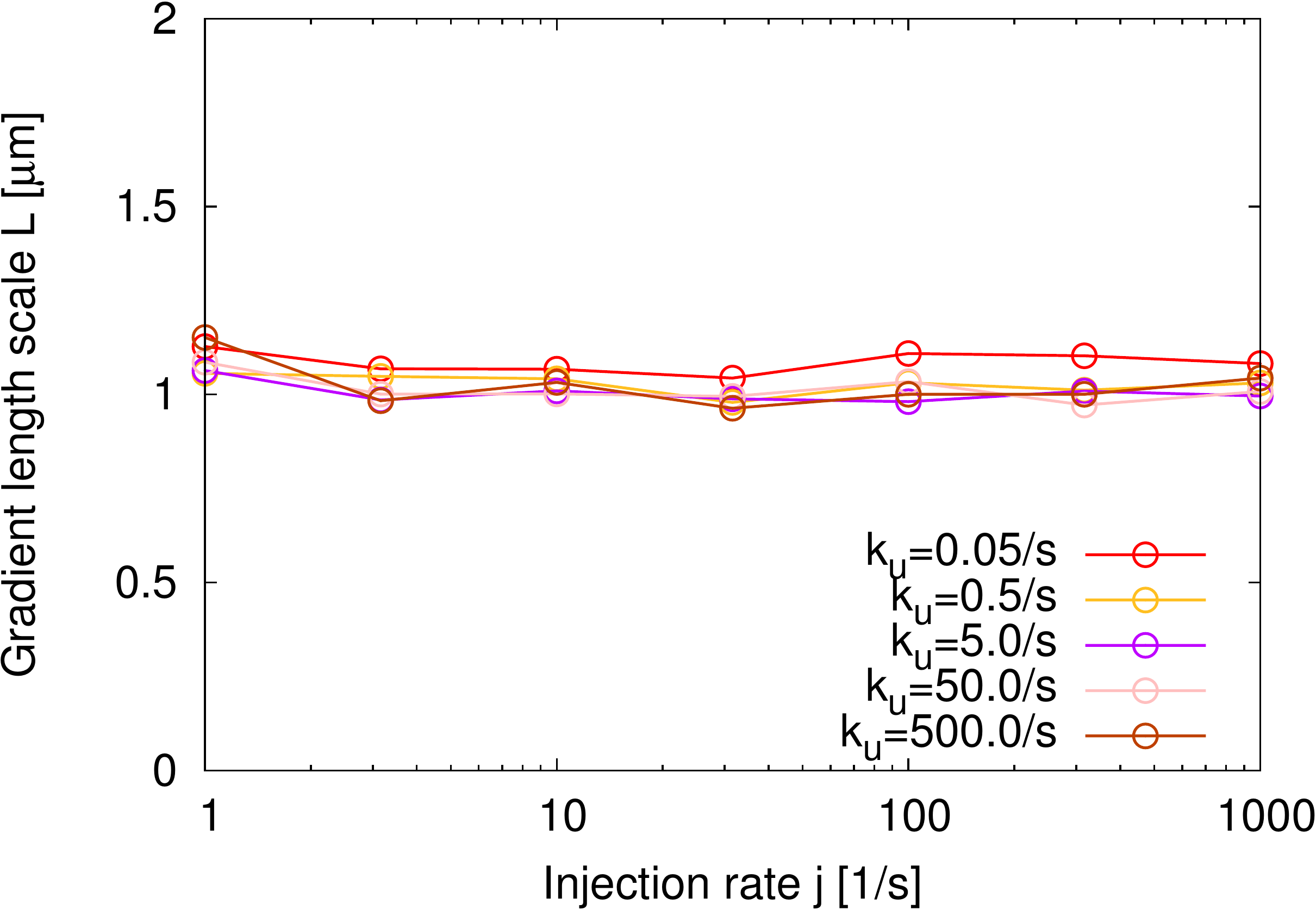}
  }\\  
  \vspace{1EM}
  \subfigure[][]{
    \label{fig:cis-Gradients-Corr}
    \includegraphics[width=\GradientPlotWidth]{\PlotsDir/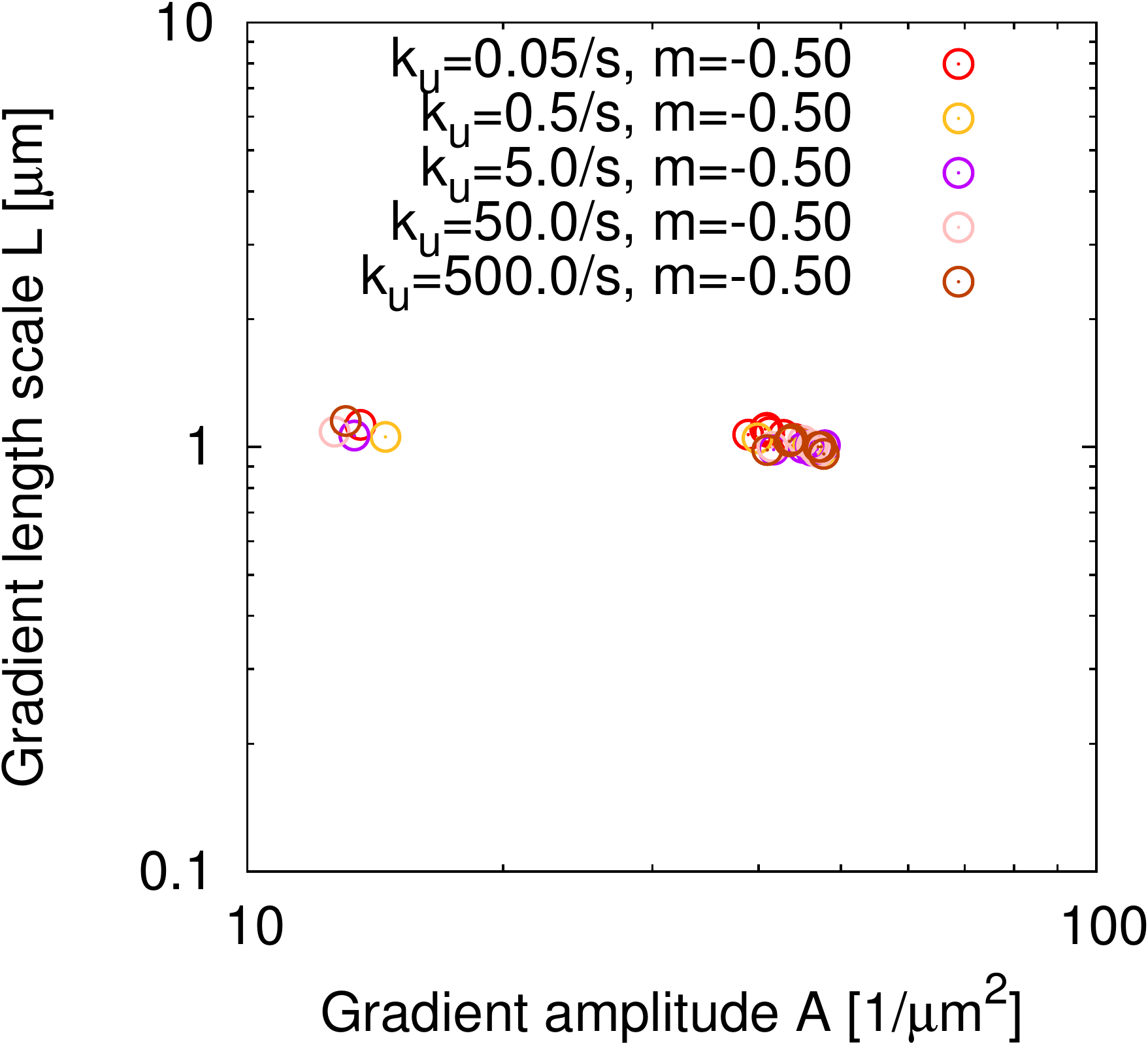}
  }
    \subfigure[][]{
    \label{fig:cis-Gradients-Noise}
    \includegraphics[width=\GradientPlotWidth]{\PlotsDir/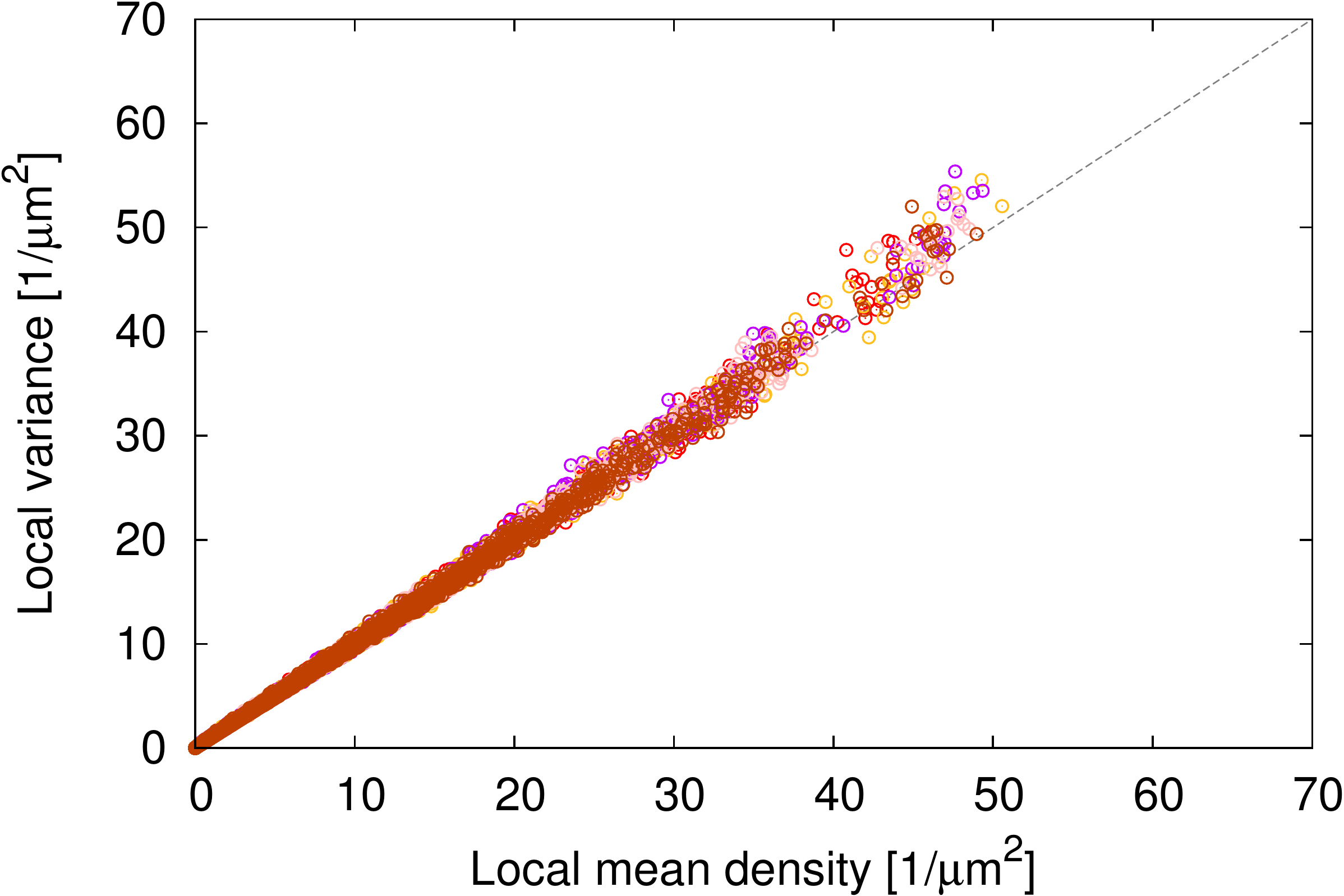}
  }  
\caption{
  \textbf{Gradient characteristics as a function of key parameters (cis-model).}
  Gradient amplitude \subfigref{\subref{fig:cis-Gradients-Ampl}} and length scale $L$ \subfigref{\subref{fig:cis-Gradients-Length}},
  obtained by fitting an exponential function to stationary radial profiles,
  plotted against the injection rate $j$ at the gradient origin for different unbinding rates $k_{\rm u}$ of the fully phosphorylated Pom1.
  Lines are guidelines for the eye.
  \subfigref{\subref{fig:cis-Gradients-Corr}} Log-log plot of the gradient length $L$ vs. gradient amplitude $A$ for the same data.  
  \subfigref{\subref{fig:cis-Gradients-Noise}} Variance vs. mean for the local total membrane-bound Pom1 density,
  combining the values of all histogram bins into one scatter plot;
  point colors correspond to the colors in \subfigref{\subref{fig:cis-Gradients-Ampl}} - \subfigref{\subref{fig:cis-Gradients-Corr}} (different values of $k_{\rm u}$).
  \label{fig:cis-Gradients}
}
\end{figure}

For comparison, we also repeated the simulations of Pom1 gradient formation for the cis-autophosphorylation model,
in which each membrane-bound Pom1 particle can autonomously autophosphorylate itself, without the need to encounter another Pom1 particle.
Again, we varied the Pom1 injection rate $j$ and the rate of unbinding upon complete autophosphorylation, $k_{\rm u}$.
The results for the cis-model are presented in Fig.~\ref{fig:cis-Gradients}.
Panels \subref{fig:cis-Gradients-Ampl} and \subref{fig:cis-Gradients-Length} of the figure show 
that--as long as the injection rate is high enough ($j\gtrsim 3/s$)--neither the gradient amplitude $A$ nor its length scale $L$
depend significantly on $j$ or $k_{\rm u}$;
consequently, as seen in panel \subref{fig:cis-Gradients-Corr}, in contrast to the trans-model, the anticorrelation between $A$ and $L$ is virtually non-existent.
This is not surprising, since in the cis-model the density-dependent amplification of local autophosphorylation that forms the basis of the buffering effect in the trans-model is not present.
Rather, the subsequent autonomous phosphorylation steps of the Pom1 particles in the cis-model constitute a cascade of events with exponentially distributed times,
at the end of which unbinding can occur; the total time until unbinding thus is a gamma-distributed random variable.
The benefit of the cis-autophosphorylation therefore is diametrally different as the one of trans-phosphorylation:
cis-autophosphorylation mainly acts as a precise ``timer'' for the unbinding events,
which keeps the distances travelled by the particles sharply distributed
and thus makes the resulting gradient shapes extraordinarily robust to variations of the other parameters.
To end with, panel \subref{fig:cis-Gradients-Noise} of Fig.~\ref{fig:cis-Gradients} shows that also the cis-model operates in the Poissonian noise regime;
this is equally unsurprising, because all processes involved in the gradient formation are first-order and Poissonian.

\clearpage
\bibliographystyle{styleTomek-v2}
\bibliography{General,GFRD,Intro,StochSim,SpatialEffects,Pom1,GapGenes}

\end{document}